\documentclass[12pt]{article}
\pdfoutput=1

\usepackage{graphicx,psfrag,epsf,xcolor}
\usepackage{amsmath,amssymb,amsfonts}
\usepackage{array}
\usepackage{cite}
\usepackage{rotating}
\usepackage{slashed, cancel}
\usepackage{scalerel,stackengine}
\usepackage{float}
\usepackage{subcaption}
\usepackage{caption}
\definecolor{cBlue}{RGB}{0,110,191}
\usepackage[pdftitle={Multiple collinear limits of scattering amplitudes},hypertexnames=true]{hyperref}
\hypersetup{bookmarksnumbered,colorlinks,
    linkcolor={black},
    citecolor={cBlue},
    urlcolor={cBlue}}
\numberwithin{equation}{section}

\usepackage{bm}
\usepackage{booktabs}
\usepackage{colortbl}
\usepackage{enumerate}
\usepackage{latexsym}
\usepackage{lscape}
\usepackage{multirow}
\usepackage{psfrag}

\def\beq{\begin{equation}}
\def\eeq{\end{equation}}
\def\beqa{\begin{eqnarray}}
\def\eeqa{\end{eqnarray}}

\def\be{\begin{equation}}
\def\ee{\end{equation}}
\def\bea{\begin{eqnarray}}
\def\eea{\end{eqnarray}}

\def\nn{\nonumber}

\newcommand\Eqn[1]     {Eq.\,(\ref{#1})}
\newcommand\Eqns[2]    {Eqs.\,(\ref{#1}) and~(\ref{#2})}

\newcommand{\bei}{\begin{itemize}}
\newcommand{\eei}{\end{itemize}}

\def\T{{\bf T}}

\def\DR{\mathcal{D}^R}

\newcommand{\as}{\alpha_s}

\newcommand{\eps}{\epsilon}

\def\MM{\mathcal{M}}

\def\SP{{\bf Sp}}

\def\HH{\mathcal{H}} 
\def\Hhard{\mathcal{H}} 

\def\Z{{\bf Z}} 

\definecolor{darkgreen}{rgb}{0.0,0.6,0.0}
\definecolor{cPurple}{RGB}{93,35,125}

\begin{document}
\allowdisplaybreaks
	
\begin{titlepage}
		
\begin{flushright}
{\small
BONN-TH-2025-25\\
\today
}
\end{flushright}
		
\vskip0.7cm
\begin{center}
{\Large \bf\boldmath
Infrared singularities and the collinear limits 
of multi-leg scattering amplitudes}
\end{center}
		
\vspace{0.3cm}
\begin{center}
{\sc Claude Duhr},$^{a}$
{\sc Einan Gardi},$^{b}$
{\sc Sebastian Jaskiewicz},$^{c}$ 
\\[0.05cm]
{\sc Jonas L\"ubken},$^{b,d}$
and {\sc Leonardo Vernazza}$^{e}$\\[6mm]

{\it $^a$ Bethe Center for Theoretical Physics,
Universit\"{a}t Bonn,\\ Wegelerstrasse 10, 53115 Bonn, Germany
\\[0.2cm]}
{\it $^b$ Higgs Centre for Theoretical Physics,
School of Physics and Astronomy, \\
The University of Edinburgh, Edinburgh EH9 3FD, United Kingdom 
\\[0.2cm]}
{\it $^c$ Albert Einstein Center for Fundamental Physics,
Institut f\"ur Theoretische Physik, Universit\"at Bern,
Sidlerstrasse 5, CH-3012 Bern, Switzerland 
\\[0.2cm]}
{\it $^d$ Institut f\"{u}r Theoretische Physik, 
Universit\"{a}t M\"{u}nster, Wilhelm-Klemm-Stra\ss e 9,
D-48149 M\"{u}nster, Germany
\\[0.2cm]}
{\it $^e$ INFN, Sezione di Torino, Via P. Giuria 1, 
I-10125 Torino, Italy}
\end{center}
		
\vspace{0.4cm}
\begin{abstract}
\vskip0.2cm\noindent

Scattering amplitudes are expected to 
admit a factorised structure in special 
kinematic limits, such as the Regge, soft and collinear limits. However, less is known about 
the precise  mechanisms through which
factorisation of $n$-particle
scattering amplitudes is realised 
at high perturbative orders, where more complex structures arise. Starting with the soft anomalous
dimension, in this work we investigate the multi-particle
collinear limits of massless amplitudes at three-
and four-loop orders.
Using colour conservation and rescaling symmetry, we show how strict collinear factorisation of multiple massless 
final-state coloured particles is realised, 
and provide results for the corresponding splitting amplitude soft anomalous dimensions. In particular,  we demonstrate through four loops that the conditions on the structure of the soft anomalous dimension that are required by strict collinear factorisation in all two-particle collinear limits, are sufficient to guarantee such factorisation also in any multiple collinear limit. 
Then, assuming that strict collinear factorisation of massless partons holds also for amplitudes containing massive coloured particles, we derive new constraints on the soft anomalous dimension from multi-collinear limits. 
\end{abstract}
\end{titlepage}


\tableofcontents

\section{Introduction}
\label{sec:introduction}

The study of scattering amplitudes in non-abelian gauge theories at high orders in the perturbative expansion is critical for the success of the precision physics program at the LHC and further advancing our understanding of quantum  field theory itself. Particularly important in this endeavour is gaining insight into the quantities that govern the all-order behaviour of the scattering amplitudes. A prototypical example is the \emph{soft anomalous dimension}, which dictates their long-distance singularity structure. This quantity admits a high degree of universality: it is independent of the details of the short-distance process, and to some degree on the specific gauge theory considered. All-order insight can be obtained through investigations of scattering amplitudes in various kinematic limits. Indeed, the study of scattering processes at high energy, the so-called Regge limit, predates QCD \cite{Veneziano:1968yb,Collins:1977jy,Mandelstam:1965zz,Gribov:1973iv} and, to this day, continues to provide valuable data about the intricate structure of gauge-theory amplitudes \cite{Falcioni:2021buo,Falcioni:2021dgr,DelDuca:2019tur}.  

Investigations of kinematic limits are especially attractive as scattering amplitudes in these specific configurations often exhibit a factorised structure. In other words, in certain limits, the scattering amplitudes can be written in terms of universal functions, which do not depend on the details of the particular scattering process. These universal functions can be calculated once and for all and applied to a multitude of other contexts. 
The soft anomalous dimension, for instance, can be used to predict the infrared (IR) singularity structure for a specific scattering process and in this way serve to check explicit new calculations.
Furthermore, the universal functions which arise in such limits typically depend on fewer kinematic variables compared to the full scattering process, e.g.\ they do not depend on the kinematics of non-coloured  particles involved in the scattering. The analytic structure of these functions is thus simpler, and they can be calculated to high orders in perturbation theory. 
Moreover, more general quantities can be constrained using the knowledge gathered from complementary kinematic limits. Flagship examples are the bootstrap approach 
applied to ${\cal N}=4$ supersymmeytric Yang-Mills amplitudes in the planar limit, see e.g. Refs.~\cite{Dixon:2011pw,Caron-Huot:2016owq,Caron-Huot:2019vjl}, or to the soft anomalous dimension in QCD~\cite{Almelid:2017qju,Becher:2019avh,Falcioni:2021buo} (see  Refs.~\cite{Gardi:2009qi,Becher:2009qa,Dixon:2009ur,Ahrens:2012qz} for earlier work in this direction).
The latter application will be of direct relevance here. These papers demonstrated the possibility of determining the soft anomalous dimension for $n$-point massless scattering, starting from a general ansatz in terms of iterated integrals, and  constraining their coefficients using symmetries and information from further kinematic limits. 

\begin{figure}[t]
	\begin{center}
 \includegraphics[width=0.32\textwidth]{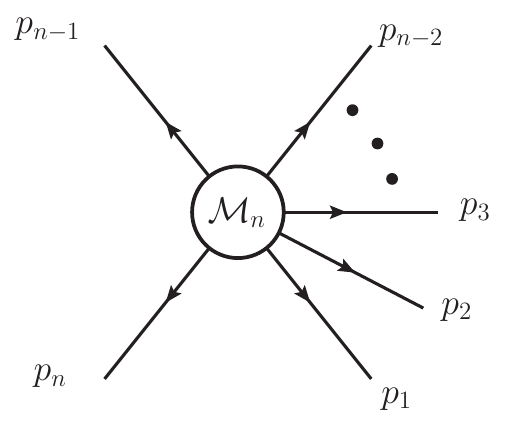} 
			\includegraphics[width=0.32\textwidth]{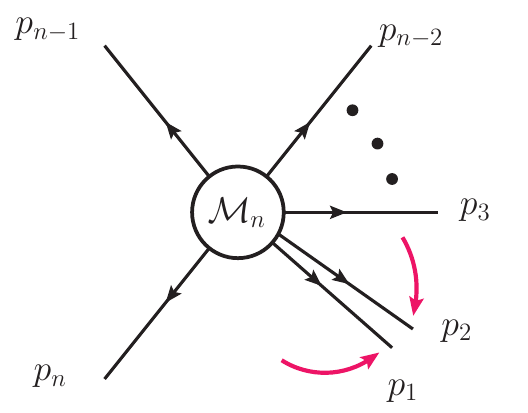} 
   		\includegraphics[width=0.33\textwidth]{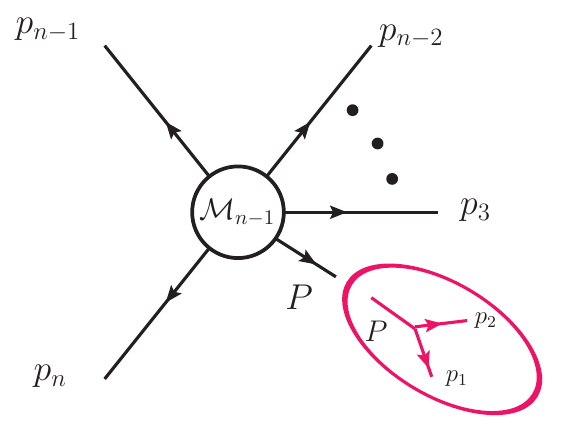}
	\end{center}
	\caption{(left) Generic $n$-point scattering amplitude with each of the external particles widely separated, i.e. $ p_i\cdot p_j$ is parametrically large. (middle) Situation depicting the approach of a two-particle collinear limit, where the angle between two of the $n$ scattering partons becomes small. (right) Diagram showing a timelike collinear limit where strict collinear factorisation is expected to hold. The object circled in red represents the splitting amplitude  which depends only on the degrees of freedom of the particles becoming collinear, i.e. it is completely factorised from the rest of the process, which is precisely the $(n-1)$-leg on-shell amplitude, $\mathcal{M}_{n-1}$.  }
\label{fig:m-collinearTL-introduction}
\end{figure}

A kinematic limit with a particularly rich structure is the collinear limit 
of (strictly) massless  quarks and gluons.
Starting from a generic $n$-point scattering amplitude where the external partons are widely separated, the collinear limit is characterised by the kinematic configuration where the angles between two or more of the external massless partons become vanishingly small. This situation is schematically depicted in Fig.~\ref{fig:m-collinearTL-introduction}.  
In this limit, the QCD scattering amplitude develops an additional singularity related to the parent parton being nearly on shell and admits a factorised structure. 

There are two qualitatively different cases of collinear limits. The first and simpler of the two, is the case in which the partons becoming collinear all belong to either the final or all to the initial state. This case is referred to as the \emph{timelike} collinear limit, since the squared sum of all four momenta is timelike. In the timelike collinear limit, the $n$-point amplitude factorises into a universal function, known as the \emph{splitting amplitude}, which depends only on the momenta, colour, and spin of the partons becoming collinear, times a lower-point amplitude which contains information about the rest of the process \cite{Berends:1988zn,Bern:1995ix,Kosower:1999xi,Mangano:1990by,Catani:2011st,Feige:2014wja}. In this case, the amplitude is said to admit \emph{strict collinear factorisation}. 
Factorisation theorems for the timelike two-particle collinear limit and the corresponding splitting amplitudes have been studied at tree level in Refs.~\cite{Altarelli:1977zs,Berends:1987me,Mangano:1990by}, at one loop in Refs.~\cite{Bern:1998sc,Bern:1999ry,Bern:1993qk,Bern:1994zx,Bern:1995ix,Kosower:1999rx, Sborlini:2013jba}, at two loops in Refs.~\cite{Badger:2004uk,Bern:2004cz,Duhr:2014nda}, and recently at three loops in Ref.~\cite{Guan:2024hlf}. Timelike collinear factorisation has been discussed in the context of the soft-anomalous dimension in~Refs.~\cite{Becher:2009qa,Dixon:2009ur,Ahrens:2012qz,Becher:2019avh}.

The second case, known as the \emph{spacelike} collinear limit, is where the partons becoming collinear belong in part to the initial and in part to the final state. The strict collinear factorisation is known to be broken in this case, which means that separation of the amplitude into different terms may still be possible, however these would generally depend on the degrees of freedom of all external partons, not only on the collinear subset \cite{Catani:2011st}, i.e.\ they are not universal.
This is often referred to as violation of (strict) collinear factorisation, or collinear factorisation breaking, and it has been further investigated in Refs.~\cite{Forshaw:2012bi,Schwartz:2017nmr,Schwartz:2018obd}.  
Recent work also considered these effects in more general limits, e.g.\ when two or more separate sets of particles become spacelike collinear~\cite{Cieri:2024ytf,Duhr:2025lyg}.
The violation of strict collinear factorisation also manifests itself in 
the appearance of so-called super-leading logarithms in certain observables, such as those involving a veto on the radiation in a gap between jets~\cite{Forshaw:2008cq,Becher:2021zkk}.
In this context, it was recently demonstrated explicitly how factorisation of parton-density functions in gap between jets cross-sections is restored, despite the presence of colour coherence breaking complex phases (due to strict collinear factorisation violation effects) in the hard-scattering amplitudes~\cite{Becher:2024kmk}.
Very recently results for the spacelike collinear limit have been obtained at the two-loop order~\cite{Henn:2024qjq} from a massless five-gluon amplitude in ${\cal N}=4$ supersymmetric Yang-Mills theory.

In this article, we explore multi-particle collinear limits of $n$-point scattering amplitudes up to fourth-order in the strong coupling expansion. We focus our investigations on the first of the cases described above, that is the timelike collinear limit for $m$ massless final-state partons becoming collinear. In this limit, an on-shell $n$-point amplitude admits strict collinear factorisation, i.e. it factorises, to all orders in perturbation theory, into a $1\to m$ \emph{splitting amplitude} times an on-shell $(n-m+1)$-point amplitude. 
Results for the multi-particle collinear limit have been given at tree level in Refs.~\cite{Catani:1999ss,Campbell:1997hg,Catani:1998nv,DelDuca:1999iql,Birthwright:2005ak,Birthwright:2005vi}; partial results at one loop are available in Refs.~\cite{Catani:2003vu,Badger:2015cxa}. 
The triple collinear limit has also been studied  in Refs.~~\cite{Sborlini:2014mpa,Sborlini:2014kla,Czakon:2022fqi,Dhani:2023uxu,Craft:2023aew} and four-particle collinear limit in Refs.~\cite{DelDuca:2019ggv,DelDuca:2020vst}.

The IR singularity structure in the timelike collinear limit in dimensional regularisation is governed by the so-called \emph{splitting amplitude soft anomalous dimension}. This object is related to the soft anomalous dimensions of the $n$-point and $(n-m+1)$-point scattering amplitudes and is the primary quantity of interest in this work. Its definition embodies both said splitting amplitude factorisation and IR factorisation, making it highly constrained even at high multiplicities and high loop orders. 

We begin our study from the soft anomalous dimension governing the IR structure of $n$-point \emph{massless} scattering amplitudes.
In this case the three-loop soft anomalous dimension is known based on explicit computations~\cite{Almelid:2015jia,Gardi:2016ttq,Almelid:2016lrq}; the result has been further checked by comparing it against the singularity structure of three-loop four-point amplitudes in ${\cal N}=4$ supersymmetric Yang-Mills theory~\cite{Henn:2016jdu} and in QCD~\cite{Caola:2021rqz} and it was also recovered by bootstrap in Ref.~\cite{Almelid:2017qju}.
Moreover, the 
general form of the soft anomalous dimension at four loops, which is dictated by nonabelian exponentiation~\cite{Gardi:2013ita}, colour conservation, Bose symmetry as well as rescaling invariance properties~\cite{Becher:2009cu,Becher:2009qa,Gardi:2009qi,Gardi:2009zv}, has been determined in Ref.~\cite{Becher:2019avh}, where constrains from the two-particle collinear limit were also derived.
Further constraints on its  kinematically-dependent functions  have been derived based on the Regge limit in Ref.~\cite{Falcioni:2021buo}.  These constraints are not sufficient to fully determine these functions~\cite{Maher:2023jqy}. 

Here we start with the general form of the soft anomalous dimension for $n$-particle scattering~\cite{Almelid:2015jia,Almelid:2017qju,Becher:2019avh,Falcioni:2021buo} and study the soft anomalous dimension of the multi-particle splitting amplitude   through four loops. Our derivation makes manifest the way in which
strict collinear factorisation is realised at these
high perturbative orders. We show that through an interplay between colour conservation and rescaling invariance of the soft anomalous dimension, the conditions that emerge from all two-particle collinear limits are sufficient to guarantee the strict collinear factorisation property also in generic multi-particle collinear limits. In particular, they are sufficient to derive the form of the relevant multi-particle splitting amplitude soft anomalous dimension, and demonstrate that it only depends on the set of particles becoming collinear. 

Less is known about the soft
anomalous dimension for $n$-point scattering amplitudes (beyond two loops) 
when some of the coloured particles are \emph{massive}. 
These objects are of direct interest for resummation of large logarithms in cross sections in the case of top production (possibly, in association with jets). Here we restrict our attention to the case where only one of the external coloured particles is massive, while allowing for an arbitrary number of massless coloured particles. We shall refer to this as the one-mass soft anomalous dimension. 
In this case, the general form of the three-loop soft anomalous dimension was derived in Ref.~\cite{Liu:2022elt}. The same paper also computed explicitly one of the functions appearing there, 
namely the one associated with interactions between the   massive particle and two massless ones.
The contributions due to interactions between the massive particle and three massless particles remained undetermined in Ref.~\cite{Liu:2022elt}, but they were recently computed using a different method in Ref.~\cite{GZ-TBP}, thus completing the determination of this three-loop soft anomalous dimension with any number of massless particles.

Requiring strict collinear factorisation in the two-particle collinear limit, Ref.~\cite{Liu:2022elt} derived new constraints on the one-mass soft anomalous dimension functions. Here we analyse the triple-collinear limit and show that it yields an independent constraint of these functions, namely one that does not follow from the two-particle collinear limits. Interestingly, the same constraint arises from another limit, namely the limit where the massive particle becomes massless. These constraints are satisfied by the computation in Ref.~\cite{GZ-TBP}, where they provide a boundary condition as well as a check of the result.

The outline of this article is as follows: in Section~\ref{sec:IRdivergences} we review the structure of the infrared singularities of $n$-point scattering amplitudes. We define the soft anomalous dimension and give the state-of-the-art three-loop results in the massless~\cite{Almelid:2015jia} and massive~\cite{Liu:2022elt} cases, as well as the general structure of this object at four loops for the massless case~\cite{Becher:2019avh}. 
In Section~\ref{sec:factorisationincollinearlimits}, we focus on the collinear limits of amplitudes and define the splitting amplitude soft anomalous dimension for $m$ partons becoming collinear. We also study the kinematic dependence of particles in the multi-particle collinear limit and their associated degrees of freedom. 
Sections~\ref{sec:massless} and~\ref{sec:massive} contain the main results of this work. In Section~\ref{sec:massless}, we compute the multi-particle collinear splitting amplitude soft anomalous dimensions at the three- and four-loop order for the case where only massless external partons participate in the $n$-point scattering. 
In Section~\ref{sec:massive}, we consider the case where in addition to $n$ massless coloured particles participating in the scattering, a massive coloured particle is also present. We confirm the result of Ref.~\cite{Liu:2022elt} regarding the constraint on the soft anomalous dimension arising from two-particle strict collinear factorisation, and then derive a new constraint using the three-particle collinear limit. 
We conclude and provide an outlook in Section~\ref{sec:conclusion}. The paper also has six appendices where we provide all explicit derivations.


\section{Infrared divergences in scattering amplitudes}
\label{sec:IRdivergences}
We begin by considering a scattering amplitude $\MM_n$, for $n$ massless coloured partons, and any number of non-coloured particles (massless or massive). 
The momentum of the $i^{{\,\rm{th}}}$ parton is denoted by $p_i$. For the purpose of this section, we take the amplitude at \emph{fixed angle}, such that all invariants between the external partons, $p_i\cdot p_j$, are large with respect to the QCD confinement scale. 
It is well known that such scattering amplitudes 
develop infrared singularities, originating 
from loop integrations 
when the loop momenta 
become soft or collinear to one of the external 
partons. Infrared divergences are known to factorise, 
i.e., they can be collected in a \emph{universal} multiplicative factor 
$\Z_n$ times a finite reminder ${\cal H}_n$, known as the
\emph{hard function}, which contains only finite terms
in the limit $\eps \to 0$. The hard function ${\cal H}_n$ is process-dependent. It can be shown that an $n$-point UV renormalised scattering amplitude $\MM_n$ takes the following form \cite{Becher:2009qa,Gardi:2009zv,Gardi:2016ttq}
\beq \label{IRfacteq}
\MM_n \left(\{p_i\},\mu, \as (\mu^2),\epsilon \right) \, = \, 
{\bf Z}_n \left(\{p_i\},\mu_f, \as (\mu_f^2), \epsilon \right)
\Hhard_n \left(\{p_i\},\mu_f, \mu, \as (\mu^2) \right).
\eeq
Here the parameter $\mu$ represents the UV renormalisation 
scale, while $\mu_f$ is the scale at which infrared 
singularities are regularised. Dependence on the scales enters
explicitly, and implicitly through the running coupling, as
indicated in \Eqn{IRfacteq}. The most important result 
concerning infrared factorisation is that the renormalisation 
factor ${\bf Z}_n$ obeys a renormalisation group equation 
\cite{Catani:1998bh,Sterman:2002qn,Aybat:2006mz,Aybat:2006wq,Gardi:2009qi,Becher:2009qa,Becher:2009cu,Gardi:2009zv,Becher:2019avh,Almelid:2015jia,Almelid:2017qju,Dixon:2009ur,Magnea:2021fvy,Beneke:2017ztn,Beneke:2018rbh}
\be\label{rgeZ}
\frac{d}{d\ln\mu_f}{\bf Z}_n \left(\{p_i\},\mu_f, \as (\mu_f^2),\epsilon  \right) 
=- {\bf Z}_n \left(\{p_i\},\mu_f, \as (\mu_f^2),\epsilon  \right)  
{\bf \Gamma}_n \left(\{p_i\},\mu_f, \as(\mu_f^2) \right),
\ee
which formally defines the \emph{soft anomalous dimension} ${\bf \Gamma}_n$. 
The solution to \Eqn{rgeZ}
can be expressed in terms of a 
path-ordered exponential of the  
soft anomalous dimension as follows
\beq \label{RGsol}
{\bf Z}_n \left(\{p_i\},\mu_f, \as (\mu_f^2),\epsilon  \right) \, = \,  
{\cal P} \exp \left\{ -\frac{1}{2}\int_0^{\mu_f^2} \frac{d \lambda^2}{\lambda^2}\,
{\bf \Gamma}_n \left(\{p_i\},\lambda, \as(\lambda^2) \right) \right\}\,,
\eeq
where ${\cal P}$ is the path-ordering operator. The soft anomalous dimension is itself finite, and the IR singularities contained in $ {\bf Z}_n $ are generated by performing the integral over the scale $\lambda$.

The invariance of the unfactorised amplitude on the left-hand side of Eq.~(\ref{IRfacteq}) implies that the same anomalous dimension controls the evolution of the hard function ${\cal H}_n$, namely
\begin{equation}
\label{RG_H}
\frac{d}{d\ln\mu_f} \Hhard_n \left(\{p_i\},\mu_f, \mu, \as (\mu^2) \right)={\bf \Gamma}_n \left(\{p_i\},\mu_f, \as(\mu_f^2) \right) \Hhard_n \left(\{p_i\},\mu_f, \mu, \as (\mu^2) \right)\,.
\end{equation}

Much effort has been devoted to the study of the structure
of the  soft anomalous dimension over the years. 
In this section, we introduce the notation and draw attention
to features of ${\bf \Gamma}_n$
which are important for our investigations.  For a recent review on the topic  we recommend~\cite{Agarwal:2021ais}. 
The soft anomalous dimension for scattering of massless 
partons ($p_i^2=0$), introduced in \Eqn{rgeZ}, depends on kinematic invariants 
formed between the external partons 
\begin{eqnarray}\label{eq:s_ij}
(-s_{ij}) = 2 |p_i \cdot p_j | e^{-i \pi \lambda_{ij}}
\end{eqnarray}
with 
$\lambda_{ij} = 1$ if partons $i$ and $j$ both belong to either 
the initial or the final state and $\lambda_{ij} = 0$ otherwise. Owing to the rescaling invariance of Wilson lines, correlators of Wilson lines are typically functions of \emph{conformally invariant cross-ratios} (CICRs)~\cite{Gardi:2009qi,Becher:2009qa}, defined as\footnote{The rescaling invariance is broken by IR divergences, leading to an additional dependence in $\ln(-s_{ij})$.}
\begin{eqnarray}\label{eq:CICR}
\rho_{ijkl} = \frac{(-s_{ij})(-s_{kl})}{(-s_{ik})(-s_{jl})}  \,. 
\end{eqnarray}
In what follows, we also use the logarithms of CICRs denoted by 
\begin{eqnarray}\label{eq:lnCICR}
    \beta_{ijkl} = \ln \rho_{ijkl}
    =\ln \left( \frac{(-s_{ij})(-s_{kl})}{(-s_{ik})(-s_{jl})} 
    \right)\,.
\end{eqnarray} 

The anomalous dimension 
is an operator 
in colour space depending on $\{\T_i\}_{i=1}^{n}$, which is
the set of colour charge operators~\cite{Catani:1996jh,Catani:1996vz,Catani:1998bh}  
in an arbitrary representation
associated with the external partons.
The definition of  ${\bf \Gamma}_n$ is subject to colour 
conservation 
\be
\left(\sum_{i=1}^{n}\T_i\right)\MM_n= \left(\sum_{i=1}^{n}\T_i\right)\Hhard_n= 0\,,
\ee
where $\Hhard_n$ is the hard function defined
in \Eqn{IRfacteq}.
The structure of the soft anomalous dimension is strongly 
constrained due to Bose symmetry, non-Abelian
exponentiation, and rescaling invariance. We point
the reader to Section~7~of Ref.~\cite{Falcioni:2021buo}
for an in-depth discussion of the various constraints
on the general form of ${\bf \Gamma}_n$. 
For us, the starting point is the parametrisation of
the soft anomalous dimension given in Eq.~(7.13) of Ref.~\cite{Falcioni:2021buo}. Namely, we write 
${\bf \Gamma}_n$ as 
\begin{eqnarray}\label{eq:adm-param}
    {\bf{ \Gamma}}_{n}\left(\{s_{ij}\},\lambda,\alpha_s(\lambda^2) \right)
     &=& {\bf{ \Gamma}}^{{\rm{dip.}}}_{n}\left(\{s_{ij}\},\lambda,\alpha_s  \right)
     + {\bf{ \Gamma}}_{n,4{\rm{T}}-3{\rm{L}}}(\alpha_s)
     + {\bf{ \Gamma}}_{n,4{\rm{T}}-4{\rm{L}}}(\{\beta_{ijkl}\},\alpha_s)
    \nonumber\\ && 
    + {\bf{ \Gamma}}_{n,{\rm{Q}}4{\rm{T}}-2,3{\rm{L}}}(\{s_{ij} \},\lambda,\alpha_s)  
    + {\bf{ \Gamma}}_{n,{\rm{Q}}4{\rm{T}}-4{\rm{L}}}(\{\beta_{ijkl} \} ,\alpha_s)  
    \\ \nonumber && 
    + {\bf{ \Gamma}}_{n,5{\rm{T}}-4{\rm{L}}}(\{\beta_{ijkl} \} ,\alpha_s)  
        + {\bf{ \Gamma}}_{n,5{\rm{T}}-5{\rm{L}}}(\{\beta_{ijkl} \} ,\alpha_s)
        + \mathcal{O}(\alpha_s^5)\,,
\end{eqnarray}
where the subscript on each of the terms includes: the number
of partons $n$, Q if the contribution is related to 
a colour structure involving a totally symmetric combination of four colour charge operators, a number followed by T to indicate the 
number generators present in that term, and lastly, a number
followed by L which indicates the number of distinct external legs 
(partons) that are connected together in this term. 
For concreteness, the dipole part of the soft anomalous 
dimension, which involves only pair-wise interactions between 
the partons and constitutes the complete result through
two loops, is given by 
\begin{eqnarray}\label{eq:dipole}
    {\bf \Gamma}_{n}^{\rm dip.}\left(\{s_{ij}\},\lambda, \as \right)=
-\frac{1}{2}\gamma_{K}  (\as) \, \, \sum_{i<j} 
\ln \left(\frac{-s_{ij}}{\lambda^2}\right) \, \T_i \cdot \T_j 
\,+\, \sum_i \gamma_i (\as) {\mathbf{1}}\,.
\end{eqnarray}
The function ${\gamma}_{K}(\alpha_s)$
is the (lightlike) cusp anomalous dimension
\cite{Korchemsky:1985xj,Korchemsky:1985xu,Korchemsky:1987wg}, 
and $\gamma_i (\as)$ 
represents the field anomalous dimension corresponding 
to the parton $i$, which governs collinear singularities \cite{Moch:2005tm,Falcioni:2019nxk,Dixon:2017nat}. 
The anomalous dimensions which enter in \Eqn{eq:dipole}
are known to four loops in QCD \cite{Boels:2017ftb,Boels:2017skl,Moch:2017uml,Grozin:2017css,Henn:2019swt,vonManteuffel:2020vjv,Agarwal:2021zft}.
The second and third term in \Eqn{eq:adm-param}
start contributing at three loops,
and all other remaining terms 
start contributing at four loops. We note that 
the last two terms (which contain five colour generators)
could be excluded from ${\bf{ \Gamma}}_{n}$
following the argument made in Ref.~\cite{Vladimirov:2017ksc}.
However, we keep them in the 
general parametrisation to check whether independent 
constraints arise for these objects from considerations
of the multi-particle collinear limits. We now introduce 
each term in \Eqn{eq:adm-param}. We make use of the
notation used in Refs.~\cite{Falcioni:2021buo,Becher:2019avh},
although we choose to write the sums over the subsets
of particles entering these terms in an ordered form, 
rather than fully symmetric as in the aforementioned references,
which turns out to be useful for considerations of multi-particle
collinear limits. 

The terms starting at the three-loop order, 
$ {\bf{ \Gamma}}_{n,4{\rm{T}}-3{\rm{L}}}(\alpha_s)$ and
     $ {\bf{ \Gamma}}_{n,4{\rm{T}}-4{\rm{L}}}(\{\beta_{ijkl}\},\alpha_s)$  
in \Eqn{eq:adm-param}, were explicitly computed in Ref.~\cite{Almelid:2015jia}. They involve the colour 
and kinematic degrees of freedom of subsets of 
three or four partons. 
These terms are given by
\begin{eqnarray}\label{eq:Gammaf}
&&\hspace{-5.77cm}{\bf{ \Gamma}}^{}_{n,4{\rm{T}}-3{\rm{L}}}(\alpha_s) = 2 f(\alpha_s) \, 
\sum_{i=1}^n \sum_{\substack{1\leq j < k \leq n,\\ j,k\neq i}} 
\mathcal{T}_{iijk} \,,
\\ \label{eq:GammaF}
{\bf{ \Gamma}}^{}_{n,4{\rm{T}}-4{\rm{L}}}
(\{\beta_{ijkl} \},\alpha_s) = \,8\,
\sum_{1\leq i<j<k<l \leq n} \Big[ &&\!\!\!\!\!
\mathcal{T}_{iklj} 
\,{\cal F}  (\beta_{ikjl},\beta_{iljk}) \\ \nn
&&\hspace{-2cm}+\,\,\mathcal{T}_{ijlk} 
\,{\cal F}  (\beta_{ijkl},\beta_{ilkj}) 
+\,\mathcal{T}_{ijkl} 
\,{\cal F} (\beta_{ijlk},\beta_{iklj}) \Big] \,.
\end{eqnarray}
The colour structure in the above terms involves four generators
\begin{eqnarray}
\label{Tauijkl}
        \mathcal{T}_{ijkl} = f^{ade}f^{bce}\left\{ 
        \T^a_i,\T^b_j,\T^c_k,\T^d_l
        \right\}_+ \,,
 \end{eqnarray}
where the curly brackets with a `$+$'- subscript 
denote symmetrisation, defined as 
\begin{eqnarray}
    \left\{ 
        \T^{a_1}_i,\T^{a_2}_j,\ldots ,\T^{a_n}_l
        \right\}_+
        \equiv 
        \frac{1}{n!} \sum_{\pi\in S_n}
         \T^{a_{\pi(1)}}_i
         \T^{a_{\pi(2)}}_j\ldots \T^{a_{\pi(n)}}_l\,,
\end{eqnarray}
where the sum is over all permutations of the indices. 
The symmetrisation effectively only  applies to generators 
connected to the same line, because generators attached to distinct 
lines commute. The functions $f(\alpha_s)$
and 
$\mathcal{F}(\beta_{ijkl},\beta_{ilkj};\alpha_s)$
can be calculated order by order in perturbation theory, 
\begin{eqnarray}\label{eq:f-constant}
        f(\alpha_s) = \left(\frac{\alpha_s}{\pi} \right)^3 f^{(3)} +
        \left(\frac{\alpha_s}{\pi} \right)^4 \sum_R f_R^{(4)} + \mathcal{O}(\alpha_s^5)\,,
 \end{eqnarray}
 and 
 \begin{eqnarray}\label{eq:F-function}
 \hspace{-0.5cm}
         \mathcal{F}(\beta_{ijkl},\beta_{ilkj};\alpha_s)
         = \left(\frac{\alpha_s}{\pi} \right)^3 \mathcal{F}^{(3)}(\beta_{ijkl},\beta_{ilkj} )
        +\left(\frac{\alpha_s}{\pi} \right)^4 \sum_R \mathcal{F}_R^{(4)}(\beta_{ijkl},\beta_{ilkj} )
        + \mathcal{O}(\alpha_s^5)\,.
  \end{eqnarray}
In \Eqn{eq:f-constant}
the $f^{(3)}$ and $f^{(\ell)}_R$ are (transcendental) constants and in 
\Eqn{eq:F-function} the 
$\mathcal{F}^{(3)}$ and $\mathcal{F}^{(\ell)}_R$ are (transcendental) functions
of the CICRs defined in Eqs.~\eqref{eq:CICR} and~\eqref{eq:lnCICR}. 
The constant  $ f^{(3)}$ at three loops 
is given by   \cite{Almelid:2015jia}
\begin{eqnarray}\label{eq:fa3loop}
        f^{(3)} = \frac{1}{4} (\zeta_5 + 2 \zeta_2\zeta_3 ) \,.
\end{eqnarray}
For the three-loop 
kinematic function
$\mathcal{F}^{(3)}(\beta_{ijkl},\beta_{iklj})$,
we first express 
the CICRs in terms of variables $z_{ijkl}$
and $\bar{z}_{ijkl}$. We have  
\begin{eqnarray}\label{eq:CICR-z-zbar}
        \rho_{ijkl} = z_{ijkl}\bar{z}_{ijkl}\,, \quad\quad\quad 
        \rho_{ilkj} =(1-z_{ijkl})(1-\bar{z}_{ijkl})\,,
\end{eqnarray}
using which the function ${\cal F}^{(3)}$ takes the following form 
\begin{eqnarray}\label{eq:calFdef}
       {\cal F}^{(3)} (\beta_{ijkl},\beta_{ilkj}) = \frac{1}{32} \Big[F(1-z_{ijkl}) -F(z_{ijkl}) \Big] \,.
\end{eqnarray}
In turn, $F(z)$ is \cite{Almelid:2015jia}
\begin{eqnarray}\label{eq:F_SVHPL}
F(z) = \mathcal{L}_{10101}(z) + 2 \zeta_2 \left[\mathcal{L}_{001}(z) + \mathcal{L}_{100}(z)
\right]\,,        
\end{eqnarray}
where the functions ${\cal L}_w(z)$ are Brown's 
single-valued harmonic polylogarithms 
(SVHPLs)~\cite{CRMATH_2004__338_7_527_0,Dixon:2012yy,Brown:2013gia} 
in which $w$ is a word made out of $0$'s and $1$'s.
The function $F$ implicitly depends on $\bar z$ as well, 
but it is initially defined in the part of the Euclidean region 
where $\bar z = z^*$, where it is single-valued. One may 
then analytically continue the function beyond this region, 
treating $z$ and $\bar{z}$ as independent variables \cite{Almelid:2015jia,Almelid:2016lrq,Almelid:2017qju}. 

The four-loop results for functions $f$
and ${\cal F}$ have not yet been computed, and their direct computation is beyond what is achievable with present techniques.
However, several constraints on these quantities have been derived and a bootstrap programme to determine them has been initiated~\cite{Falcioni:2021buo}. As discussed in the introduction, the present work is motivated in part by this.

We note that the antisymmetry under $z\to 1-z$ in Eq.~(\ref{eq:calFdef}) is a manifestation of a general property of the function ${\cal F}$
in Eq.~(\ref{eq:GammaF})
under swapping its two arguments 
\begin{equation}
\label{calFAntisymm}
{\cal F}  (\beta_{ikjl},\beta_{iljk}) = - {\cal F}  (\beta_{iljk},\beta_{ikjl})\,,
\end{equation}
which is linked by Bose symmetry to the antisymmetry of the corresponding colour structure defined in Eq.~(\ref{Tauijkl}), under swapping $k$ and $l$, namely $\mathcal{T}_{iklj} = -\mathcal{T}_{ilkj}$, which stems from the antisymmetry of the three-gluon vertex. This of course applies to ${\cal F}$ at all loop orders.

For our analysis it is useful to define 
\begin{eqnarray} \label{eq:quardupoleABdef}
            {\bf{ \Gamma}}_{n,4{\rm{T}}-4{\rm{L}}}(\{\beta_{ijkl} \},\alpha_s)
&=& 4
\sum_{1\leq i<j<k<l \leq n} \, {\bf a}_{ijkl} (\{\beta\}) \,,
\end{eqnarray}
where
\bea \label{eq:quadrupole-ab-defshu} 
{\bf a}_{ijkl}(\{\beta\}) &=&  \nn
2\Big[ 
\mathcal{T}_{iklj} 
\,{\cal F}  (\beta_{ikjl},\beta_{iljk}) +\,\,\mathcal{T}_{ijlk} 
\,{\cal F}  (\beta_{ijkl},\beta_{ilkj}) 
+\,\mathcal{T}_{ijkl} 
\,{\cal F} (\beta_{ijlk},\beta_{iklj}) \Big]
\\
 &=&
{\cal F}_{ijkl}^{\rm A}(\{\beta\}) \,{\cal T}_{iklj} 
+ {\cal F}_{ijkl}^{\rm S}(\{\beta\}) \Big( {\cal T}_{ijlk}
+ {\cal T}_{ijkl} \Big)  \,,
\eea
with the functions ${\cal F}_{ijkl}^{\rm A}(\{\beta\})$
and ${\cal F}_{ijkl}^{\rm S}(\{\beta\}) $ being combinations 
of the kinematic functions ${\cal F}$ in \Eqn{eq:GammaF}
\begin{subequations} 
    \label{quadrupole-AS-defsh} 
    \begin{align}
{\cal F}_{ijkl}^{\rm A}(\{\beta\}) &=
{\cal F}(\beta_{ijkl},\beta_{ilkj}) - {\cal F}(\beta_{ijlk},\beta_{iklj}) + 2{\cal F}(\beta_{ikjl},\beta_{iljk})\,, \\[0.1cm]
{\cal F}_{ijkl}^{\rm S}(\{\beta\}) &= {\cal F}(\beta_{ijkl},\beta_{ilkj}) +  {\cal F}(\beta_{ijlk},\beta_{iklj})\,,
\end{align}
\end{subequations} 
which are, respectively, antisymmetric and symmetric with respect to swapping the two velocities $\beta_k$ and $\beta_l$ (or, equivalently, $\beta_i$ and $\beta_j$).  The functions 
${\cal F}_{ijkl}^{\rm A}$ and ${\cal F}_{ijkl}^{\rm S}$ 
correspond respectively to the functions $F_a$ and $F_b$ 
introduced in Eq.~(6.26) of Ref.~\cite{Almelid:2017qju}. 
This representation can be obtained from \Eqn{eq:GammaF}
with the help of the following relation between the colour factors defined in Eq.~(\ref{Tauijkl}):
\begin{equation}
\label{JacobiTau}\mathcal{T}_{ijlk}=
\mathcal{T}_{iklj}+
\mathcal{T}_{ijkl}\,,
\end{equation}
which follows directly from the Jacobi identity 
\begin{equation}\label{eq:Jacobi}
 f^{ace}f^{bde} =  f^{abe}f^{cde} +  f^{ade}f^{bce}   \,.
\end{equation}

The remaining terms in \Eqn{eq:adm-param}
begin at the four-loop order.  Adopting the
convention of Ref.~\cite{Falcioni:2021buo} (which differs
by a factor $1/2$ in the definition of $g_R$ with
respect to Eq.~(42) of Ref.~\cite{Becher:2019avh}),
the quartic term with dependence on four colour
generators and attaching to two and three legs
can be written as follows~\cite{Becher:2019avh}
\begin{eqnarray}\label{eq:Q4T-23L}
        {\bf{ \Gamma}}^{}_{n,{\rm{Q}}4{\rm{T}}-2,3{\rm{L}}}(\{s_{ij} \},\lambda,\alpha_s) &=& 
        - \sum_R g_R(\alpha_s) \bigg[\sum_{1\leq i<j  \leq n}\Big(  { \bf{\cal{ D}}}^R_{iijj} 
        +  { \bf{\cal{ D}}}^R_{iiij}+  { \bf{\cal{ D}}}^R_{jjji}\Big)\ell_{ij} 
    \nonumber \\ &&\hspace{4cm}
       +   \sum_{i=1}^n\sum_{\substack{ 1\leq j < k \leq n\\ j,k \neq i } }{ \bf{\cal{ D}}}^R_{jkii}\ell_{jk}
        \bigg] \,,
\end{eqnarray}
where 
$\ell_{ij} = \ln\left(\frac{-s_{ij}}{\lambda^2}\right)$  
and 
\begin{eqnarray}\label{eq:calDijkl}
    { \bf{\cal{ D}}}^R_{ijkl} =  
    \frac{1}{4!} \sum_{\sigma \in S_4}
   {\rm{Tr}}_R\left(T^{\sigma(a)}T^{\sigma(b)}T^{\sigma(c)}
   T^{\sigma(d)}\right) \T^a_i\T^b_j\T^c_k\T^d_l \, .
\end{eqnarray}
The function $g_R(\alpha_s)$ is the coefficient of the quartic Casimir within the lightlike cusp anomalous dimension.\footnote{See Eqs.~(2.28) and~(7.26) in Ref.~\cite{Falcioni:2021buo} for the precise notation.} It begins at four loops; at this order it has been calculated in Refs.~\cite{Henn:2019swt,Huber:2019fxe,vonManteuffel:2020vjv}. 
Next, we have the ${\bf{ \Gamma}}_{n,{\rm{Q}}4{\rm{T}}-4{\rm{L}}}(\{\beta_{ijkl} \} ,\alpha_s)$
term which depends on the quartic Casimir operator in \Eqn{eq:calDijkl} as follows:
\begin{eqnarray}\label{eq:Q4T-4L}
{\bf{ \Gamma}}^{}_{n,{\rm Q}4{\rm{T}}-4{\rm{L}}}
(\{\beta_{ijkl} \},\alpha_s) &=& \,24\,
\sum_{R} \sum_{1\leq i<j<k<l \leq n} 
\DR_{ijkl} 
\,{\cal G}_R (\beta_{ijlk},\beta_{iklj}) \,.
\end{eqnarray}
Due to complete permutation symmetry of  $\DR_{ijkl}$, the function ${\cal{G}}_R$ satisfies
\begin{eqnarray}\label{eq:symGR}
        {\cal{G}}_R(x_1,x_2)=        {\cal{G}}_R(x_2,x_1) = 
            {\cal{G}}_R(x_1-x_2,-x_2) = {\cal{G}}_R(x_2-x_1,-x_1) \,.
\end{eqnarray}
The explicit form of this function is currently unknown.
Lastly, we have the terms containing five colour generators 
\begin{eqnarray}\label{eq:H_5T-4L}
    {\bf{ \Gamma}}^{}_{n,5{\rm{T}}-4{\rm{L}}}(\{\beta_{ijkl} \},\alpha_s) &=& \,2\,
\sum_{i=1}^n
\sum_{ \substack{ 1\leq j<k<l \leq n \\  j,k,l \neq i }} \Big[  
\mathcal{T}_{iklji}\mathcal{H}_1(\beta_{ikjl},\beta_{iljk};\alpha_s) \\[-0.1cm] \nn
&&\hspace{1.5cm}+\,  \mathcal{T}_{ijlki}\mathcal{H}_1(\beta_{ijkl},\beta_{ilkj};\alpha_s)
+\,\mathcal{T}_{ijkli}\mathcal{H}_1(\beta_{ijlk},\beta_{iklj};\alpha_s) \Big] \,,
\end{eqnarray}
and
\begin{eqnarray}\label{eq:H_5T-5L}
{\bf{ \Gamma}} ^{ }_{n,5{\rm{T}}-5{\rm{L}}}(\{\beta_{ijkl} \},\alpha_s)   
&=& 8 \sum_{m=1}^n   \sum_{\substack{1\leq i<j<k<l\leq n\\ i,j,k,l\neq m} } 
\Big[   \mathcal{T}_{ikljm}
\mathcal{H}_2(\beta_{iklj},\beta_{ikml},\beta_{ilmk},\beta_{kimj},\beta_{kjmi};
\alpha_s) \nonumber \\[-0.1cm] 
&&\hspace{1.5cm}+\, \mathcal{T}_{ijlkm}
\mathcal{H}_2(\beta_{ijlk},\beta_{ijml},\beta_{ilmj},\beta_{jimk},\beta_{jkmi};
\alpha_s) \nonumber \\[0.3cm] 
&&\hspace{1.5cm}+\, \mathcal{T}_{ijklm}
\mathcal{H}_2(\beta_{ijkl},\beta_{ijmk},\beta_{ikmj},\beta_{jiml},\beta_{jlmi};
\alpha_s)\Big] \,,
\end{eqnarray}
where
\begin{eqnarray}
\label{eq:cT5}
\mathcal{T}_{ijklm} &=& i f^{adf}f^{bcg}f^{efg}
\left\{\mathbf{T}^a_i,\mathbf{T}^b_j,\mathbf{T}^c_k,
\mathbf{T}^d_l, \mathbf{T}^e_m \right\}_+\,.
\end{eqnarray}
The kinematic functions $\mathcal{H}_1$ and $\mathcal{H}_2$
starting at the four-loop order have not been computed but, as previously mentioned, they are thought to vanish following the argument of Ref.~\cite{Vladimirov:2017ksc}. We shall not apply this argument here, so as to explore what other constraints we can deduce regarding these.

Some information about the unknown functions 
${\cal{G}}_R$, $\mathcal{H}_1$, $\mathcal{H}_2$, and four-loop
contributions to $f$ and ${\cal{F}}$ has been obtained through
analysis of specific kinematic limit such as Regge limit 
\cite{DelDuca:2011ae,Falcioni:2021buo}
and two-particle collinear limits \cite{Becher:2019avh,Almelid:2017qju,Dixon:2009ur,Becher:2009qa}. 
We will return to the question whether new data 
can be gathered through analysis of {\emph{multi-particle}} 
collinear for these massless functions in Section~\ref{sec:massless}.

For the purposes of studying multi-particle collinear  
limits in the presence of a massive coloured particle carried out 
in Section~\ref{sec:massive}, we also 
require the corresponding soft anomalous dimension terms.
We use the parametrisation of Ref.~\cite{Liu:2022elt}, and in the following the capital index $I$ denotes the one additional massive coloured
particle with mass~$m_I$. 
To the terms in the anomalous dimension given in 
\Eqn{eq:adm-param},
we add the following structures
\begin{eqnarray}\label{eq:adm-massive}
{\bf{ \Gamma}}\left(\{ p\}, m_I,\lambda,\alpha_s \right)
&=& \gamma_{K}(\alpha_s) \sum_i \T_I\cdot \T_i 
\ln \frac{m_I\, \lambda}{(-s_{Ij})} 
+ \gamma^I(\alpha_s){\mathbf{1}}
+2   \sum_{1\leq i< j \leq n} {\mathcal{T}}_{IIij}
F_{{\rm{h}}2}\left(r_{ijI},\alpha_s \right)
\nonumber \\ && \hspace{-0.5cm}
+  \sum_{1\leq i< j< k \leq n} {\bf a}^h_{ijkI}(\{r\})
\,,
\end{eqnarray}
where $\T_I$ is the colour generator of the massive particle $I$,
and its anomalous dimension~$\gamma^{I}$ is known to three 
loops~\cite{Korchemsky:1987wg,Korchemsky:1991zp,Kidonakis:2009ev,Grozin:2014hna,Grozin:2015kna,Bruser:2019yjk}, and we defined 
\begin{equation}
\label{eq:Massive-aijkI}
{\bf a}^h_{ijkI}(\{r\})  = 2\,\Big[{\mathcal{T}}_{ijkI}
F_{{\rm{h}}3}\left(r_{ijI},r_{ikI},r_{jkI}\right)
+ {\mathcal{T}}_{jikI}
F_{{\rm{h}}3}\left(r_{jiI},r_{jkI},r_{ikI} \right)
+{\mathcal{T}}_{kjiI}
F_{{\rm{h}}3}\left(r_{kjI},r_{kiI},r_{jiI}\right)\Big],
\end{equation}
where we have left the dependence of $F_{{\rm{h}}3}$ functions on $\alpha_s$ implicit. Note the first index in every ${\cal T}$ is the same as the repeated index in the first two arguments, while the two middle indices (with respect to which this ${\cal T}$ is antisymmetric) provide the second index in $r$ in the first and second arguments. 
In App.~\ref{sec:Fh3Constraints} we show that we can also write (we substitute $(i,j,k)\to (1,2,3)$) :
\begin{equation}
\label{eq:Massive-aiIjk}
\begin{split}
{\bf a}^h_{iIjk}(\{r\})
=\,& 2\,\Big[{\mathcal{T}}_{Ikji}
F_{{\rm{h}}3}\left(r_{ijI},r_{ikI},r_{jkI}\right)
+ {\mathcal{T}}_{Ikij}
F_{{\rm{h}}3}\left(r_{jiI},r_{jkI},r_{ikI} \right)
+{\mathcal{T}}_{Iijk}
F_{{\rm{h}}3}\left(r_{kjI},r_{kiI},r_{jiI}\right)\!\Big]
\\
=&\, {\mathcal{T}}_{Ikji}
F^{\rm A}_{{\rm{h}}3, iIjk}\left(\{r\} \right)
+ \left[{\mathcal{T}}_{Iikj}+{\mathcal{T}}_{Iijk}\right]
F^{\rm S}_{{\rm{h}}3, iIjk}\left(\{r\} \right) \,,
\end{split}
\end{equation}
\begin{subequations}
with
\label{eq:Fh3quadrupole-A-main} 
\begin{align}
\label{eq:Fh3quadrupole-A-main-a} 
F^{\rm A}_{{\rm{h}}3, {iIjk}}
(\{r\})
&\equiv
F_{{\rm{h}}3}\left(r_{kjI},r_{kiI},r_{jiI} \right)  
-F_{{\rm{h}}3}\left(r_{jkI},r_{jiI},r_{ikI}\right)
+2F_{{\rm{h}}3}\left(r_{ijI},r_{ikI},r_{jkI}  \right)\,,
\\
\label{eq:Fh3quadrupole-A-main-b} 
F^{\rm S}_{{\rm{h}}3, {iIjk}}
(\{r\})&\equiv 
F_{{\rm{h}}3}\left(r_{kjI},r_{kiI},r_{jiI}\right)
+F_{{\rm{h}}3}\left(r_{jkI},r_{jiI},r_{ikI}
  \right)\,.
\end{align}
\end{subequations}
Since \Eqn{eq:Massive-aijkI} shares the same colour structure with its massless counterpart in \Eqn{eq:GammaF}, also the kinematically-dependent functions  
$F_{{\rm{h}}3}$ and ${\cal F}$ have some properties in common. In particular, $F_{{\rm{h}}3}$ has a similar antisymmetry property to \Eqn{calFAntisymm} upon swapping its first two arguments,
\begin{equation}
\label{Fh3Antisymm}
F_{{\rm{h}}3}\left(r_{ijI},r_{ikI},r_{jkI}\right)= -F_{{\rm{h}}3}\left(r_{ikI}, r_{ijI},r_{jkI}\right)\,. 
\end{equation}
Moreover, also the Jacobi identity (\ref{JacobiTau}) becomes important for understanding these functions.

We note that since we are only considering one additional massive 
particle, no angle-dependent cusp contributions, nor
tripole structures ${\cal{T}}_{ijk}=if^{abc}(\T^a_i\T^b_j\T^c_k)_+$
appear in~\Eqn{eq:adm-massive}, since
non-vanishing contributions of this type (which arise first at two-loop order) must connect 
at least two massive
particles~\cite{Becher:2009kw,Ferroglia:2009ep}. The conformal cross ratio describing correlations between three coloured particles,
of which exactly one is massive, is given by 
\begin{eqnarray}\label{eq:r-kin}
    r_{ijI} = \frac{p_i\cdot p_j\, p_I^2}{2 p_i\cdot p_I\, p_j\cdot p_I }\,,
\end{eqnarray}
and the kinematic variables corresponding to quadrupole
correlations (last two lines of \Eqn{eq:adm-massive})
can be written in terms of three linearly independent 
CICRs: $r_{ijI}$, $r_{ikI}$, and~$r_{jkI}$, since one can form any 
other conformal ratio using these. 
The function $ F_{{\rm{h}}2}\left(r_{ijI},\alpha_s \right)$,
connecting the massive particle and two massless ones,
was calculated in Ref.~\cite{Liu:2022elt}, but
the $ F_{{\rm{h}}3}\left(r_{ijI},r_{ikI},r_{jkI},\alpha_s \right)$, 
connecting the massive particle and \emph{three} massless ones,
remained unknown. This missing ingredient has been
computed very recently and will be published soon~\cite{GZ-TBP}.
There are known constraints on these
functions arising from two-particle collinear limits and
small-mass expansion. In Section~\ref{sec:massive}, we 
study whether multi-particle collinear limits provide
additional information about these structures. 


\section{Factorisation in collinear limits}
\label{sec:factorisationincollinearlimits}
Having introduced the soft anomalous dimension, in this section 
we turn our attention to the central topic of this
work, i.e., the collinear limits of $n$-point 
scattering amplitudes. Taking the collinear limit 
of an $n$-point amplitude amounts to relaxing the 
fixed-angle condition for the amplitude stated 
at the very beginning of the previous section.
We now allow for a subset of invariants 
$p_i\cdot p_j$ to become parametrically 
small and investigate the factorisation 
properties of $\MM_n$ in this limit.

\begin{figure}[t]
\begin{center}
    \includegraphics[width=0.65\textwidth]{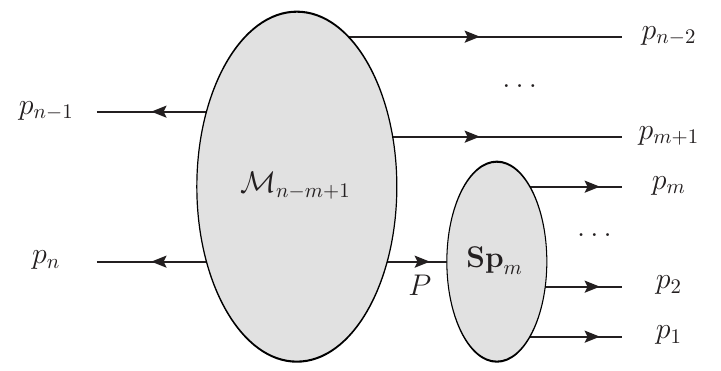}
\end{center}
    \caption{Labelling of momenta for the 
        configuration of $m$-particle timelike 
        collinear limit. The initial-state particle 
        momenta are labelled $p_{n-1}$ and $p_n$. 
        All other particles are in the final state.}
\label{m-collinearTL}
\end{figure}

The \emph{timelike} $m$-parton collinear 
limit is defined by the kinematic configuration 
in which $m$ partons, all either in the initial 
or all in the final state, become collinear, as 
shown explicitly for the latter case in 
Fig.~\ref{m-collinearTL}. In this limit the 
amplitude $\MM_n(p_1,\ldots p_n;\mu)$ becomes 
singular as $P^2\to 0$, where $P$ is the total momentum of the collinear particles, $P= p_1 + \ldots + p_m$.  
The $n$-point amplitude factorises 
into an $(n-m+1)$-point amplitude times a 
splitting amplitude, denoted~$\SP_{m}$, which captures the single pole at $P^2= 0$ and only depends on the momenta and colour of the $m$ collinear 
particles \cite{Berends:1988zn,Mangano:1990by,Bern:1995ix,Kosower:1999xi}, 
\bea\label{TL-collinear-limit} 
\MM_n(p_1,\ldots p_m, \{p_i\}_{\text{rest}};\mu) 
&\stackrel{p_1 \parallel p_2 \parallel \ldots \parallel p_m}{\longrightarrow}&
\SP_{m}(p_1, \ldots p_m;\mu) \,\MM_{n-m+1}(P,\{p_i\}_{\text{rest}};\mu)\,, 
\eea
where we denote dependence on the momenta of the rest of the process (the non-collinear particles) by $\{p_i\}_{\text{rest}}=p_{m+1},\ldots p_n$.
The flavour of the parent parton is uniquely fixed from the $m$ particles from the collinear set.
Note that the left- and right-hand sides differ by power-suppressed terms in the limit $P^2\to0$. 
Similarly to momenta, also in terms of colour, the amplitude 
$\MM_{n-m+1}(P,\{p_i\}_{\text{rest}};\mu)$ depends 
only on the \emph{total} colour charge $\T_P$ of the set of collinear particles, $\T_P = \T_1 + \ldots + \T_m$ (rather than on their individual charges) and on the colour charges of the rest of the process. \Eqn{TL-collinear-limit} is written in 
colour-space notation 
\cite{Catani:1996jh,Catani:1996vz,Catani:1998bh}: 
amplitudes are represented as vectors in colour 
space. The amplitudes $\MM_n$ and $\MM_{n-m+1}$ 
live respectively in the $n$- and $(n-m+1)$-parton 
colour space. Consequently, the splitting amplitude 
is given initially as a matrix in the 
colour space of $n \times (n-m+1)$-partons. Upon 
substituting $\T_P = \T_1 + \ldots + \T_m$, 
however, one can use the $n$-parton colour 
space only.
The fact that the splitting amplitude $\SP_m$ 
depends only on the momenta and colours of the 
partons becoming collinear constitutes a 
relatively strong statement of universality. 
For this reason, the timelike collinear 
factorisation is also referred to as 
\emph{strict collinear factorisation} 
\cite{Catani:2011st}. 

It turns out that infrared factorisation 
is useful to study collinear factorisation, and vice versa: strict collinear factorisation constrains the structure of infrared factorisation. This is 
because the amplitudes $\MM_n$ and $\MM_{n-m+1}$, 
as well as the splitting amplitude $\SP_m$ 
in \Eqn{TL-collinear-limit}, are infrared 
divergent. Owing to collinear factorisation, these divergences are related.  Both $\MM_n$ and $\MM_{n-m+1}$,  
in \Eqn{TL-collinear-limit} respectively,
admit an infrared factorisation as in 
\Eqn{IRfacteq}, which means that their 
infrared divergences can be determined 
starting from a soft anomalous dimension 
for $n$- and $(n-m+1)$-particle scattering 
process. Collinear factorisation then states that in the limit where $m$ of the $n$ partons are collinear, the soft anomalous dimension of the splitting amplitude is given by the difference between the soft anomalous dimensions of the $n$-parton amplitude and that of the $(n-m+1)$ one. Since the soft anomalous dimension of the splitting amplitude can only depend on the degrees of freedom on the $m$ collinear partons, this relation amounts to a strong constraint on these quantities.
We review the definitions of the 
relevant objects and outline the derivation of the relations between them in the next section.  


\subsection{The splitting amplitude soft anomalous dimension}
\label{sec:factorisationincollinearlimits-sub}

As a consequence of the factorisation structure in 
\Eqn{TL-collinear-limit}, the infrared singularities 
of the timelike splitting amplitude are given to all 
orders in terms of the so-called splitting amplitude 
soft anomalous dimension, which can be related to the soft anomalous dimensions of amplitudes.
Its derivation has been 
discussed for instance in 
Ref.~\cite{Becher:2009qa,Dixon:2009ur,Catani:2011st} 
and more recently in Ref.~\cite{Almelid:2017qju}. 
Here we review it since this relation is key to our investigation. 

The starting point of this derivation is the observation that the 
infrared factorisation formula in \Eqn{IRfacteq} 
applies to both the $n$- and the $(n-m+1)$-point 
scattering amplitudes\footnote{We note that the $\MM_{n-m+1}$ amplitude in 
\Eqn{TL-collinear-limit} is evaluated at $P^2=0$, 
hence it does obey the usual IR factorisation 
formula in \Eqn{IRfacteq} for massless amplitudes.} in \Eqn{TL-collinear-limit}:
\begin{align} 
\label{IRfacteqn}
\MM_n \left(p_1,\ldots p_m,\{p_i\}_{\text{rest}},\mu  \right) \, &= \, 
{\bf Z}_n \left(p_1,\ldots p_m,\{p_i\}_{\text{rest}},\mu_f  \right)
\Hhard_n \left(p_1,\ldots p_m,\{p_i\}_{\text{rest}},\mu_f, \mu  \right),
\\[0.1cm]
  \label{IRfacteqnm1}
\MM_{n-m+1} \left(P,\{p_i\}_{\text{rest}},\mu  \right) \, &= \, 
{\bf Z}_{n-m+1} \left(P,\{p_i\}_{\text{rest}},\mu_f  \right)
\Hhard_{n-m+1} \left(P,\{p_i\}_{\text{rest}},\mu_f, \mu \right),
\end{align}
where the singularities in $\epsilon\to 0$ are fully contained in the factors $Z_n$ and $Z_{n-m+1}$, respectively, which are in turn governed by the soft anomalous dimension according to Eq.~(\ref{RGsol}). The hard functions
${\cal H}_n$ and ${\cal H}_{n-m+1}$
are finite for $\epsilon\to 0$.
Furthermore, collinear factorisation in the limit where $p_i \cdot p_j \to 0$, 
with $i,j = 1,\ldots ,m$ -- all of which are massless final-state particles -- applies also to the hard function
${\cal H}_n$, which factorises as
\be\label{H-TL-collinear-limit} 
\Hhard_n(p_1,\ldots p_m, \{p_i\}_{\text{rest}},\mu_f,\mu) 
\stackrel{p_1 \parallel p_2 \parallel \ldots \parallel p_m}{\longrightarrow}
\SP_{\HH,m}(p_1, \ldots p_m;\mu_f,\mu) \,
 \Hhard_{n-m+1}(P,\{p_i\}_{\text{rest}},\mu_f,\mu)\,,
\ee
in full analogy with the corresponding amplitude
in Eq.~(\ref{TL-collinear-limit}).
In Eq.~(\ref{H-TL-collinear-limit}) we introduced the hard splitting amplitude $\SP_{\HH,m}$, 
which defines 
the collinear behaviour of the hard amplitude ${\cal H}_n$. Given that both $\Hhard_n$ and $\Hhard_{n-m+1}$ 
are finite for $\eps \to 0$, $\SP_{\HH,m}$ must 
be finite for $\eps \to 0$ as well, (although divergent for $p_i \cdot p_j \to 0$). 
Putting the infrared factorisation properties in Eqs.~\eqref{IRfacteqn} and \eqref{IRfacteqnm1} together with the collinear factorisation properties in Eqs.~\eqref{TL-collinear-limit} and 
\eqref{H-TL-collinear-limit}, we  
obtain a relation between 
$\SP_{\HH,m}$ and $\SP_m$:
\begin{equation}
    \begin{split}
\label{SPvsSPH} 
\SP_{\HH,m}&(p_1, \ldots p_m;\mu_f,\mu) =\\
&={\bf Z}^{-1}_n(p_1,\ldots p_m, \{p_i\}_{\text{rest}};\mu_f) \,
\SP_m(p_1, \ldots p_m;\mu)  
{\bf  Z}_{n-m+1}(P, \{p_i\}_{\text{rest}};\mu_f).
\end{split}\end{equation}
We note that this equation encodes extremely non-trivial 
cancellations between the rest-of-the-process partons on 
the right-hand side, in such a way that the left-hand 
side contains information only about the partons 
becoming collinear. We continue by differentiating 
\Eqn{SPvsSPH} with respect to the logarithm of the 
infrared factorisation scale, and using the 
renormalisation group equation for ${\bf Z}_n$, 
\Eqn{rgeZ}. We then we arrive at 
\bea \label{diffSPH1}\nn
 \frac{d}{d\ln\mu_f} \SP_{\HH,m}&&\!\!\!(p_1, \ldots p_m;\mu_f,\mu)\\ \nn
&&= {\bf \Gamma}_n (p_1,\ldots p_m, \{p_i\}_{\text{rest}};\mu_f) 
\,\SP_{\HH,m}(p_1, \ldots p_m;\mu_f,\mu) \\
&&\quad\quad -\, \SP_{\HH,m}(p_1, \ldots p_m;\mu_f,\mu) 
\, {\bf   \Gamma}_{n-m+1}(P, \{p_i\}_{\text{rest}};\mu_f)\,,
\eea 
which is obtained by noticing that the splitting 
function $\SP_m(p_1, \ldots p_m;\mu)$ does not 
depend on the factorisation scale $\mu_f$ and 
therefore $\frac{d}{d\ln\mu_f} \SP_m(p_1, \ldots p_m;\mu) = 0$. 

In analogy with Eq.~(\ref{RG_H}), we further define the anomalous dimension of the $m$-parton (timelike) splitting function 
$\SP_{\HH,m}(p_1, \ldots p_m;\mu_f)$
 by 
\be \label{diffSPH2}
\frac{d}{d\ln\mu_f} \SP_{\HH,m}(p_1, \ldots p_m;\mu_f,\mu)
= {\bf \Gamma}_{\SP,m} (p_1,\ldots p_m;\mu_f) 
\,\SP_{\HH,m}(p_1, \ldots p_m;\mu_f,\mu)\,.
\ee 
We now discuss how ${\bf \Gamma}_{\SP,m}$ can be determined based on Eq.~(\ref{diffSPH1}). 

Based on the definition of the hard splitting amplitude in~Eq.~(\ref{H-TL-collinear-limit}), both the left- and right-hand side of  
Eq.~(\ref{SPvsSPH}), and hence also of Eq.~(\ref{diffSPH1}), must be understood as colour-space operators acting on the hard function
$\Hhard_{n-m+1}(P,\{p_i\}_{\text{rest}},\mu_f,\mu)$, which lives in the colour space of the rest of the process along with the parent particle $P$, and is colour conserving.
Hence, upon acting on $\Hhard_{n-m+1}$, the generators admit colour conservation:
\begin{equation}
\label{colour_conservation_rest}
\sum_{i=m+1}^n \T_i\, 
\Hhard_{n-m+1}(P,\{p_i\}_{\text{rest}},\mu_f,\mu)
=-\T_P \Hhard_{n-m+1}(P,\{p_i\}_{\text{rest}},\mu_f,\mu)\,,
\end{equation}
where $\T_i$ for $i>m$ are colour generators of individual non-collinear partons in the rest of the process, while
$\T_P$ is the colour generator  of parent particle $P$ of the collinear set.

Consider now the term in the last line of \Eqn{diffSPH1}. The anomalous dimension operator ${\bf \Gamma}_{n-m+1}(P, \{p_i\}_{\text{rest}};\mu_f)$ there contains $\T_i$ for $i>m$, i.e., colour generators of the rest of the process, as well as the colour charge $\T_P$ of the parent particle. In turn, $\SP_{\HH,m}(p_1, \ldots p_m;\mu_f,\mu)$ only depends on the collinear particles, $i\leq m$. This object, just like an amplitude, must be colour conserving, 
\be \label{T1T2TP}
\,\SP_{\HH,m}(p_1, \ldots p_m;\mu_f,\mu)\,\T_P=
\sum_{i=1}^m
\T_i \, \SP_{\HH,m}(p_1, \ldots p_m;\mu_f,\mu) 
\, .
\ee 
The commutativity property\footnote{The reason why the colour generators on the right-hand side of Eq.~(\ref{T1T2TP}) is written to the left of $\SP_{\HH,m}$ can be understood from 
Eq.~(\ref{H-TL-collinear-limit}): $\SP_{\HH,m}$ acts (to the right) on the colour space of the rest of the process, and it produces an object in the colour space of the original $n$-parton amplitude. Therefore, expressing $\T_P$ in terms of the collinear parton generators is appropriate only when these are placed to the left of $\SP_{\HH,m}$.} in Eq.~(\ref{T1T2TP}) is consistent with the fact that $\T_P$ on the 
left-hand side of Eq.~(\ref{T1T2TP}) may also be written in terms of the generators $\T_i$ for $i>m$ (as implied by Eq.~(\ref{colour_conservation_rest})), and each of these trivially commutes with $\SP_{\HH,m}$.

Given Eq.~(\ref{T1T2TP}), and 
since all other generators in ${\bf   \Gamma}_{n-m+1}(P, \{p_i\}_{\text{rest}};\mu_f)$ trivially commute with $\SP_{\HH,m}(p_1, \ldots p_m;\mu_f,\mu)$, we may just as well write the term in the last line of \Eqn{diffSPH1} as
\begin{equation}
    \left.{\bf   \Gamma}_{n-m+1}(P, \{p_i\}_{\text{rest}};\mu_f)\right\vert_{\T_P\to \sum_{i=1}^m \T_i}\SP_{\HH,m}(p_1, \ldots p_m;\mu_f,\mu) 
\,, 
\end{equation}
and hence \Eqn{diffSPH1} allows us to identify the anomalous dimension of the splitting amplitude defined in Eq.~(\ref{diffSPH2}) as~\cite{Becher:2009qa,Dixon:2009ur,Catani:2011st,Almelid:2017qju}: 
\begin{align}\label{GammaSPdef} 
\begin{split}
{\bf \Gamma}_{\SP,m} (p_1,\ldots p_m;\mu_f)
&={\bf \Gamma}_n (p_1,\ldots p_m, p_{m+1},\ldots p_n;\mu_f) 
\\&\hspace*{30pt} -\,{\bf \Gamma}_{n-m+1}(P, p_{m+1},
\ldots p_n;\mu_f)|_{\T_P\to \sum_{i=1}^m \T_i}\,.
\end{split}
\end{align}
Note that we implicitly imply that in the right-hand side we also perform the replacement $P\to\sum_{i=1}^mp_i$.
Thus, in the timelike collinear limit, infrared 
divergences are obtained entirely from the knowledge 
of the splitting amplitude soft anomalous dimension.
The soft anomalous dimension of the splitting 
amplitude is therefore the basic object 
describing the infrared singularities of the 
splitting amplitude, and it can be obtained
starting from the soft anomalous dimensions
of the $n$- and $(n-m+1)$-scattering amplitudes. 
This is the strategy we will adopt in Sections~\ref{sec:massless} 
and~\ref{sec:massive} to derive multi-particle
collinear limits for amplitudes with massless and 
massive particles in external states.
It is remarkable that, while the two anomalous dimensions on the right-hand side of Eq.~(\ref{GammaSPdef}) depend in a complicated way on all partons involved in the respective scattering amplitudes, in the collinear limit their difference depends exclusively on the collinear set. The way any dependence on the rest of the process cancels out is very interesting and will be investigated below. 

For convenience, we define separate contributions to the 
$m$-particle splitting amplitude soft anomalous dimension
according to the different types of structures appearing on 
the right-hand side of~\Eqn{eq:adm-param} that start to contribute
at different orders in the perturbative expansion.
First, we define the $m$-particle splitting amplitude 
soft anomalous dimension for the dipole terms in Eq.~(\ref{eq:dipole}) as
\bea \label{eq:GammaSPdefDip} \nn
{\bf \Gamma}_{\SP,m}^{\rm dip.} (p_1,\ldots p_m;\mu_f)
&=&{\bf \Gamma}_n^{\rm dip.} (p_1,\ldots p_m, p_{m+1},\ldots p_n;\mu_f) \\
&&-\,{\bf \Gamma}_{n-m+1}^{\rm dip.}(P, p_{m+1},\ldots p_n;\mu_f)|_{\T_P\to \sum_{i=1}^m \T_i}.
\eea
For the terms starting at three loops, given in Eqs.~(\ref{eq:Gammaf}) and~(\ref{eq:quardupoleABdef}), we define 
\begin{eqnarray}
\label{eq:GammaSpQuadm} 
{\bf{ \Gamma}}_{\SP, m}^{4{\rm{T}}}(p_1,\ldots p_m;\mu_f) &=& 
{\bf{ \Gamma}}_{n,4{\rm{T}}-3{\rm{L}}}(\alpha_s)
+ {\bf{ \Gamma}}_{n,4{\rm{T}}-4{\rm{L}}}(\{\beta_{ijkl}\},\alpha_s)
\nonumber \\  && \hspace{-2.3cm} 
-  \Big(  {\bf{ \Gamma}}_{n-m+1,4{\rm{T}}-3{\rm{L}}}(\alpha_s)
+ {\bf{ \Gamma}}_{n-m+1,4{\rm{T}}-4{\rm{L}}}(\{\beta_{ijkl}\},\alpha_s)\Big) 
\Big|_{\T_P\to \sum_{i=1}^m \T_i}.
\end{eqnarray}
Next, we define the corresponding object for terms 
beginning at four loops, containing colour structures
related to the quartic Casimir, given in Eqs.~(\ref{eq:Q4T-23L}) and (\ref{eq:Q4T-4L}):
\begin{eqnarray}
\label{eq:GammaSpQ4Tm}
{\bf{ \Gamma}}_{\SP, m}^{{\rm Q}4{\rm{T}}}(p_1,\ldots p_m;\mu_f) &=& 
{\bf{ \Gamma}}_{n,{\rm{Q}}4{\rm{T}}-2,3{\rm{L}}}(\{s_{ij} \},\mu_f,\alpha_s)
+{\bf{ \Gamma}}_{n,{\rm Q}4{\rm{T}}-4{\rm{L}}}(\{\beta_{ijkl} \},\alpha_s) 
\nonumber \\ && \hspace{-4cm}
- \Big({\bf{ \Gamma}}_{n-m+1,{\rm{Q}}4{\rm{T}}-2,3{\rm{L}}}(\{s_{ij} \},\mu_f,\alpha_s)
+{\bf{ \Gamma}}_{n-m+1,{\rm Q}4{\rm{T}}-4{\rm{L}}}(\{\beta_{ijkl} \},\alpha_s)\Big) 
\Big|_{\T_P\to \sum_{i=1}^m \T_i},
\end{eqnarray}
and lastly we introduce the terms beginning 
at four loops, which contain five colour generators, given in Eqs.~(\ref{eq:H_5T-4L}) and~(\ref{eq:H_5T-5L}):
\begin{eqnarray}
\label{eq:GammaSp5Tm}
{\bf{ \Gamma}}_{\SP, m}^{5{\rm{T}}}(p_1,\ldots p_m;\mu_f) &=&
{\bf{ \Gamma}}_{n,5{\rm{T}}-4{\rm{L}}}(\{\beta_{ijkl} \},\alpha_s) 
+{\bf{ \Gamma}}_{n,5{\rm{T}}-5{\rm{L}}}(\{\beta_{ijkl} \},\alpha_s) 
\nonumber \\ 
&& \hspace{-3.2cm}
- \Big(  {\bf{ \Gamma}}_{n-m+1,5{\rm{T}}-4{\rm{L}}}(\{\beta_{ijkl} \},\alpha_s)
+ {\bf{ \Gamma}}_{n-m+1,5{\rm{T}}-5{\rm{L}}}(\{\beta_{ijkl} \},\alpha_s) 
\Big) |_{\T_P\to \sum_{i=1}^m \T_i}\,.
\end{eqnarray}
Then, the complete  ${\bf \Gamma}_{\SP,m}$ of \Eqn{GammaSPdef} is
given by
\begin{align}
\label{eq:GammaSPdef-shand} 
\begin{split}
{\bf \Gamma}_{\SP,m} (p_1,\ldots p_m;\mu_f)
= &\,\,{\bf \Gamma}_{\SP,m}^{\rm dip.} (p_1,\ldots p_m;\mu_f)+ 
{\bf{ \Gamma}}_{\SP, m}^{4{\rm{T}}}(p_1,\ldots p_m;\mu_f) 
\\ &+ 
 {\bf{ \Gamma}}_{\SP, m}^{{\rm Q}4{\rm{T}}}(p_1,\ldots p_m;\mu_f) +
{\bf{ \Gamma}}_{\SP, m}^{5{\rm{T}}}(p_1,\ldots p_m;\mu_f)
+ \mathcal{O}(\alpha_s^5),
\end{split}
\end{align}
where $\mathcal{O}(\alpha_s^5)$ indicates the 
higher-order structures contributing to the soft
anomalous dimension in \Eqn{eq:adm-param}.


\subsection{Kinematics}
\label{sec:kinematics}

In order to analyse the behaviour of
kinematic functions and the relevant
variables that these depend on in  
the approach to the collinear limit,
it is important to have an explicit 
parametrisation of the momenta of the
particles becoming collinear. 
For the parametrisation we choose light-cone
coordinates, with the vectors $n^\mu_-$ and
$n^\mu_+$ which are lightlike and fulfil
$n_-\cdot n_+=2$. The momenta of the particles 
in the collinear limit are then given by
\begin{eqnarray}\label{eq:coll-momenta}
p_i^\mu &=& 
x_iEn^{\mu}_-+p_{i,\perp}^\mu
-\frac{p_{i,\perp}^2}{4x_iE} n^\mu_+ \,,
\end{eqnarray}
for $i=1,\ldots,m$, with $0<x_1,\ldots, x_m<1$ 
and $x_1+\ldots+x_m=1$. The label $\perp$ denotes 
the direction perpendicular to both $n_-^{\mu}$ 
and $n_+^{\mu}$ and $p_{i,\perp}^2\leq 0$. 
In this notation, the scalar product between two parton momenta is
\beq
p_i\cdot p_j=\frac12 \left(-p_{i,\perp}^2
\frac{x_j}{x_i}  
-p_{j,\perp}^2 
\frac{x_i}{x_j} 
\right) + p_{i,\perp} \cdot p_{j,\perp}\,,
\eeq
and one may verify that the external partons are all strictly on shell, $p_i^2=0$.
In contrast, the invariant mass of the $m$ collinear particles is timelike. Indeed, using the fact that 
%
\begin{align}
P^\mu = E n^\mu_- +\sum_{i=1}^m\left(p_{i,\perp}^\mu
-\frac{p_{i,\perp}^2}{4x_iE}\right)\,,
\end{align}
we find
\begin{align}
P^2 = \left(\sum_{i=1}^m p_{i,\perp}\right)^2 
- \sum_{i=1}^m\frac{p_{i,\perp}^2}{x_i} 
\,>0\,.
\end{align}
The collinear limit corresponds to taking all 
$p_{i\perp} \to 0$ simultaneously, so $P$ approaches the lightcone. Due to rescaling invariance and Bose 
symmetry, the corrections to the 
dipole formula of the soft anomalous dimension 
can only depend on the kinematic  via CICRs defined in \Eqn{eq:CICR}.
It is known that in the two-particle collinear 
limit the relevant CICRs turn to a constant value. 
However, a clear understanding of the multi-particle 
collinear dynamics and the appropriate degrees of 
freedom are missing in the literature. We fill this gap in this section. In the 
following subsections we first briefly review 
the two-particle collinear limit, and then analyse multi-particle collinear limits.  


\subsubsection{Two-particle collinear limit}
\label{sec:kin-two-part-col}

For two particles in the collinear sector, 
we find two different behaviours 
of the logarithms of the cross ratios that 
are given in \Eqn{eq:CICR-z-zbar}. 
Starting with the definition of a CICR in
\Eqn{eq:CICR} and taking its logarithm, as in
\Eqn{eq:lnCICR}, we can
choose without the loss of generality 
the collinear pair to be partons $i=1$ and $j=2$. 
The logarithm of the first CICR in \Eqn{eq:CICR-z-zbar} is given by 
\begin{align}
\beta_{12kl} =& \ln \rho_{12kl} = 
\ln\frac{(-p_1\cdot p_2)(-p_k\cdot p_l)}{(-p_1\cdot p_k)(-p_2\cdot p_l)}\,,
\end{align}
which turns to $-\infty$ as $p_1$ and $p_2$ 
become collinear. We also remark
that the order of the rest-of-the-process 
partons can be interchanged in the collinear limit. Explicitly,
we use parametrisation in \Eqn{eq:coll-momenta} to show
\begin{align}\label{eq:CICR1-2pc-ij}
\begin{split}
\beta_{12lk} = \ln 
\frac{(-p_1\cdot p_2)(-p_l\cdot p_k)}{(-x_1P\cdot p_l)(-x_2P\cdot p_k)}  
+ \mathcal O(p_\perp)
= \ln \frac{(-p_1\cdot p_2)(-p_k\cdot p_l)}{(-x_1P\cdot p_k)(-x_2P\cdot p_l)} 
+ \mathcal O(p_\perp)
=   \beta_{12kl}\, .
\end{split}
\end{align}
The logarithm 
of the second CICR given in \Eqn{eq:CICR-z-zbar} has the following behaviour in the limit that partons $i=1$ and $j=2$ are collinear
\begin{align}
\begin{split}
\beta_{1kl2} &= 
\ln\frac{(-p_1\cdot p_k)(-p_l\cdot p_2)}{(-p_1\cdot p_l)(-p_k\cdot p_2)} \\ 
&= \ln \frac{(-x_1P\cdot p_k)(-x_2P\cdot p_l)}{(-x_1P\cdot p_l)(-x_2P\cdot p_k)} 
+ \mathcal O(p_\perp)\overset{p_\perp\to 0}{\longrightarrow} 0\,,
\end{split}
\end{align}
where in the second step we made use of 
the parametrisation given above. 
We can also show that for a CICR involving  
only one of the particles from the collinear subset, the  
momentum fraction dependence of the particular
collinear particle in fact cancels out, 
leaving behind information only on the
direction, which is the same as for the parent parton. As an 
example we consider CICR with the collinear particle 
dependence in the first index position
\begin{align}\label{eq:resclinv}
\beta_{1jlk} &= 
\ln \frac{(-x_1 P\cdot p_j)(- p_l\cdot p_k)}
{(-x_1 P\cdot p_l)(-p_j\cdot p_k)}   
+ \mathcal O(p_\perp)
\\ &= \ln \frac{(-  P\cdot p_j)(- p_l\cdot p_k)}
{(-  P\cdot p_l)(-p_j\cdot p_k)}  
+ \mathcal O(p_\perp)
=\beta_{Pjlk} + \mathcal O(p_\perp)  .
\end{align}
The same is true for collinear particles appearing in the 
remaining positions $j,l,$ and $k$. 
Any CICR involving one parton from the collinear subset
along with three rest-of-the-process partons is equal, up to power corrections, to the CICR with the index of the collinear particle replaced by an index corresponding to
the parent parton. 
 

\subsubsection{Three-particle collinear limit}
\label{sec:kin-three-part-col}

\begin{figure}
	\centering
    \includegraphics[width=0.45\linewidth]{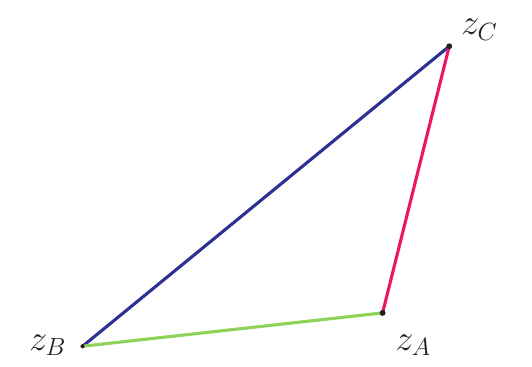}
	\caption{Geometric representation of the conformal cross ratios involving three collinear partons.
    The colouring of the sides of the triangle corresponds
    to the CICRs given in \Eqn{eq:trianglecrossratio}.}
	\label{fig:trianglecrossratio}
\end{figure}

The case of three particles becoming collinear contains
richer dynamics as both the numerator and denominator
of the CICR contain products of collinear momenta.
This fact renders the way in which this limit is
approached relevant. We now choose the three particles 
becoming collinear to be  $1,2$, and $3$. We consider 
first the scalar products between them.
We denote the collinear subset by the letters $(A,B,C)$, 
which can be any of the collinear partons $(1,2,3)$ 
in this limit. We then have
\begin{align}
p_A\cdot p_B  = -\frac{x_Ap_{B,\perp}^2}{2x_B}
-\frac{x_Bp_{A,\perp}^2}{2x_A}
+p_{A,\perp}\cdot p_{B,\perp}\,.
\end{align}
Since the CICRs can connect at most four particle momenta, 
in the three-particle collinear limit it can at most 
contain dependence on one rest-of-the-process particle
which we choose here to be $i$. Explicitly, we then have 
\begin{equation}\label{eq:3collptbeta}
\begin{split}
\rho_{iCAB}&=\frac{(-\frac{x_Ap_{B,\perp}^2}{2x_B}-\frac{x_Bp_{A,\perp}^2}{2x_A}+p_{A,\perp}\cdot p_{B,\perp})( x_CEn\cdot p_i+p_{C,\perp}\cdot p_i -\frac{p_{C,\perp}^2}{4x_CE}\bar n\cdot p_i)}{(-\frac{x_Cp_{B,\perp}^2}{2x_B}-\frac{x_Bp_{C,\perp}^2}{2x_C}+p_{C,\perp}\cdot p_{B,\perp})( x_AEn\cdot p_i+p_{A,\perp}\cdot p_i -\frac{p_{A,\perp}^2}{4x_AE}\bar n\cdot p_i)}\\
&=\frac{(-\frac{p_{B,\perp}^2}{2x_B}-\frac{x_Bp_{A,\perp}^2}{2x_A^2}+\frac{1}{x_A}p_{A,\perp}\cdot p_{B,\perp}) }{(-\frac{p_{B,\perp}^2}{2x_B}-\frac{x_Bp_{C,\perp}^2}{2x_C^2}+\frac{1}{x_C}p_{C,\perp}\cdot p_{B,\perp})} +\mathcal{O}(p_{\perp})\,.
\end{split}
\end{equation}
It is important to note that as the collinear limit
is approached,
the dependence of this CICR on
the rest-of-the-process parton
$i$ is power suppressed.
After dividing both the numerator and denominator by $x_B$,
we can replace the two-dimensional $\perp$ vectors by points in the complex plane. We define 
\begin{align}
z_A= \frac{p_{A,x}+i p_{A,y}}{x_A} \,.
\end{align}
Then the combinations appearing in \Eqn{eq:3collptbeta}
can be written in terms of $z$ variable
and its complex conjugate
\begin{align}
\frac{p_{A,\perp}^2}{x_A^2}= z_A\bar z_A\qquad \text{ and } \qquad\frac{p_{A,\perp}}{x_A}\cdot \frac{p_{B,\perp}}{x_B} = \frac{1}{2}(z_A\bar z_B+\bar z_A z_B)\,.
\end{align}
We therefore define
\begin{equation}
z_{CAB} \equiv 
\frac{\textcolor{darkgreen}{z_B-z_A}}{\textcolor{blue}{z_B-z_C}}\,,\qquad 
z_{BAC}\equiv 
\frac{\textcolor{red}{z_C-z_A}}{\textcolor{blue}{z_C-z_B}}, 
\end{equation}
which satisfy the relation 
\begin{equation}
\label{BAC_CAB}
    z_{BAC} = 1-z_{CAB}\,.
\end{equation}
Then for the CICRs in the three-particle collinear limit we obtain
\begin{align}
\rho_{iCAB}=\left|z_{CAB}\right|^2 = \left|\frac{\textcolor{darkgreen}{z_B-z_A}}{\textcolor{blue}{z_B-z_C}}\right|^2, \qquad\rho_{iBAC}=\left|z_{BAC}\right|^2=\left|\frac{\textcolor{red}{z_C-z_A}}{\textcolor{blue}{z_C-z_B}}\right|^2 ,\label{eq:trianglecrossratio}
\end{align}
where $\rho_{iCAB}$ directly follows from Eq.~(\ref{eq:3collptbeta}) and 
the expression for $\rho_{iBAC}$ can be obtained by swapping of the indices $B\leftrightarrow C$.
From this analysis it is clear that 
the CICRs become squares of ratios of distances in the complex plane as represented in Fig.~\ref{fig:trianglecrossratio} (the colours in Eq.~\eqref{eq:trianglecrossratio} refer to the sides of the triangle in the figure). 
The ratios are invariant under rescaling and under rotations. Given that all three points in the complex plane are distinct, we can choose one of the differences to equal 1, for instance: $z_C-z_B=1$.
The restriction to distinct points in the complex plane corresponds to the momenta not being exactly collinear. 
For any given value of the cross ratio $\rho_{iBAC}$, we know that
\begin{align}(\sqrt{\rho_{iBAC}}-1)^2\leq\rho_{iCAB}\leq (\sqrt{\rho_{iBAC}}+1)^2,\label{eq:crossratiocondition}
\end{align}
where the upper and lower bound correspond the angle at $z_C$ being zero or $\pi$, respectively. Now, the three-particle collinear limit $p_A\parallel p_B\parallel p_C$ can be taken with fixed values for $\rho_{iBAC}$ and $\rho_{iCAB}$. This corresponds to taking the angles of the triangle fixed while scaling it by a factor that approaches zero.
    
We can also approach the limit in a non-simultaneous way,~i.e. by first taking two out of the three of the particles to be collinear to each other followed by the third being collinear to the pair.
Geometrically, this is represented by one of the sides of the triangle becoming arbitrarily short, and then rescaling the triangle by a vanishing factor. In the first step the angles are not fixed.

\subsubsection{Four- and higher-particle collinear limits}
\label{sec:kin-four-part-col}
Since at the three- and four-loop order the relevant kinematic variables 
are the CICRs governing the behaviour of the functions that depend on either
four or five particle indices, see the soft anomalous dimension terms
in Eqs.~\eqref{eq:Q4T-4L}, \eqref{eq:H_5T-4L} and \eqref{eq:H_5T-5L}, there are two
cases to consider for the situation when four or more particles become collinear. 

The first case is when the indices of the CICRs are saturated
by the partons becoming collinear. In this instance, 
we have for example 
\begin{eqnarray}
\rho_{ABCD} = \frac{\left(- \frac{x_A\, p_{B\perp}^2}{2 x_B }  - \frac{x_B\, p_{A\perp}^2}{2 x_A } + p_{A,\perp} \cdot p_{B,\perp}  \right)\left(- \frac{x_C\, p_{D,\perp}^2}{2 x_D }  - \frac{x_D\, p_{C,\perp}^2}{2 x_C } + p_{C,\perp} \cdot p_{D,\perp}  \right) }{\left(- \frac{x_A\, p_{C\perp}^2}{2 x_C }  - \frac{x_C\, p_{A\perp}^2}{2 x_A } + p_{A,\perp} \cdot p_{C,\perp}  \right)\left(- \frac{x_B\, p_{D\perp}^2}{2 x_D }  - \frac{x_D\, p_{B\perp}^2}{2 x_B } + p_{B,\perp} \cdot p_{D,\perp}  \right) },
\end{eqnarray}
and there is
no relative hierarchy between the $p_{i\perp}$ appearing in the 
scalar products making up the CICR. Therefore, within the collinear
subset, there is no simplification and CICRs are not constrained to
take on only a subset of possible values, i.e., they keep the freedom
to exhibit full richness of kinematics within the collinear sector. 

The second case is when the kinematic functions depend on a number 
of CICRs which carry the indices of five particles in total, see
the $\mathcal{H}_2$ terms in \Eqn{eq:H_5T-5L}. In the four-particle
collinear limit, the situation resembles the one described in 
Section~\ref{sec:kin-three-part-col}. Namely, at least three of
the four indices carried by the CICRs in $\mathcal{H}_2$ will
be from the collinear subset and the dependence on the fourth one from the 
rest-of-the-process parton will scale out. Therefore, the
relevant geometric picture for the kinematic variables is the
same as the one described Section~\ref{sec:kin-three-part-col}. For 
these terms, starting from five-particle collinear limit and on, the 
situation is identical to the case of the four-particle
collinear limit of terms which depend only on the degrees of 
freedom of four partons. Namely, the indices will simply be saturated
by the collinear particles and no simplification of the
kinematic behaviour occurs. \\[-0.3cm]

With the geometric insight into the behaviour of the 
kinematic variables in the 
collinear limits obtained in the above sections, we are now ready to 
consider the splitting amplitude soft anomalous 
dimensions. We first carry out the analysis for amplitudes with
massless external partons and then for the case where one of 
the external lines is massive. 

\section{Multi-particle collinear limits: massless case}
\label{sec:massless}

In this section, we calculate the splitting
amplitude soft anomalous dimension 
in multi-particle collinear limits
for the case where all the external
legs of the scattering amplitude are
massless. We only consider the case of timelike splitting, where all of the collinear particles are in the final state.
We start from the parametrisation 
given in \Eqn{eq:adm-param} and
follow the strategy presented in
Section~\ref{sec:factorisationincollinearlimits-sub}. 
Namely, in order to calculate the splitting
amplitude soft anomalous dimension in the
$m$-particle collinear limit, we consider the
difference between the soft anomalous
dimensions of $n$ and $(n-m+1)$
particle amplitudes. The complexity naturally 
rises with the order in perturbation theory 
at which the structures first appear.

We begin with the lowest-order
contributions encapsulated by 
the first term that enters in 
\Eqn{eq:adm-param}, which is the so-called
dipole term~${\bf{ \Gamma}}^{{\rm{dip.}}}_{n}\left(\{s_{ij}\},\lambda,\alpha_s  \right)$.
Since it contains only pairwise
interactions, it is simple to derive the form
of the splitting amplitude soft anomalous dimension
for any number of particles becoming collinear.
We provide the derivation App.~\ref{app:dipoles}.
The main result is the master formula for $m$ particles
becoming collinear. It can be found in \Eqn{eq:GammaSpDipm}, 
and is given here again for convenience of the reader: 
\begin{eqnarray}
\label{GammaSpDipm}
{\bf \Gamma}_{\SP,m}^{\rm dip.} (p_1,\ldots,p_m;\mu_f)
&&  = \frac{1}{2}\gamma_{K}   \sum_{i}^m
\ln \left(x_i\right) \, \T_i \cdot \T_P  
-\frac{1}{2}\gamma_{K}  \sum_{1\leq i< j \leq m}   
\ln \left(\frac{-s_{ij}}{\mu_f^2}\right) \, \T_i \cdot \T_j
\nonumber \\ && \hspace{0.4cm} 
 - \gamma_P(\as) \, 
\,+\, \sum_{i}^m  \gamma_i(\as)  \,,
\end{eqnarray}
where $\T_P = \T_1 + \ldots + \T_m$, $x_i$ is the fraction of the parent
momentum $P$ carried by particle $i$ in the collinear set as defined in Eq.~(\ref{eq:coll-momenta}),
and $\gamma_P(\as)$ is the collinear anomalous dimension for the parent parton. 

In the following subsections, we consider the two-, three-, and four-particle collinear limits, each time discussing the contributing 
terms as the order perturbation theory increases. 

\subsection{Two-particle collinear limit}
\label{sec:two-particle-collinear}

We begin with the two-particle collinear limit, 
which corresponds to the configuration 
presented in Fig.~\ref{m-collinearTL} with $m = 2$. 
The collinear particles are parametrised through
\Eqn{eq:coll-momenta} with $i=1,2$. 
The splitting amplitude soft anomalous dimension
in the two-particle collinear limit due to dipole terms
can easily be obtained
from \Eqn{GammaSpDipm} by setting 
$m=2$. Its explicit form is
reported in \Eqn{GammaSpDip2} for completeness. 

Next, we have the terms that first appear 
at the three-loop order as corrections
to the dipole formula: 
${\bf{ \Gamma}}_{n,4{\rm{T}}-3{\rm{L}}}(\alpha_s)$ and 
${\bf{ \Gamma}}_{n,4{\rm{T}}-4{\rm{L}}}(\{\beta_{ijkl}\},\alpha_s)$. 
The explicit form of these terms is of course
known \cite{Almelid:2015jia}. Moreover, the 
splitting amplitude soft anomalous dimension 
for the two-particle collinear limit has also
been derived. In fact, demanding that 
strict collinear factorisation holds in this 
limit has yielded constraints which were used
along with constraints with the Regge limit 
in Ref.~\cite{Almelid:2017qju} to bootstrap these
terms. The constraints on the kinematic functions
arise because this object is \emph{universal}, i.e.,
it is the same regardless of the total number of 
partons $n$ in the amplitude from which it has 
been obtained. For this 
purpose, it was sufficient 
to determine the two-particle
splitting soft anomalous dimension 
entirely from the case $n= 3$, 
and to analyse the constraints 
arising from considering the case $n = 4$
as was done in Ref.~\cite{Almelid:2017qju}.  
Using the notation in \Eqn{eq:GammaSpQuadm}
with $m=2$, the result is  
\be \label{GammaSpQuad2dN}
{\bf{ \Gamma}}_{\SP, 2}^{4{\rm{T}}}(p_1,p_2;\mu_f)
= -\frac{3}{4} f(\alpha_s) \Big( 
 C_A^2 \T_1\cdot \T_2 + 8 { \cal T }_{1122}
 \Big),
\ee
where  
\begin{align}\label{eq:calD}
\mathcal{T}_{1122}  = \frac14 f^{ade}f^{bce}
\left(\T^a_1 \T^b_1 + \T^b_1 \T^a_1\right)
\left(\T^c_2 \T^d_2 + \T^d_2 \T^c_2\right).
\end{align}
The colour factor in Eq.~(\ref{eq:calD}) is given in \Eqn{Tauijkl}, specialised to two partons.
The corresponding constraints on the anomalous dimension functions in \Eqn{quadrupole-AS-defsh} are given by 
\be\label{eq:twocoll-constraint1c-main}
 \Big( 4{\cal F}^{\rm S}_{12kl}(\{\beta_{12kl}\}) -2f(\alpha_s) \Big)\bigg|_{p_1|| p_2} = 2f(\alpha_s) \,,
\ee
and 
\be \label{eq:twocoll-constraint2Hybrid-main}
{\cal F}^{\rm A}_{12kl}(\{\beta_{12kl}\}) \Big|_{p_1|| p_2}  =0\,,
\ee
where $1,2$ are indices of collinear particles and 
$k,l$ correspond to rest-of-the-process partons, 
i.e., $k=3$ and $l=4$ in Ref.~\cite{Almelid:2017qju}. 
In App.~\ref{app:2partcol}, we extend 
the analysis presented in Ref.~\cite{Almelid:2017qju}
by showing that \Eqn{GammaSpQuad2dN} can indeed be  
obtained from generic $n$-point amplitudes,
and no additional constraints can arise just 
from considering two-particle collinear limits 
between higher-point amplitudes, 
such as $ {\bf{ \Gamma}}_{5}- {\bf{ \Gamma}}_{4}$, $ {\bf{ \Gamma}}_{6}- {\bf{ \Gamma}}_{5}$,  and so on.
This gives us the opportunity to show 
that consistency with the collinear limit requires a non-trivial 
interplay between the kinematic-dependent term 
${\bf{ \Gamma}}_{n,4{\rm{T}}-4{\rm{L}}}
(\{\beta_{ijkl}\},\alpha_s)$
and the kinematic-independent 
one ${\bf{ \Gamma}}_{n,4{\rm{T}}-3{\rm{L}}}(\alpha_s)$,
which in turn can be traced back to 
an interesting interplay between kinematics and the colour algebra, which is ultimately a consequence of gauge invariance. The way this is realised is rather remarkable, and we refer the reader to App.~\ref{app:2partcol} to appreciate the details.
The pinnacle of this derivation is that a kinematic relation in Eq.~(\ref{eq:twocoll-constraint1}) must hold, for a stuffle identity to be formed relating the colour structures in Eq.~(\ref{eq:stuffle-2.p.c.c}), which then enables the use of colour conservation to show that ${\bf{ \Gamma}}_{\SP, 2}^{4{\rm{T}}}$ derived from an $n$-point amplitude is indeed universal. Here we note that a stuffle identity is an identity allowing one to write a product of two sums in terms of nested sums,
\begin{equation}
\sum_{1\le i\le r}a_i\,\sum_{1\le j\le r}b_j = \sum_{1\le i<j\le r}a_ib_j+\sum_{1\le j<i\le r}b_ja_i + \sum_{1\le i\le r}a_ib_i\,.
\end{equation}

While the function ${\cal F}$ and the constant $f(\alpha_s)$ are only known at three loops, the constraints on these components of the soft anomalous dimension in 
\Eqns{eq:twocoll-constraint1c-main}{eq:twocoll-constraint2Hybrid-main}
are valid to higher orders~\cite{Becher:2019avh}, and their potential role in bootstrapping the four-loop contributions, alongside the Regge limit, has been discussed in~Refs.~\cite{Falcioni:2021buo,Maher:2023jqy}.

Two-particle collinear limits 
were also calculated for the terms 
in the soft anomalous dimension associated with the quartic Casimir, Eqs.~(\ref{eq:Q4T-23L}) and (\ref{eq:Q4T-4L}), which
start to contribute at the four-loop order~\cite{Becher:2019avh}.
Demanding that strict collinear factorisation holds
leads to constraints on the form of the kinematic 
functions ${\cal G}_R$ appearing at four loops.  
We perform the derivation starting from 
$n$-point amplitudes using our method in 
App.~\ref{app:2partcol-fourloop}. 
Starting with the quartic terms,
and using the 
definitions for the 
splitting amplitude soft anomalous 
dimensions in \Eqn{eq:GammaSpQ4Tm} 
with $m=2$, in the end one finds that (see \Eqn{eq:6.78eg})
\begin{eqnarray}\label{eq:6.78eg-main}
{\bf{ \Gamma}}_{\SP, 2}^{{\rm Q}4{\rm{T}}}(p_1, p_2;\mu_f)
&=& \sum_R g_R \Big[\,
 \Big({\bf{\cal{D}}}^R_{1111}
+ 2 { \bf{\cal{D}}}^R_{1112}  
+ 3 { \bf{\cal{D}}}^R_{1122} 
+ 2 { \bf{\cal{ D}}}^R_{1222}\Big)\ln({x_1}) 
\nonumber \\[-0.1cm] 
&& \hspace{1.1cm}
+ \Big({ \bf{\cal{ D}}}^R_{2222}
+ 2 { \bf{\cal{ D}}}^R_{2111}        
+ 3 { \bf{\cal{ D}}}^R_{2211}  
+ 2 { \bf{\cal{ D}}}^R_{2221}\Big)\ln(x_2)   
\nonumber \\[0.1cm]  
&& \hspace{1.1cm}
-\Big(   2{ \bf{\cal{ D}}}^R_{1112}
+  3{ \bf{\cal{ D}}}^R_{1122}  
+  2{ \bf{\cal{ D}}}^R_{2221}\Big)\ell_{12}   
+2 \sum_{\substack{ 3\leq   k < l \leq n  } }  
{\cal{D}}_{12kl}^R   \beta_{12kl}  \, \Big] 
\nonumber       \\     &&  \hspace{0.7cm}
+\,24\, \sum_{R} \sum_{3\leq k<l \leq n}    
\DR_{12kl} 
\,{\cal G}_R  (\beta_{12kl},\beta_{1lk2}).
\end{eqnarray}
From this expression we read off the constraints
that can be formed on the unknown function ${\cal G}_R$
in terms of $g_R$ by demanding that the splitting amplitude
soft anomalous dimension does not depend on the
rest of the process,
\begin{eqnarray}\label{eq:GRconstraint}
     \lim_{\beta_{12kl}\to-\infty}   \,{\cal G}_R  (\beta_{12kl},0)       = -\frac{ g_R}{12}       \beta_{12kl} \,.
\end{eqnarray}
This reproduces the result obtained in Ref.~\cite{Becher:2019avh},
accounting for the factor of $1/2$ in the definition of $g_R$ as 
noted above \Eqn{eq:Q4T-23L}. Implementing the constraint in \Eqn{eq:GRconstraint}
leads to the final expression for ${\bf{ \Gamma}}_{\SP, 2}^{{\rm Q}4{\rm{T}}}$
which takes the form
\begin{eqnarray}
\label{eq:GammaSpQ4T2Res}
{\bf{ \Gamma}}_{\SP, 2}^{{\rm Q}4{\rm{T}}}(p_1, p_2;\mu_f) 
&=& 
\sum_R g_R\, \Big[\,
 \Big({\bf{\cal{D}}}^R_{1111}
+ 2 { \bf{\cal{D}}}^R_{1112}  
+ 3 { \bf{\cal{D}}}^R_{1122} 
+ 2 { \bf{\cal{ D}}}^R_{1222}\Big)\ln({x_1}) \nonumber \\[-0.2cm] && \hspace{1.2cm}
+ \Big({ \bf{\cal{ D}}}^R_{2222}
+ 2 { \bf{\cal{ D}}}^R_{2111}        
+ 3 { \bf{\cal{ D}}}^R_{2211}  
+ 2 { \bf{\cal{ D}}}^R_{2221}
\Big)\ln(x_2)  
\nonumber \\ &&\hspace{1.2cm}
-\Big(2{ \bf{\cal{ D}}}^R_{1112} 
+  3{ \bf{\cal{ D}}}^R_{1122}  
+  2{ \bf{\cal{ D}}}^R_{2221}\Big)\ell_{12} 
\,\Big] \,.
\end{eqnarray}

Next, in App.~\ref{sec:2-coll_Quintic_terms} we have performed the 
derivation of the contribution to the two-particle collinear splitting amplitude
soft anomalous dimension due to terms of the soft anomalous dimension 
that depend on five colour generators and unknown kinematic functions
$\mathcal{H}_1$ and $\mathcal{H}_2$ parametrised in Eqs.~\eqref{eq:H_5T-4L} and
\eqref{eq:H_5T-5L}. 
We recall that these terms are thought to vanish following the argument of Ref.~\cite{Vladimirov:2017ksc}; nonetheless, it is interesting to examine how constrained they are by collinear factorisation.

The relevant object is ${\bf{ \Gamma}}_{\SP, 2}^{5{\rm{T}}}$
as defined in \Eqn{eq:GammaSp5Tm} with $m=2$, for which we obtain 
(see \Eqn{eq:5.212z} in App.~\ref{sec:2-coll_Quintic_terms} and the derivation in 
preceding equations)
\begin{eqnarray}\label{eq:5.212z-main}
&& {\bf{ \Gamma}}_{\SP, 2}^{5{\rm{T}}}(p_1, p_2;\mu_f)= 
2  \sum_{k=3} \sum_{\substack{l=3 \\ l\neq k}}  \,
\Big[ \,\mathcal{T}_{12kl1} -\,  \mathcal{T}_{12klk}+\,\mathcal{T}_{21kl2} -\mathcal{T}_{21klk}\Big]
\mathcal{H}_1(\beta_{12kl},0  )
\\\nonumber && \hspace{0.6cm} 
+\sum_{ (k,l,m) \neq 1,2}
\Big[ \mathcal{T}_{12klm}- \mathcal{T}_{12lkm}\Big] 
\Big(   K_+(\beta_{1kml},\beta_{1lmk}, \beta_{12kl}  ) +  K_-(\beta_{1kml},\beta_{1lmk}, \beta_{12kl}  )  \Big), 
\end{eqnarray}
where 
\begin{eqnarray}\label{eq:Kplusminus-main}
K_{\pm}(\beta_{1kml},\beta_{1lmk}, \beta_{12kl})&\equiv &\Big(
4\mathcal{H}_2(- {\beta_{1kml}},- \beta_{12kl} - \beta_{1lmk} ,- \beta_{12kl} - \beta_{1lmk} ,0,
\beta_{12kl} + \beta_{1kml}  )
\nonumber \\ &&  \hspace{-3.9cm}
- \mathcal{H}_1( - \beta_{1lmk}, \beta_{1kml} -  \beta_{1lmk}  )
+2\mathcal{H}_2(\beta_{12kl}, \beta_{12kl} + \beta_{1lmk} ,0,\beta_{12kl} + \beta_{1kml},0  ) \Big)
\nonumber \\ &&  \hspace{-3.9cm}     
\pm \Big(  4\mathcal{H}_2(- {\beta_{1lmk}},- \beta_{12kl} - \beta_{1kml} ,
- \beta_{12kl} - \beta_{1kml} ,0,\beta_{12kl} + \beta_{1lmk}  )
\nonumber \\ &&   \hspace{-3.9cm}
- \mathcal{H}_1( - \beta_{1kml}, \beta_{1lmk} -  \beta_{1kml}  )
+2\mathcal{H}_2(\beta_{12kl}, \beta_{12kl} + \beta_{1kml} ,0,\beta_{12kl} + \beta_{1lmk},0  ) \Big).
\end{eqnarray}
The functions $K_{+}(\beta_{1kml},\beta_{1lmk}, \beta_{12kl})$
and $K_{-}(\beta_{1kml},\beta_{1lmk}, \beta_{12kl})$ are manifestly 
symmetric and  antisymmetric under interchange of their first two argument,
or equivalently interchange of indices $k\leftrightarrow l$, respectively.
The contribution due to $K_{+}$ vanishes in \Eqn{eq:5.212z-main} since it is 
multiplying a manifestly antisymmetric colour structure under the 
interchange of indices  $k\leftrightarrow l$. 

The first constraint originates from the top line of \Eqn{eq:5.212z-main}. The constraint is that
\begin{eqnarray}\label{eq:H1constraint}
    \lim_{\beta_{12kl}\to-\infty} \mathcal{H}_1(\beta_{12kl},0)=0.
\end{eqnarray}
This condition is stronger than simply independence of $\mathcal{H}_1(\beta_{12kl},0)$
in the collinear limit of the kinematics of the rest of the process. It is necessary, because if the kinematic function did not vanish, but merely
became independent of rest-of-the-process momenta, it would not have been possible to bring
the left-over colour terms to depend only on the collinear partons. 
The conclusion is therefore that the $ \mathcal{H}_1(\beta_{12kl},0)$ function must vanish in the collinear limit.  
The second constraint is that
\begin{eqnarray}\label{eq:Kminus-constraint}
K_{-}(\beta_{1kml},\beta_{1lmk}, \beta_{12kl})&\stackrel{p_1 \parallel p_2}{\longrightarrow}& 0
\end{eqnarray}
in the collinear limit. This constraint simply states that the
combination of ${\cal{H}}_1$ and ${\cal{H}}_2$ functions in top
two lines of \Eqn{eq:Kplusminus-main} must be antisymmetric under
the exchange of $k \leftrightarrow l$. Both of these constraints
have previously been reported in Ref.~\cite{Becher:2019avh}.
The result of this considerations is that in fact 
the contribution to the two-particle collinear splitting amplitude 
soft anomalous dimension due to the terms with 5-index structure vanishes
\begin{eqnarray}
\label{eq:GammaSp5T2Res}
{\bf{ \Gamma}}_{\SP, 2}^{5{\rm{T}}}(p_1, p_2;\mu_f) &=& 0\,.
\end{eqnarray}
This is of course consistent with the conclusion of
Ref.~\cite{Vladimirov:2017ksc} that ${\cal H}_i$ vanish identically, but by itself it does not imply that. This concludes our discussion of the two-particle splitting amplitude soft anomalous dimensions.

It is apparent that strict factorisation of amplitudes in two-particle
collinear limits have proven useful in providing
data for constraining thus-far-unknown functions
governing the structure of IR singularities in 
$n$-point amplitudes. In what follows, 
we study the extent to which multi-particle
collinear limits can provide additional information. 
We found that the universality of soft singularities of collinear splitting amplitudes is realised through remarkably intricate cancellations of the dependence of the rest of the process, involving an interplay of the kinematics and the colour space. We will now explore how this generalises for multi-collinear limits. In this process we will also determine the structure of the soft anomalous dimensions of the corresponding multi-parton splitting amplitudes.

\subsection{Three-particle collinear limit}
\label{sec:three-particle-collinear}

Next, we consider the three-particle collinear limit, 
which corresponds to the configuration 
drawn in Fig.~\ref{m-collinearTL} with $m = 3$. 
The collinear particles are parametrised through
\Eqn{eq:coll-momenta} with $i=1,2,3$. 
The contribution due to the dipole terms in Eq.~(\ref{eq:dipole})
to the splitting amplitude soft anomalous dimension can be obtained from \Eqn{GammaSpDipm} by setting $m=3$. The structure of the dipole terms is simple and naturally satisfies factorisation properties of the amplitude in collinear limits upon using colour conservation. The derivation is presented in App.~\ref{app:dipoles}, see in particular the explicit result for ${\bf \Gamma}_{\SP,3}^{\rm dip.}$ in \Eqn{GammaSpDip3}. 

\subsubsection{Terms starting at three loops}
\label{sec:three-particle-collinear-threeloop}

The first non-trivial considerations pertain to 
the contributions to the soft anomalous 
dimension which begin at three loops and
are captured by ${\bf{ \Gamma}}_{\SP, m}^{4{\rm{T}}}(p_1,\ldots p_m;\mu_f)$
in~\Eqn{eq:GammaSpQuadm} with $m=3$. 
Concretely, the
contribution to the splitting amplitude 
soft anomalous dimension from the terms starting at three-loops in case of the triple-collinear  limit is captured by
\begin{eqnarray}
\label{eq:GammaSpQuad3} 
{\bf{ \Gamma}}_{\SP, 3}^{4{\rm{T}}}(p_1,p_2, p_3;\mu_f) &=& 
   {\bf{ \Gamma}}_{n,4{\rm{T}}-3{\rm{L}}}(\alpha_s)
     + {\bf{ \Gamma}}_{n,4{\rm{T}}-4{\rm{L}}}(\{\beta_{ijkl}\},\alpha_s)
\nonumber \\  && \hspace{-2.3cm} 
-  \Big(  {\bf{ \Gamma}}_{n-2,4{\rm{T}}-3{\rm{L}}}(\alpha_s)
     + {\bf{ \Gamma}}_{n-2,4{\rm{T}}-4{\rm{L}}}(\{\beta_{ijkl}\},\alpha_s)\Big) 
    \Big|_{\T_P\to \sum_{i=1}^3 \T_i}.
\end{eqnarray}
This quantity must depend only
on the momenta and colour of the three particles becoming 
collinear, and  
should be, in particular, independent of the index $n$ appearing on the right-hand side. This means that in order to  calculate its form we can choose a 
specific value for $n$.
In case of ${\bf{ \Gamma}}_{\SP, 3}^{4{\rm{T}}}$
the simplest non-trivial case to consider is $n =4$, 
for which only the terms in the first line in Eq.~(\ref{eq:GammaSpQuad3}) contribute: 
\begin{eqnarray}
    \label{eq:GammaSpQuad3b} 
{\bf{ \Gamma}}_{\SP, 3}^{4{\rm{T}}}(p_1,p_2, p_3;\mu_f) &=& 
   {\bf{ \Gamma}}_{4,4{\rm{T}}-3{\rm{L}}}(\alpha_s)
     + {\bf{ \Gamma}}_{4,4{\rm{T}}-4{\rm{L}}}(\{\beta_{ijkl}\},\alpha_s)\,,
\end{eqnarray}
given that ${\bf{ \Gamma}}_{2,4{\rm{T}}-3{\rm{L}}}(\alpha_s)= 0$ and $ {\bf{ \Gamma}}_{2,4{\rm{T}}-4{\rm{L}}}(\{\beta\},\alpha_s) = 0$, since it is not possible to construct these structures with only two independent particles.  
In this case,
we need to evaluate
\Eqn{eq:GammaSpQuad3b} subject to colour conservation, 
i.e., $\T_1 + \T_2 + \T_3 + \T_4 = 0$. 
Starting from the parametrisation of these terms in 
Eqs.~\eqref{eq:Gammaf} and \eqref{eq:quardupoleABdef},
and applying colour conservation along with the
identities in App.~\ref{colourID},  we find that
${\bf{ \Gamma}}_{\SP, 3}^{4{\rm{T}}}$ 
can be written as
\bea \label{GammaSpQuad3d}  \nn
{\bf{ \Gamma}}_{\SP, 3}^{4{\rm{T}}}(p_1,p_2, p_3,\mu_f) &=&
- \frac{3 }{4}f(\alpha_s) \Big[ C_A^2 \Big( 
\T_1\cdot \T_2 +\T_1\cdot \T_3
+\T_2\cdot \T_3  \Big)
\nn \\& & \hspace{-.2cm}
+\,8\Big( \mathcal{T}_{1122}
+\mathcal{T}_{1133}
+\mathcal{T}_{2233} \Big) 
+\,\frac83  \Big(  \mathcal{T}_{1123} 
+ \mathcal{T}_{2213} + \mathcal{T}_{3312} \Big)\Big]
\nn \\ && \hspace{-.2cm}
+\,8\,\Big( \mathcal{T}_{1123}  
-\mathcal{T}_{2213} \Big) 
\,{\cal F}  (\beta_{1324},\beta_{1423})
 +\,8\Big(\mathcal{T}_{1123} 
-\mathcal{T}_{3312} \Big)
\,{\cal F}  (\beta_{1234},\beta_{1432}) 
\nn \\ && \hspace{-.2cm}
+\,8\Big(\mathcal{T}_{2213}  - \mathcal{T}_{3312} \Big)
\,{\cal F} (\beta_{1243},\beta_{1342})\,. 
\eea
As can be seen in the last three terms of \Eqn{GammaSpQuad3d}, after application of colour conservation the colour operators multiplying the 
momentum dependent functions ${\cal F}(\{\beta\})$
can be expressed in terms of the same colour operators ${\cal T }_{iijk}$ associated with the constant term $f(\alpha_s)$. 

It is evident from \Eqn{GammaSpQuad3d} that the colour factors 
satisfy the requirement that the splitting amplitude soft anomalous 
dimension must depend only on the degrees of freedom of the collinear particles
1, 2 and 3. However, at first sight the momentum-dependent functions 
${\cal F}(\{\beta\})$ still contain information about the 4$^{{\rm{th}}}$
particle, which is  not becoming collinear to the others. If this were the case,
consistency with strict collinear factorisation would require that in the 
three-particle collinear limit
the function ${\cal F}(\{\beta\}) \stackrel{!}{=}0$. Such a condition would
constitute an additional constraint on the analytic structure of infrared 
divergences, in addition 
to the constraints already obtained from the two-particle collinear limit, 
and given in Eqs.~\eqref{eq:twocoll-constraint1c-main} and~\eqref{eq:twocoll-constraint2Hybrid-main}.
However, since three-loop results for ${\cal F}(\{\beta\}) $ are known, it can be checked
that in general these functions do not vanish in the three-particle collinear limit. Rather, 
consistency with strict collinear factorisation is guaranteed for a different reason: 
inspecting the kinematic dependence of the functions 
${\cal F}(\{\beta\})$ 
in the collinear limit, as we have
carefully done 
in Section~\ref{sec:kin-three-part-col}, 
one realises that
the dependence on the 
rest-of-the-process parton, here with momentum~$p_4$,
cancels in the CICRs which capture the kinematics. More precisely, we see from Eq.~\eqref{eq:trianglecrossratio} that ${\cal F}(\{\beta\})$ is independent of $p_4$, and in fact only depends on the complex variables $z_{ABC}$ with $\{A,B,C\}=\{1,2,3\}$. At this point we make the following observation. From Eqs.~\eqref{eq:calFdef} and ~\eqref{eq:F_SVHPL}, we know that (at least) at three loops, in general kinematics the function ${\cal F}^{(3)}(\{\beta\})$ can be expressed in terms of SVHPLs depending on the complex variables $z_{ijkl}$ defined in Eq.~\eqref{eq:CICR-z-zbar}. We can now give another interpretation of the complex variables $z_{ABC}$ from Eq.~\eqref{eq:trianglecrossratio}: they can be identified with the collinear limit of the complex cross ratios $z_{iABC}$. We can make this explicit at three loops by inserting Eq.~\eqref{eq:calFdef} into Eq.~\eqref{GammaSpQuad3d},
\bea \label{GammaSpQuad3d_2}  \nn
{\bf{ \Gamma}}_{\SP, 3}^{4{\rm{T}}(3)}(p_1,p_2, p_3,\mu_f) &=&
- \frac{3 }{4}f^{(3)} \Big[ C_A^2 \Big( 
\T_1\cdot \T_2 +\T_1\cdot \T_3
+\T_2\cdot \T_3  \Big)
\nn \\& & \hspace{-2cm}
+\,8\Big( \mathcal{T}_{1122}
+\mathcal{T}_{1133}
+\mathcal{T}_{2233} \Big) 
+\,\frac83  \Big(  \mathcal{T}_{1123} 
+ \mathcal{T}_{2213} + \mathcal{T}_{3312} \Big)\Big]
\nn \\ && \hspace{-2cm}
+\,\frac{1}{4}\, \mathcal{T}_{1123}   
\,\Big(F(z_{132}) - F(z_{231})+F(z_{123}) - F(z_{321})\Big)
\nn \\ && \hspace{-2cm}
 +\,\frac{1}{4}\mathcal{T}_{2213} 
\,\Big(F(z_{213}) - F(z_{312})+F(z_{231}) - F(z_{132})\Big)
\nn \\ && \hspace{-2cm}
+\,\frac{1}{4}\mathcal{T}_{3312}
\,\Big(F(z_{312}) - F(z_{213})+F(z_{321}) - F(z_{123})\Big)\,, 
\eea
where we used the relation in Eq.~(\ref{BAC_CAB}).
This representation makes manifest the fact that the splitting function anomalous dimension depends only on the momenta of the three collinear particles. 


While it is easiest to obtain the form of the three-particle splitting 
amplitude soft anomalous dimension from the specific $n = 4$ case, 
as in \Eqn{eq:GammaSpQuad3b}, this result
is still based on the 
assumption that ${\bf{ \Gamma}}_{\SP, 3}^{4{\rm{T}}}$, as
defined in \Eqn{eq:GammaSpQuad3}, is universal, and in particular independent of $n$. This remains to be verified. 
In order to prove that this is indeed the case, we are 
required to calculate the right-hand side of 
\Eqn{eq:GammaSpQuad3} for generic $n$.
This fully generic calculation is enabled by
the approach we already employed
to evaluate the 
${\bf{ \Gamma}}_{\SP, 2}^{4{\rm{T}}}$
for any $n$, as detailed in App.~\ref{app:2partcol}.
Since the evaluation of multi-particle collinear limits
is more involved and the fact that we will subsequently extend these considerations to higher loops and include massive final
state particles (see Section~\ref{sec:massive}),
we find it instructive 
to present the calculation in detail. 
   
We start by considering the momentum-dependent 
part of 
${\bf{ \Gamma}}_{\SP, 3}^{4{\rm{T}}}$, 
which is given
by the difference ${\bf{ \Gamma}}_{n,4{\rm{T}}-4{\rm{L}}}(\{\beta\},\alpha_s) - {\bf{ \Gamma}}_{n-2,4{\rm{T}}-4{\rm{L}}}(\{\beta\},\alpha_s)$.
We proceed by
singling out the contribution of 
the collinear particles 1, 2 and 
3 and splitting the sum in 
\Eqn{eq:quardupoleABdef} for the amplitudes with $n$ and $n-2$ external partons as follows:
\begin{subequations} \label{Aterms-3part}
\bea \label{An-3part}   \nn
{\bf{ \Gamma}}_{n,4{\rm{T}}-4{\rm{L}}}(\{\beta\})
&=& 4 \bigg[
\textcolor{orange}{
\sum_{4\leq i<j<k<l \leq n} \, {\bf a}_{ijkl}^{ }(\{\beta\})}
+\textcolor{red}{\sum_{4\leq j<k<l \leq n} \, {\bf a}_{1jkl}^{ }(\{\beta\}) }
\\ 
&&\hspace{1.0cm}
+\,\textcolor{red}{\,\sum_{4\leq j<k<l \leq n} \, {\bf a}_{2jkl}^{ }(\{\beta\})}
+\,\textcolor{red}{\sum_{4\leq j<k<l \leq n} \, {\bf a}_{3jkl}^{ }(\{\beta\})} \nn \\ 
&&\hspace{1.0cm}
+\,\sum_{4\leq k<l \leq n} \, {\bf a}_{12kl}^{ }(\{\beta\})+\sum_{4\leq k<l \leq n} \, {\bf a}_{13kl}^{ }(\{\beta\}) \nn\\ 
&&\hspace{1.0cm}
+\sum_{4\leq k<l \leq n} \, {\bf a}_{23kl}^{ }(\{\beta\})
+
\textcolor{darkgreen}{
\sum_{4\leq  l \leq n} \, {\bf a}_{123l}^{ }(\{\beta\})}
\bigg], \\[0.1cm] \label{Anm1-3part}
{\bf{ \Gamma}}_{n-2,4{\rm{T}}-4{\rm{L}}}(\{\beta\})
&=& 4  \bigg[\textcolor{orange}{
\sum_{4\leq i<j<k<l \leq n} \, {\bf a}_{ijkl}^{ }(\{\beta\})}
+\textcolor{red}{\sum_{4\leq j<k<l \leq n} \, {\bf a}_{Pjkl}^{ }(\{\beta\})} \bigg].
\eea
\end{subequations} 
Taking the difference, it is clear that the terms which do not 
involve any of the collinear or parent partons cancel each other
(\textcolor{orange}{orange} terms). Using $\T_P=\T_1+\T_2+\T_3$
we also see that the terms involving one collinear particle 
cancel with the second term in ${\bf{ \Gamma}}_{n-2,4{\rm{T}}-4{\rm{L}}}(\{\beta\})$ (\textcolor{red}{red} terms).  
What remains are terms involving two collinear particles along with two from the rest of the process (black terms), which are similar to those appearing in the two-particle collinear limits,
and the last term in ${\bf{ \Gamma}}_{n,4{\rm{T}}-4{\rm{L}}}(\{\beta\})$ (\textcolor{darkgreen}{green}) which genuinely captures the three-particle collinear limit kinematics. We note that the dependence on the rest-of-the-process momenta cancels in the last term (\textcolor{darkgreen}{green}), and this contribution is a function only of the complex variables $z_{ABC}$ with $\{A,B,C\}=\{1,2,3\}$.

We now turn our attention to the constant terms in ${\bf{ \Gamma}}_{\SP, 3}^{4{\rm{T}}}$. Here again we use the method of splitting the sums contained within $ {\bf{ \Gamma}}_{n,4{\rm{T}}-3{\rm{L}}} $ and $ {\bf{ \Gamma}}_{n-2,4{\rm{T}}-3{\rm{L}}} $ (\emph{cf.}~\Eqn{BtermsHybrid}):
\begin{subequations}\label{eq:Bterms-3part}
\bea\nn
{\bf{ \Gamma}}_{n,4{\rm{T}}-3{\rm{L}}}(\alpha_s) 
&=& 2  f(\alpha_s)  
\Bigg\{ \textcolor{orange}{
\sum_{i = 4}^{n}\sum_{\substack{4\leq j<k \leq n \\ j,k\neq i}}
\, {\cal T}_{iijk}}
+\textcolor{blue}{
\sum_{ 4\leq j<k \leq n  } \Big(
{\cal T}_{11jk} +{\cal T}_{22jk} +{\cal T}_{33jk}  \Big)}
\\[-0.1cm]  &&\hspace{-0.5cm}
+\,\textcolor{red}{\sum_{i = 4}^{n}
\sum_{\substack{k = 4, \\ k\neq i}}^{n} 
\Big({\cal T}_{ii1k} +{\cal T}_{ii2k} +{\cal T}_{ii3k} \Big)}
+ \sum_{i = 4}^{n} \Big({\cal T}_{112i} + {\cal T}_{221i} 
+ {{\cal T}_{ii12}} + {\cal T}_{113i}
\nn \\[-0.1cm] &&\hspace{-0.5cm} 
+ {\cal T}_{331i} + {{\cal T}_{ii13}}  + {\cal T}_{223i}
+ {\cal T}_{332i} + {{\cal T}_{ii23}}    \Big) 
+ \textcolor{darkgreen}{ {\cal T}_{1123}
+ {\cal T}_{2213} + {\cal T}_{3312} }
\Bigg\}, \label{eq:Bterms-3partA}
\\[0.1cm] \label{eq:Bterms-3partB}
{\bf{ \Gamma}}_{n-2,4{\rm{T}}-3{\rm{L}}}(\alpha_s) &=&
2 f(\alpha_s)\Bigg\{\textcolor{orange}{\sum_{i = 4}^{n}
\sum_{\substack{4\leq j<k \leq n \\ j,k\neq i}} \,
{\cal T}_{iijk}}
+\textcolor{blue}{
\sum_{\substack{4\leq j<k \leq n }} {\cal T}_{PPjk}}
+\textcolor{red}{\sum_{i = 4}^{n} 
\sum_{\substack{k = 4, \\ k\neq i}}^{n} 
{\cal T}_{iiPk}}\Bigg\} .
\eea
\end{subequations}
As for the kinematic-dependent terms discussed above, the black terms in Eq.~(\ref{eq:Bterms-3partA}) contain information 
about pairs of particles out of the three becoming collinear, 
and the \textcolor{darkgreen}{green} terms contain information 
about all three of the collinear particles. 
When the difference between Eqs.~(\ref{eq:Bterms-3partA}) and~(\ref{eq:Bterms-3partB}) is taken, the \textcolor{orange}{orange} and \textcolor{red}{red} contributions cancel each other exactly. Whereas the \textcolor{blue}{blue} contributions leave some terms behind. This is because upon 
substituting for the total charge of the parent parton $\T_P=\T_1+\T_2+\T_3$ in Eq.~(\ref{eq:Bterms-3partB}) we obtain 
\begin{eqnarray}\label{eq:teal-threepartcol}
{\cal T}_{PPjk} &=&   {\cal T}_{11jk} 
+{\cal T}_{22jk}+{\cal T}_{33jk} 
+{\cal T}_{12jk}+{\cal T}_{12kj}
+{\cal T}_{13jk}+{\cal T}_{13kj}
+{\cal T}_{23jk}+{\cal T}_{23kj}\,,
\end{eqnarray}
which includes mixed terms involving two of the collinear particles, which are absent in Eq.~(\ref{eq:Bterms-3partA}). 
Bringing together the terms from Eqs.~\eqref{Aterms-3part} and \eqref{eq:Bterms-3part}
according to \Eqn{eq:GammaSpQuad3} gives the following expression 
\bea\label{GammaSpQuad3eHybrid} \nn 
{\bf{ \Gamma}}_{\SP, 3}^{4{\rm{T}}}(p_1,p_2, p_3)
&=&   \sum_{4\leq k<l \leq n} \bigg[
\, 4 {\cal F}^{\rm A}_{12kl}(\{\beta\}) \,{\cal T}_{1kl2} +
\Big( 4 {\cal F}^{\rm S}_{12kl}(\{\beta\}) 
\textcolor{blue}{ -2f(\alpha_s)} \Big) \Big( {\cal T}_{12lk}
+ {\cal T}_{12kl} \Big)  \bigg]
\nn \\ && \hspace{-0.35cm} 
+\sum_{4\leq k<l \leq n} \bigg[
\, 4 {\cal F}^{\rm A}_{13kl}(\{\beta\}) \,{\cal T}_{1kl3} +
\Big( 4 {\cal F}^{\rm S}_{13kl}(\{\beta\}) 
\textcolor{blue}{ -2f(\alpha_s)} \Big) \Big( {\cal T}_{13lk}
+ {\cal T}_{13kl} \Big)  \bigg]
\nn \\ && \hspace{-0.35cm} 
+ \sum_{4\leq k<l \leq n} \bigg[
\, 4 {\cal F}^{\rm A}_{23kl}(\{\beta\}) \,{\cal T}_{2kl3} +
\Big( 4 {\cal F}^{\rm S}_{23kl}(\{\beta\}) 
\textcolor{blue}{ -2f(\alpha_s)} \Big) \Big( {\cal T}_{23lk}
+ {\cal T}_{23kl} \Big)  \bigg]
\nn \\  &&\hspace{-0.35cm}
+ 2f(\alpha_s) \bigg[ 
\sum_{i = 4}^{n} \Big( 
{\cal T}_{112i} + {\cal T}_{221i} + {{\cal T}_{ii12}} 
+ {\cal T}_{113i}+ {\cal T}_{331i} + {{\cal T}_{ii13}}
+ {\cal T}_{223i} + {\cal T}_{332i} 
\nn \\[-0.1cm] &&\hspace{1.5cm}  
+ {{\cal T}_{ii23}}   \Big) 
+ \textcolor{darkgreen}{ {\cal T}_{1123}}
+ \textcolor{darkgreen}{{\cal T}_{2213}} 
+\textcolor{darkgreen}{ {\cal T}_{3312}} \bigg]
+4 \textcolor{darkgreen}{
\sum_{4\leq  l \leq n} \, {\bf a}_{123l}^{ }(\{\beta\})}\,,
\eea
where we see the emergence of familiar structures, 
ones that we have already encountered in the analysis 
of two-particle collinear limit in~App.~\ref{app:2partcol-threeloop}. 
For instance, we know from two-particle collinear limits 
that the antisymmetric functions ${\cal F}^{\rm A}_{abkl}(\{\beta\})$
vanish in the limit where particles $a,b$ are collinear and $k,l$
are rest-of-the-process partons (see \Eqn{eq:twocoll-constraint2Hybrid-main}). 

Using colour conservation~\eqref{colour_conservation_rest}, 
namely the fact that the rightmost generators in every term 
satisfy $\sum_{i=1}^{n}\T_i=0$, one finds:
\begin{eqnarray}\label{eq:calT-AABi-3pc}
    \sum_{i=4}^{n}  \, {\cal T}_{aabi}  &\stackrel{m.c.c}{=}&
    - \frac{1}{8}C_A^2 \T_a\cdot \T_b -
 \mathcal{T}_{aabb} - \mathcal{T}_{aabc}\,,
\end{eqnarray}
where the $m.c.c$ label above the equal sign
reminds us that this relation holds for the 
application of colour conservation with only 
massless external particles present in the 
process. Moreover, we see that by satisfying 
the first constraint from the two-particle 
collinear limit, \Eqn{eq:twocoll-constraint1c-main}, 
for each pair of particles out of the three 
collinear ones, 
\be\label{threecoll-constraint1}
\Big( 4 {\cal F}^{\rm S}_{abkl}(\{\rho\}) 
-2f(\alpha_s) \Big)\bigg|_{p_a|| p_b} \!\!\!\! = 2 f(\alpha_s)\,, 
\qquad (a,b) = (1,2)\,, \,\,\, (1,3)\,, \,\, {\rm or} \,\,(2,3)\,, 
\ee
it is possible to form stuffle products 
\begin{eqnarray}\label{eq:stuffleGen}
2\sum_{4\leq k<l \leq n}  \Big( {\cal T}_{ablk} + {\cal T}_{abkl} \Big)
+2\sum_{i=4}^{n}{\cal T}_{abii} = \sum_{k=4}^{n}\sum_{l=4}^{n} 
\big( {\cal T}_{ablk} +  {\cal T}_{abkl} \big)\,,
\end{eqnarray}
which enable the application of colour conservation 
to eliminate dependence on rest-of-the-process partons 
from these terms. By applying $\sum_{i=1}^{n}\T_i=0$ 
to the rightmost index for the first sum and then the 
second, we can show that 
\begin{eqnarray}\label{eq:mcolcons-triplecollinear-2a}
\sum_{k=4}^{n}\sum_{l=4}^{n}  \mathcal{T}_{abkl}
&&  \stackrel{m.c.c}{=}
- {\cal{T}}_{aabb} - \frac{1}{8}C_A^2 
\T_a\cdot \T_b - {\cal T}_{aabc} 
+ {\cal T}_{abcc}- {\cal T}_{bbac} \,,
\end{eqnarray}
where $(a,b,c)$ are the collinear particles 
$(1,2,3)$, $(3,1,2)$, $(2,3,1)$.

With the above considerations,
and using definition in \Eqn{eq:GammaF}
for the kinematic-dependent term containing all three collinear particles, we find 
\bea\label{eq:m3partcol-pen} \nn 
{\bf{ \Gamma}}_{\SP, 3}^{4{\rm{T}}}(p_1,p_2, p_3)
&=& 
-\frac{3}{4}f(\alpha_s)\left[ C_A^2 \T_1\cdot \T_2 +8\mathcal{T}_{1122}\right]
- \frac{3}{4} f(\alpha_s)\left[C_A^2 \T_1\cdot \T_3
+8\mathcal{T}_{1133} \right]
\nn \\  &&\hspace{-1.2cm} 
- \frac{3}{4} f(\alpha_s) \left[C_A^2 \T_2\cdot \T_3 
+8\mathcal{T}_{2233} \right]
- 4f(\alpha_s) \Big[ \mathcal{T}_{1123}
+ \mathcal{T}_{2213} + \mathcal{T}_{3312}
\Big] \\ \nn   && \hspace{-1.2cm} +    \,8\,
\sum_{4\leq l \leq n} \Big[  \mathcal{T}_{13l2} 
\,{\cal F}  (\beta_{132l},\beta_{1l23}) +\,\mathcal{T}_{12l3} 
\,{\cal F}  (\beta_{123l},\beta_{1l32}) 
+\,\mathcal{T}_{123l} \,{\cal F} (\beta_{12l3},\beta_{13l2}) \Big]   .
\eea
We see that the first three terms 
are identical to the two-particle collinear 
limit result in \Eqn{GammaSpQuad2dN} for each 
pair of particles out of the three collinear ones. 
The constant terms containing colour generators of all three collinear particles are new in the 
three-particle collinear limit, but of course pose no issue for strict collinear factorisation,
because they only depend on the degrees of freedom
of the collinear particles. 
Most interesting is the appearance of the
kinematic terms in the last line, together 
with a sum over all rest-of-the-process 
partons. However, as we have already discussed
above and in Section~\ref{sec:kin-three-part-col},
the dependence on the rest-of-the-process 
parton in a cross ratio containing also three collinear particles scales out. This allows us to take this term out of the sum and apply colour conservation  
to $\mathcal{T}$ structures. For example 
\begin{eqnarray} \label{eq:3partcol-kin-c.c}
\sum_{4\leq l \leq n} 
\mathcal{T}_{123l}  
&&\stackrel{m.c.c}{=} \mathcal{T}_{2213}  - \mathcal{T}_{3312},
\end{eqnarray}
as shown in \Eqn{eq:5.231}, and similar for the remaining
two colour structures,
which yields the final result
\bea\label{eq:m3partcol-fin} \nn 
{\bf{ \Gamma}}_{\SP, 3}^{4{\rm{T}}}(p_1,p_2, p_3)
&=&
-\frac{3}{4}f(\alpha_s)\left[ C_A^2 \T_1\cdot \T_2 +8\mathcal{T}_{1122}\right]
- \frac{3}{4} f(\alpha_s)\left[C_A^2 \T_1\cdot \T_3
+8\mathcal{T}_{1133} \right]
\nn \\  &&\hspace{0cm} 
- \frac{3}{4} f(\alpha_s) \left[C_A^2 \T_2\cdot \T_3 
+8\mathcal{T}_{2233} \right]
- 4f(\alpha_s) \bigg[ \mathcal{T}_{1123}+ \mathcal{T}_{2213} 
+ \mathcal{T}_{3312} \bigg]
\nn \\ && 
+\,8\,\Big( \mathcal{T}_{1123}  
-\mathcal{T}_{2213} \Big) 
\,{\cal F}  (\beta_{132l},\beta_{1l23}) +\,8\Big(\mathcal{T}_{1123} 
-\mathcal{T}_{3312} \Big)
\,{\cal F}  (\beta_{123l},\beta_{1l32}) 
\nn \\ && +\,8\Big(\mathcal{T}_{2213}  - \mathcal{T}_{3312} \Big)
\,{\cal F} (\beta_{12l3},\beta_{13l2})\,.
\eea
This result is identical to the one obtained
directly from $n=4$ case given in \Eqn{GammaSpQuad3d} showing that indeed the three-particle collinear splitting amplitude
soft anomalous dimension is a universal object.
We also see that due to properties of CICRs we
were able to render the result consistent with
expectations from strict collinear factorisation,
and no additional constraints arise from this
kinematic configuration on top of the ones
obtained in the two-particle collinear limit. 
The fact that the three-particle collinear limit
should yield no additional constraints at the 
three-loop order has been noted in App.~B of
Ref.~\cite{Ahrens:2012qz}. However, this work assumed
that a constant term does not appear in the soft anomalous dimension at three loops, so it did not provide the complete result for the splitting amplitude soft anomalous dimension. 

Summarising, \Eqn{eq:m3partcol-fin}, which constitutes one of the new contributions of the present work, provides the
beyond-dipole three-loop contribution to the splitting amplitude
soft anomalous dimension for the case of three final-state particles becoming collinear. 
In \Eqn{eq:m3partcol-fin} 
we identify a contribution, proportional
to $\frac{3}{4}f(\as)$, which corresponds 
to a generalisation of the two-particle
collinear splitting amplitude soft anomalous 
dimension and accounts for the fact that
now there are three pairwise
combinations of particles becoming collinear
($1||2$, $1||3$, and $2||3$).
There are however new contributions, not present in case 
of the two-particle collinear limit, given by the constant
terms proportional to $4f(\as)$ in the second
line of \Eqn{eq:m3partcol-fin}, as well as the
terms containing kinematic dependence in the
last two lines. Due to
the properties of the CICRs discussed in Section~\ref{sec:kin-three-part-col},
such terms do not in fact depend on the rest of the process momenta, and are hence allowed to appear in the splitting amplitude soft anomalous dimension without
violating strict collinear factorisation. Such kinematic-dependent terms can appear when there are at least three collinear particles.

\subsubsection{Terms starting at four loops}
\label{sec:three-particle-collinear-fourloop}
We now turn our attention to the calculation of the 
four-loop contributions to the 
splitting amplitude soft anomalous dimension 
in the three-particle collinear limit. 
Namely, we need to evaluate 
${\bf{ \Gamma}}_{\SP, m}^{{\rm Q}4{\rm{T}}}(p_1,\ldots p_m;\mu_f)$
defined in \Eqn{eq:GammaSpQ4Tm} and 
${\bf{ \Gamma}}_{\SP, m}^{5{\rm{T}}}(p_1,\ldots p_m;\mu_f)$
defined in \Eqn{eq:GammaSp5Tm} for $m=3$. We will consider these
in turn. \\

\noindent 
{\bf Symmetric four-generator terms:} The object we are interested in is 
\begin{eqnarray}
\label{eq:GammaSpQ4T3}
{\bf{ \Gamma}}_{\SP, 3}^{{\rm Q}4{\rm{T}}}(p_1, p_2, p_3;\mu_f) 
&=& {\bf{ \Gamma}}_{n,{\rm{Q}}4{\rm{T}}-2,3{\rm{L}}}
(\{s_{ij} \},\mu_f,\alpha_s)
+{\bf{ \Gamma}}_{n,{\rm Q}4{\rm{T}}-4{\rm{L}}}
(\{\beta_{ijkl} \},\alpha_s) \nonumber \\ 
&& \hspace{-3.2cm}
- \Big({\bf{ \Gamma}}_{n-2,{\rm{Q}}4{\rm{T}}-2,3{\rm{L}}}
(\{s_{ij} \},\mu_f,\alpha_s)
+{\bf{ \Gamma}}_{n-2,{\rm Q}4{\rm{T}}-4{\rm{L}}}
(\{\beta_{ijkl} \},\alpha_s)\Big) 
\Big|_{\T_P\to \sum_{i=1}^3 \T_i},
\end{eqnarray}
involving the four-loop contributions to the splitting amplitude soft anomalous dimension associated with the terms involving a totally symmetric combination of four colour generators, Eqs.~\eqref{eq:Q4T-23L} and \eqref{eq:Q4T-4L}. The calculation proceeds in an 
analogous way to the computation 
performed above for the terms starting at lower
perturbative order.
Without loss of generality, we choose 
particles 1, 2, and 3 to be collinear to parent 
parton $P$ and use parametrisation in \Eqn{eq:coll-momenta}. 
For the anomalous dimensions on the right-hand side
of \Eqn{eq:GammaSpQ4T3}, we split the sums appearing in 
their explicit expressions, Eqs.~\eqref{eq:Q4T-23L} and \eqref{eq:Q4T-4L}, into terms depending only on the rest-of-the-process partons, subsets of collinear particles and rest-of-the-process partons, and terms which depend only on the
particles becoming collinear. We perform the
subtraction between ${\bf{ \Gamma}}_{n}$
and ${\bf{ \Gamma}}_{n-2}$ terms, 
apply colour conservation, and simplify the resultant
expressions. Derivation following the outlined steps
is performed in App.~\ref{app:3partcol-fourloop-Q4T},  
and the result of the manipulations can be found in 
\Eqn{eq:GammaSpQ4T3Res-app}. We report 
it here after minor simplifications using properties kinematic variables in the collinear limit. 
We find that 
${\bf{ \Gamma}}_{\SP, 3}^{{\rm Q}4{\rm{T}}}$ is given by
\begin{eqnarray}
\label{eq:GammaSpQ4T3Res}
&&{\bf{ \Gamma}}_{\SP, 3}^{{\rm Q}4{\rm{T}}}(p_1, p_2, p_3;\mu_f) 
= \sum_R g_R \bigg[   
{ \bf{\cal{ D}}}^R_{1111}\ln(x_1)+  { \bf{\cal{ D}}}^R_{2222}\ln(x_2)
+  { \bf{\cal{ D}}}^R_{3333}\ln(x_3)
\nonumber \\ &&   \hspace{4.2cm}
+ \Big( 
2 { \bf{\cal{ D}}}^R_{1112}+3 { \bf{\cal{ D}}}^R_{1122}
+ 2 { \bf{\cal{ D}}}^R_{1222} \Big)  
\Big(    \ln(x_1x_2) - \ell_{12}\Big) 
\nonumber    \\ &&  \hspace{4.2cm}
+\Big(  2{ \bf{\cal{ D}}}^R_{1113}   +3 { \bf{\cal{ D}}}^R_{1133}  
+ 2 { \bf{\cal{ D}}}^R_{1333}\Big) \Big( \ln(x_1x_3) - \ell_{13}\Big)
\nonumber    \\ && \hspace{4.2cm}
+\Big(  2 { \bf{\cal{ D}}}^R_{2223}   + 3{ \bf{\cal{ D}}}^R_{2233} 
+2 { \bf{\cal{ D}}}^R_{2333}\Big) \Big( \ln(x_2x_3)- \ell_{23}\Big)  
 \nonumber \\ &&     \hspace{4.2cm}
+   2 \sum_{\substack{ 4\leq j < k \leq n  } }  \Big(
{\color{blue} 
{\cal{D}}_{jk12}^R  \beta_{12jk}
}
+   
{\color{red} 
{\cal{D}}_{jk13}^R   \beta_{13jk}
}
+   
{\color{darkgreen} 
 {\cal{D}}_{jk23}^R  \beta_{23jk}
}
 \Big)
 \bigg] 
\nonumber \\ && \nn  \hspace{1.5cm}
+\, 24   \sum_{R}
\sum_{4\leq j<k \leq n} \hspace{-0.2cm}
\Big(
{\color{blue} \,\DR_{12jk} 
\,{\cal G}_R  (\beta_{12jk},0)  }
+
{\color{red} 
\,\DR_{13jk} 
\,{\cal G}_R  (\beta_{13jk},0)   
}
+   
{\color{darkgreen}  
\DR_{23jk} 
\,{\cal G}_R  (\beta_{23jk},0)
} \Big)
\\ && \nn  \hspace{1.5cm}
+ 2 \sum_R g_R \Big(   { \bf{\cal{ D}}}^R_{1123}
+ { \bf{\cal{ D}}}^R_{1223}
+ { \bf{\cal{ D}}}^R_{1233} \Big)\Big( 2 \ln(x_1x_2x_3)
- \ell_{12}-\ell_{13}-\ell_{23} \Big)
\\  &&  \hspace{1.5cm}- \,24\,  \sum_{R}
\,\Big[ \,\DR_{1123}+\DR_{1223} 
+\DR_{1233} \Big] {\cal G}_R  (\beta_{123l},\beta_{1l32})\,.
\end{eqnarray}
In the first four lines, we recognise the 
two-particle collinear limit contributions
$ {\bf{ \Gamma}}_{\SP, 2}^{{\rm Q}4{\rm{T}}}$ found in 
\Eqn{eq:GammaSpQ4T2Res} for each 
pair of the three particles becoming collinear. The fifth and sixth lines are also related to the two-particle collinear 
limits by the virtue that they cancel upon imposing the two-particle collinear
limits constraints derived previously by demanding 
that strict collinear factorisation holds. We see that the constraint of \Eqn{eq:GRconstraint} is generalised here to apply to any two out of the three collinear particles; this is highlighted in Eq.~(\ref{eq:GammaSpQ4T3Res}) by a different colour for each pair. 
We observe that no new form of a constraint arises here in the three-particle collinear limit. Indeed, the last two lines represent genuine new additions to the splitting amplitude 
soft anomalous dimension that appear in the three-particle collinear
limit as opposed to the two-particle collinear limit. We see that they naturally obey strict collinear factorisation by depending only on the degrees of freedom of the collinear particles. Similarly to the three-particle collinear
limit kinematics in terms that start at the three-loop order, 
we observe the seeming dependence on a rest-of-the-process 
parton through CICRs such as $\beta_{123l}$. However, as discussed 
in Section~\ref{sec:kin-three-part-col}, investigation of the properties
of the CICRs shows that this seeming dependence 
scales out, and $\beta_{123l}$ type objects are entirely independent of the rest of the process. 

Implementing the constraints obtained in the two-particle collinear
limit investigations yields the following final result for the splitting amplitude anomalous dimension contribution:
\begin{eqnarray}
\label{eq:GammaSpQ4T3ResF}
&&{\bf{ \Gamma}}_{\SP, 3}^{{\rm Q}4{\rm{T}}}(p_1, p_2, p_3;\mu_f) 
= \sum_R g_R \bigg[   
{ \bf{\cal{ D}}}^R_{1111}\ln(x_1)+  { \bf{\cal{ D}}}^R_{2222}\ln(x_2)
+  { \bf{\cal{ D}}}^R_{3333}\ln(x_3)
\nonumber \\ &&   \hspace{4.2cm}
+ \Big( 
2 { \bf{\cal{ D}}}^R_{1112}+3 { \bf{\cal{ D}}}^R_{1122}
+ 2 { \bf{\cal{ D}}}^R_{1222} \Big)  
\Big(    \ln(x_1x_2) - \ell_{12}\Big) 
\nonumber    \\ &&  \hspace{4.2cm}
+\Big(  2{ \bf{\cal{ D}}}^R_{1113}   +3 { \bf{\cal{ D}}}^R_{1133}  
+ 2 { \bf{\cal{ D}}}^R_{1333}\Big) \Big( \ln(x_1x_3) - \ell_{13}\Big)
\nonumber    \\ && \hspace{4.2cm}
+\Big(  2 { \bf{\cal{ D}}}^R_{2223}   + 3{ \bf{\cal{ D}}}^R_{2233} 
+2 { \bf{\cal{ D}}}^R_{2333}\Big) \Big( \ln(x_2x_3)- \ell_{23}\Big)  
\nonumber \\ &&  \hspace{4.2cm}
+ 2  \Big(   { \bf{\cal{ D}}}^R_{1123}
+ { \bf{\cal{ D}}}^R_{1223}
+ { \bf{\cal{ D}}}^R_{1233} \Big)\Big( 2 \ln(x_1x_2x_3)
- \ell_{12}-\ell_{13}-\ell_{23} \Big) \bigg]
\nonumber \\  &&  \hspace{4.2cm}- \,24\,  \sum_{R}
\,\Big[ \,\DR_{1123}+\DR_{1223} 
+\DR_{1233} \Big] {\cal G}_R  (\beta_{123l},\beta_{1l32})\,.
\end{eqnarray}
We note that in the last term the dependence on the momentum $p_l$ of the rest-of-the-process particle cancels, and this contribution only depends on the complex variables $z_{ABC}$ from Eq.~\eqref{eq:trianglecrossratio}.

\noindent 
{\bf Five-generator terms:}
We now turn our attention to the second type of structure potentially entering the soft anomalous dimension for the first time at the
four-loop order, namely the terms containing five colour generators 
in Eqs.~\eqref{eq:H_5T-4L} and \eqref{eq:H_5T-5L}. 
As discussed below \Eqn{eq:dipole}, there exist arguments why this type of 
contribution is not present in the soft anomalous dimension in
the first place. 
We have already seen in Section~\ref{sec:two-particle-collinear}, following Ref.~\cite{Becher:2019avh}, that certain constrains on these contributions arise from two-particle collinear limits, which are however not strong enough to exclude them altogether. Here we are interested to explore 
whether multi-particle collinear limits can provide additional information on these terms, going beyond the two-particle collinear limit constraints. 

We are interested in ${\bf{ \Gamma}}_{\SP, m}^{5{\rm{T}}}$ defined
in \Eqn{eq:GammaSp5Tm} with index $m$ set to $m=3$.
Explicitly, we compute the 
following object
\begin{eqnarray}
\label{eq:GammaSp5T3}
{\bf{ \Gamma}}_{\SP, 3}^{5{\rm{T}}}(p_1, p_2, p_3;\mu_f) &=&
{\bf{ \Gamma}}_{n,5{\rm{T}}-4{\rm{L}}}(\{\beta_{ijkl} \},\alpha_s) 
+{\bf{ \Gamma}}_{n,5{\rm{T}}-5{\rm{L}}}(\{\beta_{ijkl} \},\alpha_s) 
\nonumber \\ 
&& \hspace{-3.2cm}
- \Big(  {\bf{ \Gamma}}_{n-2,5{\rm{T}}-4{\rm{L}}}(\{\beta_{ijkl} \},\alpha_s)
+ {\bf{ \Gamma}}_{n-2,5{\rm{T}}-5{\rm{L}}}(\{\beta_{ijkl} \},\alpha_s) 
\Big) |_{\T_P\to \sum_{i=1}^3 \T_i}\,.
\end{eqnarray}
Starting from $n$ and $n-2$ soft anomalous dimensions, the calculation proceeds following
the same strategy as for the 
previous terms and it is presented in detail in App.~\ref{app:3partcol-fourloop-5T}.
Namely, for the anomalous dimensions on the right-hand side
of \Eqn{eq:GammaSp5T3}, we separate the sums appearing in 
these terms (see Eqs.~\eqref{eq:H_5T-4L} and \eqref{eq:H_5T-5L}) into 
pieces that depend solely on the rest-of-the-process partons,
subsets of collinear particles and rest-of-the-process
partons, and ones depending only on the collinear partons. 
Due to the length of the expressions, we implement the
constraints obtained in the two-particle collinear
limit, Eqs.~\eqref{eq:H1constraint} and \eqref{eq:Kminus-constraint},
for each pair out of the three collinear particle at an intermediate stage of the derivation, and continue only with genuinely new terms containing dependence on all three collinear particles.

The critical step in the derivation is given in \Eqn{eq:5.308-c4},
and is copied here to facilitate the following discussion.
Following the steps described above, we find that ${\bf{ \Gamma}}_{\SP, 3}^{5{\rm{T}}}$
is given by  
\begin{eqnarray}\label{eq:5.308-c4-main}
&&\hspace{-0.1cm} {\bf{ \Gamma}}_{\SP, 3}^{5{\rm{T}}}(p_1, p_2, p_3) =
2  \sum_{l=4}^{n}  \Big(\mathcal{T}_{13l21} 
-\mathcal{T}_{23l12} +\,\mathcal{T}_{312l3} \Big)\mathcal{H}_1(\beta_{132l},\beta_{1l23})
+ 2 \sum_{l=4}^{n} \Big( \mathcal{T}_{12l31}
\\[-0.25cm] && \hspace{+0.4cm}
-\mathcal{T}_{32l13} +\mathcal{T}_{213l2} \Big)\mathcal{H}_1(\beta_{123l},\beta_{1l32})
+2  \sum_{l=4}^{n} \Big(\mathcal{T}_{21l32}
- \mathcal{T}_{31l23} +\mathcal{T}_{123l1} \Big)\mathcal{H}_1(\beta_{12l3},\beta_{13l2})
\nonumber \\[-0.25cm] && \hspace{0.4cm}
+8 {\sum_{k=4}^{n}   \sum_{\substack{l =4 \\   l\neq k} }^ {n} }
\Big[ \mathcal{H}_2(\beta_{23kl},\beta_{1k23},\beta_{132k},\beta_{1l32},\beta_{123l})
+\mathcal{H}_2(\beta_{231l},\beta_{23k1},\beta_{21k3},\beta_{23lk},0 )
\nonumber \\[-0.6cm] && \hspace{0.4cm}
+\mathcal{H}_2(\beta_{13l2},\beta_{132k},\beta_{1k23},\beta_{13lk},0 ) \Big] \mathcal{T}_{23kl1}
+2 \sum_{l=4}^{n}\big(\mathcal{H}_1(\beta_{132l},\beta_{1l23}) 
- \mathcal{H}_1(\beta_{123l},\beta_{1l32})\big) \mathcal{T}_{23ll1}
\nonumber \\[-0.25cm] && \hspace{0.4cm}
+8 {\sum_{k=4}^{n}   \sum_{\substack{l =4 \\   l\neq k} }^ {n} }  
\Big[  \mathcal{H}_2(\beta_{13kl},\beta_{13k2},\beta_{12k3},\beta_{132l},\beta_{1l23} )
+\mathcal{H}_2(\beta_{13l2},\beta_{132k},\beta_{1k23},\beta_{13lk},0 ) 
\nonumber \\[-0.6cm] && \hspace{0.4cm}
+\mathcal{H}_2(\beta_{123l},\beta_{12k3},\beta_{13k2},\beta_{12lk},0 ) \Big] \mathcal{T}_{31kl2} 
+2 \sum_{l=4}^{n}\big(\mathcal{H}_1(\beta_{132l},\beta_{1l23})
-\mathcal{H}_1(\beta_{13l2},\beta_{12l3})\big)\mathcal{T}_{31ll2}
\nonumber \\[-0.25cm] && \hspace{0.4cm}
+8 {\sum_{k=4}^{n}   \sum_{\substack{l =4 \\   l\neq k} }^ {n} } 
\Big[\mathcal{H}_2(\beta_{12kl},\beta_{123k},\beta_{1k32},\beta_{12l3},\beta_{13l2} )
+\mathcal{H}_2(\beta_{123l},\beta_{12k3},\beta_{13k2},\beta_{12lk},0 )
\nonumber \\[-0.6cm] \nonumber && \hspace{0.5cm}
+\mathcal{H}_2(\beta_{231l},\beta_{23k1},\beta_{21k3},\beta_{23lk},0 )\Big]\mathcal{T}_{12kl3}
+2 \sum_{l=4}^{n}  \big( \mathcal{H}_1(\beta_{12l3},\beta_{13l2})
-\mathcal{H}_1(\beta_{123l},\beta_{1l32})\big)\mathcal{T}_{12ll3} \,.
\end{eqnarray}
At this point, we recognise that we have reached a step in the 
calculation that is analogous to the one in which constraints were 
obtained in the derivation of two-particle collinear limit 
for the terms starting at three loops. 
Namely, in order to apply colour conservation to the generators 
of the non-collinear partons in two index positions, we are 
required to form stuffle products. Here, analogously to 
\Eqn{eq:stuffleGen}, we need  
\begin{eqnarray}\label{eq:NEWstuffle3-main}
2{\sum_{k=4}^{n}   \sum_{\substack{l =4 \\   k\neq l} }^ {n} }   {\cal T}_{abklc}  
+2\sum_{l=4}^{n}{\cal T}_{abllc} = \sum_{k=4}^{n}\sum_{l=4}^{n} 
\big( {\cal T}_{abklc} +  {\cal T}_{ablkc} \big)\,,
\end{eqnarray}
for cyclic permutations of $(a,b,c)=(1,2,3)$. 
To combine the colour terms in this way, we must 
demand that their coefficients can be identified 
with each other in the strict collinear limit.
This reasoning, as was the case for two-particle 
collinear limit at three loops (see 
\Eqn{GammaSpQuad2eHybrid} and the discussion 
below), yields a seemingly new constraint
that the ${\cal{H}}_1$ and ${\cal{H}}_2$ 
kinematic functions have to satisfy
in the three-particle collinear limit:
\begin{eqnarray}\label{eq:Kminus-3coll}
 && \hspace{-0.9cm}  4\mathcal{H}_2(\beta_{abkl},\beta_{abck},\beta_{akcb},\beta_{ablc},\beta_{aclb} )
+4\mathcal{H}_2(\beta_{abcl},\beta_{abkc},\beta_{ackb},\beta_{ablk},0 )
\nonumber \\  && \hspace{-0.95cm}
+\,\, 4 \mathcal{H}_2(\beta_{bcal},\beta_{bcka},\beta_{bakc},\beta_{bclk},0 )
=   \mathcal{H}_1(\beta_{ablc},\beta_{aclb})
-\mathcal{H}_1(\beta_{abcl},\beta_{alcb}).
\end{eqnarray}
This condition appears to be independent of the 
constraints of~\Eqns{eq:H1constraint}{eq:Kminus-constraint} 
we derived from the two-particle collinear limit in
Section~\ref{sec:two-particle-collinear} and 
App.~\ref{sec:2-coll_Quintic_terms}. In particular,
the new relation in \Eqn{eq:Kminus-3coll} seems to 
be distinct from \Eqn{eq:Kplusminus-main} that indeed 
includes combinations of ${\cal{H}}_1$ and~${\cal{H}}_2$ 
functions, but with different coefficients. Moreover, 
here for the first time we see a constraint containing 
non-trivial dependence of the ${\cal{H}}_1$ and~${\cal{H}}_2$ 
kinematic functions on the three collinear 
particles~$(a,b,c)=(1,2,3)$, along with the cyclic 
permutations. However, we now recall that the 
constraints of~\Eqn{eq:Kminus-constraint} applies 
with different assignments of the parton indices, 
and from the point of view of the functional 
dependence of kinematic functions on the CICRs, 
it is equivalent to consider a configuration where 
three of the particles become collinear and one is 
from the rest of the process, and a configuration 
where three partons are from the rest of the process 
along with one from the collinear set. 
The constraint on~$K_-(\beta_{1kml},\beta_{1lmk}, \beta_{12kl})$
in \Eqn{eq:Kminus-constraint} has been obtained 
for the latter case, where partons~$k, l$, and $m$ 
belonged to the rest-of-the-process set. However, 
if we replace the particles $k, l$, and $m$ by 
the three collinear partons, i.e. $k\to b$, $l\to a$, 
and $m\to c$, and simultaneously replace the collinear 
indices $1$ and $2$ by two partons $k$ and $l$ from 
the rest-of-the-process, 
 the function $K_-$ in \Eqn{eq:Kplusminus-main} becomes: 
\begin{eqnarray}\label{eq:Kminus-3pcABC}
K_-(\beta_{kbca},\beta_{kacb},\beta_{klba})&=&\Big(  
4\mathcal{H}_2(\beta_{kcba}, \beta_{kcab}, \beta_{kbac},\beta_{abkl},0)
\nonumber \\ && \hspace{-3.6cm} - \mathcal{H}_1(\beta_{kcab},\beta_{kbac})
+2 \mathcal{H}_2(\beta_{abkl},\beta_{kcba},\beta_{kabc},\beta_{kcab},\beta_{kbac})\Big)
\nonumber \\ &&  \hspace{-3.6cm}     
-\Big(  4\mathcal{H}_2( \beta_{kcab},\beta_{kcba},\beta_{kabc},\beta_{abkl},0)
- \mathcal{H}_1(\beta_{kcba},\beta_{kabc})
\nonumber \\ &&  \hspace{-3.6cm}
+2 \mathcal{H}_2(\beta_{abkl},\beta_{kcab},\beta_{kbac},\beta_{kcba},\beta_{kabc}) \Big)\,,
\end{eqnarray}
after using the identities in Eqs.~\eqref{eq:H2-sym2} and \eqref{eq:H2-sym3}. 

Now, starting with \Eqn{eq:Kminus-3coll}, we bring the ${\cal{H}}_1$ terms to the
right-hand side and upon permuting indices to start to match the structures 
in \Eqn{eq:Kminus-3pcABC} we have
\begin{eqnarray}\label{eq:E.122copy2-7}
&&  4 \mathcal{H}_2(\beta_{abkl},\beta_{kcba},\beta_{kabc},\beta_{lcab},\beta_{lbac})
+4\mathcal{H}_2(\beta_{lcba},\beta_{kcab},\beta_{kbac},\beta_{abkl},0)
\\ &&  \nonumber 
+ 4 \mathcal{H}_2(\beta_{lacb},\beta_{kabc},\beta_{kcba},\beta_{bckl},0)   
+ \mathcal{H}_1(\beta_{lbac},\beta_{lcab})
+ \mathcal{H}_1(\beta_{lcba},\beta_{labc})=0.
\end{eqnarray}
Next, working with the~${\cal{H}}_2$ function in the second line, 
we exchange~$\beta_{bckl}$ in favour of $\beta_{abkl}$ using 
$\beta_{bckl} = \beta_{abkl} + \beta_{kacb}$ and apply one of 
the symmetry properties of this function in~\Eqn{eq:H2-sym3} 
leading to
\begin{eqnarray}\label{eq:E.122copy2-9}
&&  4 \mathcal{H}_2(\beta_{abkl},\beta_{kcba},\beta_{kabc},\beta_{lcab},\beta_{lbac})
+4\mathcal{H}_2(\beta_{lcba},\beta_{kcab},\beta_{kbac},\beta_{abkl},0)
\\ &&  \nonumber 
 - 4 \mathcal{H}_2(-\beta_{lacb},\beta_{kcba},\beta_{kabc}, \beta_{abkl} + \beta_{kacb} -\beta_{kabc}+\beta_{kcba},\beta_{lacb}-\beta_{kabc}+\beta_{kcba} )
\\ &&  \nonumber    +   
\mathcal{H}_1(\beta_{lbac},\beta_{lcab})
 +\, \mathcal{H}_1(\beta_{lcba},\beta_{labc})=0\,.
\end{eqnarray}
The middle term can be simplified using 
$\beta_{lacb}-\beta_{kabc}+\beta_{kcba} =0$. 
We split the first~${\cal{H}}_2$ term and 
simplify to yield
\begin{eqnarray}\label{eq:E.122copy2-11}
&&  -   \mathcal{H}_1(\beta_{lcab},\beta_{lbac})
+4\mathcal{H}_2(\beta_{lcba},\beta_{kcab},\beta_{kbac},\beta_{abkl},0)
\nonumber    \\ && 
+ 2 \mathcal{H}_2(\beta_{abkl},\beta_{kcba},\beta_{kabc},\beta_{lcab},\beta_{lbac})
\nonumber    \\ && 
+\, \mathcal{H}_1(\beta_{lcba},\beta_{labc} )
- 4 \mathcal{H}_2(\beta_{lcab},\beta_{kcba},\beta_{kabc}, \beta_{abkl} ,0) 
\\ &&  \nonumber   
+ 2 \mathcal{H}_2(\beta_{abkl},\beta_{kcba},\beta_{kabc},\beta_{lcab},\beta_{lbac}) =0\,.
\end{eqnarray}
In the final step, we use another symmetry property of ${\cal{H}}_2$ 
given in~\Eqn{eq:H2-sym1} to rewrite the last term
of the above equation. Then, recalling that the rest-of-the-process
parton in a CICR  with three collinear partons scales out, we find
\begin{eqnarray}\label{eq:E.122copy2-13}
&&  - \mathcal{H}_1(\beta_{kcab},\beta_{kbac})
+4\mathcal{H}_2(\beta_{kcba},\beta_{kcab},\beta_{kbac},\beta_{abkl},0)
\nonumber    \\ && 
+ 2 \mathcal{H}_2(\beta_{abkl},\beta_{kcba},\beta_{kabc},\beta_{kcab},\beta_{kbac})
\nonumber    \\ &&
+\, \mathcal{H}_1(\beta_{kcba},\beta_{kabc})
-4
\mathcal{H}_2(\beta_{kcab},\beta_{kcba},\beta_{kabc}, \beta_{abkl} ,0) 
\\ &&  \nonumber   
- 2 \mathcal{H}_2(\beta_{abkl},\beta_{kcab},\beta_{kbac},\beta_{kcba},\beta_{kabc}) =0\,,
\end{eqnarray}
which brings the seemingly new constraint in \Eqn{eq:Kminus-3coll}
to the same form as \Eqn{eq:Kminus-3pcABC}, meaning that in this case 
as well, it is the two-particle collinear limit that provides constraints
on the form of the functions in the soft anomalous dimension, and no new 
constrains appear upon considering three-particle collinear limits. 

The upshot of the above discussion in terms of a result for the 
three-particle collinear splitting amplitude soft anomalous dimension
of the five generator terms is that it is given by the top two lines of 
\Eqn{eq:5.308-c4-main} where a single rest-of-the-process 
parton index is free and the double summed terms after
the stuffle relations in \Eqn{eq:NEWstuffle3-main} have been formed by
enforcing the constraint on the kinematic functions.
Then, colour conservation can be applied  to give the following result 
\begin{eqnarray}\label{eq:5.308-c4-main-res}
&&\hspace{-0.1cm} {\bf{ \Gamma}}_{\SP, 3}^{5{\rm{T}}}(p_1, p_2, p_3) =
2\mathcal{H}_1(\beta_{132k},\beta_{1k23})
\Big(  
\frac{C_A^2}{12} \mathcal{T}_{123}
-2  \mathcal{T}_{11231}
+2  \mathcal{T}_{22132}
+ 4 \mathcal{T}_{11322} \Big)
\nonumber \\ && \hspace{2.9cm}
+  2 \mathcal{H}_1(\beta_{213k},\beta_{2k31})
\Big(  
\frac{C_A^2}{12} \mathcal{T}_{231}
-2  \mathcal{T}_{22312}
+2  \mathcal{T}_{33213}
+ 4 \mathcal{T}_{22133} \Big)
\nonumber \\ && \hspace{2.9cm}
+  2 \mathcal{H}_1(\beta_{321k},\beta_{3k12})
\Big(  
\frac{C_A^2}{12} \mathcal{T}_{312}
-2  \mathcal{T}_{33123}
+2  \mathcal{T}_{11321}
+ 4 \mathcal{T}_{33211} \Big).
\end{eqnarray}

\subsection{Four and higher - particle collinear limits}
\label{sec:four-particle-collinear}
In this section, we move to consider the four-particle
collinear limit. We also discuss the form that
the splitting amplitude anomalous dimension takes as
the number of particles becoming collinear increases
beyond four. 
The four-particle collinear limit
corresponds to the configuration 
depicted in Fig.~\ref{m-collinearTL} with $m = 4$ 
and the collinear particles are parametrised using
\Eqn{eq:coll-momenta} with $i=1,2,3$, and $4$. 

For the dipole contributions,
we have already presented the result for the splitting amplitude anomalous dimension for any number of 
particles, $m$, becoming collinear in \Eqn{GammaSpDipm}. 
Non-dipole contributions are considered in the following two subsections.

\subsubsection{Terms starting at three loops}
\label{sec:four-particle-collinear-threeloop}

For the terms in the soft anomalous dimension 
that start at three loops, the case of four 
collinear particles is 
a borderline case: it is the limit with the highest possible number of collinear 
particles where a new type of behaviour can arise in the splitting amplitude. This is because at three loops at most four external
particles can be connected by gluon webs. 
Therefore, the last possible new type of contribution
to the splitting amplitude anomalous dimension
is one where all the exchanges happen between 
the particles becoming collinear. This contribution is therefore trivially consistent
with strict collinear factorisation. 
It follows that for these terms no new constraints can arise from limits involving four or more collinear particles, beyond those obtained in the two-particle collinear limit. For completeness, we now explicitly consider the four-particle
collinear limit. 

For the case of four collinear particles, the
relevant object of 
interest is~${\bf{ \Gamma}}_{\SP, 4}^{4{\rm{T}}}$
defined through \Eqn{eq:GammaSpQuadm} with $m=4$.
Concretely, it is given by
\begin{eqnarray}
\label{eq:GammaSpQuad4} 
{\bf{ \Gamma}}_{\SP, 4}^{4{\rm{T}}}(p_1,p_2, p_3,p_4;\mu_f) &=& 
   {\bf{ \Gamma}}_{n,4{\rm{T}}-3{\rm{L}}}(\alpha_s)
     + {\bf{ \Gamma}}_{n,4{\rm{T}}-4{\rm{L}}}(\{\beta_{ijkl}\},\alpha_s)
\nonumber \\  && \hspace{-2.3cm} 
-  \Big(  {\bf{ \Gamma}}_{n-3,4{\rm{T}}-3{\rm{L}}}(\alpha_s)
     + {\bf{ \Gamma}}_{n-3,4{\rm{T}}-4{\rm{L}}}(\{\beta_{ijkl}\},\alpha_s)\Big) 
    \Big|_{\T_P\to \sum_{i=1}^4 \T_i}.
\end{eqnarray}
We compute ${\bf{ \Gamma}}_{\SP, 4}^{4{\rm{T}}}$
as for previous cases, namely, directly
from the definition above using the expressions 
for soft anomalous dimensions with $n$ and $n-3$
partons appearing on the right-hand side of 
\Eqn{eq:GammaSpQuad4}. We follow our procedure and
choose the particles $1,2,3$, and~$4$ to be the 
collinear ones, split the sums in the soft 
anomalous dimension accordingly,
perform the subtractions, and impose the 
two-particle collinear limit constraints 
for each pair out of the four becoming collinear. 
These manipulations are carried out in  
App.~\ref{app:4partcol} leading to the
expression in \Eqn{GammaSpQuad4eHybrid-g},
which we copy here for convenience
\bea\label{GammaSpQuad4eHybrid-g-main-cp} \nn 
{\bf{ \Gamma}}_{\SP, 4}^{4{\rm{T}}}(p_1,p_2, p_3,p_4)
&=&  - \frac{3}{4} f(\alpha_s)
\big( C_A^2 \T_1\cdot \T_2 + 8{\cal{T}}_{1122} \big)
- \frac{3}{4}f(\alpha_s)
\big(   C_A^2 \T_1\cdot \T_3+8{\cal{T}}_{1133} \big)
\nn \\ && \hspace{-0.0cm} 
- \frac{3}{4}f(\alpha_s)  
\big( C_A^2 \T_1\cdot \T_4  - 8{\cal{T}}_{1144}  \big)
 - \frac{3}{4}f(\alpha_s) 
\big(    C_A^2 \T_2\cdot \T_3- 8{\cal{T}}_{2233} \big)
\nn \\ && \hspace{-0.0cm} 
 - \frac{3}{4}f(\alpha_s)
\big(  C_A^2 \T_2\cdot \T_4- 8{\cal{T}}_{2244} \big)
- \frac{3}{4}f(\alpha_s) 
\big(  C_A^2 \T_3\cdot \T_4- 8{\cal{T}}_{3344} \big)
\nn \\  &&\hspace{-0.0cm}
- 4f(\alpha_s) \Big[ 
  \mathcal{T}_{1123}+\mathcal{T}_{1124}
+ \mathcal{T}_{1134}+\mathcal{T}_{2213}
+ \mathcal{T}_{2214}+\mathcal{T}_{2234}
\nn \\  &&\hspace{1.4cm}
+\mathcal{T}_{3312}+\mathcal{T}_{3314}
+\mathcal{T}_{3324}+\mathcal{T}_{4412}
+\mathcal{T}_{4413}+\mathcal{T}_{4423} \Big]
\nn \\[-0.1cm] &&\hspace{-3.9cm} 
+4 \textcolor{black}{
\sum_{5\leq  l \leq n} \,\big( {\bf a}_{123l}^{ }(\{\beta\})
+ {\bf a}_{124l}^{ }(\{\beta\})
+ \, {\bf a}_{134l}^{ }(\{\beta\})
+ \, {\bf a}_{234l}^{ }(\{\beta\}) \big)}
+4 \textcolor{darkgreen}{
 \, {\bf a}_{1234}^{ }(\{\beta\})}\,.
\eea
This expression resembles the corresponding 
one found in the three-particle collinear limit case given
in~\Eqn{eq:m3partcol-pen}, with the same type of 
terms appearing for subsets of two or three out of
the four collinear particles. Highlighted in
\textcolor{darkgreen}{green} for emphasis,
the expression above contains in addition the genuinely new type of term
for the splitting amplitude soft anomalous dimension that
contains the kinematics of all four collinear particles. 

It remains to apply colour conservation to the summed over 
terms in the last line of~\Eqn{GammaSpQuad4eHybrid-g-main-cp}. 
For this, we cannot use the result for similar terms encountered
in the three-particle collinear limit case, see Eq.~\eqref{eq:3partcol-kin-c.c},
as the sum now starts with index 5. Therefore, new terms containing colour
of four collinear partons appear through colour conservation. 
We find 
\begin{eqnarray} \label{eq:4partcol-kin-c.c}
\sum_{5\leq l \leq n} \mathcal{T}_{123l}  
&&\stackrel{m.c.c}{=} 
\mathcal{T}_{2213}  - \mathcal{T}_{3312}
- \mathcal{T}_{1234} ,
\end{eqnarray}
which is 
derived in \Eqn{eq:5.231-four}. Using similar identities for 
the different subsets of collinear particles in the end 
we arrive at the following result
\bea\label{eq:GammaSpQuad4eHybrid-i} \nn 
&&{\bf{ \Gamma}}_{\SP, 4}^{4{\rm{T}}}(p_1,p_2, p_3,p_4)
= -\frac{3}{4} f(\alpha_s)
\left[ C_A^2 \T_1\cdot \T_2+ 8{\cal{T}}_{1122}\right]
- \frac{3}{4}f(\alpha_s)
\left[ C_A^2 \T_1\cdot \T_3+ 8{\cal{T}}_{1133}\right]
\nn \\ &&  \hspace{3.8cm}
- \frac{3}{4}f(\alpha_s)  
\left[C_A^2 \T_1\cdot \T_4+ 8 {\cal{T}}_{1144} \right]
- \frac{3}{4}f(\alpha_s) 
\left[C_A^2 \T_2\cdot \T_3 +8{\cal{T}}_{2233}\right]
\nn \\ &&  \hspace{3.8cm}
- \frac{3}{4}f(\alpha_s)
\left[C_A^2 \T_2\cdot \T_4 +8{\cal{T}}_{2244} \right]
- \frac{3}{4}f(\alpha_s) 
\left[C_A^2 \T_3\cdot \T_4  +8{\cal{T}}_{3344}\right]
\nn \\  &&\hspace{3.8cm}
- 4f(\alpha_s) \Big[ 
 \mathcal{T}_{1123}+\mathcal{T}_{1124}
+\mathcal{T}_{1134}+\mathcal{T}_{2213}
+\mathcal{T}_{2214}+\mathcal{T}_{2234}
\nn \\  &&\hspace{5.25cm}
+\mathcal{T}_{3312}+\mathcal{T}_{3314}
+\mathcal{T}_{3324}+\mathcal{T}_{4412}
+\mathcal{T}_{4413}+\mathcal{T}_{4423}
\Big]
\nn \\[-0.0cm]  &&\hspace{0cm} 
+ \,8
\big(\mathcal{T}_{1123}  
-\mathcal{T}_{2213}  -\mathcal{T}_{1342} \big)
\,{\cal F}  (\beta_{132l},\beta_{1l23})  
+\,8\big(\mathcal{T}_{1123} 
-\mathcal{T}_{3312} 
-\mathcal{T}_{1243}  \big)
\,{\cal F}  (\beta_{123l},\beta_{1l32}) 
\nn \\[-0.0cm]  &&\hspace{0cm} 
+\,8\big( \mathcal{T}_{2213}  - \mathcal{T}_{3312}
- \mathcal{T}_{1234}  \big)
\,{\cal F} (\beta_{12l3},\beta_{13l2}) 
+    \,8 \big( \mathcal{T}_{1124}  
-\mathcal{T}_{2214}  -\mathcal{T}_{1432}\big)
\,{\cal F}  (\beta_{142l},\beta_{1l24}) 
\nn \\[-0.0cm]  &&\hspace{0cm} 
+\,8  \big(\mathcal{T}_{1124} 
-\mathcal{T}_{4412} 
-\mathcal{T}_{1234} \big)
\,{\cal F}  (\beta_{124l},\beta_{1l42}) 
+\,8  \big(\mathcal{T}_{2214}  - \mathcal{T}_{4412}
- \mathcal{T}_{1243}  \big)
\,{\cal F} (\beta_{12l4},\beta_{14l2}) 
\nn \\[-0.0cm]  &&\hspace{0cm} 
+    \,8
\big(\mathcal{T}_{1134}  
-\mathcal{T}_{3314}  -\mathcal{T}_{1423} \big)
\,{\cal F}  (\beta_{143l},\beta_{1l34}) 
+\,8\big(\mathcal{T}_{1134} 
-\mathcal{T}_{4413} 
-\mathcal{T}_{1324}\big)
\,{\cal F}  (\beta_{134l},\beta_{1l43}) 
\nn\\[-0.0cm]  &&\hspace{0cm}
+\,8 \big(\mathcal{T}_{3314}  - \mathcal{T}_{4413}
- \mathcal{T}_{1342} \big)
\,{\cal F} (\beta_{13l4},\beta_{14l3}) 
+    \,8
\big(\mathcal{T}_{2234}  
-\mathcal{T}_{3324}  -\mathcal{T}_{2413} \big)
\,{\cal F}  (\beta_{243l},\beta_{2l34})
\nn \\[-0.0cm]  &&\hspace{0cm} 
+\,8\big(\mathcal{T}_{2234} 
-\mathcal{T}_{4423} 
-\mathcal{T}_{2314}\big)
\,{\cal F}  (\beta_{234l},\beta_{2l43}) 
+\,8 \big(\mathcal{T}_{3324}  - \mathcal{T}_{4423}
- \mathcal{T}_{2341}  \big)
\,{\cal F} (\beta_{23l4},\beta_{24l3})
\nn\\[-0.0cm]  &&\hspace{0cm}
+    \,8\, \Big[ 
\mathcal{T}_{1342} 
\,{\cal F}  (\beta_{1324},\beta_{1423}) +\,\mathcal{T}_{1243} 
\,{\cal F}  (\beta_{1234},\beta_{1432}) 
+\,\mathcal{T}_{1234} 
\,{\cal F} (\beta_{1243},\beta_{1342}) \Big]\,.
\eea
We note that the dependence on the momentum $p_l$ of the rest-of-the-process particle cancels, and there is dependence only on the complex variables $z_{ABC}$ from Eq.~\eqref{eq:trianglecrossratio}.
From this point on, for the terms starting at three-loop order, 
there can be no more new types of terms contributing to the 
splitting amplitude soft anomalous dimension for $m$ larger 
than 4. As argued at the beginning of this section, this fact follows
because in this contribution to the soft anomalous dimension at most 
four partons can be connected. 
Therefore, there is no possibility for new interesting structures 
to arise in a five-particle or higher collinear limit. The result
for higher-particle collinear limit splitting amplitude anomalous
dimensions will resemble the one found in \Eqn{eq:GammaSpQuad4eHybrid-i}
with sums over the subsets of the particles becoming collinear.

\subsubsection{Terms starting at four loops}
\label{sec:four-particle-collinear-fourloop}
In this section, we turn our attention to the four-
and higher-particle collinear limits of the terms 
that first appear at four loops. 
The discussion is divided into two parts,
first we consider the terms related to totally symmetric combinations of four generators 
and subsequently the five-generator terms.

The situation for terms involving totally symmetric combinations of four generators is not more
involved than the analysis of terms contributing first
at three loops, since they too can connect to at most four partons. Therefore, as in the previous case, for the
four-particle collinear limit the new type of terms contributing
to the splitting amplitude soft anomalous dimension are the ones which
saturate the particle indices with the collinear particles.
Naturally, the results also contain combinatorics of the lower-particle collinear limit results, for each subset of the two and
three particles becoming collinear. This leads to quite lengthy,
but not particularly illuminating expressions, including contributing
terms of the type encountered in \Eqn{eq:GammaSpQ4T3Res}, but extended 
to include all subsets of the particles becoming collinear.
We do not find it useful to explicitly write out these expressions,
but their structure is clear. For higher-particle collinear
limits it only remains to take care of the combinatorics for the same
type of terms appearing for subsets of collinear particles.
Since these terms are automatically consistent with strict collinear
factorisation, no additional constraints on the functions governing the behaviour of these terms can be obtained
from multi-particle collinear limits.  

Lastly, we make comments regarding the five-generator terms. 
This sector has the lengthiest expressions, but the situation can
be understood without performing a  lengthy derivation. 
Moreover, since this sector is believed to not contribute
to the soft anomalous dimension, we choose not to include explicit results for this piece.
The derivation proceeds along the same lines as performed in
this article thus far. The point is that for the four-particle
collinear limit, the situation resembles the structure encountered in
the considerations of the three-particle collinear limit of terms
starting at three loops. Namely, the kinematic-dependent functions depend only on
CICRs which contain four parton indices. 
For terms that contain information on
five particles, since four are collinear,
there can be at most one index involving a parton from the rest of the process.
Therefore, the dependence on this parton in fact scales out, as argued in Section~\ref{sec:kin-three-part-col}. It follows that just as in the 
three-particle collinear limit for terms starting at three loops, these
functions can then be taken out of the sum over the rest-of-the-process
partons, and then colour conservation can be applied on the remaining one
rest-of-the-process parton index to render these terms compatible with strict
collinear factorisation. For five and higher 
collinear limits, 
the situation then turns to the one described in the paragraph above.
Now, the indices are saturated by the particles
becoming collinear, so the functions depending on five collinear
particles appear in the result, but no new structures can appear beyond those. 
Only the combinatorics needs to be taken care of for even higher-particle collinear limits.

\subsection{Summary for the massless case}
\label{sec:massless-summary}
In this section, we focused on generic scattering amplitudes
with $n$ massless external legs and calculated the splitting
amplitude soft anomalous dimensions in timelike multi-particle collinear limits.
The starting point for our calculations are the definitions given
in Section~\ref{sec:factorisationincollinearlimits-sub}. In particular, 
the splitting amplitude soft anomalous dimensions for any number 
of particles becoming collinear written as differences of 
$n$ and $(n-m+1)$-point soft anomalous dimensions in 
Eqs.~\eqref{eq:GammaSPdefDip}, \eqref{eq:GammaSpQuadm},
\eqref{eq:GammaSpQ4Tm} and \eqref{eq:GammaSp5Tm}, 
for the dipole terms, terms starting at three loops, and the 
four and five generator terms starting at four loops, respectively. 
The complete $m$-particle collinear limit splitting amplitude
up to four loops is then given by the sum of these according
to~\Eqn{eq:GammaSPdef-shand}.  

Due to the simplicity of dipole terms, we have written down 
a general result for any number $m$ of particles becoming collinear
at the beginning of Section~\ref{sec:massless} in~\Eqn{GammaSpDipm}.
Subsequently, in 
Section~\ref{sec:two-particle-collinear},~\ref{sec:three-particle-collinear},
and~\ref{sec:four-particle-collinear} we have studied the two-, three-,
four- and higher-particle collinear limits, respectively, where we discussed the terms that start contributing to the soft anomalous
dimension at the three- and four-loop order in separate subsections. Corresponding appendices contain
details of the derivations performed to arrive at the results presented in
these sections. 

The two-particle collinear results obtained from fixed $n$
amplitudes are known in the literature, both for the terms starting at three and
four loops. In our analysis, we have demonstrated the universality of the 
two-particle splitting amplitude soft anomalous dimension
by showing how the calculation can be organised as to obtain 
${\bf \Gamma}_{\SP,2}$ truly from amplitudes with any number of legs. 
From our derivation it is transparent how constraints on the functions
appearing in the soft anomalous dimension arise by demanding that strict collinear
factorisation holds. It is then also clear that no additional constraints arise 
by considering the same limits for amplitudes with higher number of legs. 

The splitting amplitude anomalous dimensions for 
three-, four- and higher-particle collinear limits
worked out in Section~\ref{sec:three-particle-collinear}
and~\ref{sec:four-particle-collinear} for terms starting
at three and four loops are new results of this work. 
We have observed how through the interplay between kinematics
and colour terms strict collinear factorisation is 
delicately preserved for high perturbative order terms. 
As a corollary, we have discovered that no additional 
constraints on the soft anomalous dimension, to the ones obtained
in the two-particle collinear limit, arise from three-, four- and higher-particle collinear limits up to four loops. 

Collecting all the expressions according to~\Eqn{eq:GammaSPdef-shand}
constitutes one of the main results of this section: it encodes, 
up to four loops, the structure of infrared divergences 
of the splitting amplitude, for two-, three-, four- and higher-particle  
collinear limit in the timelike case. The results for the splitting
amplitude soft anomalous dimensions for  $m$ particles becoming 
collinear can be used as consistency checks for higher order 
multi-leg amplitude calculations.  


\section{Multi-particle collinear limits: massive case}
\label{sec:massive}

We now extend our considerations beyond scattering 
amplitudes with only massless external partons and consider amplitudes
with $n$ massless partons -- out of which $m$ final-state ones become collinear -- and one additional massive coloured particle, in either the initial or final state. We recall that, as in the purely massless case considered in the previous section, the amplitude may contain additional non-coloured particles (these do not change the long-distance singularities, and hence do not influence our considerations). 
The relevant soft anomalous dimension for the one-mass case is given by
the sum of the terms in~Eqs.~\eqref{eq:adm-param} and~\eqref{eq:adm-massive}.
We recall that the parametrisation in~\Eqn{eq:adm-massive} of soft anomalous dimension contributions associated with a single massive particle and any number of massless ones, first presented in~\cite{Liu:2022elt}, involves two new terms, with two associated kinematically-dependent functions,
$F_{{\rm{h}}2}$ and $F_{{\rm{h}}3}$. These describe, respectively, soft singularities emanating from the interaction between the massive particle and two and three massless ones. The former, which depends on a single cross ratio~$r_{ijI}$, defined in Eq.~(\ref{eq:r-kin}),
and was computed in Ref.~\cite{Liu:2022elt}, while the latter, which depends on three independent cross ratios, is the final ingredient needed for the three-loop soft anomalous dimension in the one-mass case (with any number of massless partons) was computed very recently~\cite{GZ-TBP}.

The expression for the two-particle collinear splitting amplitude  
anomalous dimension in this case has also been obtained in Ref.~\cite{Liu:2022elt}.
Moreover, the same paper determined constraints due to strict collinear factorisation in the two-particle collinear limit
and due to the small-mass expansion, which amount to 
relations between the three-loop functions $F_{{\rm{h}}2}$ and $F_{{\rm{h}}3}$ 
and their massless counterparts $f$ and ${\cal F}$ of Eqs.~(\ref{eq:Gammaf}) and~(\ref{eq:GammaF}). 

In this section, our first aim is to determine the splitting amplitude soft
anomalous dimension for both the two- and three-particle collinear limits, for amplitudes involving a single massive particle. In this process we reproduce the results of~\cite{Liu:2022elt} for the constraints that arise in the two-particle collinear limit and obtain new constraints for the three-particle collinear limit.
Secondly, we wish to understand the detailed mechanism responsible for the validity
of strict collinear factorisation. It is interesting to utilise the methods
developed for the massless case and explore any differences and necessary adaptations
required in the case with one additional massive leg. 

To keep the analysis as compact as possible, we define the combination of the massless and massive anomalous dimensions 
terms starting at three loops as
\begin{subequations}
\label{quardupoleABdefshv}
  \bea 
{\bf A}_{n+I}(\{\beta\},\{r\}) &=& 4
\sum_{1\leq i<j<k<l \leq n} \, {\bf a}_{ijkl} (\{\beta\})
+2 \sum_{1\leq i< j< k \leq n} {\bf a}^h_{ijkI}(\{r\}), 
\label{quardupoledefshvA}
\\ 
\label{quardupoleAdefshvB}
{\bf B}_{n+I}(\{r\}) &=& 2f(\alpha_s)    
\sum_{i=1}^n \sum_{\substack{1\leq j < k \leq n,\\ j,k\neq i}} 
\,\mathcal{T}_{iijk}
+ 2   \sum_{1\leq i< j \leq n} {\mathcal{T}}_{IIij}
F_{{\rm{h}}2}\left(r_{ijI},\alpha_s \right),
\eea  
\end{subequations}
where  ${\bf a}_{ijkl} (\{\beta\})$ 
are given in~\Eqn{eq:quadrupole-ab-defshu}
and ${\bf a}^h_{ijkI}(\{r\})$ in~\Eqn{eq:Massive-aijkI}.

Analogously to \Eqn{GammaSPdef}, the three-loop $m$-particle collinear splitting amplitude
soft anomalous dimension is given by the difference of the soft anomalous dimensions of two amplitudes: the one for $n$ massless plus one heavy particle 
and the one for $(n-m+1)$ massless plus one heavy particle. These are given by~\Eqn{quardupoleABdefshv}
that capture the interactions of both massive and massless particles
\begin{eqnarray}
\label{eq:GammaSpQuadmMassive} 
{\bf{ \Gamma}}_{\SP, m}^{4{\rm{T}}}(p_1,\ldots p_m;\mu_f) &=& 
{\bf A}_{n+I}(\{\beta\},\{r\})
+{\bf B}_{n+I}(\{r\})
\nonumber \\  && \hspace{-1.4cm} 
-\Big( {\bf A}_{n-m+1+I}(\{\beta\},\{r\})
+{\bf B}_{n-m+1+I}(\{r\}) \Big) 
\Big|_{\T_P\to \sum_{i=1}^m \T_i}.
\end{eqnarray}
Since the massive particle cannot become part of the collinear
subset, we expect that the results for ${\bf{ \Gamma}}_{\SP, m}^{4{\rm{T}}}$
derived here will coincide with the ones obtained in the massless section. However, the mechanism how this happens is interesting to explore, in particular since contributions from massive coloured particles are linked with the purely massless ones through colour conservation, which was a key element proving the universality of the splitting amplitude anomalous dimension in the purely massless case. We will see that indeed, new constraints emerge on the interactions between the massive particle and the massless ones through colour conservation, as soon as strict collinear factorisation is assumed.

We start in Section~\ref{sec:two-particle-collinear-massive}
by reviewing the considerations and  
discussing the two-particle collinear limit results.
The derivation of the splitting amplitude soft anomalous 
dimension for the three-particle collinear limit is 
performed in Section~\ref{sec:three-particle-collinear-massive}.

\subsection{Two-particle collinear limit}
\label{sec:two-particle-collinear-massive}
We start as in the massless case, choosing without loss of generality 
partons 1 and 2 to make up the collinear subset out of the $n$ massless 
and one massive coloured particles of the amplitude. The two collinear 
particles are parametrised using~\Eqn{eq:coll-momenta} with $m=2$.
We compute the two-particle collinear splitting amplitude according 
to~\Eqn{eq:GammaSpQuadmMassive} with $m=2$. Starting with the 
terms that contain $F_{{\rm{h}}3}$, we split the sum over 
both of the massless and massive particles in the amplitudes 
in the following way

\begin{subequations} \label{AtermsHybridb}
\bea \label{AnHybridb} \nn
&&\hspace{-0.5cm}{\bf A}_{n+I}(\{\beta\},\{r\}) = 4 \bigg[\,
\textcolor{orange}{\sum_{3\leq i<j<k<l \leq n} 
{\bf a}_{ijkl}}
+\hspace{-0.1cm}\textcolor{red}{\sum_{3\leq j<k<l \leq n} 
{\bf a}_{1jkl}} 
+\hspace{-0.1cm}\textcolor{red}{\sum_{3\leq j<k<l \leq n} 
{\bf a}_{2jkl}}
+\hspace{-0.1cm}\sum_{3\leq k<l \leq n}  {\bf a}_{12kl}^{ }\bigg]
\nonumber \\[0.1cm] && \hspace{2.5cm}
+\, 2 \bigg[\, \textcolor{orange}{    \sum_{3\leq i< j< k \leq n}
{\bf a}^h_{ijkI}} +\hspace{-0.1cm} \textcolor{red}{\sum_{3\leq   j< k \leq n} {\bf a}^h_{1jkI}}
+ \hspace{-0.1cm}\textcolor{red}{   \sum_{3\leq   j< k \leq n} {\bf a}^h_{2jkI} } 
+ \sum_{3\leq   k \leq n} {\bf a}^h_{12kI} \bigg],
\\[0.2cm] \label{Anm1Hybridb}
&&\hspace{-0.5cm}{\bf A}_{n-1+I}(\{\beta\},\{r\}) = 
4 \bigg[\,\textcolor{orange}{\sum_{3\leq i<j<k<l \leq n} 
{\bf a}_{ijkl}^{ }}
+\textcolor{red}{\sum_{3\leq j<k<l \leq n}
{\bf a}_{Pjkl}^{ }} \bigg]
\nonumber \\ && \hspace{2.9cm}+ \,  2 \bigg[ \,
\textcolor{orange}{\sum_{3\leq i< j< k \leq n}
{\bf a}^h_{ijkI}} +\textcolor{red}{\sum_{3\leq   j< k \leq n}{\bf a}^h_{PjkI} } \bigg] ,
\eea
\end{subequations} 
where we have suppressed the arguments on the right-hand side to keep
the focus on the structure. 
We have marked in \textcolor{orange}{orange} the terms that 
cancel each other immediately when taking the difference 
as they only depend on the rest-of-the-process partons. 
Furthermore, upon using 
rescaling invariance and the fact that $\T_P = \T_1 + \T_2$, we observe that 
the terms marked in  \textcolor{red}{red} cancel each other. These considerations lead to a simple expression for the difference of the above two equations
\be \label{deltaA-twocollHybridb}
{\bf A}_{n+I}(\{\beta\},\{r\}) - {\bf A}_{n-1+I}(\{\beta\},\{r\})\Big|_{p_1||p_2} =
4 \sum_{3\leq k<l \leq n} \, {\bf a}_{12kl}
+ 2   \sum_{3\leq   k \leq n} {\bf a}^h_{12kI},
\ee
We now turn our attention to the ${\bf B}_{n+I}(\{r\})$ terms in~\Eqn{quardupoleABdefshv}. 
In this case, we again use the method of splitting the sums contained within these contributions to find the following starting expressions 
\begin{subequations}  \label{BtermsHybridb}
\bea  \label{BnHybridb} 
{\bf B}_{n+I}(\{r\}) &=& 2  f(\alpha_s)  \Bigg\{ 
\textcolor{orange}{
\sum_{i = 3}^{n}\sum_{\substack{3\leq j<k \leq n \\ j,k\neq i}}
\, {\cal T}_{iijk}}
+\textcolor{blue}{
\sum_{ 3\leq j<k \leq n  } \Big(
{\cal T}_{11jk} +{\cal T}_{22jk}  \Big)} \\[-0.1cm] 
&&\hspace{2.0cm}
+\,\textcolor{red}{\sum_{i = 3}^{n}\sum_{\substack{k = 3, \\ k\neq i}}^{n} 
\Big({\cal T}_{ii1k} +{\cal T}_{ii2k} \Big)}
+\sum_{i = 3}^{n} \Big( 
{\cal T}_{112i} + {\cal T}_{221i} + 
\textcolor{black}{{\cal T}_{ii12} } \Big) \Bigg\} 
\nonumber  \\[+0.1cm] && \hspace{-0.1cm}
+\, {\textcolor{orange}{2   \sum_{3\leq i< j \leq n}
{\mathcal{T}}_{IIij} F_{{\rm{h}}2}\left(r_{ijI},\alpha_s \right)}}
\,+\, {\textcolor{red}{    2  \sum_{3\leq j \leq n}
{\mathcal{T}}_{II1j}F_{{\rm{h}}2}\left(r_{1jI},\alpha_s \right) }}
\nonumber \\ &&  \hspace{-0.1cm}
+\, {\textcolor{red}{   2   \sum_{3\leq j \leq n}
{\mathcal{T}}_{II2j}F_{{\rm{h}}2}\left(r_{2jI},\alpha_s \right)}}
\,+\, 2{\mathcal{T}}_{II12}
F_{{\rm{h}}2}\left(r_{12I},\alpha_s \right), \nn
\\[0.2cm] \label{Bnm1Hybridb}
{\bf B}_{n-1+I}(\{r\}) &=& 2 f(\alpha_s)\Bigg\{
\textcolor{orange}{\sum_{i = 3}^{n}
\sum_{\substack{3\leq j<k \leq n \\ j,k\neq i}} \, {\cal T}_{iijk}}
\, +\,\textcolor{blue}{
\sum_{\substack{3\leq j<k \leq n }} {\cal T}_{PPjk}  }
+\textcolor{red}{\sum_{i = 3}^{n} 
\sum_{\substack{k = 3, \\ k\neq i}}^{n} {\cal T}_{iiPk}}\Bigg\} 
 \\ && \nn \hspace*{2cm}
+\, {\textcolor{orange}{  2   \sum_{3\leq i< j \leq n}
{\mathcal{T}}_{IIij} F_{{\rm{h}}2}\left(r_{ijI},\alpha_s \right) }}
+ {\textcolor{red}{  2   \sum_{3\leq j \leq n}{\mathcal{T}}_{IIPj}
F_{{\rm{h}}2}\left(r_{PjI},\alpha_s \right) }}.
\eea
\end{subequations}
We use the same colour coding as above, so in the difference
the \textcolor{orange}{orange} and \textcolor{red}{red} contributions cancel each other exactly. The terms in \textcolor{blue}{blue}, for which there is partial cancellation, have already been considered in the massless case, so we can directly use \Eqn{eq:teal-twopartcol} to arrive at 
\bea \label{deltaB-twocollHybridb} \nn
{\bf B}_{n+I}(\{r\}) - {\bf B}_{n-1+I}(\{r\})\Big|_{p_1||p_2} &=&
2 f(\alpha_s)  \bigg[  \textcolor{blue}{-\sum_{3\leq j<k \leq n} \Big( {\cal T}_{12jk} +{\cal T}_{12kj}\Big) }
\\ &&\hspace{-1.8cm}
+\, \sum_{i=3}^{n} \Big( 
\textcolor{black}{{\cal T}_{ii12}} + {\cal T}_{112i} + {\cal T}_{221i} \Big)\bigg]
+ 2{\mathcal{T}}_{II12} F_{{\rm{h}}2}\left(r_{12I},\alpha_s \right).
\eea
Upon summing~Eqs.~\eqref{deltaA-twocollHybridb} and \eqref{deltaB-twocollHybridb} 
according to \Eqn{eq:GammaSpQuadmMassive} with $m=2$
we obtain the two-particle collinear
splitting amplitude soft anomalous dimension, which reads
\bea\label{GammaSpQuad2eHybridb} \nn 
&& {\bf{ \Gamma}}_{\SP, 2}^{4{\rm{T}}}(p_1,p_2;\mu_f)
= \hspace{-0.1cm}\sum_{3\leq k<l \leq n} \bigg[
4 {\cal F}^{\rm A}_{12kl}(\{\beta\}) \,{\cal T}_{1kl2} +
\Big( 4 {\cal F}^{\rm S}_{12kl}(\{\beta\}) \textcolor{blue}{ -2f(\alpha_s)} \Big) \Big( 
\textcolor{blue}{{\cal T}_{12lk} + {\cal T}_{12kl}} \Big)  \bigg] 
\nn \\  &&\hspace{1.5cm}
+ 2f(\alpha_s) \sum_{i=3}^{n} \left(
\textcolor{black}{{\cal T}_{ii12}} 
+ {\cal T}_{112i} + {\cal T}_{221i} \right)
+ 2 \sum_{3\leq   k \leq n}
{\bf a}^h_{12kI} + 2 {\mathcal{T}}_{II12}F_{{\rm{h}}2}\left(r_{12I},\alpha_s \right),
\eea
where we have made use of the definition in~\Eqn{eq:quadrupole-ab-defshu}
for the massless kinematic functions. We see that in addition to the appearance
of the massive terms, the massless structures
encountered in~\Eqn{GammaSpQuad2eHybrid} are reproduced. For the latter,
we again need to form the stuffle product in \Eqn{eq:stuffle-2.p.c.c}
such that the usual massless two-particle collinear 
constraint in~\Eqn{eq:twocoll-constraint1c-main} naturally reappears here. 
Implementing the latter as well as~\Eqn{eq:twocoll-constraint2Hybrid-main}, 
we find:
\bea\label{eq:GammaSpQuad2-1-M} \nn 
&&{\bf{ \Gamma}}_{\SP, 2}^{4{\rm{T}}}(p_1,p_2;\mu_f)
=f(\alpha_s)\sum_{k=3}^{n}\sum_{l=3}^{n} 
\Big( {\cal T}_{12lk}
+ {\cal T}_{12kl} \Big)  
+ 2 f(\alpha_s) \sum_{i=3}^{n} \left(  {\cal T}_{112i} + {\cal T}_{221i} \right)
\nn \\  &&\hspace{3.5cm}
+ \, 2   \sum_{3\leq   k \leq n}
{\bf a}^h_{12kI} + 2 {\mathcal{T}}_{II12}F_{{\rm{h}}2}\left(r_{12I},\alpha_s \right).
\eea
After this step, we apply colour conservation. Care is needed, since now 
colour conservation includes the massive particle. Therefore, the results
of application of colour conservation change with respect to the massless case. 
For instance, instead of Eqs.~\eqref{eq:T-2-colcons1} and \eqref{eq:T-2-colcons2}, we have
\begin{eqnarray}\label{eq:T-2-colcons1-M}
\sum_{k=3}^{n}\sum_{l=3}^{n}  \mathcal{T}_{12kl} 
&\stackrel{M.c.c}{=}&-  \frac{1}{8}C_A^2\T_1\cdot \T_2-\mathcal{T}_{1122}
- {\cal T}_{12II}  -\sum_{k=3}^n \left( {\cal T}_{12kI} +  {\cal T}_{21kI}  \right)\,,
\end{eqnarray}
and
\bea\label{eq:T-2-colcons2-M}
\sum_{i=3}^{n}  \, {\cal T}_{112i} &\stackrel{M.c.c}{=}&
-\frac{1}{8}C_A^2\T_1\cdot \T_2 - \mathcal{T}_{1122} - \mathcal{T}_{112I}\,,
\eea
where the $M.c.c$ label in the above equations 
serves as a reminder that this relation 
holds for the application of colour conservation
for the case where the amplitude contains one
massive coloured particle. 
After using the two identities (\ref{eq:T-2-colcons1-M}) and (\ref{eq:T-2-colcons2-M}) in 
\Eqn{eq:GammaSpQuad2-1-M}, we still end up with a redundant set of colour operators. To bring it to a proper basis we make use of the identity 
\begin{eqnarray}
\label{eq:M-col-basis}
{\cal T}_{12II} = \frac{1}{2}\left( 
{\cal T}_{221I} +  {\cal T}_{112I}  
\right) -\frac{1}{2}\sum_{k=3}^n \left( 
{\cal T}_{12kI} +  {\cal T}_{21kI}  
\right),
\end{eqnarray}
to remove the ${\cal T}_{221I}$ and ${\cal T}_{112I}$ colour structures in favour of ones which have no repeated index of light partons. The result for the splitting amplitude soft anomalous dimension in this basis then becomes:
\bea\label{eq:GammaSpQuad2-2-M} \nn 
&&   {\bf{ \Gamma}}_{\SP, 2}^{4{\rm{T}}}(p_1,p_2;\mu_f)
= -\frac{3}{4} f(\alpha_s)  \Big(C_A^2\T_1\cdot \T_2+8\mathcal{T}_{1122} \Big)
\hspace{-0.1cm} -4 f(\alpha_s)\hspace{-0.1cm}\sum_{k=3}^n
\left( {\cal T}_{12kI} +  {\cal T}_{21kI}  \right)
\nn \\  && \hspace{2.6cm}+ \,
2   \sum_{   k=3 }^n {\mathcal{T}}_{12kI}
F_{{\rm{h}}3}\left(r_{12I},r_{1kI},r_{2kI},\alpha_s \right)
+  2   \sum_{   k=3 }^n {\mathcal{T}}_{21kI}
F_{{\rm{h}}3}\left(r_{21I},r_{2kI},r_{1kI},\alpha_s \right)
\nn \\  &&\hspace{2.6cm}
 -\, 6f(\alpha_s) {\cal T}_{12II} 
 +\, 2 {\mathcal{T}}_{12II}F_{{\rm{h}}2}\left(r_{12I},\alpha_s \right),
\eea
where we have reinstated the explicit form of ${\bf a}^h_{12kI}$
from~\Eqn{eq:Massive-aijkI}, already removing the term 
$F_{{\rm{h}}3}\left(r_{k2I},r_{k1I},r_{21I},\alpha_s \right)$,
which vanishes by itself in the
$1||2$ collinear limit due to the antisymmetry\footnote{The symmetry properties of $F_{{\rm{h}}3}$ can be read off from~\Eqn{eq:Massive-aijkI} using those of the colour factor in~\Eqn{Tauijkl}.} of $F_{{\rm{h}}3}$ with respect to swapping its first two arguments, and the fact that these two arguments become identical in this limit.

Having brought~\Eqn{eq:GammaSpQuad2-2-M} to a basis of independent colour structures, we can now impose the requirement of strict collinear factorisation: structures that carry colour or kinematic information on the rest of the process must conspire to cancel. This leads to two independent constraints on 
the $F_{{\rm{h}}2}$ and $F_{{\rm{h}}3}$ functions:
\begin{eqnarray}\label{eq:2pc-Massive-constraints}
F_{{\rm{h}}2}\left(r_{12I},\alpha_s \right)\Big|_{p_1|| p_2} = 3 f(\alpha_s)\,,
\quad \qquad
F_{{\rm{h}}3}\left(r_{12I},r_{1kI},r_{2kI},\alpha_s \right)\Big|_{p_1|| p_2}
= 2 f(\alpha_s)\,,
\end{eqnarray}
The constraints in \Eqn{eq:2pc-Massive-constraints} agree with the ones reported
in Eq.~(9) of Ref.~\cite{Liu:2022elt}.
It is evident
that only the first term in \Eqn{eq:GammaSpQuad2-2-M} survives, and it reproduces the result for the two-particle collinear splitting amplitude soft anomalous dimension
obtained previously, see~\Eqn{GammaSpQuad2dN}. 
In conclusion, we have seen that, subject to Eq.~(\ref{eq:2pc-Massive-constraints}), the universality property of the 
splitting function soft anomalous dimension in the two particle collinear limit
holds independently of the features of the rest of the process, including whether or not it contains a heavy coloured particle. 

\subsection{Three-particle collinear limit}
\label{sec:three-particle-collinear-massive}
We now turn our attention to the three-particle collinear limit.
We take partons~1, 2, and 3 as the ones becoming collinear
and parametrise them using~\Eqn{eq:coll-momenta}.
The three-particle collinear splitting amplitude is defined according 
to~\Eqn{eq:GammaSpQuadmMassive} with \hbox{$m=3$}. We start our
considerations with the ${\bf A}_{n+I}(\{\beta\},\{r\})$ and 
${\bf A}_{n-2+I}(\{\beta\},\{r\})$ terms, splitting the 
sums in the soft anomalous dimensions similarly to 
the way it is carried out in Eqs.~\eqref{AnHybridb} and 
\eqref{Anm1Hybridb}, but now isolating the three 
collinear partons instead of just a collinear pair. 
Taking the difference, as dictated by~\Eqn{eq:GammaSpQuadmMassive},
we obtain for the ${\bf A}_{n+I}(\{\beta\},\{r\})$ type terms
the following expression 
\begin{eqnarray}  \label{eq:deltaA-threecoll-M}
&& {\bf A}_{n+I}(\{\beta\},\{r\}) - {\bf A}^{ }_{n-2+I}(\{\beta\},\{r\})\Big|_{1||2||3} 
\hspace{-0.1cm}= 4 \bigg[\sum_{3\leq k<l \leq n} \hspace{-0.1cm} {\bf a}_{12kl}
+\sum_{4\leq k<l \leq n} \hspace{-0.1cm} {\bf a}_{13kl} 
+\sum_{4\leq k<l \leq n} \hspace{-0.1cm} {\bf a}_{23kl}
\nn\\  &&  \hspace{2.4cm}
+\textcolor{darkgreen}{
\sum_{4\leq  l \leq n}  {\bf a}_{123l}}\bigg] 
+ 2 \bigg[    \,\, \sum_{4\leq   k \leq n} {\bf a}^h_{12kI} 
+ \sum_{4\leq   k \leq n} {\bf a}^h_{13kI}
+ \sum_{4\leq   k \leq n} {\bf a}^h_{23kI}
+ \textcolor{darkgreen}{{\bf a}^h_{123I}} \bigg]\,,
\end{eqnarray}
which is analogous to 
\Eqn{deltaA-twocollHybridb} in the two-particle collinear limit.
Similarly for the ${\bf B}_{n+I}(\{r\})$ terms we find in the three-particle collinear
limit
\begin{eqnarray}\label{deltaB-threecollHybridI}
&&{\bf B}_{n+I}(\{r\})-{\bf B}_{n-2+I}(\{r\}) = 
2 f(\alpha_s)  \Bigg\{ 
\textcolor{blue}{
-\hspace{-0.2cm}\sum_{4\leq j<k \leq n} \Big( {\cal T}_{12jk} +{\cal T}_{12kj}\Big)}
\textcolor{blue}{
-\hspace{-0.2cm}\sum_{4\leq j<k \leq n} \Big( {\cal T}_{13jk} +{\cal T}_{13kj}\Big) }
\nn\\[-0.1cm]  &&\hspace{2cm}\,
\textcolor{blue}{
-\sum_{4\leq j<k \leq n} \Big( {\cal T}_{23jk} +{\cal T}_{23kj}\Big) }
+\sum_{i = 4}^{n} \Big({\cal T}_{112i} + {\cal T}_{221i} + {\cal T}_{ii12} 
+ {\cal T}_{113i}+ {\cal T}_{331i} + {\cal T}_{ii13}
\nn \\[-0.1cm] &&\hspace{2cm}  
+ {\cal T}_{223i} + {\cal T}_{332i} + {\cal T}_{ii23}   \Big) 
+ \textcolor{darkgreen}{ {\cal T}_{1123}}
+ \textcolor{darkgreen}{ {\cal T}_{2213}}
+ \textcolor{darkgreen}{ {\cal T}_{3312} }
\Bigg\} \nonumber \\ && \hspace{2cm}  
+ 2\textcolor{black}{ {\mathcal{T}}_{II12}
F_{{\rm{h}}2}\left(r_{12I},\alpha_s \right) }
+2\textcolor{black}{ {\mathcal{T}}_{II13}
F_{{\rm{h}}2}\left(r_{13I},\alpha_s \right) }
+2\textcolor{black}{     {\mathcal{T}}_{II23}
F_{{\rm{h}}2}\left(r_{23I},\alpha_s \right)}\,,
\end{eqnarray} 
which is analogous to 
\Eqn{deltaB-twocollHybridb}. 
In both of the above equations we recognise structures that are similar to those we encountered in the case of the two-particle
collinear limit. Here they emerge for each pair out of the three collinear particles. On top of that,
we have marked in \textcolor{darkgreen}{green} the new
terms appearing for the first time in the three-particle
collinear limit. Summing Eqs.~\eqref{eq:deltaA-threecoll-M}
and \eqref{deltaB-threecollHybridI}
according to \Eqn{eq:GammaSpQuadmMassive}
gives an expression for the three-particle collinear splitting amplitude soft
anomalous dimension. The structure is very close to the expression 
in~\Eqn{GammaSpQuad3eHybrid} obtained in the three-particle collinear
case for massless amplitudes, only with additional $I$ dependent-terms
found in the two equations above. Imposing then two-particle collinear constraints for each pair
out of the three collinear particles, allows us to form the 
stuffle products precisely of the form found in~\Eqn{eq:stuffleGen}. 
Carrying out these steps yields the following expression
\bea\label{GammaSpQuad3eHybrids2I} \nn 
&&\hspace{-0.5cm}{\bf{ \Gamma}}_{\SP, 3}^{4{\rm{T}}}(p_1,p_2,p_3;\mu_f)
=  f(\alpha_s) \sum_{k=4}^{n}\sum_{l=4}^{n} 
\big( {\cal T}_{12lk} +  {\cal T}_{12kl} \big)
+f(\alpha_s)   \sum_{k=4}^{n}\sum_{l=4}^{n} 
\big( {\cal T}_{13lk} +  {\cal T}_{13kl} \big)  
\nn \\ && \hspace{3.1cm}
+f(\alpha_s)   \sum_{k=4}^{n}\sum_{l=4}^{n} 
\big( {\cal T}_{23lk} +  {\cal T}_{23kl} \big)
+ 2f(\alpha_s) \bigg[ 
\sum_{i = 4}^{n} \Big( 
{\cal T}_{112i} + {\cal T}_{221i}  
\nn \\[-0.1cm]  &&\hspace{4.1cm}
+ {\cal T}_{113i}+ {\cal T}_{331i}
+  {\cal T}_{223i} + {\cal T}_{332i}    \Big)   
+ \textcolor{darkgreen}{ {\cal T}_{1123}}
+ \textcolor{darkgreen}{{\cal T}_{2213}} 
+ \textcolor{darkgreen}{{\cal T}_{3312} }
\bigg]
\nonumber \\ && \hspace{1.7cm}
+4\textcolor{darkgreen}{
\sum_{4\leq  l \leq n}  {\bf a}_{123l}}
+ 2   \sum_{4\leq   k \leq n}  {\bf a}^h_{12kI}
+ 2\sum_{4\leq   k \leq n} {\bf a}^h_{13kI}
+ 2  \sum_{4\leq   k \leq n} {\bf a}^h_{23kI}
+ 2\, \textcolor{darkgreen}{{\bf a}^h_{123I}} 
\nonumber \\ &&   \hspace{1.7cm} 
+2\,\textcolor{black}{ {\mathcal{T}}_{II12}
F_{{\rm{h}}2}\left(r_{12I},\alpha_s \right) }
+ 2\,\textcolor{black}{ {\mathcal{T}}_{II13}
F_{{\rm{h}}2}\left(r_{13I},\alpha_s \right) }
+ 2\,\textcolor{black}{ {\mathcal{T}}_{II23}
F_{{\rm{h}}2}\left(r_{23I},\alpha_s \right)}\,.
\eea
In the next step, we apply colour conservation
to the massless terms with double sums and the ones with a single index. 
As in the case of the two-particle collinear limit calculation above,
the results for application of colour conservation differs from 
the corresponding massless ones in Eqs.~\eqref{eq:calT-AABi-3pc} 
and \eqref{eq:mcolcons-triplecollinear-2a}
due to the presence of the coloured massive particle. 
The relevant colour-conservation relations are
\begin{eqnarray}
\sum_{i=4}^{n}  \, {\cal T}_{aabi}  &\stackrel{M.c.c}{=}&
- \frac{1}{8}C_A^2 \T_a\cdot \T_b -\mathcal{T}_{aabb} 
- \mathcal{T}_{aabc}- \mathcal{T}_{aabI}\,,
\end{eqnarray}
and
\begin{equation}\label{eq:stuffle-M.c.c-Gen}
\sum_{k=4}^{n}\sum_{l=4}^{n}  \mathcal{T}_{abkl}
\stackrel{M.c.c}{=}
- \frac{1}{8}C_A^2 \T_a\cdot \T_b- {\cal{T}}_{aabb} 
- {\cal T}_{aabc} + {\cal T}_{abcc} - {\cal T}_{bbac}   
-  {\cal T}_{abII} -\sum_{k=4}^n \left( 
{\cal T}_{abkI} +  {\cal T}_{bakI}  \right)\,,
\end{equation}
where $\{a,b,c\}$ can be replaced by any of the three permutations of the collinear partons $\{1,2,3\}$.
As before, the~$M.c.c.$ label denotes that the colour conservation
equality holds in the presence of a massive particle. Compared to 
Eqs.~\eqref{eq:T-2-colcons1-M} and \eqref{eq:T-2-colcons2-M} we note the
appearance of terms depending on the third particle becoming collinear,
since the sums on the left-hand side now start from the fourth particle. 
Implementing these, we find
\bea\label{GammaSpQuad3eHybrids2Iv2-3} \nn 
&&{\bf{ \Gamma}}_{\SP, 3}^{4{\rm{T}}}(p_1,p_2,p_3;\mu_f)
= - \frac{3}{4}f(\alpha_s)\Big(C_A^2 \T_1\cdot \T_2+ 8{\cal{T}}_{1122} \Big)
- \frac{3}{4}f(\alpha_s)\Big(C_A^2 \T_1\cdot \T_3+ 8{\cal{T}}_{1133} \Big)
\nn \\ &&  \hspace{3.2cm}
- \frac{3}{4}f(\alpha_s)\Big( C_A^2 \T_2\cdot \T_3+8{\cal{T}}_{2233} \Big)
- 4f(\alpha_s) \Big[ \textcolor{darkgreen}{ {\cal T}_{1123}}
+ \textcolor{darkgreen}{ {\cal T}_{2213}} 
+ \textcolor{darkgreen}{ {\cal T}_{3312}} \Big]
\nn \\  && \hspace{3.2cm}
- 2f(\alpha_s)\bigg[\mathcal{T}_{112I}+\mathcal{T}_{221I} 
+{\cal T}_{12II}+\sum_{k=4}^n\left({\cal T}_{12kI}+{\cal T}_{21kI}\right)\bigg]
\nn \\ && \hspace{3.2cm}
- 2f(\alpha_s) \bigg[\mathcal{T}_{113I} + \mathcal{T}_{331I} 
+ {\cal T}_{13II}+\sum_{k=4}^n\left({\cal T}_{13kI}+{\cal T}_{31kI}\right) \bigg]
\nn \\ && \hspace{3.2cm}
- 2f(\alpha_s) \bigg[\mathcal{T}_{223I} + \mathcal{T}_{332I}
+ {\cal T}_{23II}+\sum_{k=4}^n\left({\cal T}_{23kI}+{\cal T}_{32kI}\right)\bigg]
\nonumber \\ &&  \hspace{1.8cm}     
+4\textcolor{darkgreen}{
\sum_{4\leq  l \leq n}  {\bf a}_{123l}} 
+ 2 \bigg[    \,\,
\sum_{4\leq   k \leq n} {\bf a}^h_{12kI}
+\sum_{4\leq   k \leq n} {\bf a}^h_{13kI}
+\sum_{4\leq   k \leq n}{\bf a}^h_{23kI}
+\textcolor{darkgreen}{{\bf a}^h_{123I}} \bigg]
\nonumber \\ && \hspace{1.8cm}  
+2\,\textcolor{black}{ {\mathcal{T}}_{II12}
F_{{\rm{h}}2}\left(r_{12I},\alpha_s \right) }
+ 2\,\textcolor{black}{ {\mathcal{T}}_{II13}
F_{{\rm{h}}2}\left(r_{13I},\alpha_s \right) }
+ 2\,\textcolor{black}{ {\mathcal{T}}_{II23}
F_{{\rm{h}}2}\left(r_{23I},\alpha_s \right)} \,.
\eea
We see that already here, the massless two-particle collinear structures
involving pairs of collinear particles and the constant parts of the massless three-particle
collinear expression have emerged (\emph{cf.}~\Eqn{eq:m3partcol-fin}). However, \Eqn{GammaSpQuad3eHybrids2Iv2-3} is still written in terms of a redundant set of colour structures. 
We proceed by using
the same colour-conservation identity as in \Eqn{eq:M-col-basis} to eliminate the terms with repeated indices of  collinear partons, terms of the type 
${\cal T}_{221I}$ and ${\cal T}_{112I}$. In the present context we must extract from the sum over $k$ in~\Eqn{eq:M-col-basis} the term corresponding to the third collinear parton, so we get 
\begin{eqnarray}
\label{IdentityForThreeCollinear}
\left( {\cal T}_{221I} +  {\cal T}_{112I}  \right) =
2  {\cal T}_{12II} + \big({\cal T}_{123I} +  {\cal T}_{213I} \big)
+\sum_{k=4}^n \left( {\cal T}_{12kI} +  {\cal T}_{21kI}   \right).
\end{eqnarray}
Two similar identities can be derived by permutations
of the set of collinear partons $(1,2,3)$ in \Eqn{IdentityForThreeCollinear}.
Using these three relations, and writing out the explicit forms of~${\bf a}^h_{12kI}$, ${\bf a}^h_{13kI}$, and
${\bf a}^h_{23kI}$ according to~\Eqn{eq:Massive-aijkI}, we find the following expression
\bea\label{GammaSpQuad3eHybrids2Iv2-5} \nn 
&&{\bf{ \Gamma}}_{\SP, 3}^{4{\rm{T}}}(p_1,p_2,p_3;\mu_f)
= - \frac{3}{4}f(\alpha_s)\Big(C_A^2 \T_1\cdot \T_2+ 8{\cal{T}}_{1122} \Big)
- \frac{3}{4}f(\alpha_s)\Big(C_A^2 \T_1\cdot \T_3+ 8{\cal{T}}_{1133} \Big)
\nn \\ &&  \hspace{3.9cm}
- \frac{3}{4}f(\alpha_s)\Big( C_A^2 \T_2\cdot \T_3+8{\cal{T}}_{2233} \Big)
- 4f(\alpha_s) \Big[ \textcolor{darkgreen}{ {\cal T}_{1123}}
+ \textcolor{darkgreen}{ {\cal T}_{2213}} 
+ \textcolor{darkgreen}{ {\cal T}_{3312}} \Big]
\nn \\  && \hspace{2.0cm} 
\textcolor{red}{-6f(\alpha_s)   {\cal T}_{12II}} 
\textcolor{orange}{
 -\,\, 4 f(\alpha_s) \sum_{k=4} \left( 
{\cal T}_{12kI} +
{\cal T}_{21kI}  \right)}
-  2f(\alpha_s) \big(\textcolor{teal}{{\cal T}_{123I}} +  \textcolor{teal}{{\cal T}_{213I}} \big)
\nonumber \\   &&  \hspace{2cm}     
- 6f(\alpha_s) {\cal T}_{13II} -4f(\alpha_s)\sum_{k=4} \left( 
{\cal T}_{13kI} + {\cal T}_{31kI}   \right)   
 - 2f(\alpha_s) \big(\textcolor{teal}{{\cal T}_{132I}} + \textcolor{teal}{ {\cal T}_{312I}} \big) 
 \nonumber \\    &&  \hspace{2cm}     
- 6f(\alpha_s)   {\cal T}_{23II} -4f(\alpha_s)  \sum_{k=4} \left( 
{\cal T}_{23kI} +  {\cal T}_{32kI}   \right)
-  2f(\alpha_s) \big(\textcolor{teal}{{\cal T}_{231I} }+  \textcolor{teal}{ {\cal T}_{321I} }\big) 
\nn \\ && \hspace{0.8cm}     
+\, \textcolor{orange}{ 2   \sum_{4\leq   k \leq n}  {\mathcal{T}}_{12kI}
F_{{\rm{h}}3}\left(r_{12I},r_{1kI},r_{2kI},\alpha_s \right)}
+ \textcolor{orange}{ 2   \sum_{4\leq   k \leq n} {\mathcal{T}}_{21kI}
F_{{\rm{h}}3}\left(r_{21I},r_{2kI},r_{1kI},\alpha_s \right)}
\nn \\ && \hspace{0.8cm}     
+ \, 2   \sum_{4\leq   k \leq n} {\mathcal{T}}_{13kI}
F_{{\rm{h}}3}\left(r_{13I},r_{1kI},r_{3kI},\alpha_s \right)
+  2   \sum_{4\leq   k \leq n} {\mathcal{T}}_{31kI}
F_{{\rm{h}}3}\left(r_{31I},r_{3kI},r_{1kI},\alpha_s \right)
\nn \\ && \hspace{0.8cm}     
+ \,  2   \sum_{4\leq   k \leq n} {\mathcal{T}}_{23kI}
F_{{\rm{h}}3}\left(r_{23I},r_{2kI},r_{3kI},\alpha_s \right)
+  2   \sum_{4\leq   k \leq n} {\mathcal{T}}_{32kI}
F_{{\rm{h}}3}\left(r_{32I},r_{3kI},r_{2kI},\alpha_s \right)
\nonumber \\ && \hspace{0.8cm}  
+\,4\textcolor{darkgreen}{
\sum_{4\leq  l \leq n}  {\bf a}_{123l}} 
+ 2\,\textcolor{darkgreen}{ {\bf a}^h_{123I}}  
+\textcolor{red}{ 2\,{\mathcal{T}}_{II12}
F_{{\rm{h}}2}\left(r_{12I},\alpha_s \right) }
+ 2\,\textcolor{black}{ {\mathcal{T}}_{II13}
F_{{\rm{h}}2}\left(r_{13I},\alpha_s \right) }
\nonumber \\ && \hspace{0.8cm} 
+ \,2\,\textcolor{black}{ {\mathcal{T}}_{II23}
F_{{\rm{h}}2}\left(r_{23I},\alpha_s \right)} \,.
\eea
In the above, we can identify the two independent 
structures appearing in the derivation two-particle
collinear limit splitting amplitude in the massive
case in~\Eqn{eq:GammaSpQuad2-2-M}, which we have 
marked respectively in~\textcolor{red}{red} and~\textcolor{orange}{orange} for particles 
1 and 2 as an example. These parts recover
the constraints found in~\Eqn{eq:2pc-Massive-constraints}
for each pair out of the three particles becoming collinear.
We can immediately implement these constraints which we
do in the following. It is interesting to focus on the 
new type of constant term that contains dependence on
all three collinear particles as well as the massive particle, which we marked in 
\textcolor{teal}{blue}.
We notice that these terms are spurious and in fact cancel among themselves
as ${\mathcal{T}_{132I}}=-{\mathcal{T}_{123I}}$,
${\mathcal{T}_{312I}}=-{\mathcal{T}_{321I}}$,
${\mathcal{T}_{231I}}=-{\mathcal{T}_{213I}}$.
We then find 
\bea\label{GammaSpQuad3eHybrids2Iv2-7} \nn 
&&{\bf{ \Gamma}}_{\SP, 3}^{4{\rm{T}}}(p_1,p_2,p_3;\mu_f)
= - \frac{3}{4}f(\alpha_s)\Big(C_A^2 \T_1\cdot \T_2+ 8{\cal{T}}_{1122} \Big)
- \frac{3}{4}f(\alpha_s)\Big(C_A^2 \T_1\cdot \T_3+ 8{\cal{T}}_{1133} \Big)
\nn \\ &&  \hspace{3.9cm}
- \frac{3}{4}f(\alpha_s)\Big( C_A^2 \T_2\cdot \T_3+8{\cal{T}}_{2233} \Big)
- 4f(\alpha_s) \Big[ \textcolor{darkgreen}{ {\cal T}_{1123}}
+ \textcolor{darkgreen}{ {\cal T}_{2213}} 
+ \textcolor{darkgreen}{ {\cal T}_{3312}} \Big]
\nn \\ && \hspace{0.6cm}
+    \,8\,
\sum_{4\leq l \leq n} \Big[ 
\textcolor{darkgreen}{ \mathcal{T}_{13l2} 
\,{\cal F}  (\beta_{132l},\beta_{1l23})} +\textcolor{darkgreen}{ \,\mathcal{T}_{12l3} 
\,{\cal F}  (\beta_{123l},\beta_{1l32}) }
+\,\textcolor{darkgreen}{ \mathcal{T}_{123l} 
\,{\cal F} (\beta_{12l3},\beta_{13l2}) }\Big] 
\\ &&\nonumber \hspace{0.6cm}
+\, 2\textcolor{darkgreen}{    {\mathcal{T}}_{123I}
    F_{{\rm{h}}3}\left(r_{12I},r_{13I},r_{23I} \right)}
+2\textcolor{darkgreen}{   {\mathcal{T}}_{213I}
    F_{{\rm{h}}3}\left(r_{21I},r_{23I},r_{13I} \right)}
+2\textcolor{darkgreen}{      {\mathcal{T}}_{312I}
    F_{{\rm{h}}3}\left(r_{31I},r_{32I},r_{21I}\right)}\,,
\eea
where we have written the kinematically-dependent terms
involving a massive particle using~\Eqn{eq:Massive-aijkI}. Similarly, we have also written out explicitly the 
kinematically-dependent massless terms $ {\bf a}_{123l}$ 
using \Eqn{eq:GammaF}, as was done in \Eqn{eq:m3partcol-pen}. 
Colour conservation can be applied to these terms as before, but
now the resulting expression differs from~\Eqn{eq:3partcol-kin-c.c},
because the massive coloured particle enters through colour conservation. 
Instead of the expression in 
\Eqn{eq:3partcol-kin-c.c}, in the presence of a massive particle we find
\begin{eqnarray} \label{eq:3partcol-kin-M.c.c}
  \sum_{4\leq l \leq n} 
\mathcal{T}_{123l}  
\stackrel{M.c.c. }{=} \mathcal{T}_{2213}  - \mathcal{T}_{3312} -
\mathcal{T}_{123I}.
\end{eqnarray}
Using similar results for the remaining two colour structures to 
implement colour conservation, we arrive at the following result
\bea\label{eq:M-GammaSpQuad3-9} \nn 
&&{\bf{ \Gamma}}_{\SP, 3}^{4{\rm{T}}}(p_1,p_2,p_3;\mu_f)
= - \frac{3}{4}f(\alpha_s)\Big(C_A^2 \T_1\cdot \T_2+ 8{\cal{T}}_{1122} \Big)
- \frac{3}{4}f(\alpha_s)\Big(C_A^2 \T_1\cdot \T_3+ 8{\cal{T}}_{1133} \Big)
\nn \\ &&  \hspace{3.9cm}
- \frac{3}{4}f(\alpha_s)\Big( C_A^2 \T_2\cdot \T_3+8{\cal{T}}_{2233} \Big)
- 4f(\alpha_s) \Big[ \textcolor{black}{ {\cal T}_{1123}}
+ \textcolor{black}{ {\cal T}_{2213}} 
+ \textcolor{black}{ {\cal T}_{3312}} \Big]
\nn \\ && \hspace{2.0cm}
+\,8\, \Big(\mathcal{T}_{1123} -\mathcal{T}_{2213} \Big)
\,{\cal F}  (\beta_{132l},\beta_{1l23}) 
+\,8\,\Big(\mathcal{T}_{1123}  -\mathcal{T}_{3312} \Big) 
\,{\cal F}  (\beta_{123l},\beta_{1l32}) 
\nonumber\\ &&\hspace{2.0cm} \nonumber
+\,8\Big(\mathcal{T}_{2213}  - \mathcal{T}_{3312} \Big)
\,{\cal F} (\beta_{12l3},\beta_{13l2})
\nonumber \\ &&\hspace{2cm}
+ 2{\mathcal{T}}_{123I}\Big(
F_{{\rm{h}}3}\left(r_{12I},r_{13I},r_{23I},\alpha_s \right)
-4{\cal F} (\beta_{12l3},\beta_{13l2}) \Big)
\nonumber \\ && \hspace{2cm}
+ 2{\mathcal{T}}_{213I}\Big( 
F_{{\rm{h}}3}\left(r_{21I},r_{23I},r_{13I},\alpha_s \right)
-4{\cal F}  (\beta_{123l},\beta_{1l32}) \Big)
\nonumber \\ && \hspace{2cm}
+ 2{\mathcal{T}}_{312I} \Big(
F_{{\rm{h}}3}\left(r_{31I},r_{32I},r_{21I},\alpha_s \right)
 - 4  \,{\cal F}  (\beta_{132l},\beta_{1l23})  \Big).
\eea
We see that in the first four lines of this expression we recover
the result for the three-particle collinear splitting soft amplitude anomalous dimension that can be found in~\Eqn{eq:m3partcol-fin}. Indeed, in strict
collinear factorisation, this object must be universal and hence not 
depend on the precise nature of the rest-of-the-process particles, in particular whether they are massless or massive. Therefore, from 
\Eqn{eq:M-GammaSpQuad3-9} we can obtain constraints on the $F_{{\rm{h}}3}$
functions by demanding that the last three lines vanish:
\begin{align}
\label{TripleCollinearMassiveConstraints}
\begin{split}
0\,\,&=\,\,\Bigg\{2{\mathcal{T}}_{123I}\Big(
F_{{\rm{h}}3}\left(r_{12I},r_{13I},r_{23I},\alpha_s \right)
-4{\cal F} (\beta_{12l3},\beta_{13l2}) \Big)
 \\ & \hspace{.7cm}
+ 2{\mathcal{T}}_{213I}\Big( 
F_{{\rm{h}}3}\left(r_{21I},r_{23I},r_{13I},\alpha_s \right)
-4{\cal F}  (\beta_{123l},\beta_{1l32}) \Big)
 \\ &\hspace{.7cm}
\left.+ 2{\mathcal{T}}_{312I} \Big(
F_{{\rm{h}}3}\left(r_{31I},r_{32I},r_{21I},\alpha_s \right)
 - 4  \,{\cal F}  (\beta_{132l},\beta_{1l23})  \Big)\Bigg\}\right\vert_{p_1||p_2||p_3}\,.
\end{split}
\end{align}
Before we formulate the constraints which follow from \Eqn{TripleCollinearMassiveConstraints}, it is useful to 
understand the dependence of the functions on the kinematic
variables. At first glance they may appear incompatible,  since in one case they include only massless CICRs 
and in the second case the $r$ variables given in~\Eqn{eq:r-kin}
with a massive particle index $I$ and two massless collinear particles. 
As described in Section~\ref{sec:kin-three-part-col}, the CICRs carrying
indices of three collinear partons contain rich kinematic structure
of the approach of the three-particle collinear limit. On the other hand,
the variables  $r_{abI}\to 0$ for $a$ and $b$ becoming collinear.  However, 
we see that indeed for the  $F_{{\rm{h}}3}$ functions the correct
parametrisation in this limit is in fact through ratios of the $r_{abI}$ variables, which remain finite.
In general, we have that
\begin{eqnarray}
\label{r_CICRs}
\frac{r_{ijI}}{r_{jkI}} = 
\frac{p_i\cdot p_j \, p_I^2  }{p_i\cdot p_I \, p_j\cdot p_I}
\frac{p_j\cdot p_I \, p_k\cdot p_I}{p_j\cdot p_k \,p_I ^2 }
= \frac{p_i\cdot p_j \,p_I\cdot p_k   }{p_i\cdot p_I \, p_j\cdot p_k }
= \rho_{ijIk} \,,
\end{eqnarray}
and
\begin{eqnarray}
\frac{r_{ikI}}{r_{jkI}}= 
\frac{p_i\cdot p_k \, p_I^2   }{p_i\cdot p_I \, p_k\cdot p_I}
\frac{p_j\cdot p_I \, p_k\cdot p_I}{p_j\cdot p_k \, p_I^2  }
= \frac{p_i\cdot p_k \,p_I\cdot p_j  }{p_i\cdot p_I \,p_j\cdot p_k}
= \rho_{ikIj}\,.
\end{eqnarray}
Therefore, we can express $F_{{\rm{h}}3}$ in terms of alternative variables as follows
\begin{eqnarray} 
    F_{{\rm{h}}3}\left( \beta_{abIc}, \beta_{acIb} \,;r_{bcI} \right)\equiv F_{{\rm{h}}3}\left(r_{abI},r_{acI},r_{bcI}  \right)   
    ,  
\end{eqnarray}
where we used \Eqn{r_CICRs} to relate the variables, rescaling the first two arguments of $F_{{\rm{h}}3}$ by the third, to obtain the usual CICRs and then taking their logarithms.
Since \emph{in the collinear limit}, $r_{bcI}\to 0$,
this allows us to express $F_{{\rm{h}}3}$ using the same two variables as the massless counterpart, and  \Eqn{TripleCollinearMassiveConstraints}
yields the following constraint
\begin{eqnarray}\label{eq:M-3pc-constraint}
 \lim_{p_a||p_b||p_c} 
F_{{\rm{h}}3}\left(r_{abI},r_{acI},r_{bcI}  \right) 
 =
\lim_{r_{bcI}\to 0} F_{{\rm{h}}3}\left( \beta_{abIc}, \beta_{acIb} \,;r_{bcI} \right)   = {4}{\cal F} (\beta_{ablc},\beta_{aclb})\Big\rvert_{p_a||p_b||p_c}\,,
\end{eqnarray}
for cyclic permutations of $(a,b,c) = ( 1,2,3)$. This 
result is the first case where considerations of the multi-particle collinear
limit provide additional constraining information on the functions 
appearing in the soft anomalous dimension beyond what could be 
deduced through the two-particle collinear limit.

We note the three colour factors appearing in the constraint equation \Eqn{TripleCollinearMassiveConstraints} are in fact related by the
Jacobi identity given in~\Eqn{JacobiTau}, which may be written as
\begin{equation}
\label{JacobiI123}
\mathcal{T}_{213I}=
\mathcal{T}_{123I}
+\mathcal{T}_{312I}\,.
\end{equation}
Therefore, since the colour terms are not all independent, the expression in~\Eqn{TripleCollinearMassiveConstraints}
amounts to two rather than three independent constraints on the kinematic functions. In~App.~\ref{sec:Fh3Constraints}, we show that the antisymmetric and symmetric parts of $F_{h3}$, defined respectively in Eqs.~\eqref{eq:Fh3quadrupole-A-main-a} and \eqref{eq:Fh3quadrupole-A-main-b},   are related in the three-particle collinear limit to the respective combinations of the purely massless ${\cal F}$, in \Eqn{quadrupole-AS-defsh}. Namely, the final form of the constraints are obtained in \Eqn{TripleCollinearMassiveConstraintsFinalFormLim}, which we state here for convenience 
\begin{subequations}
\label{eq:TripleCollinearMassiveConstraintsFinalFormLim-main}
    \begin{align}
\lim_{p_1||p_2||p_3}F^{\rm A}_{{\rm{h}}3, {1I23}}\left(\{r\} \right)
&=\left.
4{\cal F}^{\rm A}_{1l23}  (\{\beta\}) \right\vert_{p_1||p_2||p_3}\,,
\\
\lim_{p_1||p_2||p_3}F^{\rm S}_{{\rm{h}}3, {1I23}}\left(\{r\} \right)
&=\left.4{\cal F}^{\rm S}_{1l23}  (\{\beta\})\right\vert_{p_1||p_2||p_3}\,.
\end{align}
\end{subequations}

\subsubsection*{Connection with the small-mass limit}
An interesting connection exists between the three-particle 
collinear limit of soft singularities of amplitudes with a massive external leg 
and the small-mass limit of the same, i.e., taking the
limit where the extra massive leg $I$ becomes lightlike.
Namely, the behaviour of the variables $r_{abI}$ in the
three-particle collinear limit of the soft anomalous dimension is indistinguishable from
the one encountered in considerations of the small-mass expansion
limit. The point is that, albeit for different reasons, in both cases the variables
$r_{abI}\to 0$. We consider now the explicit expression for 
$r_{abI}$ given in~\Eqn{eq:r-kin}.
In the three-particle collinear limit case, the numerator of 
$r_{abI}$ turns to zero due to the vanishing scalar product between the
collinear momenta, $p_a\cdot p_b\to 0$. Conversely, in the small-mass
limit case  the numerator of the $r_{abI}$ variables also turns to 
zero, but this time due to $p_I^2\to 0$.  The resultant 
behaviour of the~variables is indistinguishable from each other.
Therefore, the relation between~$F_{{\rm{h}}3}$
and functions describing massless kinematics $ {\cal F}$
obtained in the three-particle collinear limit and the small-mass
limit must coincide:
\begin{equation}
\label{masslessTripleCollinearlimit}
\lim_{p_a||p_b||p_c} 
F_{{\rm{h}}3}\left(r_{abI},r_{acI},r_{bcI}  \right) =
\lim_{p_I^2\to 0} F_{{\rm{h}}3}\left(r_{abI},r_{acI},r_{bcI}  \right) = {4}{\cal F} (\beta_{ablc},\beta_{aclb})\Big\rvert_{p_a||p_b||p_c}.
\end{equation}
The small-mass expansion has been used to obtain the limit
\[
\lim_{p_I^2\to 0} F_{{\rm{h}}3}\left(r_{abI},r_{acI},r_{bcI}  \right)
\]
in the appendix of Ref.~\cite{Liu:2022elt}. We note that the result obtained in Eq.~(31) there has an additional constant term involving $f(\alpha_s)$. The origin of this superfluous term is that Eq.~(30) is written using an overcomplete set of colour structures, while in fact it can be shown that the terms proportional to $f(\alpha_s)$ all cancel, in an analogous way to the cancellation of the \textcolor{teal}{blue} terms in~\Eqn{GammaSpQuad3eHybrids2Iv2-5} in our triple collinear limit computation. With this accounted for, the considerations of Ref.~\cite{Liu:2022elt} agree with the result in \Eqn{masslessTripleCollinearlimit} above.

\section{Conclusions} 
\label{sec:conclusion}
Considerations of various kinematic limits, such as Regge and collinear limits, 
provide ground for stringent tests of fundamental properties
of scattering amplitudes,
as well as providing data on the functions governing 
the behaviour of the amplitudes themselves. 
At higher perturbative orders
it becomes non-trivial to demonstrate explicitly that fundamental properties, such
as strict collinear factorisation in the timelike collinear limit,
that are expected to hold to all perturbative orders indeed 
do so, due to the increasing complexity
of the functions and colour structures arising at higher loop orders. In this article, we investigated multi-particle collinear limits of multi-leg 
scattering amplitudes at three and four loops. 
The motivation is two-fold. Firstly, to determine 
and explore the precise mechanisms that are 
responsible for the way in which colour and kinematic
degrees of freedom conspire to satisfy the expected strict 
collinear factorisation of scattering amplitudes 
in multi-collinear 
limits at high loop orders. Secondly, making use of 
expected strict collinear factorisation, we studied 
whether multi-particle collinear limits can provide 
additional information on the functions appearing
in the soft anomalous dimensions themselves, on top 
of the information that is gathered from the two-particle collinear limit.

To achieve these aims, 
we calculated the 
splitting amplitude soft anomalous 
dimensions directly from general considerations of anomalous dimensions of
$n$ and $(n-m+1)$-point scattering amplitudes, 
which explicitly demonstrates their universality. 
We provide novel results for these objects in the case of  three, four and higher
particles becoming collinear at the three and four-loop order for amplitudes with massless, and additionally
one massive, coloured particles. 
We found that in the case of purely massless scattering, 
no new constraining information could be obtained
in higher than two-particle collinear limits at 
the three- and four-loop order. Remarkably, the 
colour and kinematics interplay in such a  way as
to render the results for splitting amplitude
soft anomalous dimensions in each of the cases
immediately compatible with 
strict collinear factorisation. 
On the other hand, in considerations of the 
three-particle collinear limit of an amplitude
with $n$ massless and one massive particle $I$, 
new constraints, in addition to the ones obtained
from the two-particle collinear limit, do appear.  
Interestingly, we have observed that from the 
point of view of the kinematic variables appearing in the soft anomalous dimension, the three-particle collinear limit and small-mass expansion limit
are indistinguishable, leading to a useful consistency
check of the results obtained in Ref.~\cite{Liu:2022elt}.

The consistency with our novel results 
for the three-, four- and higher-particle splitting amplitude soft anomalous dimensions 
can  be used  to test new 
results for scattering
amplitudes at high loop orders.   
In future work, it is also interesting to apply our 
systematic techniques to investigate factorisation
breaking properties in the spacelike kinematics  
at high loop orders.


\vspace{20pt}
\subsubsection*{Acknowledgments} 

We thank Ze Long Liu and Nicolas Schalch for communication regarding~\cite{Liu:2022elt}, and Andrew McLeod and Zehao Zhu for useful
discussions. 

E.G., S.J., and L.V. are grateful to the Mainz Institute for Theoretical Physics (MITP) of the  
Cluster of Excellence \emph{PRISMA${}^+$} (Project ID 390831469) and
the ``Resummation, Amplitudes, and Subtraction'' workshop at the  CERN Theoretical Physics Department 
for hospitality during the completion of this work. 

S.J. and L.V. thank the ``EFT and multi-loop methods for advancing precision in collider and gravitational wave physics'' 
workshop at 
the Munich Institute for Astro-, Particle and BioPhysics (MIAPbP), which is funded by the Deutsche Forschungsgemeinschaft (DFG, German Research Foundation) under Germany's Excellence Strategy - EXC-2094 - 390783311, where parts of this work were completed.

E.G. thanks the Albert Einstein Center for Fundamental Physics (AEC) at the University of Bern for kind hospitality in the summer of 2024.
S.J. thanks the Higgs Centre for Theoretical Physics at the University of Edinburgh for hosting fruitful research visits throughout the project.

E.G.'s  research is supported by the STFC Consolidated Grant 
``Particle Physics at the Higgs Centre''. The work of C.D. is
supported by the European Research Council (ERC) under the European Union’s research and innovation programme grant agreement 101043686 (ERC Consolidator
Grant LoCoMotive). Views and opinions expressed are
however those of the author(s) only and do not necessarily reflect those of the European Union or the European
Research Council. Neither the European Union nor the
granting authority can be held responsible for them.


\appendix

\section{Multi-particle collinear limits of dipole terms}
\label{app:dipoles}
In this appendix, we obtain the $m$-particle splitting amplitude soft anomalous dimension for the dipole terms given in \Eqn{eq:dipole}.
We perform calculations directly from generic $n$-point and 
$(n-m+1)$-point soft anomalous dimensions as defined in \Eqn{eq:GammaSPdefDip}. We start with the two-particle 
collinear limit, $m=2$, 
and in the first step we separate the sum over the dipoles
in the $n$-point and 
$(n-m+1)$-point soft anomalous dimensions in the following way
\begin{eqnarray}
\label{eq:GammaSPdefDip:m2eq1}
{\bf \Gamma}_{\SP,2}^{\rm dip.} (p_1,p_2;\mu_f)
&&={\bf \Gamma}_n^{\rm dip.} (p_1, p_2, p_{3},\ldots p_n;\mu_f) 
- {\bf \Gamma}_{n-1}^{\rm dip.}(P, p_{3},\ldots p_n;\mu_f)|_{\T_P\to  \T_1+\T_2}
\\ \nonumber&& \hspace{-2.7cm}  
= \bigg[
-\frac{1}{2}\gamma_{K} \sum_{3\leq i<j\leq n} 
\ln \left(\frac{-s_{ij}}{\mu_f^2}\right) \, \T_i \cdot \T_j 
-\frac{1}{2}\gamma_{K}  \sum_{3\leq j \leq n} 
\ln \left(\frac{-s_{1j}}{\mu_f^2}\right) \, \T_1 \cdot \T_j 
\\ && \hspace{-1.9cm} \nonumber 
-\frac{1}{2}\gamma_{K}   \sum_{3\leq j\leq n} 
\ln \left(\frac{-s_{2j}}{\mu_f^2}\right) \, \T_2 \cdot \T_j
-\frac{1}{2}\gamma_{K}    
\ln \left(\frac{-s_{12}}{\mu_f^2}\right) \, \T_1 \cdot \T_2
\,+\, \sum_i \gamma_i \, {\mathbf{1}}
\bigg]
\\ && \hspace{-2.7cm} \nonumber 
- \bigg[
-\frac{1}{2}\gamma_{K} \sum_{3\leq i<j\leq n} 
\ln \left(\frac{-s_{ij}}{\mu_f^2}\right) \, \T_i \cdot \T_j 
-\frac{1}{2}\gamma_{K}  \sum_{3\leq j \leq n} 
\ln \left(\frac{-s_{Pj}}{\mu_f^2}\right) \, \T_P \cdot \T_j
\,+\, \sum_{P,3\leq i \leq n} \gamma_i \, {\mathbf{1}}
\bigg],
\end{eqnarray}
where the first two lines are occupied by the split up sum 
terms from ${\bf \Gamma}_n^{\rm dip.}$ and the last line 
by terms from ${\bf \Gamma}_{n-1}^{\rm dip.}$. We see that
the terms with sums exclusively over the rest-of-the-process partons
cancel each other exactly. For terms involving particles 
1 and 2 we use the parametrisation in \Eqn{eq:coll-momenta},
which we substitute into the definition of $(-s_{ij})$  in \Eqn{eq:s_ij}, such that we have 
\begin{eqnarray}\label{eq:s_ij_colrop}
(-s_{1j}) = 2  |p_1 \cdot p_j | e^{-i \pi \lambda_{1j}}
= 2 x_1 |P \cdot p_j | e^{-i \pi \lambda_{Pj}} = x_1 (-s_{Pj})\,,
\end{eqnarray}
up to power corrections, and a similar result with $x_1 \to x_2$ 
is obtained for $(-s_{2j})$. Using also the fact that 
$\T_P = \T_1 + \T_2$, we can simplify \Eqn{eq:GammaSPdefDip:m2eq1}
into the following form 
\begin{eqnarray}
\label{eq:GammaSPdefDip:m2eq2}
{\bf \Gamma}_{\SP,2}^{\rm dip.} (p_1,p_2;\mu_f)
&&  =  
-\frac{1}{2}\gamma_{K}  \sum_{3\leq j \leq n} 
\ln \left(x_1\right) \, \T_1 \cdot \T_j 
-\frac{1}{2}\gamma_{K}   \sum_{3\leq j\leq n} 
\ln \left(x_2\right) \, \T_2 \cdot \T_j
\nonumber \\ && \hspace{0.4cm} 
-\frac{1}{2}\gamma_{K}    
\ln \left(\frac{-s_{12}}{\mu_f^2}\right) \, \T_1 \cdot \T_2
\,+\,  \gamma_1 \, + \gamma_2 \, -
 \gamma_P \, .
\end{eqnarray}
Finally, we can use colour conservation to rewrite 
$ \sum_{3\leq j\leq n}\T_j = - \T_1 -\T_2 =-\T_P$ which 
gives a result for the dipole contribution 
to the two-particle splitting amplitude soft anomalous dimension
which is manifestly dependent only on the particles 
becoming collinear
\begin{eqnarray}
\label{GammaSpDip2}
{\bf \Gamma}_{\SP,2}^{\rm dip.} (p_1,p_2;\mu_f)
&&  =  
\frac{1}{2}\gamma_{K}   
\ln \left(x_1\right) \, \T_1 \cdot \T_P  
+\frac{1}{2}\gamma_{K}  
\ln \left(x_2\right) \, \T_2 \cdot \T_P 
\nonumber \\ && \hspace{0.4cm} 
-\frac{1}{2}\gamma_{K}    
\ln \left(\frac{-s_{12}}{\mu_f^2}\right) \, \T_1 \cdot \T_2
\,+\,  \gamma_1 \, + \gamma_2 \, -
 \gamma_P \, .
\end{eqnarray}

Next, we extend our analysis to three particles becoming collinear.
Again, starting from the definition in 
\Eqn{eq:GammaSPdefDip} (now with $m=3$), we generalise the way we split the initial 
expressions in \Eqn{eq:GammaSPdefDip:m2eq1} to the following
\begin{eqnarray}
\label{eq:GammaSPdefDip:m3eq1}
{\bf \Gamma}_{\SP,3}^{\rm dip.} (p_1,p_2,p_3;\mu_f)
&&={\bf \Gamma}_n^{\rm dip.} (p_1, \ldots , p_n;\mu_f) 
- {\bf \Gamma}_{n-2}^{\rm dip.}(P, p_{4},\ldots p_n;\mu_f)|_{\T_P\to  \T_1+\T_2+\T_3}
\\ && \hspace{-3cm}  
= \bigg[
-\frac{1}{2}\gamma_{K} \sum_{4\leq i<j\leq n} 
\ln \left(\frac{-s_{ij}}{\mu_f^2}\right) \, \T_i \cdot \T_j 
-\frac{1}{2}\gamma_{K}  \sum_{4\leq j \leq n} 
\ln \left(\frac{-s_{1j}}{\mu_f^2}\right) \, \T_1 \cdot \T_j 
\\ && \hspace{-2.2cm} \nonumber 
-\frac{1}{2}\gamma_{K}   \sum_{4\leq j\leq n} 
\ln \left(\frac{-s_{2j}}{\mu_f^2}\right) \, \T_2 \cdot \T_j
-\frac{1}{2}\gamma_{K}   \sum_{4\leq j\leq n} 
\ln \left(\frac{-s_{3j}}{\mu_f^2}\right) \, \T_3 \cdot \T_j
\\ && \hspace{-2.2cm} \nonumber 
-\frac{1}{2}\gamma_{K}     \sum_{1\leq i < j\leq 3} 
\ln \left(\frac{-s_{ij}}{\mu_f^2}\right) \, \T_i \cdot \T_j
\,+\, \sum_i \gamma_i \, {\mathbf{1}}
\bigg]
\\ && \hspace{-3cm} \nonumber 
+\,\frac{1}{2}\gamma_{K}\! \sum_{4\leq i<j\leq n} 
\ln \left(\frac{-s_{ij}}{\mu_f^2}\right) \, \T_i \cdot \T_j 
+\frac{1}{2}\gamma_{K}  \sum_{4\leq j \leq n} 
\ln \left(\frac{-s_{Pj}}{\mu_f^2}\right) \, \T_P \cdot \T_j
- \sum_{P,4\leq i \leq n} \gamma_i \, {\mathbf{1}}\,.
\end{eqnarray}
Similarly to the analysis of the two-particle 
collinear limit, the sums involving only the 
rest-of-the-process partons cancel exactly. Then, 
for the collinear momenta entering $s_{1j}, s_{2j}$, and $s_{3j}$
we use parametrisation of \Eqn{eq:coll-momenta}
with $m=3$, which leads to result of \Eqn{eq:s_ij_colrop} for
each of the invariants containing one of the particles becoming collinear. 
Next, using $\T_P \equiv \T_1 + \T_2 + \T_3$, 
allows us to simplify the expression significantly. Through application
of colour conservation in the form
\begin{equation}
 \sum_{4\leq j\leq n}\T_j =-\T_P= - \T_1 -\T_2-\T_3 \,,
\end{equation} 
we arrive at the following 
result for the dipole part of the three-particle splitting amplitude
soft anomalous dimension
\begin{eqnarray}
\label{GammaSpDip3}
{\bf \Gamma}_{\SP,3}^{\rm dip.} (p_1,p_2,p_3;\mu_f)
&&  = \frac{1}{2}\gamma_{K}   \sum_{i}^3
\ln \left(x_i\right) \, \T_i \cdot \T_P  
-\frac{1}{2}\gamma_{K}  \sum_{1\leq i< j \leq 3}   
\ln \left(\frac{-s_{ij}}{\mu_f^2}\right) \, \T_i \cdot \T_j
\nonumber \\ && \hspace{0.4cm} 
 - \gamma_P(\as) \, 
\,+    \gamma_1(\as)+    \gamma_2(\as)+    \gamma_3(\as)  \,.
\end{eqnarray}

At this point, it is straightforward to generalise the above 
considerations to any number $m$ of particles becoming collinear.
Working directly from the definition of this object in \Eqn{eq:GammaSPdefDip},
we split the sum over dipole terms into the rest-of-the-process partons  starting from $m+1$, the mixed terms with one collinear and one rest-of-the-process parton, and pairwise interaction terms among the $m$ particles becoming collinear. To simplify the expression and eliminate the dependence of the rest of the precess, we first use the parametrisation in \Eqn{eq:coll-momenta}
for the collinear partons and then apply colour conservation (\ref{colour_conservation_rest}):
\begin{equation}
\label{colour-Conserv-m}
\sum_{(m+1)\leq j\leq n}\T_j = - \T_1 - \ldots -\T_m =-\T_P\,.
\end{equation}
With this we arrive at the following result for the dipole part of the $m$-particle splitting amplitude soft anomalous dimension
\begin{eqnarray}
\label{eq:GammaSpDipm}
{\bf \Gamma}_{\SP,m}^{\rm dip.} (p_1,\ldots,p_m;\mu_f)
&&  = \frac{1}{2}\gamma_{K}   \sum_{i}^m
\ln \left(x_i\right) \, \T_i \cdot \T_P  
-\frac{1}{2}\gamma_{K}  \sum_{1\leq i< j \leq m}   
\ln \left(\frac{-s_{ij}}{\mu_f^2}\right) \, \T_i \cdot \T_j
\nonumber \\ && \hspace{0.4cm} 
 - \gamma_P(\as) \, 
\,+\, \sum_{i}^m  \gamma_i(\as)  \,,
\end{eqnarray}
where $\T_P = \T_1 + \ldots + \T_m$, $x_i$ is the fraction of the parent
momentum $P$ carried by parton $i$ of the collinear set, and $\gamma_P(\as)$ is the collinear anomalous dimension for the 
parent parton. We thus confirm that Eq.~(\ref{eq:GammaSpDipm}) depends solely on the collinear partons, consistently with strict collinear factorisation. 

\section{Two-particle collinear limits}
\label{app:2partcol}
In Ref.~\cite{Almelid:2017qju}
the two-particle collinear limit constraints on the 
kinematic functions ${\bf{ \Gamma}}^{}_{n,4{\rm{T}}-4{\rm{L}}}$ first appearing at three loops 
were derived using amplitudes with $n=4$ and $n=3$ partons. 
In this appendix, we generalise these considerations
to obtaining the two-particle collinear splitting anomalous
dimension from arbitrary $n$ and $n-1$ point amplitudes. 
In particular, this implies that no further constraints 
can arise from the two-particle collinear limit by
considering the difference between amplitudes
with higher points, i.e., using amplitudes with
$n=5$ and $n=4$ points, or with
$n=6$ and $n=5$ points, and so on, yields the same constraints 
as the ones obtained from $n=4$ and $n=3$ point amplitudes. 

\begin{figure}[t]
\begin{center}
\includegraphics[width=0.50\textwidth]{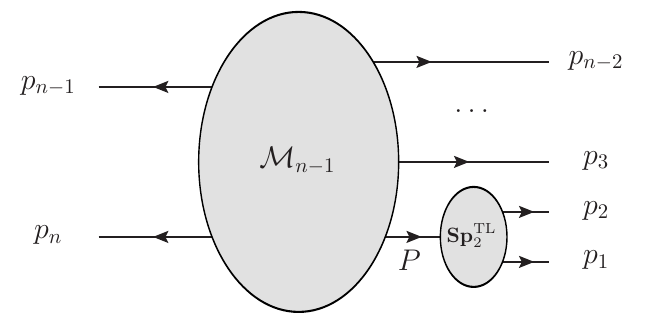}
\end{center}
\caption{Labelling of momenta for the 
configuration of two-particle timelike collinear limit.}
\label{fig:2collinearTL}
\end{figure}

In case of timelike two-particle
collinear limit we assign 
the external momenta as in Fig.~\ref{fig:2collinearTL}.
The momenta are parametrised 
as described in Section~\ref{sec:kinematics},
in particular we choose $i=1,2$ in 
\Eqn{eq:coll-momenta}.

\subsection{Terms starting at three loops}
\label{app:2partcol-threeloop}
We begin by considering the 
terms in the soft anomalous dimension 
which start at the three-loop order. 
Namely, for the 
splitting amplitude soft anomalous dimension, 
we need to evaluate ${\bf{ \Gamma}}_{\SP, m}^{4{\rm{T}}}(p_1,\ldots p_m;\mu_f)$
as defined in \Eqn{eq:GammaSpQuadm} for $m=2$. 
For convenience, the explicit object 
we are interested in is the following 
\begin{eqnarray}
\label{eq:GammaSpQuad2} 
{\bf{ \Gamma}}_{\SP, 2}^{4{\rm{T}}}(p_1, p_2;\mu_f) &=& 
   {\bf{ \Gamma}}_{n,4{\rm{T}}-3{\rm{L}}}(\alpha_s)
     + {\bf{ \Gamma}}_{n,4{\rm{T}}-4{\rm{L}}}(\{\beta_{ijkl}\},\alpha_s)
\nonumber \\  && \hspace{-2.0cm} 
-  \Big(  {\bf{ \Gamma}}_{n-1,4{\rm{T}}-3{\rm{L}}}(\alpha_s)
     + {\bf{ \Gamma}}_{n-1,4{\rm{T}}-4{\rm{L}}}(\{\beta_{ijkl}\},\alpha_s)\Big) 
     |_{\T_P\to \sum_{i=1}^2 \T_i}\,,
\end{eqnarray}
with ${\bf{ \Gamma}}_{n,4{\rm{T}}-3{\rm{L}}}$ and 
${\bf{ \Gamma}}_{n,4{\rm{T}}-4{\rm{L}}}$
given by Eqs.~(\ref{eq:Gammaf}) and (\ref{eq:quardupoleABdef}), respectively.
The quantity ${\bf{ \Gamma}}_{\SP, 2}^{4{\rm{T}}}$ is expected to depend only
on the momenta and colour of the partons becoming 
collinear, namely 1 and 2, therefore ${\bf{ \Gamma}}_{\SP, 2}^{4{\rm{T}}}$ 
must be independent of the index~$n$ appearing on the 
right-hand side of~\Eqn{eq:GammaSpQuad2}. For this reason, it is
easiest to obtain a result for \Eqn{eq:GammaSpQuad2}
by setting $n=3$. We then find that the following terms vanish: 
\hbox{${\bf{ \Gamma}}_{2,4{\rm{T}}-3{\rm{L}}}=0$} and  
\hbox{${\bf{ \Gamma}}_{2,4{\rm{T}}-4{\rm{L}}}= {\bf{ \Gamma}}_{3,4{\rm{T}}-4{\rm{L}}} = 0$},  because there are not enough partons to form the 
corresponding structures.
The result for ${\bf{ \Gamma}}_{\SP, 2}^{4{\rm{T}}}$ must be then be simply given by
\begin{equation}\label{eq:GammaSpQuad2b} 
{\bf{ \Gamma}}_{\SP, 2}^{4{\rm{T}}} 
(p_1,p_2;\mu_f)
= 
 {\bf{ \Gamma}}_{3,4{\rm{T}}-3{\rm{L}}}(\alpha_s),
\end{equation}
which is 
\begin{equation} \label{eq:GammaSpQuad2c} 
  {\bf{ \Gamma}}_{3,4{\rm{T}}-3{\rm{L}}}(\alpha_s) = 2 f(\alpha_s) \left( 
 {\cal T}_{1123}+{\cal T}_{2213} + {\cal T}_{3312}
 \right) \,.
\end{equation}
The conclusion is that only the constant term contributes: the kinematic 
dependent term cannot contribute, given that it is not  possible to
form a CICR with only three independent momenta. 
\Eqn{eq:GammaSpQuad2c}
can be written
in such a way 
to show that ${\bf{ \Gamma}}_{\SP, 2}^{4{\rm{T}}}$ does indeed only 
depend on the degrees of freedom of partons 1 and 2: using  colour conservation,
$\T_1 + \T_2 + \T_3 = 0$, in \Eqn{eq:GammaSpQuad2c} 
and arranging the result by means of the colour identities
collected in App.~\ref{colourID}, one arrives at
\be \label{eq:GammaSpQuad2d}
 {\bf{ \Gamma}}_{\SP, 2}^{4{\rm{T}}} 
 = -\frac{3}{4} f(\alpha_s) \Big( 
 C_A^2 \T_1\cdot \T_2 + 8 { \cal T }_{1122}
 \Big),
\ee
where we identify ${\bf{ \Gamma}}_{\SP, 2}^{4{\rm{T}}} \equiv {\bf{ \Gamma}}_{\SP, 2}^{4{\rm{T}}} 
(p_1,p_2;\mu_f)$,
given the absence of momentum dependence. 

While it is possible to determine ${\bf{ \Gamma}}_{\SP, 2}^{4{\rm{T}}} $
entirely from the case $n= 3$, the result
in~\Eqn{eq:GammaSpQuad2d}
by itself does not guarantee that \Eqn{eq:GammaSpQuad2} is 
indeed independent of $n$. This can be verified only by 
explicitly calculating \Eqn{eq:GammaSpQuad2}, 
proving that for generic $n$ the function ${\bf{ \Gamma}}_{\SP, 2}^{4{\rm{T}}} $ 
does coincide with the result in \Eqn{eq:GammaSpQuad2d}. 
This consistency check can be turned into a constraint
on the soft anomalous dimension.
As mentioned above, calculations with $n=4$
were performed in Ref.~\cite{Almelid:2017qju}; here we extend
this analysis to generic~$n$.

We start by considering the kinematic dependent
term, which contributes to ${\bf{ \Gamma}}_{\SP, 2}^{4{\rm{T}}} $
through the difference ${\bf{ \Gamma}}_{n,4{\rm{T}}-4{\rm{L}}}(\{\beta\},\alpha_s) 
- {\bf{ \Gamma}}_{n-1,4{\rm{T}}-4{\rm{L}}}(\{\beta\},\alpha_s)$.
This quantity is best evaluated by splitting the sum in 
\Eqn{eq:quardupoleABdef} as follows:
\begin{subequations} \label{AtermsHybrid}
\bea \label{AnHybrid} \nn
{\bf{ \Gamma}}_{n,4{\rm{T}}-4{\rm{L}}}(\{\beta\}) &=& 4 \bigg[
\textcolor{orange}{ \sum_{3\leq i<j<k<l \leq n} \, {\bf a}_{ijkl}^{ }(\{\beta\})}
+\textcolor{red}{ \sum_{3\leq j<k<l \leq n} \, {\bf a}_{1jkl}^{ }(\{\beta\}) }\\ 
&&\hspace{1.0cm}
+\textcolor{red}{\,\sum_{3\leq j<k<l \leq n} \, {\bf a}_{2jkl}^{ }(\{\beta\})}
+\sum_{3\leq k<l \leq n} \, {\bf a}_{12kl}^{ }(\{\beta\}) \bigg], \\[0.1cm]
\label{Anm1Hybrid}
{\bf{ \Gamma}}_{n-1,4{\rm{T}}-4{\rm{L}}}(\{\beta\}) &=& 4  \bigg[
\textcolor{orange} {\sum_{3\leq i<j<k<l \leq n} \, {\bf a}_{ijkl}^{ }(\{\beta\})}
+\textcolor{red}{\sum_{3\leq j<k<l \leq n} \, {\bf a}_{Pjkl}^{ }(\{\beta\})} \bigg].
\eea
\end{subequations}
Taking the difference, it is evident that the first term in 
${\bf{ \Gamma}}_{n,4{\rm{T}}-4{\rm{L}}}(\{\beta\})$ cancels with the first term in 
${\bf{ \Gamma}}_{n-1,4{\rm{T}}-4{\rm{L}}}(\{\beta\})$. Furthermore, upon using 
rescaling invariance, which implies $\beta_{1jkl} = \beta_{2jkl} = \beta_{Pjkl}$
(recall that $P = p_1 + p_2$), and the identity $\T_P = \T_1 + \T_2$, one easily
observes that the second term in~\Eqn{Anm1Hybrid} cancels with the second and
third terms in~\Eqn{AnHybrid}, such that one gets
\be \label{deltaA-twocollHybrid}
{\bf{ \Gamma}}_{n,4{\rm{T}}-4{\rm{L}}}(\{\beta\}) 
- {\bf{ \Gamma}}_{n-1,4{\rm{T}}-4{\rm{L}}}(\{\beta\})\Big|_{p_1||p_2} =
4 \sum_{3\leq k<l \leq n} \, {\bf a}_{12kl}^{ }(\{\beta\}),
\ee
where the exact form of  ${\bf a}_{ijkl}(\{\beta\})$ 
can be read off from~\Eqn{eq:quadrupole-ab-defshu}. 
Considering now the constant term, we split the sum over the 
colour indices in a similar way as done in~\Eqn{AtermsHybrid} for the kinematic term: 
\begin{subequations}  \label{BtermsHybrid}
\bea  \label{BnHybrid} \nn
\!\!\!{\bf{ \Gamma}}_{n,4{\rm{T}}-3{\rm{L}}}(\alpha_s) &=& 2  f(\alpha_s)  \Bigg\{ 
\textcolor{orange}{
\sum_{i = 3}^{n}\sum_{\substack{3\leq k<l \leq n \\ k,l\neq i}}
\, {\cal T}_{iikl}}
+\textcolor{blue}{
\sum_{ 3\leq k<l \leq n  } \Big(
{\cal T}_{11kl} +{\cal T}_{22kl}  \Big)} \\[-0.1cm] 
&&\hspace{0.4cm}+\,
\textcolor{red}{\sum_{i = 3}^{n}\sum_{\substack{k = 3, \\ k\neq i}}^{n} 
\Big({\cal T}_{ii1k} +{\cal T}_{ii2k} \Big) }
+ \sum_{i = 3}^{n} \Big( 
{\cal T}_{112i} + {\cal T}_{221i} + {\cal T}_{ii12} 
\Big) \Bigg\} \,, \\[0.1cm] 
\label{Bnm1Hybrid}
\!\!\!{\bf{ \Gamma}}_{n-1,4{\rm{T}}-3{\rm{L}}}(\alpha_s) &=&
2 f(\alpha_s)  \Bigg\{   \textcolor{orange}{\sum_{i = 3}^{n}
\sum_{\substack{3\leq k<l \leq n \\ k,l\neq i}} \, {\cal T}_{iikl}}
\,+\,\textcolor{blue}{
\sum_{\substack{3\leq k<l \leq n }} {\cal T}_{PPkl}  }
+\textcolor{red}{\sum_{i = 3}^{n} \sum_{\substack{k = 3, \\ k\neq i}}^{n} {\cal T}_{iiPk}}\Bigg\}\,.
\eea
\end{subequations} 
In the two-particle collinear limit colour conservation implies $\T_P = \T_1 + \T_2$, therefore
when taking the difference, \textcolor{orange}{orange} and
\textcolor{red}{red} contributions cancel each other exactly.
On the other hand, there are left over terms originating from
the \textcolor{blue}{blue} contributions. Using $\T_P = \T_1 + \T_2$ we see that 
\begin{eqnarray}\label{eq:teal-twopartcol}
 {\cal T}_{PPjk} &=&   {\cal T}_{11jk} + {\cal T}_{22jk} +  {\cal T}_{12jk}+  {\cal T}_{12kj}\,.
\end{eqnarray}
Then
the difference between \Eqn{BnHybrid} and \Eqn{Bnm1Hybrid} 
is given by 
\be \label{deltaB-twocollHybrid}  
{\bf{ \Gamma}}_{n,4{\rm{T}}-3{\rm{L}}} 
-  {\bf{ \Gamma}}_{n-1,4{\rm{T}}-3{\rm{L}}}  
= 2 f(\alpha_s)  \bigg[  
\textcolor{blue}{-   \sum_{3\leq k<l \leq n} 
\Big( {\cal T}_{12kl} +{\cal T}_{12lk}\Big) } 
+\, \sum_{i=3}^{n} \Big( {\cal T}_{ii12} 
+ {\cal T}_{112i} + {\cal T}_{221i} \Big)\bigg].
\ee
Summing \Eqns{deltaA-twocollHybrid}{deltaB-twocollHybrid} according 
to \Eqn{eq:GammaSpQuad2} gives the 
splitting amplitude soft anomalous dimension
${\bf{ \Gamma}}_{\SP, 2}^{4{\rm{T}}} 
(p_1,p_2;\mu_f)$, which reads
\bea
\label{GammaSpQuad2eHybrid} 
\nn &&  \hspace{-0.4cm}
{\bf{ \Gamma}}_{\SP, 2}^{4{\rm{T}}} 
(p_1,p_2;\mu_f) = 
\Big[{\bf{ \Gamma}}_{n,4{\rm{T}}-3{\rm{L}}}
+ {\bf{ \Gamma}}_{n,4{\rm{T}}-4{\rm{L}}} \Big]
- \Big[{\bf{ \Gamma}}_{n-1,4{\rm{T}}-3{\rm{L}}} 
+ {\bf{ \Gamma}}_{n-1,4{\rm{T}}-4{\rm{L}}} \Big]
\nn \\  && \hspace{2.6cm}
=   \sum_{3\leq k<l \leq n} \bigg[
\, 4 {\cal F}^{\rm A}_{12kl}(\{\beta\}) \,{\cal T}_{1kl2} 
+ \Big( 4 {\cal F}^{\rm S}_{12kl}(\{\beta\}) 
\textcolor{blue}{ -2f(\alpha_s)} \Big) \Big( {\cal T}_{12lk}
+ {\cal T}_{12kl} \Big)  \bigg]
\nn \\  &&\hspace{3.1cm}
+ 2f(\alpha_s) \sum_{i=3}^{n} \left( {\cal T}_{ii12} 
+ {\cal T}_{112i} + {\cal T}_{221i} \right).
\eea
Universality of the splitting amplitude requires that this result must coincide with \Eqn{eq:GammaSpQuad2d}.
In fact, for $n=3$ we can 
immediately see that only the last line 
of the above equation can contribute and 
it gives precisely the same result as 
one in \Eqn{eq:GammaSpQuad2c}, after which colour 
conservation can easily be used to eliminate 
$\T_3$ in favour of $\T_1$ and  $\T_2$, i.e. 
$\T_3 = -\T_1 - \T_2$, yielding \Eqn{eq:GammaSpQuad2d}. However, 
for $n\geq 4$, the remaining terms in~\Eqn{GammaSpQuad2eHybrid} start to contribute
and this implies non-trivial relations
between the constant and kinematic-dependent terms,
which must conspire to produce the same 
result as for the $n=3$ case.
We notice that stuffle-type relations 
for the colour generators,
\begin{eqnarray}\label{eq:stuffle-2.p.c.c}
2\sum_{3\leq k<l \leq n}  \Big( {\cal T}_{12lk} + {\cal T}_{12kl} \Big)
+2\sum_{i=3}^{n}{\cal T}_{ii12} = \sum_{k=3}^{n}\sum_{l=3}^{n} 
\big( {\cal T}_{12lk} +  {\cal T}_{12kl} \big)\,,
\end{eqnarray}
can be formed, provided that the following condition holds for the kinematic function in the two-particle
collinear limit
\be\label{eq:twocoll-constraint1}
 \Big( 4{\cal F}^{\rm S}_{12kl}(\{\beta\}) -2f(\alpha_s) \Big)\bigg|_{p_1|| p_2} = 2f(\alpha_s) \,.
\ee 
If indeed this constraint is 
satisfied, the two-particle
collinear splitting amplitude 
soft anomalous dimension becomes 
\bea\label{eq:GammaSpQuad2-2} \nn 
&&  
{\bf{ \Gamma}}_{\SP, 2}^{4{\rm{T}}} 
(p_1,p_2;\mu_f) = 
\sum_{3\leq k<l \leq n} 
\, 4 {\cal F}^{\rm A}_{12kl}(\{\beta\}) \,{\cal T}_{1kl2} +f(\alpha_s)
 \sum_{k=3}^{n}\sum_{l=3}^{n} 
 \Big( {\cal T}_{12lk}
+ {\cal T}_{12kl} \Big)   \nn \\ 
&&\hspace{3.1cm}
+ 2f(\alpha_s) \sum_{i=3}^{n} \left(  {\cal T}_{112i} + {\cal T}_{221i} \right)
.
\eea
Application of colour conservation and the identities in App.~\ref{colourID} 
yields the following relations 
\begin{eqnarray}\label{eq:T-2-colcons1}
   \sum_{k=3}^{n}\sum_{l=3}^{n}  \mathcal{T}_{12kl} \stackrel{m.c.c}{=}
  -\frac{1}{8}C_A^2 \T_1\cdot \T_2-\mathcal{T}_{1122}\,,
\end{eqnarray}
and 
\bea  \label{eq:T-2-colcons2}
\sum_{i=3}^{n}  \, {\cal T}_{112i}  &\stackrel{m.c.c}{=}&
- \frac{1}{8}\,C_A^2 \,\T_1\cdot \T_2 -
 \mathcal{T}_{1122}\,,
\eea
where the $m.c.c$ label was defined in Eq.~\eqref{eq:calT-AABi-3pc}. 
Using these, \Eqn{eq:GammaSpQuad2-2}
can be brought into the form of
\Eqn{eq:GammaSpQuad2d}: 
\bea\label{GammaSpQuad2hHybrid} 
&&  \hspace{-0.5cm}
{\bf{ \Gamma}}_{\SP, 2}^{4{\rm{T}}} 
(p_1,p_2;\mu_f) =  
\sum_{3\leq k<l \leq n} 
\, 4 {\cal F}^{\rm A}_{12kl}(\{\beta\}) \,{\cal T}_{1kl2} 
 -\frac{3}{4}f(\alpha_s) \Big( C_A^2 \T_1\cdot \T_2
 + 8 {\cal T}_{1122} \Big) \,.
\eea
It is clear at this point that \Eqn{GammaSpQuad2hHybrid}
matches \Eqn{eq:GammaSpQuad2d}, as long as  
\be \label{eq:twocoll-constraint2}
{\cal F}^{\rm A}_{12kl}(\{\beta\}) \Big|_{p_1|| p_2}  =0\,.
\ee
\Eqns{eq:twocoll-constraint1}{eq:twocoll-constraint2} provide
constraints on the kinematic-dependent term of the soft anomalous 
dimension, and generalise to all $n$ the results obtained in \cite{Almelid:2017qju}
for the case $n = 4$ (see in particular Eqs.~(6.24) and (6.25) there). 
The indices $k$ and $l$ represent any of the non-collinear particles 
from the sub-amplitude $\MM_{n-1}$. Collinear factorisation requires 
that the soft anomalous dimension of the splitting amplitude must 
depend only on the quantum numbers of the particles becoming 
collinear, namely particles $1$ and $2$ in this case. This is the origin 
of the constraints in \Eqns{eq:twocoll-constraint1}{eq:twocoll-constraint2}, 
which can be used either to constrain the free parameters in 
the bootstrap approach, or to check consistency between infrared and 
collinear factorisation.
Since these functions have been calculated 
at the three-loop order \cite{Almelid:2015jia},
we can now
adopt the latter point of view and 
consider the exact form of the functions ${\cal F}(\beta_{ijkl},\beta_{ilkj})$ 
given in \Eqn{eq:calFdef} to investigate whether the relations 
\Eqns{eq:twocoll-constraint1}{eq:twocoll-constraint2} are indeed satisfied. 
With this purpose in mind, it is sufficient to consider the case 
$n = 4$, given that the quadrupole correction involves at 
most four gluons. We need to evaluate the 
logarithms of the 
CICRs 
$\beta_{1234}$, $\beta_{1432}$ in the limit $p_1 || p_2$, i.e.
in the kinematic configuration 
considered in Section~\ref{sec:kin-two-part-col}.
We obtain (see also Ref.~\cite{Dixon:2009ur}) 
\bea \nn
\rho_{1234} &=& z \bar z = 
\frac{p_1 \cdot p_2 \, p_3 \cdot p_4}{p_1 \cdot p_3 \, p_2 \cdot p_4}
= \frac{P^2 \, p_3 \cdot p_4}{x_1 x_2 \, P \cdot p_3 \, P \cdot p_4} \to 0, \\ 
\rho_{1432} &=& (1-z) (1- \bar z) = 
\frac{p_1 \cdot p_4 \, p_3 \cdot p_2}{p_1 \cdot p_3 \, p_2 \cdot p_4}
= \frac{P \cdot p_4 \, P \cdot p_3}{P \cdot p_3 \, P \cdot p_4} \to 1,
\eea
which implies $z \to 0$, $\bar z \to 0$ in the collinear limit. 
Evaluating the functions ${\cal F}$ in 
\Eqns{eq:twocoll-constraint1}{eq:twocoll-constraint2} in this limit 
gives
\bea \nn
{\cal F}(\beta_{1234},\beta_{1432}) 
&\stackrel{p_1\parallel p_2}{\longrightarrow}&\frac{1}{2} f(\alpha_s)\,, \\ 
{\cal F}(\beta_{1243},\beta_{1342}) 
&\stackrel{p_1\parallel p_2}{\longrightarrow}&\frac{1}{2} f(\alpha_s)\,, \\ \nn
{\cal F}(\beta_{1324},\beta_{1423}) 
&\stackrel{p_1\parallel p_2}{\longrightarrow}&0\,,
\eea
where we recall that the three-loop result for $f(\alpha_s)$ is given in \Eqn{eq:fa3loop}.
Inserting these results into the definition of ${\cal F}^{\rm A/S}$,
\Eqn{quadrupole-AS-defsh}, we immediately obtain 
\be
{\cal F}^{\rm A}_{1234}(\{\beta\}) \Big|_{p_1|| p_2}  = 0\,, 
\ee
as expected from \Eqn{eq:twocoll-constraint2}, and 
\be
{\cal F}^{\rm S}_{1234}(\{\beta\}) \Big|_{p_1|| p_2} =  f(\alpha_s)\,,
\ee
as expected from \Eqn{eq:twocoll-constraint1}. We stress 
that this is a very non-trivial result, which arises as a 
consequence of an interesting interplay between the 
kinematic  and constant parts of the soft anomalous dimension starting at three loops, i.e., a non-trivial interplay 
between kinematics and colour in the soft anomalous 
dimension, which manifests for instance in \Eqn{GammaSpQuad2eHybrid} 
through the emergence of the stuffle products in \Eqn{eq:stuffle-2.p.c.c}.

\subsection{Four-loop quartic terms}
\label{app:2partcol-fourloop}

Two-particle collinear limits 
were also considered in Ref.~\cite{Becher:2019avh}
in order to constrain the form of the kinematic 
functions appearing at four loops. 
We rederive these constraints using
the same strategy we followed above for the terms which start to contribute
at the three-loop order. 
We begin by considering the terms which
contain totally symmetric combinations of four colour generators, Eqs.~(\ref{eq:Q4T-23L}) and (\ref{eq:Q4T-4L}).
In App.~\ref{sec:2-coll_Quintic_terms} below we shall discuss those containing five generators.

In order to obtain the contribution of the quartic terms to the splitting amplitude anomalous dimension, we begin with the
definition of ${\bf{ \Gamma}}_{\SP, m}^{{\rm Q}4{\rm{T}}}(p_1,\ldots p_m;\mu_f)$ in \Eqn{eq:GammaSpQ4Tm} for $m=2$:
\begin{eqnarray}
\label{eq:GammaSpQ4T2}
{\bf{ \Gamma}}_{\SP, 2}^{{\rm Q}4{\rm{T}}}(p_1, p_2;\mu_f) &=& {\bf{ \Gamma}}_{n,{\rm{Q}}4{\rm{T}}-2,3{\rm{L}}}(\{s_{ij} \},\mu_f,\alpha_s)
+          {\bf{ \Gamma}}_{n,\rm{Q}4{\rm{T}}-4{\rm{L}}}(\{\beta_{ijkl} \},\alpha_s) 
          \nonumber \\ && \hspace{-2.9cm}
 -  \Big(  {\bf{ \Gamma}}_{n-1,{\rm{Q}}4{\rm{T}}-2,3{\rm{L}}}(\{s_{ij} \},\mu_f,\alpha_s)
+        {\bf{ \Gamma}}_{n-1,\rm{Q}4{\rm{T}}-4{\rm{L}}}(\{\beta_{ijkl} \},\alpha_s)\Big) 
    \Big|_{\T_P\to \sum_{i=1}^2 \T_i}.
\end{eqnarray}

The ${\bf{ \Gamma}}_{n,\rm{Q}4{\rm{T}}-4{\rm{L}}}$
term in the soft anomalous dimension is given in
\Eqn{eq:Q4T-4L}. We split the sum over the partons into pieces that depend on particles becoming collinear, 1 and~2, and the rest-of-the-process partons, 
\begin{eqnarray}\label{eq:Q4T-4L-a}
{\bf{ \Gamma}}_{n,\rm{Q}4{\rm{T}}-4{\rm{L}}}
(\{\beta_{ijkl} \},\alpha_s) &=& \,24\,
\sum_{R}
\sum_{3\leq i<j<k<l \leq n} \DR_{ijkl} 
\,{\cal G}_R (\beta_{ijlk},\beta_{iklj})  
\\ \nn &+&  
 \,24\,  \sum_{R}
\sum_{3\leq j<k<l \leq n} \DR_{1jkl} 
\,{\cal G}_R (\beta_{1jlk},\beta_{1klj}) 
\\ \nn &+& 
 \,24\, \sum_{R}
\sum_{3\leq j<k<l \leq n} \DR_{2jkl} 
\,{\cal G}_R (\beta_{2jlk},\beta_{2klj})  
\\ \nn &+& 
 \,24\,   \sum_{R}
\sum_{3\leq k<l \leq n} \DR_{12kl} 
\,{\cal G}_R (\beta_{12lk},\beta_{1kl2})  \,.
\end{eqnarray}
Similarly, the form of this contribution for the 
amplitude with $n-1$ legs is given by 
\begin{eqnarray}\label{eq:Q4T-4L-b}
{\bf{ \Gamma}}_{n-1,\rm{Q}4{\rm{T}}-4{\rm{L}}}
(\{\beta_{ijkl} \},\alpha_s) &=& \,24\,
\sum_{R}
\sum_{3\leq i<j<k<l \leq n} \DR_{ijkl} 
\,{\cal G}_R (\beta_{ijlk},\beta_{iklj}) 
\\ \nn &+& 
 \,24\, \sum_{R}
\sum_{3\leq j<k<l \leq n} \DR_{Pjkl} 
\,{\cal G}_R (\beta_{Pjlk},\beta_{Pklj}) \,.
\end{eqnarray}
We now take the 
difference. It is apparent that the terms in the first 
line in the above two equations cancel in the difference, since they contain the same rest-of-the-process dependence 
only, and not information on particles becoming collinear. 
Moreover, we use rescaling invariance and 
$\T_P = \T_1 + \T_2$, which means that the terms in the
second and third lines of \Eqn{eq:Q4T-4L-a} cancel
the terms in the second line of \Eqn{eq:Q4T-4L-b}.
The remaining terms are given by 
\begin{equation}\label{eq:sp2-Q4T-4L-a}
{\bf{ \Gamma}}_{n,\rm{Q}4{\rm{T}}-4{\rm{L}}}
(\{\beta_{ijkl} \},\alpha_s) 
- {\bf{ \Gamma}}_{n-1,\rm{Q}4{\rm{T}}-4{\rm{L}}}
(\{\beta_{ijkl} \},\alpha_s) 
=\,24\, \sum_{R}
\sum_{3\leq k<l \leq n} \DR_{12kl} 
\,{\cal G}_R (\beta_{12lk},\beta_{1kl2}) \,.
\end{equation}
Considering now the contributions to 
${\bf{ \Gamma}}_{\SP, 2}^{{\rm Q}4{\rm{T}}}$ from the
${{\bf\Gamma}}_{n,{\rm{Q}}4{\rm{T}}-2,3{\rm{L}}}$ and ${{\bf\Gamma}}_{n-1,{\rm{Q}}4{\rm{T}}-2,3{\rm{L}}}$ terms in the soft anomalous dimension, given in \Eqn{eq:Q4T-23L}, we split the sum over the 
colour indices similarly to~Eqs.~\eqref{eq:Q4T-4L-a} and \eqref{eq:Q4T-4L-b} in  the case of the ${\cal G}_R$  kinematic terms: 
\begin{eqnarray}\label{eq:Q4T-23L-a}
{\bf{ \Gamma}}_{n,{\rm{Q}}4{\rm{T}}-2,3{\rm{L}}}
(\{s_{ij} \},\lambda,\alpha_s) &=& 
-\sum_R g_R \bigg[\sum_{3\leq i<j  \leq n}
\Big(  { \bf{\cal{ D}}}^R_{iijj} 
+ { \bf{\cal{ D}}}^R_{iiij}+{ \bf{\cal{ D}}}^R_{jjji}\Big)\ell_{ij} 
\\ &&\hspace{-4cm}
+ \sum_{  j = 3}^n\Big(  { \bf{\cal{ D}}}^R_{11jj} 
+{ \bf{\cal{ D}}}^R_{111j}+  { \bf{\cal{ D}}}^R_{jjj1}\Big)\ell_{1j}
+ \sum_{  j = 3}^n\Big(  { \bf{\cal{ D}}}^R_{22jj} 
+{ \bf{\cal{ D}}}^R_{222j}+  { \bf{\cal{ D}}}^R_{jjj2}\Big)\ell_{2j}
 \nonumber \\ &&\hspace{-4cm}
+\Big(  { \bf{\cal{ D}}}^R_{1122} 
+{ \bf{\cal{ D}}}^R_{1112}+  { \bf{\cal{ D}}}^R_{2221}\Big)\ell_{12}
+\sum_{i=3}^n\sum_{\substack{ 3\leq j < k \leq n\\ j,k \neq i } }
{ \bf{\cal{ D}}}^R_{jkii}\ell_{jk}
+\sum_{\substack{ 3\leq j < k \leq n } } 
\Big({ \bf{\cal{ D}}}^R_{jk11}
+ { \bf{\cal{ D}}}^R_{jk22}\Big) \ell_{jk}
\nonumber \\ \nonumber &&\hspace{-4cm}
+\sum_{i=3}^n\sum_{\substack{ k=3\\  k \neq i } }^n 
\Big( { \bf{\cal{ D}}}^R_{1kii}\ell_{1k}
+ { \bf{\cal{ D}}}^R_{2kii}\ell_{2k} \Big) 
+\sum_{i=3}^n \Big({ \bf{\cal{ D}}}^R_{2i11}\ell_{2i} 
+{\bf{\cal{ D}}}^R_{1i22}\ell_{1i}
+{\bf{\cal{ D}}}^R_{12ii}\ell_{12}\Big) 
\bigg]\,,
\end{eqnarray}
and
\begin{eqnarray}\label{eq:Q4T-23L-b}
{\bf{ \Gamma}}_{n-1,{\rm{Q}}4{\rm{T}}-2,3{\rm{L}}}
(\{s_{ij} \},\lambda,\alpha_s) 
\nonumber \\ &&\hspace{-4.0cm}
= -\sum_R g_R \bigg[\sum_{3\leq i<j  \leq n}
\hspace{-0.3cm}
\Big( { \bf{\cal{ D}}}^R_{iijj} 
+{\bf{\cal{ D}}}^R_{iiij}
+{\bf{\cal{ D}}}^R_{jjji}\Big)\ell_{ij} 
+\sum_{i=3}^n\sum_{\substack{ 3\leq j < k \leq n\\ j,k \neq i}}
\hspace{-0.3cm}
{\bf{\cal{D}}}^R_{jkii}\ell_{jk}
\\ &&\hspace{-4.0cm} \nonumber 
+ \sum_{ j = 3}^n\Big(  { \bf{\cal{ D}}}^R_{PPjj} 
+{ \bf{\cal{ D}}}^R_{PPPj}
+{ \bf{\cal{ D}}}^R_{jjjP}\Big)\ell_{Pj}
+\sum_{\substack{ 3\leq j < k \leq n  } } 
\hspace{-0.2cm}
{ \bf{\cal{ D}}}^R_{jkPP}\ell_{jk}
+\sum_{i=3}^n\sum_{\substack{ k=3\\  k \neq i } }^n
{ \bf{\cal{ D}}}^R_{Pkii}\ell_{Pk} \bigg] \,.
\end{eqnarray}
The terms depending only on the rest-of-the-process
partons cancel directly in the difference between the
two terms. Next, additional cancellations occur when
we substitute $\T_P = \T_1 + \T_2$ and 
$\ell_{1j}=\ln(x_1) + \ell_{Pj}$ and 
$\ell_{2j}=\ln(x_2) + \ell_{Pj}$. Concretely, 
we make use of the following identities
\begin{eqnarray}\label{eq:D12-TP-2pc}
\DR_{jjjP} = \DR_{jjj1}+\DR_{jjj2}\,,
\qquad
\DR_{PPjj} & =& \DR_{11jj}+\DR_{22jj}
+ 2 \DR_{12jj}\,,
\end{eqnarray}
and
\begin{eqnarray}\label{eq:D3-TP-2pc}
\DR_{PPPj} & =&  \DR_{111j}+\DR_{222j}+ 3 \DR_{112j} + 3 \DR_{122j}\,,
\end{eqnarray} 
and similar ones involving two rest-of-the-process partons. 
Substituting these relations and simplifying yields
the following expression for the difference between the 
$n$ and $n-1$ soft anomalous dimension terms
\begin{eqnarray}
&&{\bf{ \Gamma}}_{n,{\rm{Q}}4{\rm{T}}-2,3{\rm{L}}}
(\{s_{ij} \},\lambda,\alpha_s) 
- {\bf{ \Gamma}}_{n-1,{\rm{Q}}4{\rm{T}}-2,3{\rm{L}}}
(\{s_{ij} \},\lambda,\alpha_s)
\nonumber \\ &&    =
- \sum_R g_R \bigg[
\sum_{  j = 3}^n\Big(  { \bf{\cal{ D}}}^R_{11jj} 
+  { \bf{\cal{ D}}}^R_{111j}
+  { \bf{\cal{ D}}}^R_{jjj1}\Big)   \ln(x_1) 
+ \sum_{  j = 3}^n\Big( { \bf{\cal{ D}}}^R_{22jj}  
+  { \bf{\cal{ D}}}^R_{222j} 
+  { \bf{\cal{ D}}}^R_{jjj2}\Big)\ln(x_2)
\nonumber \\ && \hspace{2.5cm}
+\Big(  { \bf{\cal{ D}}}^R_{1122} 
+  { \bf{\cal{ D}}}^R_{1112}
+  { \bf{\cal{ D}}}^R_{2221}\Big)\ell_{12}       
+\sum_{i=3}^n\sum_{\substack{ k=3\\  k \neq i } }^n 
\Big( { \bf{\cal{ D}}}^R_{1kii}\ln(x_1) + 
{ \bf{\cal{ D}}}^R_{2kii}\ln(x_2)  \Big)
\nonumber \\[-0.1cm] && \hspace{2.5cm}
+\sum_{i=3}^n \Big( { \bf{\cal{ D}}}^R_{2i11}\ln({x_2})  
+ { \bf{\cal{ D}}}^R_{1i22}\ln({x_1}) 
+ { \bf{\cal{ D}}}^R_{12ii}\ell_{12} \Big)
\bigg] 
\nonumber \\ && \hspace{0.6cm}+
\sum_R g_R \bigg[
\sum_{  j = 3}^n\Big(    2 {\cal{D}}^R_{12jj} 
+    2 {\cal{D}}^R_{112j} + 2 {\cal{D}}^R_{122j}  
\Big)\ell_{Pj}
+ \sum_{\substack{ 3\leq j < k \leq n  } }  
\Big(   2 {\cal{D}}_{jk12}^R  \Big)\ell_{jk}  \bigg] .
\end{eqnarray}
After these manipulations, we start applying colour 
conservation to simplify the expression. For example, using the explicit
definition for ${\bf{\cal{ D}}}^R_{ijkl}$, given 
in \Eqn{eq:calDijkl}, we can show that  
\begin{eqnarray}\label{eq:D2pc-mcc}\nonumber
\sum_{  j = 3}^n   { \bf{\cal{ D}}}^R_{111j} 
&=& d_R^{abcd}\T_1^a\T_1^b\T_1^c
\big(\T_3^d + \T_4^d +\T_5^d+ \ldots + \T_n^d\big)
\\[1ex] &=&d_R^{abcd}\T_1^a\T_1^b\T_1^c\big(-\T_1^d - \T_2^d  \big)
= -{ \bf{\cal{ D}}}^R_{1111} - { \bf{\cal{ D}}}^R_{1112} \,,
\end{eqnarray}
and 
\begin{eqnarray}\label{eq:calD1iikcc}
\sum_{i=3}^n\sum_{\substack{ k=3\\  k \neq i } }^n \,
{ \bf{\cal{ D}}}^R_{1iik}
= - \sum_{i=3}^{n} { \bf{\cal{ D}}}^R_{11ii}
- \sum_{i=3}^{n} { \bf{\cal{ D}}}^R_{12ii}  
-\sum_{i=3}^{n} { \bf{\cal{ D}}}^R_{1iii} \,.
\end{eqnarray}
The rest of the necessary identities are given in 
\Eqns{eq:calD1i22cc}{eq:calD2iikcc}. After application of 
colour conservation, additional terms cancel
and we are left with
\begin{eqnarray}\label{eq:Q4T-23L-b34} \nonumber
{\bf{ \Gamma}}_{n,{\rm{Q}}4{\rm{T}}-2,3{\rm{L}}}
(\{s_{ij} \},\lambda,\alpha_s) 
-{\bf{ \Gamma}}_{n-1,{\rm{Q}}4{\rm{T}}-2,3{\rm{L}}}
(\{s_{ij} \},\lambda,\alpha_s) && \\ 
&&\hspace{-7.4cm} =\,  
\sum_R g_R \bigg[\,
\Big({\bf{\cal{ D}}}^R_{1111}
+{\bf{\cal{ D}}}^R_{1112}
+{\bf{\cal{ D}}}^R_{1122} 
+{\bf{\cal{ D}}}^R_{1222}\Big)\ln({x_1})   
\nonumber \\ && \hspace{-5.7cm}
+\Big({\bf{\cal{ D}}}^R_{2222}
+{\bf{\cal{ D}}}^R_{2221}
+{\bf{\cal{ D}}}^R_{2211}    
+{\bf{\cal{ D}}}^R_{2111}\Big)\ln({x_2})  
\nonumber\\[0.2cm] && \hspace{-5.7cm}
-\Big({ \bf{\cal{ D}}}^R_{1112} 
+{\bf{\cal{ D}}}^R_{1122}
+{ \bf{\cal{ D}}}^R_{2221}\Big)\ell_{12}\,\bigg]
\nonumber \\ &&\hspace{-7.8cm}
+\sum_R g_R \bigg[\sum_{j = 3}^n\Big(2 {\cal{D}}_{12jj}^R  
+ 2 {\cal{D}}_{112j}^R + 2 {\cal{D}}_{122j}^R \Big)\ell_{Pj}
\nonumber \\[-0.1cm] && \hspace{-6.34cm}
+\sum_{i=3}^{n} { \bf{\cal{ D}}}^R_{12ii} 
\Big( \ln(x_1) +\ln(x_2)-\ell_{12} \Big)
+\sum_{\substack{ 3\leq j < k \leq n  } }  
2 {\cal{D}}_{jk12}^R  \ell_{jk}\bigg]. 
\end{eqnarray}
We next focus on manipulating the terms which
still depend on the rest-of-the-process partons. 
We rewrite 
\begin{eqnarray}\label{eq:logPx1x2}
2  \ell_{Pj} +\ln(x_1)+\ln(x_2) =   \ell_{1j} +  \ell_{2j}\,,
\end{eqnarray}
and 
\begin{eqnarray}\label{eq:logjk}
\ell_{jk} &=&  \beta_{12jk} +
\ell_{1j}+ \ell_{2k} - \ell_{12}\,.
\end{eqnarray} 
Then, using colour conservation we can show that
\begin{eqnarray}
-2\sum_{\substack{ 3\leq j < k \leq n  } }  
{\cal{D}}_{jk12}^R \ell_{12}   
&\stackrel{m.c.c.  }{=}&  \ell_{12}  \Big(
- {\cal{D}}_{1112}^R - 2{\cal{D}}_{1122}^R
-{\cal{D}}_{1222}^R + \sum_{i=3}^n 
{\cal{D}}_{12ii}^R  \Big)\,, 
\label{eq.6.89} \\
2\sum_{\substack{ 3\leq j < k \leq n  } }  
{\cal{D}}_{jk12}^R  \,    \ell_{1j}  
&\stackrel{m.c.c.  }{=}& \sum_{i=3}^n \ell_{1i}
\big( - {\cal{D}}_{112i}^R\,    
- {\cal{D}}_{122i}^R \,     
- {\cal{D}}_{12ii}^R \,\big)\,,
\label{eq.6.92}
\end{eqnarray}
and similarly for 
$ \sum_{\substack{ 3\leq j < k \leq n }}
{\cal{D}}_{jk12}^R \,\ell_{2k}$.
Making use of the above relations, we can write 
\Eqn{eq:Q4T-23L-b34}  as 
\begin{eqnarray}\label{eq:6.78b}
&&{\bf{ \Gamma}}_{n,{\rm{Q}}4{\rm{T}}-2,3{\rm{L}}}(\{s_{ij} \},\lambda,\alpha_s) - {\bf{ \Gamma}}_{n-1,{\rm{Q}}4{\rm{T}}-2,3{\rm{L}}}(\{s_{ij} \},\lambda,\alpha_s)
\nonumber \\ && \hspace{1.0cm} =
\sum_R g_R \bigg[\,
\Big({ \bf{\cal{ D}}}^R_{1111} 
+{ \bf{\cal{ D}}}^R_{1112}
+{ \bf{\cal{ D}}}^R_{1122} 
+ { \bf{\cal{ D}}}^R_{1222}\Big)\ln({x_1})   
\nonumber \\ && \hspace{2.5cm}
+\Big({ \bf{\cal{ D}}}^R_{2222}
+{ \bf{\cal{ D}}}^R_{2221} 
+{ \bf{\cal{ D}}}^R_{2211}
+{ \bf{\cal{ D}}}^R_{2111}\Big)\ln({x_2})  
\nonumber \\[0.2cm] &&\hspace{2.5cm}
-\Big(2 { \bf{\cal{ D}}}^R_{1112}
+3 { \bf{\cal{ D}}}^R_{1122} 
+ 2 { \bf{\cal{ D}}}^R_{2221}\Big)\ell_{12} \,\bigg] 
\nonumber \\ &&\hspace{1.0cm}
+ \sum_R g_R \bigg[ \sum_{  j = 3}^n\Big(      
2 {\cal{D}}_{112j}^R + 2 {\cal{D}}_{122j}^R  \Big)\ell_{Pj} 
+2\sum_{\substack{ 3\leq j < k \leq n  } }  
        {\cal{D}}_{12jk}^R   \beta_{12jk}  
\nonumber \\[-0.2cm] &&\hspace{2.5cm} 
+\sum_{i=3}^n \ell_{1i}\big( - {\cal{D}}_{112i}^R\,  
- {\cal{D}}_{122i}^R \,\big)
+\sum_{i=3}^n \ell_{2i}\big( - {\cal{D}}_{112i}^R\,     
- {\cal{D}}_{122i}^R \, \big)        \bigg] \,.
\end{eqnarray}
Rewriting $l_{1i} = \ln(x_1) + l_{Pi} $, $l_{2i} = \ln(x_2) + l_{Pi} $, cancelling terms, and using colour conservation, we finally arrive at
\begin{eqnarray}\label{eq:6.78e}
&&{\bf{ \Gamma}}_{n,{\rm{Q}}4{\rm{T}}-2,3{\rm{L}}}
(\{s_{ij} \},\lambda,\alpha_s) 
-   {\bf{ \Gamma}}_{n-1,{\rm{Q}}4{\rm{T}}-2,3{\rm{L}}}
(\{s_{ij} \},\lambda,\alpha_s)  
\nonumber \\ && \hspace{1cm} =
\sum_R g_R \bigg[\,
\Big({ \bf{\cal{ D}}}^R_{1111} 
+ 2 { \bf{\cal{ D}}}^R_{1112}   
+ 3   { \bf{\cal{ D}}}^R_{1122} 
+ 2  { \bf{\cal{ D}}}^R_{1222}\Big)\ln({x_1})   
\nonumber \\ && \hspace{2.5cm}
+ \Big({ \bf{\cal{ D}}}^R_{2222}
+ 2 { \bf{\cal{ D}}}^R_{2111}
+ 3 { \bf{\cal{ D}}}^R_{2211}  
+ 2 { \bf{\cal{ D}}}^R_{2221} \Big)\ln(x_2)  
\nonumber \\[0.2cm] &&\hspace{2.5cm}
-\Big(   2{ \bf{\cal{ D}}}^R_{1112}
+  3{ \bf{\cal{ D}}}^R_{1122}  
+  2{ \bf{\cal{ D}}}^R_{2221}\Big)\ell_{12}   
+2\sum_{\substack{ 3\leq j < k \leq n  } }  
{\cal{D}}_{12jk}^R   \beta_{12jk} 
 \bigg] \,.
\end{eqnarray}
Combing this now with the splitting anomalous dimension
coming from the ${\bf{ \Gamma}}_{n,\rm{Q}4{\rm{T}}-4{\rm{L}}}$ sector, 
obtained in \Eqn{eq:sp2-Q4T-4L-a}, we have 
\begin{eqnarray}\label{eq:6.78eg}
{\bf{ \Gamma}}_{\SP, 2}^{{\rm Q}4{\rm{T}}}(p_1, p_2;\mu_f)    
&=& {\bf{ \Gamma}}_{n,{\rm{Q}}4{\rm{T}}-2,3{\rm{L}}}(\{s_{ij} \},\mu_f,\alpha_s)
-{\bf{ \Gamma}}_{n-1,{\rm{Q}}4{\rm{T}}-2,3{\rm{L}}}(\{s_{ij} \},\mu_f,\alpha_s)
\nonumber \\[0.2cm] &&
+\,{\bf{ \Gamma}}_{n,\rm{Q}4{\rm{T}}-4{\rm{L}}}(\{\beta_{ijkl} \},\alpha_s) 
-{\bf{ \Gamma}}_{n-1,\rm{Q}4{\rm{T}}-4{\rm{L}}}(\{\beta_{ijkl} \},\alpha_s)
\nonumber \\[0.2cm] &=& 
\sum_R g_R \bigg[ \,
\Big({ \bf{\cal{ D}}}^R_{1111} 
+ 2{ \bf{\cal{ D}}}^R_{1112}     
+ 3{ \bf{\cal{ D}}}^R_{1122} 
+ 2{ \bf{\cal{ D}}}^R_{1222}\Big)\ln({x_1})   
\nonumber \\[-0.1cm] && \hspace{1.2cm}
+ \Big({ \bf{\cal{ D}}}^R_{2222}
+ 2{ \bf{\cal{ D}}}^R_{2111}        
+ 3{ \bf{\cal{ D}}}^R_{2211}  
+ 2{ \bf{\cal{ D}}}^R_{2221} \Big)\ln(x_2)   
\nonumber \\[0.3cm] &&\hspace{1.2cm}
-\Big( 2{ \bf{\cal{ D}}}^R_{1112}
+ 3{ \bf{\cal{ D}}}^R_{1122}  
+ 2{ \bf{\cal{ D}}}^R_{2221}\Big)\ell_{12}   
+2\sum_{\substack{ 3\leq   k < l \leq n  } }  
{\cal{D}}_{12kl}^R   \beta_{12kl} 
 \bigg] 
\nonumber \\ && 
+\,24\, \sum_{R} \sum_{3\leq k<l \leq n}    
\DR_{12kl} \,{\cal G}_R  (\beta_{12kl},\beta_{1lk2})\,.
\end{eqnarray}
In order for the splitting anomalous dimension to be consistent with 
strict collinear factorisation and not depend on the rest-of-the-process
partons, we must demand that
\begin{eqnarray}
{\cal G}_R  (\beta_{12kl},\beta_{1lk2})\Big|_{p_1|| p_2} =
{\cal G}_R  (\beta_{12kl}\to-\infty,\beta_{1lk2}\to 0)
= -\frac{ g_R}{12}\beta_{12kl}  \,.
\end{eqnarray}
Accounting for the factor of $1/2$ in the definition of $g_R$, as 
noted above \Eqn{eq:Q4T-23L},
these constraints are in agreement with the ones
first obtained in Eq.~(76) of Ref.~\cite{Becher:2019avh}
and reported in \Eqn{eq:GRconstraint} above.

\subsection{Calculation of \texorpdfstring{${\bf{ \Gamma}}_{\SP, 2}^{5{\rm{T}}}$}{}}
\label{sec:2-coll_Quintic_terms}

We now turn our attention to the calculation of the 
splitting amplitude soft anomalous dimension of the 
terms connecting five partons, which is the
second type of term that starts to contribute at the 
four-loop order. Concretely, here we compute 
${\bf{ \Gamma}}_{\SP, m}^{5{\rm{T}}}(p_1,\ldots p_m;\mu_f)$
as defined in \Eqn{eq:GammaSp5Tm}
with $m=2$. The object we are interested in is given by
\begin{eqnarray}
\label{eq:GammaSp5T2}
{\bf{ \Gamma}}_{\SP, 2}^{5{\rm{T}}}(p_1, p_2;\mu_f) &=&
    {\bf{ \Gamma}}_{n,5{\rm{T}}-4{\rm{L}}}(\{\beta_{ijkl} \},\alpha_s) + 
     {\bf{ \Gamma}}_{n,5{\rm{T}}-5{\rm{L}}}(\{\beta_{ijkl} \},\alpha_s) 
       \nonumber \\ && \hspace{-2.4cm}
         -   \Big(  {\bf{ \Gamma}}_{n-1,5{\rm{T}}-4{\rm{L}}}(\{\beta_{ijkl} \},\alpha_s)
     + {\bf{ \Gamma}}_{n-1,5{\rm{T}}-5{\rm{L}}}(\{\beta_{ijkl} \},\alpha_s) \Big) |_{\T_P\to \sum_{i=1}^2 \T_i}\,.
\end{eqnarray}
We proceed as in the earlier cases, and we 
split the sums appearing in ${\bf{ \Gamma}}_{n,5{\rm{T}}-5{\rm{L}}}
(\{\beta_{ijkl} \},\alpha_s)$ as parametrised  in \Eqn{eq:H_5T-5L}
in the following way
\begin{eqnarray}
{\bf{ \Gamma}}_{n,5{\rm{T}}-5{\rm{L}}}
(\{\beta_{ijkl} \},\alpha_s) &=& 
8\sum_{m=3}^n \sum_{\substack{3\leq i<j<k<l\leq n\\ i,j,k,l\neq m} }\Big[
\mathcal{T}_{ikljm}
\mathcal{H}_2(\beta_{iklj},\beta_{ikml},\beta_{ilmk},\beta_{kimj},\beta_{kjmi})
\nonumber \\[-0.6cm] &&\hspace{3.0cm}
+\, \mathcal{T}_{ijlkm}
\mathcal{H}_2(\beta_{ijlk},\beta_{ijml},\beta_{ilmj},\beta_{jimk},\beta_{jkmi})
\nonumber \\ && \hspace{3.0cm}
+\mathcal{T}_{ijklm}
\mathcal{H}_2(\beta_{ijkl},\beta_{ijmk},\beta_{ikmj},\beta_{jiml},\beta_{jlmi})\Big]
\nonumber \\[0.1cm] && 
+8 \sum_{\substack{3\leq i<j<k<l\leq n } } 
\Big[ \mathcal{T}_{iklj1}
\mathcal{H}_2(\beta_{iklj},\beta_{ik1l},\beta_{il1k},\beta_{ki1j},\beta_{kj1i}  ) 
\nonumber \\[-0.2cm] && \hspace{2.8cm}
+ \mathcal{T}_{ijlk1}
\mathcal{H}_2(\beta_{ijlk},\beta_{ij1l},\beta_{il1j},\beta_{ji1k},\beta_{jk1i}  )
\nonumber \\[0.2cm] && \hspace{2.8cm}
+\mathcal{T}_{ijkl1}
\mathcal{H}_2(\beta_{ijkl},\beta_{ij1k},\beta_{ik1j},\beta_{ji1l},\beta_{jl1i}  )    
\nonumber \\[0.2cm] && \hspace{2.8cm}
+\mathcal{T}_{iklj2}
\mathcal{H}_2(\beta_{iklj},\beta_{ik2l},\beta_{il2k},\beta_{ki2j},\beta_{kj2i}  ) 
\nonumber \\[0.2cm] && \hspace{2.8cm}
+\mathcal{T}_{ijlk2}
\mathcal{H}_2(\beta_{ijlk},\beta_{ij2l},\beta_{il2j},\beta_{ji2k},\beta_{jk2i}  )
\nonumber \\[0.2cm] && \hspace{2.8cm}
+\mathcal{T}_{ijkl2}
\mathcal{H}_2(\beta_{ijkl},\beta_{ij2k},\beta_{ik2j},\beta_{ji2l},\beta_{jl2i}  ) \Big]   
\nonumber \\[-0.2cm] &&
+8\sum_{m=3}^n   \sum_{\substack{3\leq  j<k<l\leq n\\ j,k,l\neq m} } \Big[
\mathcal{T}_{1kljm}
\mathcal{H}_2(\beta_{1klj},\beta_{1kml},\beta_{1lmk},\beta_{k1mj},\beta_{kjm1}  ) 
\nonumber \\[-0.5cm] && \hspace{3.0cm}
+\mathcal{T}_{1jlkm}
\mathcal{H}_2(\beta_{1jlk},\beta_{1jml},\beta_{1lmj},\beta_{j1mk},\beta_{jkm1}  )
\nonumber \\[0.2cm] && \hspace{3.0cm}
+\mathcal{T}_{1jklm}
\mathcal{H}_2(\beta_{1jkl},\beta_{1jmk},\beta_{1kmj},\beta_{j1ml},\beta_{jlm1}  )   
\nonumber \\[0.2cm] && \hspace{3.0cm}
+\mathcal{T}_{2kljm}
\mathcal{H}_2(\beta_{2klj},\beta_{2kml},\beta_{2lmk},\beta_{k2mj},\beta_{kjm2}  ) 
\nonumber \\[0.2cm] && \hspace{3.0cm}
+\mathcal{T}_{2jlkm}
\mathcal{H}_2(\beta_{2jlk},\beta_{2jml},\beta_{2lmj},\beta_{j2mk},\beta_{jkm2}  )
\nonumber \\[0.2cm] && \hspace{3.0cm}
+\mathcal{T}_{2jklm}
\mathcal{H}_2(\beta_{2jkl},\beta_{2jmk},\beta_{2kmj},\beta_{j2ml},\beta_{jlm2}  ) \Big]  
\nonumber \\[0.1cm] &&
+8 \sum_{\substack{3\leq  j<k<l\leq n } } \Big[ 
\mathcal{T}_{2klj1}
\mathcal{H}_2(\beta_{2klj},\beta_{2k1l},\beta_{2l1k},\beta_{k21j},\beta_{kj12}  ) 
\nonumber \\[-0.2cm] && \hspace{2.4cm}
+ \mathcal{T}_{2jlk1}
\mathcal{H}_2(\beta_{2jlk},\beta_{2j1l},\beta_{2l1j},\beta_{j21k},\beta_{jk12}  )
\nonumber \\[0.2cm] && \hspace{2.4cm}
+\mathcal{T}_{2jkl1}
\mathcal{H}_2(\beta_{2jkl},\beta_{2j1k},\beta_{2k1j},\beta_{j21l},\beta_{jl12}  )
\nonumber \\[0.2cm] && \hspace{2.4cm}
+\mathcal{T}_{1klj2}
\mathcal{H}_2(\beta_{1klj},\beta_{1k2l},\beta_{1l2k},\beta_{k12j},\beta_{kj21}  ) 
\nonumber \\[0.2cm] && \hspace{2.4cm}
+ \mathcal{T}_{1jlk2}
\mathcal{H}_2(\beta_{1jlk},\beta_{1j2l},\beta_{1l2j},\beta_{j12k},\beta_{jk21}  )
\nonumber \\[0.2cm] && \hspace{2.4cm}
+\mathcal{T}_{1jkl2}
\mathcal{H}_2(\beta_{1jkl},\beta_{1j2k},\beta_{1k2j},\beta_{j12l},\beta_{jl21}  ) \Big]
 \\[-0.1cm] &&
+ 8\sum_{m=3}^n   \sum_{\substack{3\leq  k<l\leq n\\  k,l\neq m} }\Big[
\mathcal{T}_{1kl2m}
\mathcal{H}_2(\beta_{1kl2},\beta_{1kml},\beta_{1lmk},\beta_{k1m2},\beta_{k2m1}  ) 
\nonumber \\[-0.5cm] && \hspace{2.7cm}
+\mathcal{T}_{12lkm}
\mathcal{H}_2(\beta_{12lk},\beta_{12ml},\beta_{1lm2},\beta_{21mk},\beta_{2km1}  )
\nonumber \\[0.2cm] && \hspace{2.7cm}
+\mathcal{T}_{12klm}
\mathcal{H}_2(\beta_{12kl},\beta_{12mk},\beta_{1km2},\beta_{21ml},\beta_{2lm1}  )
\Big]. \nonumber
\end{eqnarray}
The ${\cal{H}}_2$ functions appearing on the right-hand side 
have the $\alpha_s$ dependence kept implicit, as will be done 
for the terms appearing below. 
The corresponding $n-1$ soft anomalous dimension is split
in a similar fashion
\begin{eqnarray}
{\bf{ \Gamma}}_{n-1,5{\rm{T}}-5{\rm{L}}}(\{\beta_{ijkl} \},\alpha_s) 
&& \nonumber \\ && \hspace{-4.0cm}
=\, 8 \sum_{m=3}^n  \sum_{\substack{3\leq i<j<k<l\leq n\\ i,j,k,l\neq m} }
\Big[\, \mathcal{T}_{ikljm}
\mathcal{H}_2(\beta_{iklj},\beta_{ikml},\beta_{ilmk},\beta_{kimj},\beta_{kjmi}  )
\nonumber \\[-0.5cm] && \hspace{-0.5cm}+ 
\mathcal{T}_{ijlkm}
\mathcal{H}_2(\beta_{ijlk},\beta_{ijml},\beta_{ilmj},\beta_{jimk},\beta_{jkmi}  )
\nonumber \\[0.2cm] && \hspace{-0.5cm}+
\mathcal{T}_{ijklm}
\mathcal{H}_2(\beta_{ijkl},\beta_{ijmk},\beta_{ikmj},\beta_{jiml},\beta_{jlmi}  ) \Big]
\nonumber \\[0.3cm] && \hspace{-4.0cm}
+ 8 \sum_{\substack{3\leq i<j<k<l\leq n } } 
\Big[ \mathcal{T}_{ikljP}
\mathcal{H}_2(\beta_{iklj},\beta_{ikPl},\beta_{ilPk},\beta_{kiPj},\beta_{kjPi}  ) 
\nonumber \\[-0.2cm] && \hspace{-1.5cm}
+ \mathcal{T}_{ijlkP}
\mathcal{H}_2(\beta_{ijlk},\beta_{ijPl},\beta_{ilPj},\beta_{jiPk},\beta_{jkPi}  )
\nonumber \\[0.1cm] && \hspace{-1.5cm}
+ \mathcal{T}_{ijklP}
\mathcal{H}_2(\beta_{ijkl},\beta_{ijPk},\beta_{ikPj},\beta_{jiPl},\beta_{jlPi}  )
\Big]
\nonumber \\[-0.1cm] && \hspace{-4.0cm}
+ 8 \sum_{m=3}^n   \sum_{\substack{3\leq  j<k<l\leq n\\  j,k,l\neq m} }\,\,\,
\Big[ \mathcal{T}_{Pkljm}
\mathcal{H}_2(\beta_{Pklj},\beta_{Pkml},\beta_{Plmk},\beta_{kPmj},\beta_{kjmP}  ) 
\nonumber \\[-0.5cm] && \hspace{-1.0cm}+ 
\mathcal{T}_{Pjlkm}
\mathcal{H}_2(\beta_{Pjlk},\beta_{Pjml},\beta_{Plmj},\beta_{jPmk},\beta_{jkmP}  )
\nonumber \\[0.1cm] && \hspace{-1.0cm}+
\mathcal{T}_{Pjklm}
\mathcal{H}_2(\beta_{Pjkl},\beta_{Pjmk},\beta_{Pkmj},\beta_{jPml},\beta_{jlmP}  )
\Big].
\end{eqnarray}
Taking the difference between the above
two equations as dictated by~\Eqn{eq:GammaSp5T2},
we see that terms depending exclusively on the 
rest-of-the-process partons directly cancel.
More cancellations occur upon substituting 
$\T_P = \T_1 + \T_2$ and using rescaling 
invariance for the CICRs as before 
$\beta_{1jkl} = \beta_{2jkl} = \beta_{Pjkl}$.
We then find that
\begin{eqnarray}\label{eq:5T5L-2pcdiff}
{\bf{ \Gamma}}_{n,5{\rm{T}}-5{\rm{L}}}(\{\beta_{ijkl} \},\alpha_s)  
- {\bf{ \Gamma}}_{n-1,5{\rm{T}}-5{\rm{L}}}(\{\beta_{ijkl} \},\alpha_s)  
&& \nonumber \\[0.1cm] 
&&\hspace{-6.0cm}
= 8 \sum_{\substack{3\leq  j<k<l\leq n } } \Big[ 
\mathcal{T}_{2klj1}
\mathcal{H}_2(\beta_{2klj},\beta_{2k1l},\beta_{2l1k},\beta_{k21j},\beta_{kj12}  )
\nonumber \\[-0.3cm] && \hspace{-3.4cm}
+ \mathcal{T}_{2jlk1}
\mathcal{H}_2(\beta_{2jlk},\beta_{2j1l},\beta_{2l1j},\beta_{j21k},\beta_{jk12}  )
\nonumber \\[0.2cm] && \hspace{-3.4cm}
+ \mathcal{T}_{2jkl1}
\mathcal{H}_2(\beta_{2jkl},\beta_{2j1k},\beta_{2k1j},\beta_{j21l},\beta_{jl12}  )
\nonumber \\[0.2cm] && \hspace{-3.4cm}
+\mathcal{T}_{1klj2}
\mathcal{H}_2(\beta_{1klj},\beta_{1k2l},\beta_{1l2k},\beta_{k12j},\beta_{kj21}  )
\nonumber \\[0.2cm] && \hspace{-3.4cm}
+\mathcal{T}_{1jlk2}
\mathcal{H}_2(\beta_{1jlk},\beta_{1j2l},\beta_{1l2j},\beta_{j12k},\beta_{jk21}  )
\nonumber \\[0.1cm] && \hspace{-3.4cm}
+\mathcal{T}_{1jkl2}
\mathcal{H}_2(\beta_{1jkl},\beta_{1j2k},\beta_{1k2j},\beta_{j12l},\beta_{jl21}  ) \Big] 
\nonumber \\[-0.2cm] && \hspace{-6.0cm}
+ 8\sum_{m=3}^n \sum_{\substack{3\leq  k<l\leq n\\  k,l\neq m} }
\Big[ \mathcal{T}_{1kl2m}
\mathcal{H}_2(\beta_{1kl2},\beta_{1kml},\beta_{1lmk},\beta_{k1m2},\beta_{k2m1}  )
\nonumber \\[-0.5cm] && \hspace{-3.2cm}+ 
\mathcal{T}_{12lkm}
\mathcal{H}_2(\beta_{12lk},\beta_{12ml},\beta_{1lm2},\beta_{21mk},\beta_{2km1}  )
\nonumber \\[0.2cm] && \hspace{-3.2cm}+
\mathcal{T}_{12klm}
\mathcal{H}_2(\beta_{12kl},\beta_{12mk},\beta_{1km2},\beta_{21ml},\beta_{2lm1}  )
\Big]. 
\end{eqnarray}
Next, we use general symmetry properties of CICRs,
such as $\beta_{1k2l} = -  \beta_{12kl}$, and 
their behaviour in the two-particle collinear limit
(see Section~\ref{sec:kin-two-part-col}) to 
simplify the expression. We have 
\begin{eqnarray}
{\bf{ \Gamma}}_{n,5{\rm{T}}-5{\rm{L}}}(\{\beta_{ijkl} \},\alpha_s)  
- {\bf{ \Gamma}}_{n-1,5{\rm{T}}-5{\rm{L}}}
(\{\beta_{ijkl} \},\alpha_s)  \nonumber && \\[0.1cm] 
&& \hspace{-7.0cm} =\, 8 
\sum_{\substack{3\leq  j<k<l\leq n }} \Big[
\mathcal{T}_{2klj1}\mathcal{H}_2(\beta_{1klj},-\beta_{12kl},-\beta_{12kl},0,\beta_{12jk}  )
\nonumber \\[-0.2cm] && \hspace{-4.2cm}
+ \mathcal{T}_{2jlk1}
\mathcal{H}_2(\beta_{1jlk},-\beta_{12jl},-\beta_{12jl},0,\beta_{12jk} )
\nonumber \\[0.2cm] && \hspace{-4.2cm}
+ \mathcal{T}_{2jkl1}
\mathcal{H}_2(\beta_{1jkl},-\beta_{12jk},-\beta_{12jk},0,\beta_{12jl} )
\nonumber \\[0.2cm] && \hspace{-4.2cm}
+ \mathcal{T}_{1klj2}
\mathcal{H}_2(\beta_{1klj},-\beta_{12kl},-\beta_{12kl},0,\beta_{12jk} )
\nonumber \\[0.2cm] && \hspace{-4.2cm}
+ \mathcal{T}_{1jlk2}
\mathcal{H}_2(\beta_{1jlk},-\beta_{12jl},-\beta_{12jl},0,\beta_{12jk} )
\nonumber \\[0.1cm] && \hspace{-4.2cm}
+ \mathcal{T}_{1jkl2}
\mathcal{H}_2(\beta_{1jkl},-\beta_{12jk},-\beta_{12jk},0,\beta_{12jl} )\Big]
\nonumber \\[-0.1cm] && \hspace{-7.0cm}
+\, 8 \, \sum_{m=3}^n 
\sum_{\substack{3\leq  k<l\leq n\\  k,l\neq m} }\Big[
\mathcal{T}_{1kl2m}
\mathcal{H}_2(0,\beta_{1kml},\beta_{1lmk},-\beta_{12km},-\beta_{12km} )
\nonumber \\[-0.4cm] && \hspace{-4.0cm}
+ \mathcal{T}_{12lkm}
\mathcal{H}_2(\beta_{12kl},\beta_{12lm},0,\beta_{12km},0  )
\nonumber \\[0.2cm] && \hspace{-4.0cm}
+\mathcal{T}_{12klm}
\mathcal{H}_2(\beta_{12kl},\beta_{12km},0,\beta_{12lm},0  )\Big]\,. 
\end{eqnarray}
This expression can be simplified further. 
Due to Bose symmetry of the soft anomalous dimension,
we know that the function $\mathcal{H}_2$ 
has the following symmetry property \cite{Becher:2019avh}
\begin{eqnarray}\label{eq:H2-sym2}
\mathcal{H}_2(y_1,y_2,y_3,y_4,y_5) = 
-\mathcal{H}_2(-y_1,y_2-y_4+y_5,y_3+y_1-y_4+y_5,y_5,y_4)\,.
\end{eqnarray}
However, considering $\mathcal{H}_2(0,\beta_{1kml},\beta_{1lmk},-\beta_{12km},-\beta_{12km}  )$
in the collinear limit using this symmetry  we observe that the function must be equal to minus
itself as $y_1=0$ and $y_4=y_5$ in this case. Therefore, the only solution is that 
\begin{eqnarray}
\mathcal{H}_2(0,\beta_{1kml},\beta_{1lmk},-\beta_{12km},-\beta_{12km}  ) =0 \,.
\end{eqnarray}
Using this result and simplifying, we find
\begin{eqnarray}\label{eq:5T5L-2pc}
&&  {\bf{ \Gamma}}_{n,5{\rm{T}}-5{\rm{L}}}(\{\beta_{ijkl} \},\alpha_s)  
- {\bf{ \Gamma}}_{n-1,5{\rm{T}}-5{\rm{L}}}(\{\beta_{ijkl} \},\alpha_s)  
\nonumber \\ && \hspace{2.4cm}= 8
\sum_{\substack{3\leq  j<k<l\leq n }}\hspace{0.0cm}\Big[
\big(\mathcal{T}_{1klj2}+ \mathcal{T}_{2klj1}\big)
\mathcal{H}_2(\beta_{1klj},-\beta_{12kl},-\beta_{12kl},0,\beta_{12jk}  ) 
\nonumber \\[-0.3cm] && \hspace{4.75cm}
+\big(\mathcal{T}_{1jlk2}+\mathcal{T}_{2jlk1}\big)
\mathcal{H}_2(\beta_{1jlk},-\beta_{12jl},-\beta_{12jl},0,\beta_{12jk}  )
\nonumber \\ && \hspace{4.75cm}
+ \big(\mathcal{T}_{1jkl2}+ \mathcal{T}_{2jkl1}\big)
\mathcal{H}_2(\beta_{1jkl},-\beta_{12jk},-\beta_{12jk},0,\beta_{12jl}  ) \Big]
\nonumber \\[-0.3cm] &&  \hspace{2.4cm}
+ 8 \sum_{m=3}^n   \sum_{\substack{3\leq  k<l\leq n\\  k,l\neq m} } \,
\Big( \mathcal{T}_{12lkm}
\mathcal{H}_2(\beta_{12kl},\beta_{12lm},0,\beta_{12km},0  )
\nonumber \\[-0.7cm] && \hspace{5.35cm}+
\mathcal{T}_{12klm}
\mathcal{H}_2(\beta_{12kl},\beta_{12km},0,\beta_{12lm},0  ) \Big)\,.
\end{eqnarray} 
We now proceed to consider the second type of term depending on 
five colour generators 
${\bf{ \Gamma}}_{n,5{\rm{T}}-4{\rm{L}}}(\{\beta_{ijkl} \},\alpha_s)$
which is given in \Eqn{eq:H_5T-4L}.
We split the sums appearing in this contribution as follows
\begin{eqnarray} \label{eq:5.148}
{\bf{ \Gamma}}_{n,5{\rm{T}}-4{\rm{L}}}(\{\beta_{ijkl} \},\alpha_s) 
\nonumber && \\ 
&&\hspace{-4.0cm}=\, 
2 \sum_{i=3}^n\sum_{\substack{ 3\leq j < k < l\leq n\\ j,k,l \neq i } } 
\Big[\mathcal{T}_{iklji}\mathcal{H}_1(\beta_{ikjl},\beta_{iljk} )
+\mathcal{T}_{ijlki}\mathcal{H}_1(\beta_{ijkl},\beta_{ilkj} )
+\mathcal{T}_{ijkli}\mathcal{H}_1(\beta_{ijlk},\beta_{iklj} ) \Big]  
\nonumber \\[-0.3cm] &&\hspace{-4.0cm}
+  2 \sum_{\substack{ 3\leq j < k < l\leq n } } \Big[     
\mathcal{T}_{1klj1}\mathcal{H}_1(\beta_{1kjl},\beta_{1ljk} )  
+\,\mathcal{T}_{1jlk1}\mathcal{H}_1(\beta_{1jkl},\beta_{1lkj} )
+\,\mathcal{T}_{1jkl1}\mathcal{H}_1(\beta_{1jlk},\beta_{1klj} )  
\nonumber \\[-0.2cm] && \hspace{-1.7cm}
+\mathcal{T}_{2klj2}\mathcal{H}_1(\beta_{2kjl},\beta_{2ljk} ) 
+\mathcal{T}_{2jlk2}\mathcal{H}_1(\beta_{2jkl},\beta_{2lkj} )
+\mathcal{T}_{2jkl2}\mathcal{H}_1(\beta_{2jlk},\beta_{2klj} )
\Big] \nonumber \\[0.0cm] &&\hspace{-4.0cm}
+ 2  \sum_{i=3}^n\sum_{\substack{ 3\leq  k < l\leq n\\ k,l \neq i } } 
\Big[\mathcal{T}_{ikl1i}\mathcal{H}_1(\beta_{ik1l},\beta_{il1k} ) 
+\mathcal{T}_{i1lki}\mathcal{H}_1(\beta_{i1kl},\beta_{ilk1} )
+\mathcal{T}_{i1kli}\mathcal{H}_1(\beta_{i1lk},\beta_{ikl1} ) 
\nn \\[-0.4cm] &&\hspace{-1.2cm}+
\mathcal{T}_{ikl2i}\mathcal{H}_1(\beta_{ik2l},\beta_{il2k}  )
+\mathcal{T}_{i2lki}\mathcal{H}_1(\beta_{i2kl},\beta_{ilk2}  )
+\mathcal{T}_{i2kli}\mathcal{H}_1(\beta_{i2lk},\beta_{ikl2}  )\Big] 
\nonumber \\[0.1cm] &&\hspace{-4.0cm}
+ 2 \sum_{3\leq k< l\leq n}  \Big[
\mathcal{T}_{1kl21}\mathcal{H}_1(\beta_{1k2l},\beta_{1l2k}  )
+\mathcal{T}_{12lk1}\mathcal{H}_1(\beta_{12kl},\beta_{1lk2}  )
+\mathcal{T}_{12kl1}\mathcal{H}_1(\beta_{12lk},\beta_{1kl2}  ) 
\nonumber  \\[-0.2cm] &&  \hspace{-1.8cm}
+\mathcal{T}_{2kl12}\mathcal{H}_1(\beta_{2k1l},\beta_{2l1k}  )
+\mathcal{T}_{21lk2}\mathcal{H}_1(\beta_{21kl},\beta_{2lk1}  )
+\mathcal{T}_{21kl2}\mathcal{H}_1(\beta_{21lk},\beta_{2kl1}  )  \Big]
\nonumber  \\[0.1cm] &&  \hspace{-4.0cm}
+  2  \sum_{i=3} \sum_{\substack{l=3 \\ l\neq i}}    \Big[  
\mathcal{T}_{i2l1i}\mathcal{H}_1(\beta_{i21l},\beta_{il12}  )
+\mathcal{T}_{i1l2i}\mathcal{H}_1(\beta_{i12l},\beta_{il21}  )
+\mathcal{T}_{i12li}\mathcal{H}_1(\beta_{i1l2},\beta_{i2l1}  ) \Big].  
\end{eqnarray}
The corresponding anomalous dimension for $n-1$ external partons is given by
\begin{eqnarray}
{\bf{ \Gamma}}_{n-1,5{\rm{T}}-4{\rm{L}}}(\{\beta_{ijkl} \},\alpha_s)  
&& \\ && \hspace{-4.0cm}
=\,2 \sum_{i=3}^n\sum_{\substack{ 3\leq j < k < l\leq n\\ j,k,l \neq i } } 
\Big[\mathcal{T}_{iklji}\mathcal{H}_1(\beta_{ikjl},\beta_{iljk}  )
+\mathcal{T}_{ijlki}\mathcal{H}_1(\beta_{ijkl},\beta_{ilkj}  )
+\mathcal{T}_{ijkli}\mathcal{H}_1(\beta_{ijlk},\beta_{iklj}  ) \Big]  
\nonumber \\[-0.2cm]  &&\hspace{-3.9cm}
+2\hspace{-0.25cm}\sum_{\substack{ 3\leq j < k < l \leq n  } } \hspace{-0.25cm}
\Big[ \mathcal{T}_{PkljP}\mathcal{H}_1(\beta_{Pkjl},\beta_{Pljk}  )
+\mathcal{T}_{PjlkP}\mathcal{H}_1(\beta_{Pjkl},\beta_{Plkj}  )
+\mathcal{T}_{PjklP}\mathcal{H}_1(\beta_{Pjlk},\beta_{Pklj}  ) \Big]  
\nonumber \\[-0.2cm] \nn &&\hspace{-3.9cm}
+2\sum_{i=3}^n\sum_{\substack{3 \leq k < l \leq n \\ k,l \neq i}}
\hspace{-0.15cm}
\Big[\mathcal{T}_{iklPi}\mathcal{H}_1(\beta_{ikPl},\beta_{ilPk}  ) 
+\mathcal{T}_{iPlki}\mathcal{H}_1(\beta_{iPkl},\beta_{ilkP}  )
+\mathcal{T}_{iPkli}\mathcal{H}_1(\beta_{iPlk},\beta_{iklP}  ) \Big].  
\end{eqnarray}
Taking the difference between the above two equations according to 
\Eqn{eq:GammaSp5T2}, substituting $\T_P = \T_1 + \T_2$, 
and using rescaling invariance of \Eqn{eq:resclinv} leads to
\begin{eqnarray}\label{eq:Sp5T4Ld1}
&&{\bf{ \Gamma}}_{n,5{\rm{T}}-4{\rm{L}}}(\{\beta_{ijkl} \},\alpha_s) 
-{\bf{ \Gamma}}_{n-1,5{\rm{T}}-4{\rm{L}}}(\{\beta_{ijkl} \},\alpha_s)
\\[0.1cm] && = 2\sum_{3\leq k< l\leq n}  \Big(
\mathcal{T}_{1kl21}\mathcal{H}_1(\beta_{1k2l},\beta_{1l2k}  )
+\,\mathcal{T}_{12lk1}\mathcal{H}_1(\beta_{12kl},\beta_{1lk2}  )
+\,\mathcal{T}_{12kl1}\mathcal{H}_1(\beta_{12lk},\beta_{1kl2}  ) 
\nonumber  \\[-0.3cm] &&  \hspace{2cm}
+ \mathcal{T}_{2kl12}\mathcal{H}_1(\beta_{2k1l},\beta_{2l1k}  )
+\,\mathcal{T}_{21lk2}\mathcal{H}_1(\beta_{21kl},\beta_{2lk1}  )
+\,\mathcal{T}_{21kl2}\mathcal{H}_1(\beta_{21lk},\beta_{2kl1}  ) \Big)
\nonumber  \\[0.2cm] &&   \hspace{0.25cm}
+  2\,\,  \sum_{i=3} \sum_{\substack{l=3 \\ l\neq i}}\,\,    \Big(\,
\mathcal{T}_{i2l1i}\,\mathcal{H}_1(\beta_{i21l},\beta_{il12}  )
+\,  \mathcal{T}_{i1l2i}\,\mathcal{H}_1(\beta_{i12l},\beta_{il21}  )
+\mathcal{T}_{i12li}\,\mathcal{H}_1(\beta_{i1l2},\beta_{i2l1}  )\, \Big)  
\nonumber \\[-0.4cm] &&  \hspace{0.25cm}
- 2\sum_{\substack{ 3\leq j < k < l \leq n  } } 
\Big( \mathcal{T}_{1klj2}\mathcal{H}_1(\beta_{Pkjl},\beta_{Pljk}  )
+ \mathcal{T}_{2klj1}\mathcal{H}_1(\beta_{Pkjl},\beta_{Pljk}  )
+  \mathcal{T}_{1jlk2}\mathcal{H}_1(\beta_{Pjkl},\beta_{Plkj}  )
\nonumber \\[-0.3cm] \nonumber&&\hspace{2.5cm}
+  \mathcal{T}_{2jlk1}\mathcal{H}_1(\beta_{Pjkl},\beta_{Plkj}  )
+\mathcal{T}_{1jkl2 }\mathcal{H}_1(\beta_{Pjlk},\beta_{Pklj}  )
+\mathcal{T}_{ 2jkl1}\mathcal{H}_1(\beta_{Pjlk},\beta_{Pklj}  )\Big).     
\end{eqnarray}
Since the function $\mathcal{H}_1$ 
is antisymmetric under exchange of its arguments \cite{Becher:2019avh}
\begin{eqnarray}\label{eq:H1-sym1}
\mathcal{H}_1(y_1,y_2) = 
-\mathcal{H}_1(y_2,y_1),
\end{eqnarray}
and from \Eqn{eq:CICR1-2pc-ij} we know that in the collinear limit
$\beta_{1k2l}$ and  $\beta_{1l2k}$ are equal to each other, we 
deduce that  
\be
\mathcal{H}_1(\beta_{1k2l},\beta_{1l2k}  )\stackrel{p_1||p_2}{\longrightarrow}0\,,
\qquad
\mathcal{H}_1(\beta_{2k1l},\beta_{2l1k}  )\stackrel{p_1||p_2}{\longrightarrow}0\,,
\qquad
\mathcal{H}_1(\beta_{i1l2},\beta_{i2l1}  )\stackrel{p_1||p_2}{\longrightarrow}0\,.
\ee
Moreover, using general symmetry properties of the logarithms of the CICRs
such as $\beta_{21lk} = \beta_{12kl}$,
$\beta_{2kl1}=\beta_{1lk2}$, $\beta_{il21}=\beta_{12li}$
and $\beta_{i12l}=\beta_{1il2}$, and their behaviour in the collinear limit, 
such as $\beta_{1lk2} \to 0$ and $\beta_{12lk} =\beta_{12kl}$,
as discussed in Section~\ref{sec:kin-two-part-col}, we can simplify 
\Eqn{eq:Sp5T4Ld1} to 
\begin{eqnarray}\label{eq:5.165}
{\bf{ \Gamma}}_{n,5{\rm{T}}-4{\rm{L}}}(\{\beta_{ijkl} \},\alpha_s) 
- {\bf{ \Gamma}}_{n-1,5{\rm{T}}-4{\rm{L}}}(\{\beta_{ijkl} \},\alpha_s)
\nonumber && \\[0.1cm] && \hspace{-7.0cm}
=\, 2  \sum_{k=3} \sum_{\substack{l=3 \\ l\neq k}}  \, 
\Big[ \mathcal{T}_{12kl1} 
+\mathcal{T}_{21kl2} +\mathcal{T}_{kl21k}
+ \mathcal{T}_{kl12k} \Big]
\mathcal{H}_1(\beta_{12kl},0 )
\nonumber \\[-0.2cm] && \hspace{-6.8cm}
-2 \sum_{\substack{ 3\leq j < k < l \leq n  } }  
\Big[\Big(\mathcal{T}_{1klj2}+\mathcal{T}_{2klj1}\Big)
\mathcal{H}_1(\beta_{Pkjl},\beta_{Pljk}  )
\nonumber \\[-0.2cm] &&\hspace{-4.6cm}
+ \Big( \mathcal{T}_{1jlk2}+\,  \mathcal{T}_{2jlk1}\Big)
\mathcal{H}_1(\beta_{Pjkl},\beta_{Plkj}  ) 
\nonumber \\[0.1cm] &&\hspace{-4.6cm} 
+\Big(\mathcal{T}_{1jkl2 }+\,\mathcal{T}_{ 2jkl1}\Big)
\mathcal{H}_1(\beta_{Pjlk},\beta_{Pklj}  ) \Big].
\end{eqnarray}
At this point we combine \Eqn{eq:5.165}
with \Eqn{eq:5T5L-2pc} according to definition
of the splitting amplitude soft anomalous dimension
${\bf{ \Gamma}}_{\SP, 2}^{5{\rm{T}}}$ 
in \Eqn{eq:GammaSp5T2}. We then find 
\begin{eqnarray} \label{eq:5.187}
&&{\bf{ \Gamma}}_{\SP, 2}^{5{\rm{T}}}(p_1, p_2;\mu_f)= 
2  \sum_{k=3} \sum_{\substack{l=3 \\ l\neq k}}  \,
\Big[  \,   \mathcal{T}_{12kl1} + \,\mathcal{T}_{21kl2} 
+\mathcal{T}_{kl21k}+\,  \mathcal{T}_{kl12k}
\Big]\mathcal{H}_1(\beta_{12kl},0  )
\\[-0.4cm] && \hspace{0.0cm} 
+\hspace{-0.2cm}\sum_{\substack{3\leq  j<k<l\leq n } }
\bigg\{ \Big[\mathcal{T}_{1klj2}+\mathcal{T}_{2klj1}\Big] 
\Big( 8\mathcal{H}_2(\beta_{1klj},-\beta_{12kl},-\beta_{12kl},0,\beta_{12jk}  )
-2 \mathcal{H}_1(\beta_{Pkjl},\beta_{Pljk}  ) \Big)
\nonumber \\[-0.2cm] && \hspace{1.9cm}
+\Big[ \mathcal{T}_{1jlk2}+ \mathcal{T}_{2jlk1}\Big] 
\Big(  8\mathcal{H}_2(\beta_{1jlk},-\beta_{12jl},-\beta_{12jl},0,\beta_{12jk}  )
-2\mathcal{H}_1(\beta_{Pjkl},\beta_{Plkj}  )\Big)
\nonumber \\[0.0cm] && \hspace{1.9cm}
+ \Big[ \mathcal{T}_{1jkl2}+ \mathcal{T}_{2jkl1}\Big] 
\Big(  8\mathcal{H}_2(\beta_{1jkl},-\beta_{12jk},-\beta_{12jk},0,\beta_{12jl}  )
-2 \mathcal{H}_1(\beta_{Pjlk},\beta_{Pklj}  )\Big) \bigg\}
\nonumber \\[-0.2cm] \nn && \hspace{0.0cm}
+ 8 \sum_{m=3}^n   \sum_{\substack{3\leq  k<l\leq n\\  k,l\neq m} } 
\hspace{-0.2cm}\Big(\mathcal{T}_{12lkm}
\mathcal{H}_2(\beta_{12kl},\beta_{12lm},0,\beta_{12km},0  )
+\mathcal{T}_{12klm}
\mathcal{H}_2(\beta_{12kl},\beta_{12km},0,\beta_{12lm},0  )  \Big).   
\end{eqnarray}
The result for ${\bf{ \Gamma}}_{\SP, 2}^{5{\rm{T}}}$ presented here
has previously been obtained in App.~C of Ref.~\cite{Becher:2019avh}.
However, this fact is not immediately obvious, as our result is not
written in the most compact form. In order to touch base with the
result in the literature we perform the following manipulations:
we use the colour identities 
$ \mathcal{T}_{kl12k} = - \mathcal{T}_{12klk}$ and 
$ \mathcal{T}_{1jkl2}=  \mathcal{T}_{12klj}-\mathcal{T}_{12jlk}$, 
and  employ antisymmetry properties of $\mathcal{H}_1$ and $\mathcal{H}_2$
to rewrite the terms in the middle sum. We then relabel the indices
in the last sum to match the ones in the middle sum. Lastly, we remove 
the ordering present in the sums throughout the derivation 
and use the $ \mathcal{T}_{12klm} = - \mathcal{T}_{21lkm}$ colour identity
in the last term to combine the middle and last sums. We then find 
\begin{eqnarray}
&& {\bf{ \Gamma}}_{\SP, 2}^{5{\rm{T}}}(p_1, p_2;\mu_f)=  
2  \sum_{k=3} \sum_{\substack{l=3 \\ l\neq k}}  \,
\Big[\,\mathcal{T}_{12kl1} -\,  \mathcal{T}_{12klk}+\,\mathcal{T}_{21kl2}-\mathcal{T}_{21klk}\Big]
\mathcal{H}_1(\beta_{12kl},0  )
\nonumber \\[-0.2cm] && \hspace{1.4cm} 
+ \sum_{ (k,l,m) \neq 1,2}
\Big[ \mathcal{T}_{12klm}+ \mathcal{T}_{21klm}\Big] 
\Big(  8\mathcal{H}_2(\beta_{1mkl},-\beta_{12mk},-\beta_{12mk},0,\beta_{12ml}  )
\nonumber \\[-0.2cm] &&  \hspace{3.8cm} 
-2 \mathcal{H}_1(\beta_{1mlk},\beta_{1klm}  )
+4
\mathcal{H}_2(\beta_{12kl},\beta_{12km},0,\beta_{12lm},0  ) \Big), 
\end{eqnarray}
which is in agreement with Eq. (C.3) of Ref.~\cite{Becher:2019avh}.
In order to formulate the constraints, we write the functions appearing in the above
result in terms of the same variables using 
$ \beta_{12km} = \beta_{12kl} + \beta_{1lmk}$,    
$\beta_{12lm} = \beta_{12kl} + \beta_{1kml}$,
$\beta_{1mlk} = - \beta_{1lmk}$, and
$\beta_{1klm} =   \beta_{1kml} -  \beta_{1lmk}$.  
We then find 
\begin{eqnarray}\label{eq:5.212z}
&& {\bf{ \Gamma}}_{\SP, 2}^{5{\rm{T}}}(p_1, p_2;\mu_f)= 
2  \sum_{k=3} \sum_{\substack{l=3 \\ l\neq k}}  \,
\Big[ \,\mathcal{T}_{12kl1} -\,  \mathcal{T}_{12klk}+\,\mathcal{T}_{21kl2} -\mathcal{T}_{21klk}\Big]
\mathcal{H}_1(\beta_{12kl},0  )
\\\nonumber && \hspace{0.6cm} 
+\sum_{ (k,l,m) \neq 1,2}
\Big[ \mathcal{T}_{12klm}- \mathcal{T}_{12lkm}\Big] 
\Big(   K_+(\beta_{1kml},\beta_{1lmk}, \beta_{12kl}  ) +  K_-(\beta_{1kml},\beta_{1lmk}, \beta_{12kl}  )  \Big)\,, 
\end{eqnarray}
where 
\begin{eqnarray}\label{eq:Kplusminus}
K_{\pm}(\beta_{1kml},\beta_{1lmk}, \beta_{12kl})&=&\Big(
4\mathcal{H}_2(- {\beta_{1kml}},- \beta_{12kl} - \beta_{1lmk} ,- \beta_{12kl} - \beta_{1lmk} ,0,
\beta_{12kl} + \beta_{1kml}  )
\nonumber \\ &&  \hspace{-3.9cm}
- \mathcal{H}_1( - \beta_{1lmk}, \beta_{1kml} -  \beta_{1lmk}  )
+2\mathcal{H}_2(\beta_{12kl}, \beta_{12kl} + \beta_{1lmk} ,0,\beta_{12kl} + \beta_{1kml},0  ) \Big)
\nonumber \\ &&  \hspace{-3.9cm}     
\pm \Big(  4\mathcal{H}_2(- {\beta_{1lmk}},- \beta_{12kl} - \beta_{1kml} ,
- \beta_{12kl} - \beta_{1kml} ,0,\beta_{12kl} + \beta_{1lmk}  )
\nonumber \\ &&   \hspace{-3.9cm}
- \mathcal{H}_1( - \beta_{1kml}, \beta_{1lmk} -  \beta_{1kml}  )
+2\mathcal{H}_2(\beta_{12kl}, \beta_{12kl} + \beta_{1kml} ,0,\beta_{12kl} + \beta_{1lmk},0  ) \Big)\,.
\end{eqnarray}
A discussion of the constraints obtained from this contribution is carried out in the main text below 
\Eqn{eq:5.212z-main}.

\section{Three-particle collinear limits}
\label{app:3-part-col}
The three-particle collinear splitting amplitude
soft anomalous dimension for terms starting 
at three loops has explicitly been computed
in Section~\ref{sec:three-particle-collinear-threeloop}. 
In this appendix, we perform the derivation of the 
corresponding contribution for terms in the 
soft anomalous dimension entering at the four-loop
order. The results of this analysis are discussed in 
Section~\ref{sec:three-particle-collinear-fourloop} of
the main text. We consider the terms containing
four colour generators in App.~\ref{app:3partcol-fourloop-Q4T}
and five colour generators in App.~\ref{app:3partcol-fourloop-5T}.

\subsection{Four-loop quartic terms}
\label{app:3partcol-fourloop-Q4T}
In this appendix, we perform the derivation to obtain 
${\bf{ \Gamma}}_{\SP, 3}^{{\rm Q}4{\rm{T}}}$, which is
defined in \Eqn{eq:GammaSpQ4T3} in terms of the difference
of quartic contributions to
soft anomalous dimensions
for amplitudes with $n$ and $n-2$ legs.

We start with $ {\bf{ \Gamma}}_{n,\rm{Q}4{\rm{T}}-4{\rm{L}}}$, 
which is given in \Eqn{eq:Q4T-4L}, and we split the sum over
the partons similarly to how it is done in the case of the
two-particle collinear limit in \Eqn{eq:Q4T-4L-a}. Namely, 
we separate this contribution into pieces that contain 
dependence on the collinear particles 1, 2 and 3 and the rest-of-the-process partons
\begin{eqnarray}\label{eq:5.205}
{\bf{ \Gamma}}_{n,\rm{Q}4{\rm{T}}-4{\rm{L}}}
(\{\beta_{ijkl} \},\alpha_s) &=& \,24\,
\sum_{R} \sum_{4\leq i<j<k<l \leq n} \DR_{ijkl} 
\,{\cal G}_R (\beta_{ijlk},\beta_{iklj}) \nn \\ 
&&  \hspace{-4.2cm}
+ \,24\,\sum_{R}
\sum_{4\leq j<k<l \leq n} \Big[
\DR_{1jkl} 
\,{\cal G}_R (\beta_{1jlk},\beta_{1klj}) 
+ \DR_{2jkl} 
\,{\cal G}_R (\beta_{2jlk},\beta_{2klj}) 
+ \DR_{3jkl} 
\,{\cal G}_R (\beta_{3jlk},\beta_{3klj})  
\Big] \nn \\ && \hspace{-4.2cm}
+ \,24\, \sum_{R}
\sum_{4\leq k<l \leq n} \Big[ 
\DR_{12kl} 
\,{\cal G}_R (\beta_{12lk},\beta_{1kl2})  
+\DR_{13kl} 
\,{\cal G}_R (\beta_{13lk},\beta_{1kl3})
+\DR_{23kl} 
\,{\cal G}_R (\beta_{23lk},\beta_{2kl3}) \Big]
\nn \\  && \hspace{-4.2cm}
+ \,24\,  \sum_{R}
\sum_{4\leq l \leq n} 
\DR_{123l} 
\,{\cal G}_R (\beta_{12l3},\beta_{13l2}) \,,
\end{eqnarray}
and the $n-2$ amplitude with collinear particles 
1, 2 and 3 replaced by parent parton $P$ is given by 
\begin{eqnarray}\label{eq:5.206} \nn
{\bf{ \Gamma}}_{n-2,\rm{Q}4{\rm{T}}-4{\rm{L}}}
(\{\beta_{ijkl} \},\alpha_s) &=& \,24\,
\sum_{R} \sum_{4\leq i<j<k<l \leq n}  
\DR_{iklj} 
\,{\cal G}_R  (\beta_{ikjl},\beta_{iljk}) 
\\  && + \,24\,\sum_{R}
\sum_{4\leq j<k<l \leq n} \DR_{Pjkl} 
\,{\cal G}_R (\beta_{Pjlk},\beta_{Pklj}) \,.
\end{eqnarray}
Taking the difference between these two terms, 
we see that the terms depending only on the 
rest-of-the-process partons cancel immediately.
Next, using rescaling invariance and 
$\T_P = \T_1 + \T_2 + \T_3$, leads to cancellation between
the second, third, and fourth brackets of \Eqn{eq:5.205} and
the terms in the second bracket of \Eqn{eq:5.206}.
The remaining terms are given by 
\begin{eqnarray}\label{eq:5.211} \nn
{\bf{ \Gamma}}_{n,\rm{Q}4{\rm{T}}-4{\rm{L}}}
(\{\beta_{ijkl} \},\alpha_s) 
- {\bf{ \Gamma}}_{n-2,\rm{Q}4{\rm{T}}-4{\rm{L}}}
(\{\beta_{ijkl} \},\alpha_s) && \\[0.2cm] \nn 
&& \hspace{-8.0cm} = \,24\, \sum_{R}
\sum_{4\leq k<l \leq n} \Big[ 
\,\DR_{12kl} 
\,{\cal G}_R  (\beta_{12kl},\beta_{1lk2})  
+ \DR_{13kl} 
\,{\cal G}_R  (\beta_{13kl},\beta_{1lk3}) 
\\ && \hspace{-6.0cm}
+\, \DR_{23kl} 
\,{\cal G}_R  (\beta_{23kl},\beta_{2lk3})  \Big]
+ \,24\,  \sum_{R}
\sum_{4\leq l \leq n} 
\DR_{123l} 
\,{\cal G}_R  (\beta_{123l},\beta_{1l32}) \,,
\end{eqnarray}
where we have used general symmetry properties of the kinematic 
functions given in \Eqn{eq:symGR}. 
We now consider the contributions from the
${{\Gamma}}_{n,{\rm{Q}}4{\rm{T}}-2,3{\rm{L}}}$ and ${{\Gamma}}_{n-2,{\rm{Q}}4{\rm{T}}-2,3{\rm{L}}}$ terms. We again split the sum over the 
indices similarly to~Eqs.~\eqref{eq:Q4T-23L-a} and \eqref{eq:Q4T-23L-b} in the
case of the two-particle collinear limit, but now for 
the case of particles 1, 2 and 3 becoming collinear  
\begin{eqnarray}
{\bf{ \Gamma}}_{n,{\rm{Q}}4{\rm{T}}-2,3{\rm{L}}}
(\{s_{ij} \},\lambda,\alpha_s) &=& 
- \sum_R g_R(\alpha_s) \bigg[\sum_{4\leq i<j  \leq n}
\Big(  { \bf{\cal{ D}}}^R_{iijj} 
+  { \bf{\cal{ D}}}^R_{iiij} 
+  { \bf{\cal{ D}}}^R_{jjji}\Big)\ell_{ij} 
 \\ &&\hspace{-1.6cm}
+ \sum_{  j = 4}^n\Big(  { \bf{\cal{ D}}}^R_{11jj} 
+  { \bf{\cal{ D}}}^R_{111j}
+  { \bf{\cal{ D}}}^R_{jjj1}\Big)\ell_{1j}
+ \sum_{  j = 4}^n\Big(  { \bf{\cal{ D}}}^R_{22jj} 
+  { \bf{\cal{ D}}}^R_{222j}
+  { \bf{\cal{ D}}}^R_{jjj2}\Big)\ell_{2j}
\nonumber \\ &&\hspace{-1.6cm}
+ \sum_{  j = 4}^n\Big(  { \bf{\cal{ D}}}^R_{33jj} 
+  { \bf{\cal{ D}}}^R_{333j}
+  { \bf{\cal{ D}}}^R_{jjj3}\Big)\ell_{3j}
+\Big(  { \bf{\cal{ D}}}^R_{1122} 
+  { \bf{\cal{ D}}}^R_{1112}
+  { \bf{\cal{ D}}}^R_{2221}\Big)\ell_{12}
\nonumber \\ &&\hspace{-1.6cm}
+\Big(  { \bf{\cal{ D}}}^R_{1133} 
+  { \bf{\cal{ D}}}^R_{1113}
+  { \bf{\cal{ D}}}^R_{3331}\Big)\ell_{13}
+\Big(  { \bf{\cal{ D}}}^R_{2233} 
+  { \bf{\cal{ D}}}^R_{2223}
+  { \bf{\cal{ D}}}^R_{3332}\Big)\ell_{23}
\nonumber \\ &&\hspace{-1.6cm}
+   \sum_{i=4}^n\sum_{\substack{ 4\leq j < k \leq n\\ j,k \neq i } }
{ \bf{\cal{ D}}}^R_{jkii}\ell_{jk}  
+ \sum_{\substack{ 4\leq j < k \leq n } } 
\Big(   { \bf{\cal{ D}}}^R_{jk11}+ { \bf{\cal{ D}}}^R_{jk22}
+ { \bf{\cal{ D}}}^R_{jk33}\Big) \ell_{jk}
\nonumber \\ &&\hspace{-1.6cm}
+   \sum_{i=4}^n\sum_{\substack{ k=4\\  k \neq i } }^n 
\Big( { \bf{\cal{ D}}}^R_{1kii}\ell_{1k}
+ { \bf{\cal{ D}}}^R_{2kii}\ell_{2k} 
+ { \bf{\cal{ D}}}^R_{3kii}\ell_{3k} \Big) 
\nonumber    \\ &&\hspace{-1.6cm}
+  \sum_{i=4}^n \Big({ \bf{\cal{ D}}}^R_{2i11}\ell_{2i} 
+ { \bf{\cal{ D}}}^R_{1i22}\ell_{1i}
+ { \bf{\cal{ D}}}^R_{12ii}\ell_{12}
+ { \bf{\cal{ D}}}^R_{3i11}\ell_{3i} 
+ { \bf{\cal{ D}}}^R_{1i33}\ell_{1i}
+ { \bf{\cal{ D}}}^R_{13ii}\ell_{13}
\nonumber    \\ &&\hspace{-0.8cm} \nonumber
+ { \bf{\cal{ D}}}^R_{3i22}\ell_{3i} 
+ { \bf{\cal{ D}}}^R_{2i33}\ell_{2i}
+ { \bf{\cal{ D}}}^R_{23ii}\ell_{23} \Big) 
+ { \bf{\cal{ D}}}^R_{2311}\ell_{23} 
+ { \bf{\cal{ D}}}^R_{1322}\ell_{13}
+ { \bf{\cal{ D}}}^R_{1233}\ell_{12} \bigg] \,,
\end{eqnarray}
and the $n-2$ counterpart is given by
 \begin{eqnarray}
{\bf{ \Gamma}}_{n-2,{\rm{Q}}4{\rm{T}}-2,3{\rm{L}}}
(\{s_{ij} \},\lambda,\alpha_s) &=& 
- \sum_R g_R \bigg[\sum_{4\leq i<j  \leq n}
\Big(  { \bf{\cal{ D}}}^R_{iijj} 
+  { \bf{\cal{ D}}}^R_{iiij}+  { \bf{\cal{ D}}}^R_{jjji}\Big)\ell_{ij} 
\nonumber \\ &&\hspace{-1cm}
+   \sum_{i=4}^n\sum_{\substack{ 4\leq j < k \leq n\\ j,k \neq i } }
{ \bf{\cal{ D}}}^R_{jkii}\ell_{jk}
+ \sum_{  j = 4}^n\Big(  { \bf{\cal{ D}}}^R_{PPjj} 
+  { \bf{\cal{ D}}}^R_{PPPj} +  
{ \bf{\cal{ D}}}^R_{jjjP}\Big)\ell_{Pj}
\nonumber \\ &&\hspace{-1cm}
+    \sum_{\substack{ 4\leq j < k \leq n  } } 
{ \bf{\cal{ D}}}^R_{jkPP}\ell_{jk}
+   \sum_{i=4}^n\sum_{\substack{ k=4\\  k \neq i } }^n
{ \bf{\cal{ D}}}^R_{Pkii}\ell_{Pk}  \bigg] \,.
\end{eqnarray}
When taking the difference between these two terms 
to obtain the splitting amplitude soft anomalous dimension,
the terms depending only on the rest-of-the-process
partons cancel directly. 
Moreover, cancellations also occur when
we substitute $\T_P = \T_1 + \T_2 +\T_3$ and 
$\ell_{1j}=\ln(x_1) + \ell_{Pj}$,  
$\ell_{2j}=\ln(x_2) + \ell_{Pj}$, and 
$\ell_{3j}=\ln(x_3) + \ell_{Pj}$. Concretely, 
we need the generalisation of the expressions in 
Eqs.~\eqref{eq:D12-TP-2pc} and \eqref{eq:D3-TP-2pc}
to the three-particle collinear limit case 
\begin{eqnarray}\label{eq:D1-TP-3pc}
{\cal{D}}_{jjjP}^R &=& {\cal{D}}_{jjj1}^R
+{\cal{D}}_{jjj2}^R+{\cal{D}}_{jjj3}^R \,, \\ 
\label{eq:D2-TP-3pc}
{\cal{D}}_{PPjj}^R & =&  {\cal{D}}_{11jj}^R+{\cal{D}}_{22jj}^R
+ {\cal{D}}_{33jj}^R  
+ 2 {\cal{D}}_{12jj}^R
+ 2 {\cal{D}}_{13jj}^R+ 2{\cal{D}}_{23jj}^R  \,, 
\end{eqnarray}
and 
\begin{eqnarray}\label{eq:D3-TP-3pc}
{\cal{D}}_{PPPj}  & =&  {\cal{D}}_{111j}^R+{\cal{D}}_{222j}^R
+{\cal{D}}_{333j}^R
+ 3 {\cal{D}}_{112j}^R + 3 {\cal{D}}_{122j}^R
 \nonumber     \\ && 
+ 3 {\cal{D}}_{113j}^R + 3 {\cal{D}}_{133j}^R
+ 3 {\cal{D}}_{223j}^R + 3 {\cal{D}}_{233j}^R
+ 6 {\cal{D}}_{123j}^R \,,
 \end{eqnarray} 
and corresponding relations where two rest-of-the-process 
parton indices appear in the $\DR$ terms. Using
these relations and simplifying, we find for the 
difference of these two anomalous dimension terms
\begin{eqnarray}
&& {\bf{ \Gamma}}_{n,{\rm{Q}}4{\rm{T}}-2,3{\rm{L}}}
(\{s_{ij} \},\lambda,\alpha_s) 
- {\bf{ \Gamma}}_{n-2,{\rm{Q}}4{\rm{T}}-2,3{\rm{L}}}
(\{s_{ij} \},\lambda,\alpha_s)
  \\ &&  = 
- \sum_R g_R \bigg\{   
\sum_{  j = 4}^n
\Big(  { \bf{\cal{ D}}}^R_{11jj} 
+  { \bf{\cal{ D}}}^R_{111j} 
+  { \bf{\cal{ D}}}^R_{jjj1}\Big)\ln(x_1)
+ \sum_{  j = 4}^n\Big( 
{ \bf{\cal{ D}}}^R_{22jj} 
+  { \bf{\cal{ D}}}^R_{222j} 
+  { \bf{\cal{ D}}}^R_{jjj2}\Big)\ln(x_2)
\nonumber \\[-0.1cm] &&\hspace{1.9cm}
+ \sum_{  j = 4}^n\Big(  { \bf{\cal{ D}}}^R_{33jj} 
+  { \bf{\cal{ D}}}^R_{333j} 
+  { \bf{\cal{ D}}}^R_{jjj3}\Big)\ln(x_3)
+ \big(  { \bf{\cal{ D}}}^R_{1122} 
+  { \bf{\cal{ D}}}^R_{1112}
+  { \bf{\cal{ D}}}^R_{2221}
+  { \bf{\cal{ D}}}^R_{1233}\big)\ell_{12}
\nonumber \\[0.2cm]     &&   \hspace{1.8cm} 
+\big(  { \bf{\cal{ D}}}^R_{1133} 
+  { \bf{\cal{ D}}}^R_{1113} 
+  { \bf{\cal{ D}}}^R_{3331}
+  { \bf{\cal{ D}}}^R_{1223}\big)\ell_{13}
+\big(  { \bf{\cal{ D}}}^R_{2233} 
+  { \bf{\cal{ D}}}^R_{2223} 
+  { \bf{\cal{ D}}}^R_{3332} 
+  { \bf{\cal{ D}}}^R_{1123}\big)\ell_{23}  
\nonumber \\[0.2cm]     &&     \hspace{1.9cm}
+   \sum_{i=4}^n\sum_{\substack{ k=4\\  k \neq i } }^n 
\Big({ \bf{\cal{ D}}}^R_{1kii}\ln(x_1) 
+ { \bf{\cal{ D}}}^R_{2kii}\ln(x_2) 
+ { \bf{\cal{ D}}}^R_{3kii}\ln(x_3) \Big) 
\nonumber    \\[-0.2cm] &&\hspace{1.9cm}
+\sum_{i=4}^n \Big[  
\Big( { \bf{\cal{ D}}}^R_{1i22}   
+ { \bf{\cal{ D}}}^R_{1i33}\Big) \ln(x_1)
+ \Big({ \bf{\cal{ D}}}^R_{2i11}   
+ { \bf{\cal{ D}}}^R_{2i33}\Big) \ln(x_2)
\nonumber    \\ &&\hspace{2.9cm}
+ \Big({ \bf{\cal{ D}}}^R_{3i11}
+ { \bf{\cal{ D}}}^R_{3i22}\Big) \ln(x_3) 
+ { \bf{\cal{ D}}}^R_{12ii}\ell_{12}
+ { \bf{\cal{ D}}}^R_{13ii}\ell_{13} 
+ { \bf{\cal{ D}}}^R_{23ii}\ell_{23}\Big] \bigg\}
\nonumber \\ &&\hspace{0.4cm}
+  \sum_R g_R \bigg[
\sum_{  j = 4}^n\Big(
2 {\cal{D}}_{12jj}^R+ 2 {\cal{D}}_{13jj}^R+ 2{\cal{D}}_{23jj}^R     
+ 2 {\cal{D}}_{112j}^R + 2 {\cal{D}}_{122j}^R 
+ 2 {\cal{D}}_{113j}^R + 2 {\cal{D}}_{133j}^R
\nonumber     \\ && \nn \hspace{1.9cm}
+ 2 {\cal{D}}_{223j}^R + 2 {\cal{D}}_{233j}^R 
+ 6 {\cal{D}}_{123j}^R \Big)\ell_{Pj}
+\sum_{\substack{ 4\leq j < k \leq n  } }  
\Big( 2 {\cal{D}}_{jk12}^R+ 2{\cal{D}}_{jk13}^R
+ 2 {\cal{D}}_{jk23}^R \Big)\ell_{jk}  \bigg]. 
\end{eqnarray}
In this step, we also use colour conservation. 
As opposed to the two-particle collinear limit
result (see \Eqn{eq:D2pc-mcc}), the sums in relevant
terms now start from $j=4$ index. Hence, we require
\begin{eqnarray}
\sum_{  j = 4}^n   { \bf{\cal{ D}}}^R_{111j} 
&=& d_R^{abcd}\T_1^a\T_1^b\T_1^c
\big( \T_4^d +\T_5^d+ \ldots + \T_n^d\big)
\nonumber \\[1ex]
&=&d_R^{abcd}\T_1^a\T_1^b\T_1^c\big(-\T_1^d - \T_2^d - \T_3^d  \big)
= -{ \bf{\cal{ D}}}^R_{1111} - { \bf{\cal{ D}}}^R_{1112} 
- { \bf{\cal{ D}}}^R_{1113} \,,
\end{eqnarray}
and 
\begin{eqnarray}\label{eq:calD1iik-3cc}
\sum_{i=4}^n\sum_{\substack{ k=4\\  k \neq i } }^n \, 
{ \bf{\cal{ D}}}^R_{1iik}  
& = &   = - \sum_{i=4}^{n} { \bf{\cal{ D}}}^R_{11ii} 
- \sum_{i=4}^{n} { \bf{\cal{ D}}}^R_{12ii}
- \sum_{i=4}^{n} { \bf{\cal{ D}}}^R_{13ii}  
-\sum_{i=4}^{n} { \bf{\cal{ D}}}^R_{1iii} \,.
 \end{eqnarray}
Additional relations can be found in 
Eqs.~\eqref{eq:calD1i22-3cc}, \eqref{eq:calD2iik-3cc}, and
\eqref{eq:calD3iik-3cc}. Using these, performing simplifications,
and collecting according to ${\cal{D}}_{12jj}^R$, ${\cal{D}}_{13jj}^R$, and ${\cal{D}}_{23jj}^R$ 
terms, we find 
\begin{eqnarray}
{\bf{ \Gamma}}_{n,{\rm{Q}}4{\rm{T}}-2,3{\rm{L}}}
(\{s_{ij} \},\lambda,\alpha_s) 
-{\bf{ \Gamma}}_{n-2,{\rm{Q}}4{\rm{T}}-2,3{\rm{L}}}
(\{s_{ij} \},\lambda,\alpha_s) &&
\nonumber \\  && \hspace{-6.0cm}  = 
\sum_R g_R \bigg[\,   
\Big({ \bf{\cal{ D}}}^R_{1111}
+  { \bf{\cal{ D}}}^R_{1112}
+  { \bf{\cal{ D}}}^R_{1113}
+ { \bf{\cal{ D}}}^R_{1122}
+ { \bf{\cal{ D}}}^R_{1133}
\nonumber \\[-0.1cm] &&\hspace{-4.1cm}
+ { \bf{\cal{ D}}}^R_{1222} 
+ { \bf{\cal{ D}}}^R_{1233}
+ { \bf{\cal{ D}}}^R_{1322} 
+ { \bf{\cal{ D}}}^R_{1333}
\Big) \ln(x_1)
\nonumber \\[0.1cm] &&\hspace{-4.4cm}
+ \Big({ \bf{\cal{ D}}}^R_{2221}
+  { \bf{\cal{ D}}}^R_{2222}
+  { \bf{\cal{ D}}}^R_{2223}
+ { \bf{\cal{ D}}}^R_{2111}
+ { \bf{\cal{ D}}}^R_{2211}
\nonumber \\[0.1cm] &&\hspace{-4.1cm}
+ { \bf{\cal{ D}}}^R_{2311}
+ { \bf{\cal{ D}}}^R_{2133}
+ { \bf{\cal{ D}}}^R_{2233}
+ { \bf{\cal{ D}}}^R_{2333}
\Big) \ln(x_2)
\nonumber \\[0.1cm] &&\hspace{-4.4cm}
+ \Big({ \bf{\cal{ D}}}^R_{3331}
+  { \bf{\cal{ D}}}^R_{3332}
+  { \bf{\cal{ D}}}^R_{3333}
+ { \bf{\cal{ D}}}^R_{3111}
+ { \bf{\cal{ D}}}^R_{3211}
\nonumber \\[0.1cm] &&\hspace{-4.1cm}
+ { \bf{\cal{ D}}}^R_{3311}
+ { \bf{\cal{ D}}}^R_{3122}
+ { \bf{\cal{ D}}}^R_{3222}
+ { \bf{\cal{ D}}}^R_{3322} 
\Big) \ln(x_3)
\nonumber \\[0.1cm] &&\hspace{-4.4cm}
-\Big(  { \bf{\cal{ D}}}^R_{1122} 
+  { \bf{\cal{ D}}}^R_{1112}
+  { \bf{\cal{ D}}}^R_{2221}
+  { \bf{\cal{ D}}}^R_{1233}\Big)\ell_{12}
\nonumber \\[0.1cm] &&   \hspace{-4.4cm}
-\Big(  { \bf{\cal{ D}}}^R_{1133} 
+  { \bf{\cal{ D}}}^R_{1113}
+  { \bf{\cal{ D}}}^R_{3331}
+  { \bf{\cal{ D}}}^R_{1223}\Big)\ell_{13}
\nonumber \\[0.1cm] &&   \hspace{-4.4cm}
-\Big(  { \bf{\cal{ D}}}^R_{2233} 
+  { \bf{\cal{ D}}}^R_{2223}
+  { \bf{\cal{ D}}}^R_{3332}
+      { \bf{\cal{ D}}}^R_{1123}\Big)\ell_{23}  \bigg]
 \nonumber \\     &&     \hspace{-6.0cm}
+  \sum_R g_R \bigg[ 
\sum_{i=4}^{n} { \bf{\cal{ D}}}^R_{12ii}
\Big(  \ln(x_1)   +  \ln(x_2) + 2\ell_{Pi} -\ell_{12} \Big) 
\nonumber  \\[-0.2cm] &&\hspace{-4.6cm}
+ \sum_{i=4}^{n} { \bf{\cal{ D}}}^R_{13ii}
\Big( \ln(x_1)  +\ln(x_3)+ 2\ell_{Pi}-\ell_{13}\Big) 
\nonumber  \\[-0.2cm] &&\hspace{-4.6cm}
+ \sum_{i=4}^{n} { \bf{\cal{ D}}}^R_{23ii} 
\Big( \ln(x_2)  + \ln(x_3)  + 2\ell_{Pi}-\ell_{23}\Big)
\nonumber  \\[-0.0cm] &&\hspace{-4.6cm}
+\, 2 \sum_{\substack{ 4\leq j < k \leq n  } }  
\Big( {\cal{D}}_{jk12}^R+ {\cal{D}}_{jk13}^R
+ {\cal{D}}_{jk23}^R \Big)\ell_{jk}
\nonumber \\[-0.2cm] && \hspace{-4.6cm}+\, 2
\sum_{  j = 4}^n\Big(
{\cal{D}}_{112j}^R + {\cal{D}}_{122j}^R  
+ {\cal{D}}_{113j}^R + {\cal{D}}_{133j}^R
\nonumber \\[-0.2cm] && \hspace{-3.1cm}
+ {\cal{D}}_{223j}^R + {\cal{D}}_{233j}^R 
+ 3 {\cal{D}}_{123j}^R \Big)\ell_{Pj}
\bigg]\,. \label{eq:5.133}
\end{eqnarray}
Next, we can combine $\ell_{Pi}$ with logarithms of
momentum fractions $x_i$ and $x_j$ as in \Eqn{eq:logPx1x2} 
for the different pairs $(i,j) = (1,2),(2,3),(1,3)$.
For the $\ell_{jk}$ term, we will need to rewrite it 
similarly to \Eqn{eq:logjk} according to which
colour structure it is multiplying. The 
corresponding  colour conservation results, as in
Eqs.~\eqref{eq.6.89} and \eqref{eq.6.92}, now naturally also
include the third particle becoming collinear
\begin{eqnarray}
&&-2\sum_{\substack{ 4\leq j < k \leq n  } }  
{\cal{D}}_{jk12}^R       \ell_{12}    
   \stackrel{m. c.c.  }{=}  \ell_{12}  \Big(
- {\cal{D}}_{1112}^R - 2{\cal{D}}_{1122}^R  - {\cal{D}}_{1222}^R 
 \nonumber   \\[-0.6cm]  && \hspace{5.2cm} 
-   2{\cal{D}}_{1123}^R-   2{\cal{D}}_{1223}^R -   {\cal{D}}_{1233}^R 
+ \sum_{i=4}^n {\cal{D}}_{12ii}^R  \Big) , 
 \label{eq.6.89b}
\\ && \hspace{0.4cm}
2 \sum_{\substack{ 4\leq j < k \leq n  } }  
{\cal{D}}_{jk12}^R  \,    \ell_{1j}     
\stackrel{m. c.c.  }{=}
\sum_{i=4}^n \ell_{1i}\big( - {\cal{D}}_{112i}^R\,   
- {\cal{D}}_{122i}^R \,  - {\cal{D}}_{123i}^R \,  
- {\cal{D}}_{12ii}^R \,  \big), \label{eq.5.144b}
\end{eqnarray}
and similar relations involving subsets of collinear
particles $(1,3)$ and $(2,3)$. We use these, and
the rewriting $\ell_{1i} = \ln(x_1) + \ell_{Pi} $,
$\ell_{2i} = \ln(x_2) + \ell_{Pi} $, and
$\ell_{3i} = \ln(x_3) + \ell_{Pi} $ 
to arrive at 
\begin{eqnarray}
&& {\bf{ \Gamma}}_{n,{\rm{Q}}4{\rm{T}}-2,3{\rm{L}}}
(\{s_{ij} \},\lambda,\alpha_s) 
- {\bf{ \Gamma}}_{n-2,{\rm{Q}}4{\rm{T}}-2,3{\rm{L}}}
(\{s_{ij} \},\lambda,\alpha_s)
\nonumber \\ && \hspace{1cm}  = 
\sum_R g_R \bigg[   
\Big( { \bf{\cal{ D}}}^R_{1111}
+ { \bf{\cal{ D}}}^R_{1112}
+ { \bf{\cal{ D}}}^R_{1113}
+ { \bf{\cal{ D}}}^R_{1122} 
+ { \bf{\cal{ D}}}^R_{1133} 
\nonumber \\[-0.1cm]  &&\hspace{2.6cm}
+ { \bf{\cal{ D}}}^R_{1222}
+ { \bf{\cal{ D}}}^R_{1233}
+ { \bf{\cal{ D}}}^R_{1322}
+ { \bf{\cal{ D}}}^R_{1333} \Big) \ln(x_1)
\nonumber \\[0.1cm]  &&\hspace{2.6cm}
\Big({ \bf{\cal{ D}}}^R_{2221}
+ { \bf{\cal{ D}}}^R_{2222}
+ { \bf{\cal{ D}}}^R_{2223}
+ { \bf{\cal{ D}}}^R_{2111}
+ { \bf{\cal{ D}}}^R_{2211}
\nonumber \\[0.1cm]  &&\hspace{2.6cm}
+ { \bf{\cal{ D}}}^R_{2311}
+ { \bf{\cal{ D}}}^R_{2133}
+ { \bf{\cal{ D}}}^R_{2233}
+ { \bf{\cal{ D}}}^R_{2333} \Big) \ln(x_2)
\nonumber  \\[0.1cm] &&\hspace{2.6cm}
\Big({ \bf{\cal{ D}}}^R_{3331}
+ { \bf{\cal{ D}}}^R_{3332}
+ { \bf{\cal{ D}}}^R_{3333}
+ { \bf{\cal{ D}}}^R_{3111}
+ { \bf{\cal{ D}}}^R_{3211}
\nonumber \\[0.1cm]  &&\hspace{2.6cm}
+ { \bf{\cal{ D}}}^R_{3311}
+ { \bf{\cal{ D}}}^R_{3122}
+ { \bf{\cal{ D}}}^R_{3222}
+ { \bf{\cal{ D}}}^R_{3322} \Big) \ln(x_3)
\nonumber \\[0.1cm] &&\hspace{2.6cm}
-\big(  { \bf{\cal{ D}}}^R_{1122} 
+  { \bf{\cal{ D}}}^R_{1112} 
+  { \bf{\cal{ D}}}^R_{2221}
+  { \bf{\cal{ D}}}^R_{1233}\big)\ell_{12}
\nonumber \\[0.2cm]  &&     \hspace{2.6cm}
-\big(  { \bf{\cal{ D}}}^R_{1133} 
+  { \bf{\cal{ D}}}^R_{1113}
+  { \bf{\cal{ D}}}^R_{3331}
+  { \bf{\cal{ D}}}^R_{1223}\big)\ell_{13}
\nonumber \\[0.1cm]  &&     \hspace{2.6cm}
-\big(  { \bf{\cal{ D}}}^R_{2233} 
+  { \bf{\cal{ D}}}^R_{2223}
+  { \bf{\cal{ D}}}^R_{3332}
+  { \bf{\cal{ D}}}^R_{1123}\big)\ell_{23}   \bigg]
\nonumber \\[-0.1cm] && \hspace{1.4cm}
-  \sum_R g_R \bigg\{  
\Big({\cal{D}}_{1112}^R 
+ 2{\cal{D}}_{1122}^R 
+  {\cal{D}}_{1222}^R 
+ 2{\cal{D}}_{1123}^R 
+ 2{\cal{D}}_{1223}^R 
+ {\cal{D}}_{1233}^R \Big) \ell_{12} 
\nonumber \\[-0.1cm] && \hspace{2.9cm} + 
\Big({\cal{D}}_{1113}^R 
+ 2{\cal{D}}_{1133}^R 
+  {\cal{D}}_{1333}^R 
+ 2{\cal{D}}_{1123}^R 
+ 2{\cal{D}}_{1233}^R 
+  {\cal{D}}_{1223}^R \Big) \ell_{13} 
\nonumber \\[0.1cm] &&\hspace{2.9cm} +
\Big( {\cal{D}}_{2223}^R 
+ 2{\cal{D}}_{2233}^R
+  {\cal{D}}_{2333}^R 
+ 2{\cal{D}}_{1223}^R 
+ 2{\cal{D}}_{1233}^R 
+  {\cal{D}}_{1123}^R \Big) \ell_{23}
\nonumber \\ &&\hspace{2.4cm}
- \sum_{i=4}^n \bigg[ 
\Big( {\cal{D}}_{112i}^R + {\cal{D}}_{122i}^R + {\cal{D}}_{123i}^R  
+ {\cal{D}}_{113i}^R + {\cal{D}}_{123i}^R + {\cal{D}}_{133i}^R \Big)\ln(x_1)
\nonumber \\ &&\hspace{3.5cm}  
+ \Big( {\cal{D}}_{112i}^R + {\cal{D}}_{122i}^R + {\cal{D}}_{123i}^R 
+ {\cal{D}}_{123i}^R + {\cal{D}}_{223i}^R + {\cal{D}}_{233i}^R \Big) \ln(x_2)
\nonumber \\ &&\hspace{3.5cm}  
+ \Big( {\cal{D}}_{113i}^R + {\cal{D}}_{123i}^R + {\cal{D}}_{133i}^R 
+ {\cal{D}}_{123i}^R + {\cal{D}}_{223i}^R + {\cal{D}}_{233i}^R \Big) \ln(x_3)
\bigg]
\nonumber \\ &&\hspace{2.4cm}
+ 2 \sum_{\substack{ 4\leq j < k \leq n  } }   
\Big( {\cal{D}}_{jk12}^R  \beta_{12jk}
+ {\cal{D}}_{jk13}^R   \beta_{13jk}
+  {\cal{D}}_{jk23}^R  \beta_{23jk} \Big) 
\bigg\}\,. \label{eq:5.133d}
\end{eqnarray}
At this point it is straightforward to apply colour conservation
to the terms which still contain a single index sum 
over the rest-of-the-process partons. Then, regrouping the
terms we arrive at the following result
\begin{eqnarray}
{\bf{ \Gamma}}_{n,{\rm{Q}}4{\rm{T}}-2,3{\rm{L}}}
(\{s_{ij} \},\lambda,\alpha_s) 
&-& {\bf{ \Gamma}}_{n-2,{\rm{Q}}4{\rm{T}}-2,3{\rm{L}}}
(\{s_{ij} \},\lambda,\alpha_s)
\nonumber \\[0.0cm] && \hspace{-4cm}  = 
\sum_R g_R \bigg[   
{ \bf{\cal{ D}}}^R_{1111}\ln(x_1)
+  { \bf{\cal{ D}}}^R_{2222}\ln(x_2)
+  { \bf{\cal{ D}}}^R_{3333}\ln(x_3)
\nonumber \\[-0.1cm] && \hspace{-3.0cm}
+\Big(  2    { \bf{\cal{ D}}}^R_{1112}
+3 { \bf{\cal{ D}}}^R_{1122}   
+2 { \bf{\cal{ D}}}^R_{1222} \Big) 
\Big(    \ln(x_1x_2) - \ell_{12}\Big) 
\nonumber  \\[0.1cm] && \hspace{-3.0cm}
+\Big(  2{ \bf{\cal{ D}}}^R_{1113}   
+3 { \bf{\cal{ D}}}^R_{1133}  
+ 2 { \bf{\cal{ D}}}^R_{1333}\Big) 
\Big( \ln(x_1x_3) - \ell_{13}\Big)
 \nonumber  \\[0.1cm] && \hspace{-3.0cm}
+\Big(  2 { \bf{\cal{ D}}}^R_{2223}   
+ 3{ \bf{\cal{ D}}}^R_{2233}  
+2 { \bf{\cal{ D}}}^R_{2333}\Big) 
\Big( \ln(x_2x_3)- \ell_{23}\Big)
\nonumber \\[0.1cm] && \hspace{-3.0cm} 
+ 2\Big(   { \bf{\cal{ D}}}^R_{1123}
+ { \bf{\cal{ D}}}^R_{1223} 
+ { \bf{\cal{ D}}}^R_{1233} \Big)
\Big( 2 \ln(x_1x_2x_3) - \ell_{12}-\ell_{13}-\ell_{23} \Big) 
 \nonumber \\[0.1cm] && \hspace{-3.0cm}
+  2\sum_{\substack{ 4\leq j < k \leq n  } }   
\big( {\cal{D}}_{jk12}^R  \beta_{12jk}
+ {\cal{D}}_{jk13}^R   \beta_{13jk}
+ {\cal{D}}_{jk23}^R  \beta_{23jk} \Big) 
\bigg]\,. \label{eq:5.133h}
\end{eqnarray}
We observe that this result is a generalisation of 
the two-particle collinear limits in
\Eqn{eq:6.78e} for each pair of the three particles 
becoming collinear, and additional terms which
depend on degrees of freedom of all three collinear
partons. Crucially there are no terms which depend
on rest-of-the-process partons beyond the known
two-particle collinear-like terms. 
Combing the expression in \Eqn{eq:5.133h} with
and the remaining quartic terms
originating from ${\bf{ \Gamma}}_{n,\rm{Q}4{\rm{T}}-4{\rm{L}}}$
soft anomalous dimension terms in 
\Eqn{eq:5.211} we obtain
\begin{eqnarray}
\label{eq:GammaSpQ4T3Res-app}
{\bf{ \Gamma}}_{\SP, 3}^{{\rm Q}4{\rm{T}}}(p_1, p_2, p_3;\mu_f) 
&=& {\bf{ \Gamma}}_{n,{\rm{Q}}4{\rm{T}}-2,3{\rm{L}}}
(\{s_{ij} \},\mu_f,\alpha_s) -
{\bf{ \Gamma}}_{n-2,{\rm{Q}}4{\rm{T}}-2,3{\rm{L}}}
(\{s_{ij} \},\mu_f,\alpha_s)
\nonumber \\[0.1cm] && \hspace{0.0cm} 
+ {\bf{ \Gamma}}_{n,{\rm Q}4{\rm{T}}-4{\rm{L}}}
(\{\beta_{ijkl} \},\alpha_s) 
-{\bf{ \Gamma}}_{n-2,{\rm Q}4{\rm{T}}-4{\rm{L}}}
(\{\beta_{ijkl} \},\alpha_s)
\nonumber \\ 
&=& \sum_R g_R \bigg[   
{ \bf{\cal{ D}}}^R_{1111}\ln(x_1)
+  { \bf{\cal{ D}}}^R_{2222}\ln(x_2)
+  { \bf{\cal{ D}}}^R_{3333}\ln(x_3)
\nonumber \\[-0.1cm] &&  \hspace{1.0cm}
+ \Big( 2 { \bf{\cal{ D}}}^R_{1112}
+3 { \bf{\cal{ D}}}^R_{1122}
+ 2 { \bf{\cal{ D}}}^R_{1222} \Big)  
\Big(    \ln(x_1x_2) - \ell_{12}\Big) 
\nonumber  \\[0.1cm] &&  \hspace{1.0cm}
+\Big(  2{ \bf{\cal{ D}}}^R_{1113}  
+3 { \bf{\cal{ D}}}^R_{1133}  
+ 2 { \bf{\cal{ D}}}^R_{1333}\Big) 
\Big( \ln(x_1x_3) - \ell_{13}\Big)
\nonumber  \\[0.1cm] && \hspace{1.0cm}
+\Big(  2 { \bf{\cal{ D}}}^R_{2223} 
+ 3{ \bf{\cal{ D}}}^R_{2233} 
+2 { \bf{\cal{ D}}}^R_{2333}\Big) 
\Big( \ln(x_2x_3)- \ell_{23}\Big)  
\nonumber \\[0.1cm] &&   \hspace{1.0cm}
+ 2 \sum_{\substack{ 4\leq j < k \leq n  } }   
\Big({\cal{D}}_{jk12}^R  \beta_{12jk}
+{\cal{D}}_{jk13}^R   \beta_{13jk}
+ {\cal{D}}_{jk23}^R  \beta_{23jk} \Big) \bigg] 
\nonumber \\ && \nn  \hspace{1.0cm}
+24\, \sum_{R}
\sum_{4\leq k<l \leq n} \Big[ \,\DR_{12kl} 
\,{\cal G}_R  (\beta_{12kl},\beta_{1lk2})  \Big] 
\\ \nn &&  \hspace{1.0cm}
+ 24\,  \sum_{R}
\sum_{4\leq k<l \leq n} \Big[ \,\DR_{13kl} 
\,{\cal G}_R  (\beta_{13kl},\beta_{1lk3})  \Big]
\\ \nn &&  \hspace{1.0cm}
+ 24\,  \sum_{R}
\sum_{4\leq k<l \leq n} \Big[ \DR_{23kl} 
\,{\cal G}_R  (\beta_{23kl},\beta_{2lk3})  \Big]
\\ && \hspace{1.0cm}
+ 2\sum_R g_R \Big(   { \bf{\cal{ D}}}^R_{1123}
+ { \bf{\cal{ D}}}^R_{1223}
+ { \bf{\cal{ D}}}^R_{1233} \Big)
\\ && \nonumber   \hspace{3.0cm}
\times\Big( 2 \ln(x_1x_2x_3)
- \ell_{12}-\ell_{13}-\ell_{23} \Big)
\\  && \nonumber \hspace{1.0cm}
- \,24\,  \sum_{R}
\,\Big[ \,\DR_{1123}+\DR_{1223} 
+\DR_{1233} \Big] 
{\cal G}_R (\beta_{123l},\beta_{1l32})\,.
\end{eqnarray}
This is the final expression for 
${\bf{ \Gamma}}_{\SP, 3}^{{\rm Q}4{\rm{T}}}$
before we implement known
two-particle collinear limit constraints as is done in the 
main text discussion in 
Section~\ref{sec:three-particle-collinear-fourloop}.
Only the last two lines contain genuinly new terms 
appearing for the first time in the three-particle collinear
limit. The remaining terms are simply the two-particle 
collinear-like terms for each pair of the three collinear particles. 

\subsection{Calculation of \texorpdfstring{${\bf{ \Gamma}}_{\SP, 3}^{5{\rm{T}}}$}{}}
\label{app:3partcol-fourloop-5T}
Here we calculate  ${\bf{ \Gamma}}_{\SP, 3}^{5{\rm{T}}}$,
which is defined in \Eqn{eq:GammaSp5T3} in terms of the
difference of quintic contributions to soft anomalous
dimensions for amplitudes with $n$ and $n-2$ legs. We 
start by separating the sum appearing in 
the parametrisation of $ {\bf{ \Gamma}}_{n,5{\rm{T}}-5{\rm{L}}}$ 
given in \Eqn{eq:H_5T-5L} into
parts which depend on the collinear particles and the
rest-of-the-process partons 
\begin{eqnarray} \nn
{\bf{ \Gamma}}_{n,5{\rm{T}}-5{\rm{L}}}(\{\beta_{ijkl} \},\alpha_s) 
&=& 8 \sum_{m=4}^n \sum_{\substack{4\leq i<j<k<l\leq n\\ i,j,k,l\neq m}} 
\Big[ \, \mathcal{T}_{ikljm}
\mathcal{H}_2(\beta_{iklj},\beta_{ikml},\beta_{ilmk},\beta_{kimj},\beta_{kjmi}) 
\\[-0.4cm] \nn
&& \hspace{3.1cm}+\mathcal{T}_{ijlkm}
\mathcal{H}_2(\beta_{ijlk},\beta_{ijml},\beta_{ilmj},\beta_{jimk},\beta_{jkmi})
\\[0.2cm] \nn
&& \hspace{3.1cm}
+ \mathcal{T}_{ijklm}
\mathcal{H}_2(\beta_{ijkl},\beta_{ijmk},\beta_{ikmj},\beta_{jiml},\beta_{jlmi})
\, \Big] \\[0.1cm] \nn 
&+& 8 \sum_{\substack{4\leq i<j<k<l\leq n } }\,\Big[\, 
\mathcal{T}_{iklj1}
\mathcal{H}_2(\beta_{iklj},\beta_{ik1l},\beta_{il1k},\beta_{ki1j},\beta_{kj1i}) 
\\[-0.1cm] \nn 
&& \hspace{2.4cm} + \mathcal{T}_{ijlk1}
\mathcal{H}_2(\beta_{ijlk},\beta_{ij1l},\beta_{il1j},\beta_{ji1k},\beta_{jk1i})
\\[0.3cm] \nn 
&& \hspace{2.4cm} + \mathcal{T}_{ijkl1}
\mathcal{H}_2(\beta_{ijkl},\beta_{ij1k},\beta_{ik1j},\beta_{ji1l},\beta_{jl1i}) 
\\[0.3cm] \nn 
&& \hspace{2.4cm} + \mathcal{T}_{iklj2}
\mathcal{H}_2(\beta_{iklj},\beta_{ik2l},\beta_{il2k},\beta_{ki2j},\beta_{kj2i}) 
\\[0.3cm] \nn 
&& \hspace{2.4cm} + \mathcal{T}_{ijlk2}
\mathcal{H}_2(\beta_{ijlk},\beta_{ij2l},\beta_{il2j},\beta_{ji2k},\beta_{jk2i})
\\[0.3cm] \nn 
&& \hspace{2.4cm} + \mathcal{T}_{ijkl2}
\mathcal{H}_2(\beta_{ijkl},\beta_{ij2k},\beta_{ik2j},\beta_{ji2l},\beta_{jl2i})
\\[0.3cm] \nn 
&& \hspace{2.4cm} + \mathcal{T}_{iklj3}
\mathcal{H}_2(\beta_{iklj},\beta_{ik3l},\beta_{il3k},\beta_{ki3j},\beta_{kj3i}) 
\\[0.3cm] \nn 
&& \hspace{2.4cm} + \mathcal{T}_{ijlk3}
\mathcal{H}_2(\beta_{ijlk},\beta_{ij3l},\beta_{il3j},\beta_{ji3k},\beta_{jk3i})
\\[0.2cm] \nn 
&& \hspace{2.4cm} + \mathcal{T}_{ijkl3}
\mathcal{H}_2(\beta_{ijkl},\beta_{ij3k},\beta_{ik3j},\beta_{ji3l},\beta_{jl3i})
\, \Big] \\[-0.1cm] \nn 
&+& 8 \sum_{m=4}^n \sum_{\substack{4\leq  j<k<l\leq n\\ j,k,l\neq m}}
\Big[\, \mathcal{T}_{1kljm}
\mathcal{H}_2(\beta_{1klj},\beta_{1kml},\beta_{1lmk},\beta_{k1mj},\beta_{kjm1}) 
\\[-0.4cm] \nn 
&& \hspace{2.7cm} +  \mathcal{T}_{1jlkm}
\mathcal{H}_2(\beta_{1jlk},\beta_{1jml},\beta_{1lmj},\beta_{j1mk},\beta_{jkm1})
\\[0.3cm] \nn 
&& \hspace{2.7cm} + \mathcal{T}_{1jklm}
\mathcal{H}_2(\beta_{1jkl},\beta_{1jmk},\beta_{1kmj},\beta_{j1ml},\beta_{jlm1}) 
\\[0.3cm] \nn 
&& \hspace{2.7cm} + \mathcal{T}_{2kljm}
\mathcal{H}_2(\beta_{2klj},\beta_{2kml},\beta_{2lmk},\beta_{k2mj},\beta_{kjm2}) 
\\[0.3cm] \nn 
&& \hspace{2.7cm} + \mathcal{T}_{2jlkm}
\mathcal{H}_2(\beta_{2jlk},\beta_{2jml},\beta_{2lmj},\beta_{j2mk},\beta_{jkm2})
\\[0.3cm] \nn 
&& \hspace{2.7cm} + \mathcal{T}_{2jklm}
\mathcal{H}_2(\beta_{2jkl},\beta_{2jmk},\beta_{2kmj},\beta_{j2ml},\beta_{jlm2}) 
\\[0.3cm] \nn 
&& \hspace{2.7cm} + \mathcal{T}_{3kljm}
\mathcal{H}_2(\beta_{3klj},\beta_{3kml},\beta_{3lmk},\beta_{k3mj},\beta_{kjm3}) 
\\[0.3cm] \nn 
&& \hspace{2.7cm} + \mathcal{T}_{3jlkm}
\mathcal{H}_2(\beta_{3jlk},\beta_{3jml},\beta_{3lmj},\beta_{j3mk},\beta_{jkm3})
\\[0.2cm] \nn 
&& \hspace{2.7cm} + \mathcal{T}_{3jklm}
\mathcal{H}_2(\beta_{3jkl},\beta_{3jmk},\beta_{3kmj},\beta_{j3ml},\beta_{jlm3})
\, \Big] \\[0.1cm] \nn 
&+& 8 \sum_{\substack{4\leq  j<k<l\leq n } } \Big[\, 
\mathcal{T}_{2klj1}
\mathcal{H}_2(\beta_{2klj},\beta_{2k1l},\beta_{2l1k},\beta_{k21j},\beta_{kj12}) 
\\[-0.1cm] \nn 
&& \hspace{2.0cm} + \mathcal{T}_{2jlk1}
\mathcal{H}_2(\beta_{2jlk},\beta_{2j1l},\beta_{2l1j},\beta_{j21k},\beta_{jk12})
\\[0.3cm] \nn 
&& \hspace{2.0cm} + \mathcal{T}_{2jkl1}
\mathcal{H}_2(\beta_{2jkl},\beta_{2j1k},\beta_{2k1j},\beta_{j21l},\beta_{jl12})
\\[0.3cm] \nn 
&& \hspace{2.0cm} + \mathcal{T}_{1klj2}
\mathcal{H}_2(\beta_{1klj},\beta_{1k2l},\beta_{1l2k},\beta_{k12j},\beta_{kj21}) 
\\[0.3cm] \nn 
&& \hspace{2.0cm} + \mathcal{T}_{1jlk2}
\mathcal{H}_2(\beta_{1jlk},\beta_{1j2l},\beta_{1l2j},\beta_{j12k},\beta_{jk21})
\\[0.3cm] \nn 
&& \hspace{2.0cm} + \mathcal{T}_{1jkl2}
\mathcal{H}_2(\beta_{1jkl},\beta_{1j2k},\beta_{1k2j},\beta_{j12l},\beta_{jl21})
\\[0.3cm] \nn 
&& \hspace{2.0cm} +\mathcal{T}_{3klj1}
\mathcal{H}_2(\beta_{3klj},\beta_{3k1l},\beta_{3l1k},\beta_{k31j},\beta_{kj13}) 
\\[0.3cm] \nn 
&& \hspace{2.0cm} + \mathcal{T}_{3jlk1}
\mathcal{H}_2(\beta_{3jlk},\beta_{3j1l},\beta_{3l1j},\beta_{j31k},\beta_{jk13})
\\[0.3cm] \nn 
&& \hspace{2.0cm} + \mathcal{T}_{3jkl1}
\mathcal{H}_2(\beta_{3jkl},\beta_{3j1k},\beta_{3k1j},\beta_{j31l},\beta_{jl13})
\\[0.3cm] \nn 
&& \hspace{2.0cm} + \mathcal{T}_{1klj3}
\mathcal{H}_2(\beta_{1klj},\beta_{1k3l},\beta_{1l3k},\beta_{k13j},\beta_{kj31}) 
\\[0.3cm] \nn 
&& \hspace{2.0cm} + \mathcal{T}_{1jlk3}
\mathcal{H}_2(\beta_{1jlk},\beta_{1j3l},\beta_{1l3j},\beta_{j13k},\beta_{jk31})
\\[0.3cm] \nn 
&& \hspace{2.0cm} + \mathcal{T}_{1jkl3}
\mathcal{H}_2(\beta_{1jkl},\beta_{1j3k},\beta_{1k3j},\beta_{j13l},\beta_{jl31})
\\[0.3cm] \nn 
&& \hspace{2.0cm} +\mathcal{T}_{3klj2}
\mathcal{H}_2(\beta_{3klj},\beta_{3k2l},\beta_{3l2k},\beta_{k32j},\beta_{kj23})
\\[0.3cm] \nn 
&& \hspace{2.0cm} + \mathcal{T}_{3jlk2}
\mathcal{H}_2(\beta_{3jlk},\beta_{3j2l},\beta_{3l2j},\beta_{j32k},\beta_{jk23})
\\[0.3cm] \nn 
&& \hspace{2.0cm}+ \mathcal{T}_{3jkl2}
\mathcal{H}_2(\beta_{3jkl},\beta_{3j2k},\beta_{3k2j},\beta_{j32l},\beta_{jl23})
\\[0.3cm] \nn 
&& \hspace{2.0cm} + \mathcal{T}_{2klj3}
\mathcal{H}_2(\beta_{2klj},\beta_{2k3l},\beta_{2l3k},\beta_{k23j},\beta_{kj32}) 
\\[0.3cm] \nn 
&& \hspace{2.0cm} + \mathcal{T}_{2jlk3}
\mathcal{H}_2(\beta_{2jlk},\beta_{2j3l},\beta_{2l3j},\beta_{j23k},\beta_{jk32})
\\[0.2cm] \nn 
&& \hspace{2.0cm} + \mathcal{T}_{2jkl3}
\mathcal{H}_2(\beta_{2jkl},\beta_{2j3k},\beta_{2k3j},\beta_{j23l},\beta_{jl32})
\, \Big]  \\[-0.1cm] \nn 
&+& 8 \sum_{m=4}^n   \sum_{\substack{4\leq  k<l\leq n\\  k,l\neq m} }
\Big[ \, \mathcal{T}_{1kl2m}
\mathcal{H}_2(\beta_{1kl2},\beta_{1kml},\beta_{1lmk},\beta_{k1m2},\beta_{k2m1}) 
\\[-0.4cm] \nn 
&& \hspace{2.4cm} + \mathcal{T}_{12lkm}
\mathcal{H}_2(\beta_{12lk},\beta_{12ml},\beta_{1lm2},\beta_{21mk},\beta_{2km1})
\\[0.3cm] \nn 
&& \hspace{2.4cm} + \mathcal{T}_{12klm}
\mathcal{H}_2(\beta_{12kl},\beta_{12mk},\beta_{1km2},\beta_{21ml},\beta_{2lm1})
\\[0.3cm] \nn 
&& \hspace{2.4cm} + \mathcal{T}_{1kl3m}
\mathcal{H}_2(\beta_{1kl3},\beta_{1kml},\beta_{1lmk},\beta_{k1m3},\beta_{k3m1})
\\[0.3cm] \nn 
&& \hspace{2.4cm}  + \mathcal{T}_{13lkm}
\mathcal{H}_2(\beta_{13lk},\beta_{13ml},\beta_{1lm3},\beta_{31mk},\beta_{3km1})
\\[0.3cm] \nn 
&& \hspace{2.4cm} + \mathcal{T}_{13klm}
\mathcal{H}_2(\beta_{13kl},\beta_{13mk},\beta_{1km3},\beta_{31ml},\beta_{3lm1})
\\[0.3cm] \nn 
&& \hspace{2.4cm} + \mathcal{T}_{2kl3m}
\mathcal{H}_2(\beta_{2kl3},\beta_{2kml},\beta_{2lmk},\beta_{k2m3},\beta_{k3m2})
\\[0.3cm] \nn 
&& \hspace{2.4cm} +  \mathcal{T}_{23lkm}
\mathcal{H}_2(\beta_{23lk},\beta_{23ml},\beta_{2lm3},\beta_{32mk},\beta_{3km2})
\\[0.2cm] \nn 
&& \hspace{2.4cm} +\mathcal{T}_{23klm}
\mathcal{H}_2(\beta_{23kl},\beta_{23mk},\beta_{2km3},\beta_{32ml},\beta_{3lm2})
\, \Big] \\[0.1cm] \nn 
&+& 8\sum_{\substack{4\leq   k<l\leq n } }
\Big[\,  \mathcal{T}_{2kl31}
\mathcal{H}_2(\beta_{2kl3},\beta_{2k1l},\beta_{2l1k},\beta_{k213},\beta_{k312})
\\[-0.1cm] \nn 
&& \hspace{1.6cm} + \mathcal{T}_{23lk1}
\mathcal{H}_2(\beta_{23lk},\beta_{231l},\beta_{2l13},\beta_{321k},\beta_{3k12})
\\[0.3cm] \nn 
&& \hspace{1.6cm} + \mathcal{T}_{23kl1}
\mathcal{H}_2(\beta_{23kl},\beta_{231k},\beta_{2k13},\beta_{321l},\beta_{3l12})
\\[0.3cm] \nn 
&& \hspace{1.6cm} + \mathcal{T}_{1kl32}
\mathcal{H}_2(\beta_{1kl3},\beta_{1k2l},\beta_{1l2k},\beta_{k123},\beta_{k321}) 
\\[0.3cm] \nn 
&& \hspace{1.6cm} + \mathcal{T}_{13lk2}
\mathcal{H}_2(\beta_{13lk},\beta_{132l},\beta_{1l23},\beta_{312k},\beta_{3k21})
\\[0.3cm] \nn 
&& \hspace{1.6cm} + \mathcal{T}_{13kl2}
\mathcal{H}_2(\beta_{13kl},\beta_{132k},\beta_{1k23},\beta_{312l},\beta_{3l21})
\\[0.3cm] \nn 
&& \hspace{1.6cm} + \mathcal{T}_{1kl23}
\mathcal{H}_2(\beta_{1kl2},\beta_{1k3l},\beta_{1l3k},\beta_{k132},\beta_{k231}) 
\\[0.3cm] \nn 
&& \hspace{1.6cm} + \mathcal{T}_{12lk3}
\mathcal{H}_2(\beta_{12lk},\beta_{123l},\beta_{1l32},\beta_{213k},\beta_{2k31})
\\[0.2cm] \nn 
&& \hspace{1.6cm} + \mathcal{T}_{12kl3}
\mathcal{H}_2(\beta_{12kl},\beta_{123k},\beta_{1k32},\beta_{213l},\beta_{2l31})
\, \Big] \\[-0.1cm] \nn 
&+& 8 \sum_{m=4}^n   \sum_{\substack{l =4 \\   l\neq m} }^ n \,\,\,
\Big[\, \mathcal{T}_{13l2m}
\mathcal{H}_2(\beta_{13l2},\beta_{13ml},\beta_{1lm3},\beta_{31m2},\beta_{32m1}) 
\\[-0.4cm] \nn 
&& \hspace{1.8cm} + \mathcal{T}_{12l3m}
\mathcal{H}_2(\beta_{12l3},\beta_{12ml},\beta_{1lm2},\beta_{21m3},\beta_{23m1})
\\[0.2cm] 
&& \hspace{1.8cm} + \mathcal{T}_{123lm}
\mathcal{H}_2(\beta_{123l},\beta_{12m3},\beta_{13m2},\beta_{21ml},\beta_{2lm1})
\, \Big],
\end{eqnarray}
where we have kept the $\alpha_s$ dependence implicit in the 
${\cal{H}}$ functions on the right-hand side.
Doing the same for the $n-2$ soft anomalous dimension term, we have 
\begin{eqnarray} \nn
{\bf{ \Gamma}}_{n-2,5{\rm{T}}-5{\rm{L}}}(\{\beta_{ijkl} \},\alpha_s) 
&=& 8 \sum_{m=4}^n \sum_{\substack{4\leq i<j<k<l\leq n\\ i,j,k,l\neq m} } 
\Big[\, \mathcal{T}_{ikljm}
\mathcal{H}_2(\beta_{iklj},\beta_{ikml},\beta_{ilmk},\beta_{kimj},\beta_{kjmi}) 
\\[-0.4cm] \nn 
&& \hspace{3.1cm} + \mathcal{T}_{ijlkm}
\mathcal{H}_2(\beta_{ijlk},\beta_{ijml},\beta_{ilmj},\beta_{jimk},\beta_{jkmi}) 
\\[0.2cm] \nn 
&& \hspace{3.1cm} + \mathcal{T}_{ijklm}
\mathcal{H}_2(\beta_{ijkl},\beta_{ijmk},\beta_{ikmj},\beta_{jiml},\beta_{jlmi})
\, \Big] \\[-0.1cm] \nn 
&+& 8 \sum_{m=4}^n \sum_{\substack{4\leq  j<k<l\leq n\\  j,k,l\neq m}}
\Big[\, \mathcal{T}_{Pkljm}
\mathcal{H}_2(\beta_{Pklj},\beta_{Pkml},\beta_{Plmk},\beta_{kPmj},\beta_{kjmP}) 
\\[-0.4cm] \nn 
&& \hspace{2.8cm} + \mathcal{T}_{Pjlkm}
\mathcal{H}_2(\beta_{Pjlk},\beta_{Pjml},\beta_{Plmj},\beta_{jPmk},\beta_{jkmP})
\\[0.2cm] \nn 
&& \hspace{2.8cm} + \mathcal{T}_{Pjklm}
\mathcal{H}_2(\beta_{Pjkl},\beta_{Pjmk},\beta_{Pkmj},\beta_{jPml},\beta_{jlmP}) \, \Big] \\[0.0cm] \nn 
&+& 8 \sum_{\substack{4\leq i<j<k<l\leq n } }\,\, 
\Big[\, \mathcal{T}_{ikljP}
\mathcal{H}_2(\beta_{iklj},\beta_{ikPl},\beta_{ilPk},\beta_{kiPj},\beta_{kjPi}) 
\\[-0.1cm] \nn 
&& \hspace{2.5cm} + \mathcal{T}_{ijlkP}
\mathcal{H}_2(\beta_{ijlk},\beta_{ijPl},\beta_{ilPj},\beta_{jiPk},\beta_{jkPi})
\\[0.2cm]  
&& \hspace{2.5cm} + \mathcal{T}_{ijklP}
\mathcal{H}_2(\beta_{ijkl},\beta_{ijPk},\beta_{ikPj},\beta_{jiPl},\beta_{jlPi}) \, \Big]. 
\end{eqnarray}
Next, we take the difference between the above
two equations.
We see that the terms with dependence on the rest-of-the-process
partons cancel directly.
More terms cancel upon
substituting $\T_P = \T_1 + \T_2+\T_3$
and using rescaling invariance 
for the CICRs 
$\beta_{1jkl} = \beta_{2jkl}= \beta_{3jkl} = \beta_{Pjkl}$.
We obtain the following 
\begin{eqnarray} \nn
{\bf{ \Gamma}}_{n,5{\rm{T}}-5{\rm{L}}}(\{\beta_{ijkl} \},\alpha_s) 
- {\bf{ \Gamma}}_{n-2,5{\rm{T}}-5{\rm{L}}}(\{\beta_{ijkl} \},\alpha_s) 
&& \\ 
&&\hspace{-6.0cm} =\, 
8 \hspace{-0.2cm} \sum_{\substack{4\leq  j<k<l\leq n } }
\hspace{-0.2cm}\Big[\, \mathcal{T}_{2klj1}
\mathcal{H}_2(\beta_{2klj},\beta_{2k1l},\beta_{2l1k},\beta_{k21j},\beta_{kj12})
\\[-0.1cm] \nn 
&& \hspace{-3.9cm} + \mathcal{T}_{2jlk1}
\mathcal{H}_2(\beta_{2jlk},\beta_{2j1l},\beta_{2l1j},\beta_{j21k},\beta_{jk12})
\\[0.3cm] \nn 
&& \hspace{-3.9cm} + \mathcal{T}_{2jkl1}
\mathcal{H}_2(\beta_{2jkl},\beta_{2j1k},\beta_{2k1j},\beta_{j21l},\beta_{jl12})
\\[0.3cm] \nn 
&& \hspace{-3.9cm} + \mathcal{T}_{1klj2}
\mathcal{H}_2(\beta_{1klj},\beta_{1k2l},\beta_{1l2k},\beta_{k12j},\beta_{kj21}) 
\\[0.3cm] \nn 
&& \hspace{-3.9cm} + \mathcal{T}_{1jlk2}
\mathcal{H}_2(\beta_{1jlk},\beta_{1j2l},\beta_{1l2j},\beta_{j12k},\beta_{jk21})
\\[0.3cm] \nn 
&& \hspace{-3.9cm} + \mathcal{T}_{1jkl2}
\mathcal{H}_2(\beta_{1jkl},\beta_{1j2k},\beta_{1k2j},\beta_{j12l},\beta_{jl21})
\\[0.3cm] \nn 
&& \hspace{-3.9cm} + \mathcal{T}_{3klj1}
\mathcal{H}_2(\beta_{3klj},\beta_{3k1l},\beta_{3l1k},\beta_{k31j},\beta_{kj13}) 
\\[0.3cm] \nn 
&& \hspace{-3.9cm} + \mathcal{T}_{3jlk1}
\mathcal{H}_2(\beta_{3jlk},\beta_{3j1l},\beta_{3l1j},\beta_{j31k},\beta_{jk13})
\\[0.3cm] \nn 
&& \hspace{-3.9cm} + \mathcal{T}_{3jkl1}
\mathcal{H}_2(\beta_{3jkl},\beta_{3j1k},\beta_{3k1j},\beta_{j31l},\beta_{jl13})
\\[0.3cm] \nn 
&& \hspace{-3.9cm} + \mathcal{T}_{1klj3}
\mathcal{H}_2(\beta_{1klj},\beta_{1k3l},\beta_{1l3k},\beta_{k13j},\beta_{kj31}) 
\\[0.3cm] \nn 
&& \hspace{-3.9cm} + \mathcal{T}_{1jlk3}
\mathcal{H}_2(\beta_{1jlk},\beta_{1j3l},\beta_{1l3j},\beta_{j13k},\beta_{jk31})
\\[0.3cm] \nn 
&& \hspace{-3.9cm} + \mathcal{T}_{1jkl3}
\mathcal{H}_2(\beta_{1jkl},\beta_{1j3k},\beta_{1k3j},\beta_{j13l},\beta_{jl31})
\\[0.3cm] \nn 
&& \hspace{-3.9cm} + \mathcal{T}_{3klj2}
\mathcal{H}_2(\beta_{3klj},\beta_{3k2l},\beta_{3l2k},\beta_{k32j},\beta_{kj23})
\\[0.3cm] \nn 
&& \hspace{-3.9cm} + \mathcal{T}_{3jlk2}
\mathcal{H}_2(\beta_{3jlk},\beta_{3j2l},\beta_{3l2j},\beta_{j32k},\beta_{jk23})
\\[0.3cm] \nn 
&& \hspace{-3.9cm} + \mathcal{T}_{3jkl2}
\mathcal{H}_2(\beta_{3jkl},\beta_{3j2k},\beta_{3k2j},\beta_{j32l},\beta_{jl23})
\\[0.3cm] \nn 
&& \hspace{-3.9cm} + \mathcal{T}_{2klj3}
\mathcal{H}_2(\beta_{2klj},\beta_{2k3l},\beta_{2l3k},\beta_{k23j},\beta_{kj32}) 
\\[0.3cm] \nn 
&& \hspace{-3.9cm} + \mathcal{T}_{2jlk3}
\mathcal{H}_2(\beta_{2jlk},\beta_{2j3l},\beta_{2l3j},\beta_{j23k},\beta_{jk32})
\\[0.2cm] \nn 
&& \hspace{-3.9cm} + \mathcal{T}_{2jkl3}
\mathcal{H}_2(\beta_{2jkl},\beta_{2j3k},\beta_{2k3j},\beta_{j23l},\beta_{jl32})
\, \Big] \\[-0.1cm] \nn 
&& \hspace{-6.0cm}
+ 8 \sum_{m=4}^n   \sum_{\substack{4\leq  k<l\leq n\\  k,l\neq m} }
\Big[\, \mathcal{T}_{1kl2m}
\mathcal{H}_2(\beta_{1kl2},\beta_{1kml},\beta_{1lmk},\beta_{k1m2},\beta_{k2m1}) 
\\[-0.4cm] \nn 
&& \hspace{-3.3cm} + \mathcal{T}_{12lkm}
\mathcal{H}_2(\beta_{12lk},\beta_{12ml},\beta_{1lm2},\beta_{21mk},\beta_{2km1})
\\[0.3cm] \nn 
&& \hspace{-3.3cm} + \mathcal{T}_{12klm}
\mathcal{H}_2(\beta_{12kl},\beta_{12mk},\beta_{1km2},\beta_{21ml},\beta_{2lm1})
\\[0.3cm] \nn 
&& \hspace{-3.3cm} + \mathcal{T}_{1kl3m}
\mathcal{H}_2(\beta_{1kl3},\beta_{1kml},\beta_{1lmk},\beta_{k1m3},\beta_{k3m1}) 
\\[0.3cm] \nn 
&& \hspace{-3.3cm} + \mathcal{T}_{13lkm}
\mathcal{H}_2(\beta_{13lk},\beta_{13ml},\beta_{1lm3},\beta_{31mk},\beta_{3km1})
\\[0.3cm] \nn 
&& \hspace{-3.3cm} + \mathcal{T}_{13klm}
\mathcal{H}_2(\beta_{13kl},\beta_{13mk},\beta_{1km3},\beta_{31ml},\beta_{3lm1})
\\[0.3cm] \nn 
&& \hspace{-3.3cm} + \mathcal{T}_{2kl3m}
\mathcal{H}_2(\beta_{2kl3},\beta_{2kml},\beta_{2lmk},\beta_{k2m3},\beta_{k3m2})
\\[0.3cm] \nn 
&& \hspace{-3.3cm} + \mathcal{T}_{23lkm}
\mathcal{H}_2(\beta_{23lk},\beta_{23ml},\beta_{2lm3},\beta_{32mk},\beta_{3km2})
\\[0.2cm] \nn 
&& \hspace{-3.3cm} + \mathcal{T}_{23klm}
\mathcal{H}_2(\beta_{23kl},\beta_{23mk},\beta_{2km3},\beta_{32ml},\beta_{3lm2})
\, \Big] \\[-0.1cm] \nn 
&& \hspace{-6.0cm}
+ 8\sum_{\substack{4\leq   k<l\leq n } }
\Big[\, \mathcal{T}_{2kl31}
\mathcal{H}_2(\beta_{2kl3},\beta_{2k1l},\beta_{2l1k},\beta_{k213},\beta_{k312}) 
\\[-0.1cm] \nn 
&& \hspace{-4.0cm} + \mathcal{T}_{23lk1}
\mathcal{H}_2(\beta_{23lk},\beta_{231l},\beta_{2l13},\beta_{321k},\beta_{3k12})
\\[0.3cm] \nn 
&& \hspace{-4.0cm} + \mathcal{T}_{23kl1}
\mathcal{H}_2(\beta_{23kl},\beta_{231k},\beta_{2k13},\beta_{321l},\beta_{3l12})
\\[0.3cm] \nn 
&& \hspace{-4.0cm} + \mathcal{T}_{1kl32}
\mathcal{H}_2(\beta_{1kl3},\beta_{1k2l},\beta_{1l2k},\beta_{k123},\beta_{k321}) 
\\[0.3cm] \nn 
&& \hspace{-4.0cm} + \mathcal{T}_{13lk2}
\mathcal{H}_2(\beta_{13lk},\beta_{132l},\beta_{1l23},\beta_{312k},\beta_{3k21})
\\[0.3cm] \nn 
&& \hspace{-4.0cm} + \mathcal{T}_{13kl2}
\mathcal{H}_2(\beta_{13kl},\beta_{132k},\beta_{1k23},\beta_{312l},\beta_{3l21})
\\[0.3cm] \nn 
&& \hspace{-4.0cm} + \mathcal{T}_{1kl23}
\mathcal{H}_2(\beta_{1kl2},\beta_{1k3l},\beta_{1l3k},\beta_{k132},\beta_{k231}) 
\\[0.3cm] \nn 
&& \hspace{-4.0cm} + \mathcal{T}_{12lk3}
\mathcal{H}_2(\beta_{12lk},\beta_{123l},\beta_{1l32},\beta_{213k},\beta_{2k31})
\\[0.2cm] \nn 
&& \hspace{-4.0cm} + \mathcal{T}_{12kl3}
\mathcal{H}_2(\beta_{12kl},\beta_{123k},\beta_{1k32},\beta_{213l},\beta_{2l31})
\, \Big] \\[-0.1cm] \nn 
&& \hspace{-6.0cm}
+ 8 \sum_{m=4}^n \sum_{\substack{l =4 \\   l\neq m} }^ n \,\,
\Big[\, \mathcal{T}_{13l2m}
\mathcal{H}_2(\beta_{13l2},\beta_{13ml},\beta_{1lm3},\beta_{31m2},\beta_{32m1})
\\[-0.4cm] \nn 
&& \hspace{-3.9cm} + \mathcal{T}_{12l3m}
\mathcal{H}_2(\beta_{12l3},\beta_{12ml},\beta_{1lm2},\beta_{21m3},\beta_{23m1})
\\[0.3cm]  
&& \hspace{-3.9cm} + \mathcal{T}_{123lm}
\mathcal{H}_2(\beta_{123l},\beta_{12m3},\beta_{13m2},\beta_{21ml},\beta_{2lm1})
\, \Big].
\end{eqnarray}
We note that the top two sums
resemble the corresponding expression in \Eqn{eq:5T5L-2pcdiff}
for the two-particle collinear limit for each pair out of the 
three becoming collinear. The last two sums contain the new type
terms that contain information on all three collinear particles. 
We can then use symmetry properties of ${\cal{H}}_2$ such as 
\Eqn{eq:H2-sym2} and properties of logarithms of CICRs discussed in
Section~\ref{sec:kin-two-part-col} to simplify the expression. We have 
\begin{eqnarray}\label{eq:5T5L-3pc} \nn
{\bf{ \Gamma}}_{n,5{\rm{T}}-5{\rm{L}}}(\{\beta_{ijkl} \},\alpha_s)   
- {\bf{ \Gamma}}_{n-2,5{\rm{T}}-5{\rm{L}}}(\{\beta_{ijkl} \},\alpha_s)  && \\ \nn 
&& \hspace{-7.0cm} =\,
8 \sum_{\substack{4\leq  j<k<l\leq n }} \bigg\{
\Big[ \mathcal{T}_{1klj2}+\mathcal{T}_{2klj1}\Big]
\mathcal{H}_2(\beta_{Pklj},-\beta_{12kl},-\beta_{12kl},0,\beta_{12jk} ) \\[-0.3cm] \nn 
&& \hspace{-4.5cm}
+ \Big[ \mathcal{T}_{1jlk2}+ \mathcal{T}_{2jlk1}\Big]
\mathcal{H}_2(\beta_{Pjlk},-\beta_{12jl},-\beta_{12jl},0,\beta_{12jk} ) \\[0.0cm] \nn 
&& \hspace{-4.5cm}
+ \Big[\mathcal{T}_{1jkl2}+ \mathcal{T}_{2jkl1}\Big]
\mathcal{H}_2(\beta_{Pjkl},-\beta_{12jk},-\beta_{12jk},0,\beta_{12jl} ) \\[0.0cm] \nn 
&& \hspace{-4.5cm}
+\Big[\mathcal{T}_{1klj3}+\mathcal{T}_{3klj1}\Big]
\mathcal{H}_2(\beta_{Pklj},-\beta_{13kl},-\beta_{13kl},0,\beta_{13jk} ) \\[0.0cm] \nn 
&& \hspace{-4.5cm}
+\Big[ \mathcal{T}_{1jlk3}+\mathcal{T}_{3jlk1}\Big]
\mathcal{H}_2(\beta_{Pjlk},-\beta_{13jl},-\beta_{13jl},0,\beta_{13jk} ) \\[0.0cm] \nn 
&& \hspace{-4.5cm}
+ \Big[\mathcal{T}_{1jkl3}+\mathcal{T}_{3jkl1}\Big]
\mathcal{H}_2(\beta_{Pjkl},-\beta_{13jk},-\beta_{13jk},0,\beta_{13jl} ) \\[0.0cm] \nn 
&& \hspace{-4.5cm}
+\Big[ \mathcal{T}_{2klj3}+\mathcal{T}_{3klj2}\Big]
\mathcal{H}_2(\beta_{Pklj},-\beta_{23kl},-\beta_{23kl},0,\beta_{23jk} ) \\[0.0cm] \nn 
&& \hspace{-4.5cm}
+ \Big[\mathcal{T}_{2jlk3}+ \mathcal{T}_{3jlk2}\Big]
\mathcal{H}_2(\beta_{Pjlk},-\beta_{23jl},-\beta_{23jl},0,\beta_{23jk} ) \\[-0.1cm] \nn 
&& \hspace{-4.5cm}
+\Big[ \mathcal{T}_{2jkl3}+\mathcal{T}_{3jkl2}\Big]
\mathcal{H}_2(\beta_{Pjkl},-\beta_{23jk},-\beta_{23jk},0,\beta_{23jl} ) \bigg\} \\[-0.3cm] \nn 
&& \hspace{-7.0cm}
+\, 8 \sum_{m=4}^n \sum_{\substack{4\leq  k<l\leq n\\  k,l\neq m} } 
\Big[\,  \mathcal{T}_{12lkm}
\mathcal{H}_2(\beta_{12kl},\beta_{12lm},0,\beta_{12km},0 ) \\[-0.4cm] \nn 
&& \hspace{-4.2cm}
+\mathcal{T}_{12klm}
\mathcal{H}_2(\beta_{12kl},\beta_{12km},0,\beta_{12lm},0 ) \\[0.2cm] \nn 
&& \hspace{-4.2cm}
+ \mathcal{T}_{13lkm}
\mathcal{H}_2(\beta_{13kl},\beta_{13lm},0,\beta_{13km},0 ) \\[0.2cm] \nn 
&& \hspace{-4.2cm}
+\mathcal{T}_{13klm}
\mathcal{H}_2(\beta_{13kl},\beta_{13km},0,\beta_{13lm},0 ) \\[0.2cm] \nn 
&& \hspace{-4.2cm}
+ \mathcal{T}_{23lkm}
\mathcal{H}_2(\beta_{23kl},\beta_{23lm},0,\beta_{23km},0 ) \\[0.2cm] \nn 
&& \hspace{-4.2cm}
+\mathcal{T}_{23klm}
\mathcal{H}_2(\beta_{23kl},\beta_{23km},0,\beta_{23lm},0 )\, \Big] \\[-0.1cm] \nn 
&& \hspace{-7.0cm}
+\, 8\hspace{-0.1cm} \sum_{\substack{4\leq   k<l\leq n }}\hspace{-0.1cm} 
\Big[ \, \mathcal{T}_{23lk1}
\mathcal{H}_2(\beta_{23kl},\beta_{1l23},\beta_{132l},\beta_{1k32},\beta_{123k} ) \\[-0.2cm] \nn 
&& \hspace{-5.2cm} +\mathcal{T}_{23kl1}
\mathcal{H}_2(\beta_{23kl},\beta_{1k23},\beta_{132k},\beta_{1l32},\beta_{123l} ) \\[0.2cm] \nn 
&& \hspace{-5.2cm} + \mathcal{T}_{13lk2}
\mathcal{H}_2(\beta_{13kl},\beta_{132l},\beta_{1l23},\beta_{13k2},\beta_{12k3} ) \\[0.2cm] \nn 
&& \hspace{-5.2cm} +\mathcal{T}_{13kl2}
\mathcal{H}_2(\beta_{13kl},\beta_{132k},\beta_{1k23},\beta_{13l2},\beta_{12l3}) \\[0.2cm] \nn 
&& \hspace{-5.2cm} + \mathcal{T}_{12lk3}
\mathcal{H}_2(\beta_{12kl},\beta_{123l},\beta_{1l32},\beta_{12k3},\beta_{13k2} ) \\[0.1cm] \nn 
&& \hspace{-5.2cm} + \mathcal{T}_{12kl3}
\mathcal{H}_2(\beta_{12kl},\beta_{123k},\beta_{1k32},\beta_{12l3},\beta_{13l2} ) \,\Big] \\[-0.2cm] \nn 
&& \hspace{-7.0cm}
+ 8 \sum_{m=4}^n   \sum_{\substack{l =4 \\   l\neq m} }^ n 
\Big[\, \mathcal{T}_{13l2m}
\mathcal{H}_2(\beta_{13l2},\beta_{13lm},0,\beta_{132m},\beta_{1m23} ) \\[-0.4cm] \nn 
&& \hspace{-5.0cm} + \mathcal{T}_{12l3m}
\mathcal{H}_2(\beta_{12l3},\beta_{12lm},0,\beta_{123m},\beta_{1m32} ) \\[0.2cm]  
&& \hspace{-5.0cm} + \mathcal{T}_{123lm}
\mathcal{H}_2(\beta_{123l},\beta_{12m3},\beta_{13m2},\beta_{12lm},0 )\, \Big],
\end{eqnarray}
where we have used antisymmetry property of ${\mathcal{H}_2}$ \cite{Becher:2019avh}
\begin{eqnarray}\label{eq:H2-sym3}
\mathcal{H}_2(y_1,y_2,y_3,y_4,y_5)&=&
-\mathcal{H}_2(-y_1,y_3, y_2, y_4-y_2+y_3, y_5+y_1-y_2+y_3),
\end{eqnarray}
to remove terms which vanish due to this antisymmetry property such as
\begin{eqnarray}
\mathcal{H}_2(0,-\beta_{12kl},-\beta_{12kl},\beta_{13k2},\beta_{12k3}) =0\,. 
\end{eqnarray}
We now consider the contribution from 
${\bf{ \Gamma}}_{n,5{\rm{T}}-4{\rm{L}}}(\{\beta_{ijkl} \},\alpha_s)$
terms given in \Eqn{eq:H_5T-4L}.
We write this contribution in the following way
\begin{eqnarray} \nn
{\bf{ \Gamma}}_{n,5{\rm{T}}-4{\rm{L}}}(\{\beta_{ijkl} \},\alpha_s) && \\ \nn
&&\hspace{-4.0cm} =\, 
2 \sum_{i=4}^n\sum_{\substack{ 4\leq j < k < l\leq n\\ j,k,l \neq i }}
\Big[\,\mathcal{T}_{iklji}\mathcal{H}_1(\beta_{ikjl},\beta_{iljk} )
+\mathcal{T}_{ijlki} \mathcal{H}_1(\beta_{ijkl},\beta_{ilkj} )
+\mathcal{T}_{ijkli} \mathcal{H}_1(\beta_{ijlk},\beta_{iklj} )\,\Big] \\[-0.2cm] \nn 
&&\hspace{-3.6cm}
+ 2 \sum_{\substack{ 4\leq j < k < l\leq n } } 
\Big[\,\mathcal{T}_{1klj1}\mathcal{H}_1(\beta_{1kjl},\beta_{1ljk} )  
+\mathcal{T}_{1jlk1}\mathcal{H}_1(\beta_{1jkl},\beta_{1lkj} )
+\mathcal{T}_{1jkl1}\mathcal{H}_1(\beta_{1jlk},\beta_{1klj} ) \\[-0.1cm] \nn 
&&\hspace{-1.3cm}  
+\mathcal{T}_{2klj2}\mathcal{H}_1(\beta_{2kjl},\beta_{2ljk} ) 
+\mathcal{T}_{2jlk2}\mathcal{H}_1(\beta_{2jkl},\beta_{2lkj} )
+\mathcal{T}_{2jkl2}\mathcal{H}_1(\beta_{2jlk},\beta_{2klj} ) \\[0.2cm] \nn 
&&\hspace{-1.3cm} 
+\mathcal{T}_{3klj3}\mathcal{H}_1(\beta_{3kjl},\beta_{3ljk} ) 
+\mathcal{T}_{3jlk3}\mathcal{H}_1(\beta_{3jkl},\beta_{3lkj} )
+\mathcal{T}_{3jkl3}\mathcal{H}_1(\beta_{3jlk},\beta_{3klj} ) \, \Big] \\[-0.1cm] \nn 
&&\hspace{-3.6cm}
+ 2  \sum_{i=4}^n\sum_{\substack{ 4\leq  k < l\leq n\\ k,l \neq i } } 
\Big[\, \mathcal{T}_{ikl1i}\mathcal{H}_1(\beta_{ik1l},\beta_{il1k} ) 
+ \mathcal{T}_{i1lki}\mathcal{H}_1(\beta_{i1kl},\beta_{ilk1} )
+ \mathcal{T}_{i1kli}\mathcal{H}_1(\beta_{i1lk},\beta_{ikl1} ) \\[-0.3cm] \nn 
&&\hspace{-1.0cm}  
+\mathcal{T}_{ikl2i}\mathcal{H}_1(\beta_{ik2l},\beta_{il2k} )
+ \mathcal{T}_{i2lki}\mathcal{H}_1(\beta_{i2kl},\beta_{ilk2} )
+ \mathcal{T}_{i2kli}\mathcal{H}_1(\beta_{i2lk},\beta_{ikl2} ) \\[0.3cm] \nn 
&&\hspace{-1.0cm}  
+ \mathcal{T}_{ikl3i}\mathcal{H}_1(\beta_{ik3l},\beta_{il3k} )
+ \mathcal{T}_{i3lki}\mathcal{H}_1(\beta_{i3kl},\beta_{ilk3} )
+ \mathcal{T}_{i3kli}\mathcal{H}_1(\beta_{i3lk},\beta_{ikl3} ) \, \Big] \\[0.1cm] \nn 
&&\hspace{-3.6cm}
+\,2\,\sum_{4\leq k< l\leq n}\,  \Big[ \, 
\mathcal{T}_{1kl21}\mathcal{H}_1(\beta_{1k2l},\beta_{1l2k} )
+ \mathcal{T}_{12lk1}\mathcal{H}_1(\beta_{12kl},\beta_{1lk2} )
+ \mathcal{T}_{12kl1}\mathcal{H}_1(\beta_{12lk},\beta_{1kl2} ) \\[-0.1cm] \nn 
&&\hspace{-1.5cm}
+ \mathcal{T}_{2kl12}\mathcal{H}_1(\beta_{2k1l},\beta_{2l1k} )
+ \mathcal{T}_{21lk2}\mathcal{H}_1(\beta_{21kl},\beta_{2lk1} )
+ \mathcal{T}_{21kl2}\mathcal{H}_1(\beta_{21lk},\beta_{2kl1} ) \\[0.3cm] \nn 
&&\hspace{-1.5cm}
+ \mathcal{T}_{1kl31}\mathcal{H}_1(\beta_{1k3l},\beta_{1l3k} )
+ \mathcal{T}_{13lk1}\mathcal{H}_1(\beta_{13kl},\beta_{1lk3} )
+ \mathcal{T}_{13kl1}\mathcal{H}_1(\beta_{13lk},\beta_{1kl3} ) \\[0.3cm] \nn 
&&\hspace{-1.5cm}
+\mathcal{T}_{3kl13}\mathcal{H}_1(\beta_{3k1l},\beta_{3l1k} )
+\mathcal{T}_{31lk3}\mathcal{H}_1(\beta_{31kl},\beta_{3lk1} )
+\mathcal{T}_{31kl3}\mathcal{H}_1(\beta_{31lk},\beta_{3kl1} ) \\[0.3cm] \nn 
&&\hspace{-1.5cm}
+ \mathcal{T}_{2kl32}\mathcal{H}_1(\beta_{2k3l},\beta_{2l3k} )
+ \mathcal{T}_{23lk2}\mathcal{H}_1(\beta_{23kl},\beta_{2lk3} )
+ \mathcal{T}_{23kl2}\mathcal{H}_1(\beta_{23lk},\beta_{2kl3} ) \\[0.3cm] \nn 
&&\hspace{-1.5cm}
+ \mathcal{T}_{3kl23}\mathcal{H}_1(\beta_{3k2l},\beta_{3l2k} )
+ \mathcal{T}_{32lk3}\mathcal{H}_1(\beta_{32kl},\beta_{3lk2} )
+ \mathcal{T}_{32kl3}\mathcal{H}_1(\beta_{32lk},\beta_{3kl2} ) \, \Big] \\[0.1cm] \nn 
&&\hspace{-3.6cm}
+\, 2\,\, \sum_{i=4} \sum_{\substack{l=4 \\ l\neq i}}\,\,  
\Big[\,\mathcal{T}_{i2l1i}\mathcal{H}_1(\beta_{i21l},\beta_{il12} )
+ \mathcal{T}_{i1l2i}\mathcal{H}_1(\beta_{i12l},\beta_{il21} )
+ \mathcal{T}_{i12li}\mathcal{H}_1(\beta_{i1l2},\beta_{i2l1} ) \\[-0.3cm] \nn 
&&\hspace{-1.3cm}
+ \mathcal{T}_{i3l1i}\mathcal{H}_1(\beta_{i31l},\beta_{il13} )
+ \mathcal{T}_{i1l3i}\mathcal{H}_1(\beta_{i13l},\beta_{il31} )
+ \mathcal{T}_{i13li}\mathcal{H}_1(\beta_{i1l3},\beta_{i3l1} ) \\[0.2cm] \nn 
&&\hspace{-1.3cm}
+ \mathcal{T}_{i3l2i}\mathcal{H}_1(\beta_{i32l},\beta_{il23} )
+ \mathcal{T}_{i2l3i}\mathcal{H}_1(\beta_{i23l},\beta_{il32} )
+ \mathcal{T}_{i23li}\mathcal{H}_1(\beta_{i2l3},\beta_{i3l2} ) \, \Big] \\[-0.2cm] \nn 
&&\hspace{-3.6cm} +\,2 \,\,\sum_{l=4}^n\,\, \Big[\,     
\mathcal{T}_{13l21}\mathcal{H}_1(\beta_{132l},\beta_{1l23} ) 
+\,\mathcal{T}_{12l31}\mathcal{H}_1(\beta_{123l},\beta_{1l32} )
+\,\mathcal{T}_{123l1}\mathcal{H}_1(\beta_{12l3},\beta_{13l2} ) \\[-0.1cm] \nn 
&&\hspace{-2.0cm}
+\mathcal{T}_{23l12}\mathcal{H}_1(\beta_{231l},\beta_{2l13} ) 
+\mathcal{T}_{21l32}\mathcal{H}_1(\beta_{213l},\beta_{2l31} )
+\mathcal{T}_{213l2}\mathcal{H}_1(\beta_{21l3},\beta_{23l1} ) \\[0.3cm] \nn 
&&\hspace{-2.0cm}
+ \mathcal{T}_{32l13}\mathcal{H}_1(\beta_{321l},\beta_{3l12} ) 
+\mathcal{T}_{31l23}\mathcal{H}_1(\beta_{312l},\beta_{3l21} )
+\mathcal{T}_{312l3}\mathcal{H}_1(\beta_{31l2},\beta_{32l1} )  \\[0.3cm]  
&&\hspace{-2.0cm}
+\mathcal{T}_{l231l}\mathcal{H}_1(\beta_{l213},\beta_{l312} )
+\mathcal{T}_{l132l}\mathcal{H}_1(\beta_{l123},\beta_{l321} )
+\mathcal{T}_{l123l}\mathcal{H}_1(\beta_{l132},\beta_{l231} )  \, \Big].
\end{eqnarray}
The corresponding anomalous dimension for an amplitude with $n-2$
external partons is given by
\begin{eqnarray}
{\bf{ \Gamma}}_{n-2,5{\rm{T}}-4{\rm{L}}}(\{\beta_{ijkl} \},\alpha_s) && \\ \nn
&&\hspace{-4.5cm} =\, 
2 \sum_{i=4}^n\sum_{\substack{ 4\leq j < k < l\leq n\\ j,k,l \neq i } } 
\Big[\, \mathcal{T}_{iklji}\mathcal{H}_1(\beta_{ikjl},\beta_{iljk})
+\mathcal{T}_{ijlki} \mathcal{H}_1(\beta_{ijkl},\beta_{ilkj} )
+\mathcal{T}_{ijkli}\mathcal{H}_1(\beta_{ijlk},\beta_{iklj} ) \, \Big]  \\[-0.2cm] \nn
&&\hspace{-4.5cm}
+ 2 \hspace{-0.3cm}  \sum_{\substack{ 4\leq j < k < l \leq n  } } 
\Big[\, \mathcal{T}_{PkljP}\mathcal{H}_1(\beta_{Pkjl},\beta_{Pljk} )
+ \mathcal{T}_{PjlkP}\mathcal{H}_1(\beta_{Pjkl},\beta_{Plkj} )
+\mathcal{T}_{PjklP}\mathcal{H}_1(\beta_{Pjlk},\beta_{Pklj} ) \, \Big] \\[-0.2cm] \nn
&&\hspace{-4.5cm}
+ 2 \sum_{i=4}^n\sum_{\substack{4 \leq k < l \leq n \\  k,l \neq i } }
\Big[\, \mathcal{T}_{iklPi}\mathcal{H}_1(\beta_{ikPl},\beta_{ilPk} ) 
+ \mathcal{T}_{iPlki}\mathcal{H}_1(\beta_{iPkl},\beta_{ilkP} )
+ \mathcal{T}_{iPkli}\mathcal{H}_1(\beta_{iPlk},\beta_{iklP}) \, \Big]  
\end{eqnarray}
We proceed to take the difference of the above two equations in accordance with
the definition in \Eqn{eq:GammaSp5T3}. We substitute $\T_P = \T_1 + \T_2+\T_3$,
and use rescaling invariance of CICRs as demonstrated in \Eqn{eq:resclinv}.
We carry out additional simplifications using symmetry properties of 
the function $\mathcal{H}_1$ such as \Eqn{eq:H1-sym1}. Together with
\Eqn{eq:CICR1-2pc-ij} this implies that for example
$\mathcal{H}_1(\beta_{1k2l},\beta_{1l2k})~\stackrel{p_1||p_2}{\longrightarrow}~0$, 
and identical conclusions can be drawn for the other two subsets of collinear
particles from the three becoming collinear here. 
Moreover, using general symmetry properties of the logarithms of the CICRs,
and their behaviour in the collinear limit discussed in
Section~\ref{sec:kin-two-part-col}, we find
\begin{eqnarray}\label{eq:5.290} \nn
{\bf{ \Gamma}}_{n,5{\rm{T}}-4{\rm{L}}}(\{\beta_{ijkl} \},\alpha_s) 
- {\bf{ \Gamma}}_{n-2,5{\rm{T}}-4{\rm{L}}}(\{\beta_{ijkl} \},\alpha_s) && \\ \nn
&& \hspace{-7.5cm}
=\, 2  \sum_{k=4} \sum_{\substack{l=4 \\ l\neq k}} \,\,\,\,
\bigg[\,\Big(\mathcal{T}_{12kl1} +\mathcal{T}_{21kl2} 
+ \mathcal{T}_{kl21k}+ \mathcal{T}_{kl12k}\Big) \mathcal{H}_1(\beta_{12kl},0) \\[-0.5cm] \nn
&& \hspace{-5.1cm}
+\Big(\mathcal{T}_{13kl1} + \mathcal{T}_{31kl3}
+\mathcal{T}_{kl31k}+ \mathcal{T}_{kl13k}\Big) \mathcal{H}_1(\beta_{13kl},0)  \\[-0.1cm] \nn
&& \hspace{-5.1cm}
+\Big(\mathcal{T}_{23kl2}+\mathcal{T}_{32kl3}
+\mathcal{T}_{kl32k}+ \mathcal{T}_{kl23k}\Big) \mathcal{H}_1(\beta_{23kl},0) \,\bigg] \\[-0.2cm] \nn
&& \hspace{-7.5cm}
-\, 2\sum_{\substack{ 4\leq j < k < l \leq n  } } \bigg[\,
\Big(\mathcal{T}_{1klj2}+\mathcal{T}_{ 2klj1}\Big)
\mathcal{H}_1(\beta_{Pkjl},\beta_{Pljk}  )
+\Big(   \mathcal{T}_{1jlk2}+ \mathcal{T}_{2jlk1}\Big)
\mathcal{H}_1(\beta_{Pjkl},\beta_{Plkj}  ) \\[-0.2cm] \nn
&& \hspace{-5.1cm}
+\Big(\mathcal{T}_{1jkl2}+\mathcal{T}_{2jkl1}\Big)
\mathcal{H}_1(\beta_{Pjlk},\beta_{Pklj}  ) 
+\Big( \mathcal{T}_{1klj3}+\mathcal{T}_{3klj1}\Big)
\mathcal{H}_1(\beta_{Pkjl},\beta_{Pljk}  ) \\[0.0cm] \nn
&& \hspace{-5.1cm}
+\Big(\mathcal{T}_{1jlk3}+ \mathcal{T}_{3jlk1}\Big)
\mathcal{H}_1(\beta_{Pjkl},\beta_{Plkj}  )
+\Big(\mathcal{T}_{1jkl3}+\mathcal{T}_{3jkl1}\Big)
\mathcal{H}_1(\beta_{Pjlk},\beta_{Pklj}  ) \\[0.0cm] \nn
&& \hspace{-5.1cm}
+ \Big( \mathcal{T}_{2klj3}+\mathcal{T}_{3klj2}\Big)
\mathcal{H}_1(\beta_{Pkjl},\beta_{Pljk}  )
+\Big(\mathcal{T}_{2jlk3}+ \mathcal{T}_{3jlk2}\Big)\mathcal{H}_1(\beta_{Pjkl},\beta_{Plkj} ) \\[0.0cm] \nn
&& \hspace{-5.1cm}
+\Big(\mathcal{T}_{2jkl3}+\mathcal{T}_{3jkl2}\Big)\mathcal{H}_1(\beta_{Pjlk},\beta_{Pklj}  ) 
\, \bigg] \\[-0.2cm] \nn
&& \hspace{-7.5cm}  +2 \sum_{l=4}^n \bigg[\,    
\mathcal{T}_{13l21}\mathcal{H}_1(\beta_{132l},\beta_{1l23} ) 
+\mathcal{T}_{12l31}\mathcal{H}_1(\beta_{123l},\beta_{1l32} )
+\mathcal{T}_{123l1}\mathcal{H}_1(\beta_{12l3},\beta_{13l2} )  \\[-0.1cm] \nn
&& \hspace{-6.2cm}
+\mathcal{T}_{23l12}\mathcal{H}_1(\beta_{231l},\beta_{2l13}  )
+\mathcal{T}_{21l32}\mathcal{H}_1(\beta_{213l},\beta_{2l31}  )
+\mathcal{T}_{213l2}\mathcal{H}_1(\beta_{21l3},\beta_{23l1}  ) \\[0.3cm] \nn
&& \hspace{-6.2cm}
+\mathcal{T}_{32l13}\mathcal{H}_1(\beta_{321l},\beta_{3l12}  ) 
+\mathcal{T}_{31l23}\mathcal{H}_1(\beta_{312l},\beta_{3l21}  )
+\mathcal{T}_{312l3}\mathcal{H}_1(\beta_{31l2},\beta_{32l1}  ) \\[0.1cm] 
&& \hspace{-6.2cm}
+\mathcal{T}_{l231l}\mathcal{H}_1(\beta_{l213},\beta_{l312}  )
+\mathcal{T}_{l132l}\mathcal{H}_1(\beta_{l123},\beta_{l321}  )
+\mathcal{T}_{l123l}\mathcal{H}_1(\beta_{l132},\beta_{l231}  ) \, \bigg]. 
\end{eqnarray}
We now combine the result in~\Eqn{eq:5.290}
with the one in \Eqn{eq:5T5L-3pc} according to definition
of the splitting amplitude soft anomalous dimension
for three particles becoming collinear,
${\bf{ \Gamma}}_{\SP, 3}^{5{\rm{T}}}$, 
that can be found in \Eqn{eq:GammaSp5T3}. We then arrive at 
\begin{eqnarray}\label{eq:5.308}
{\bf{ \Gamma}}_{\SP, 3}^{5{\rm{T}}}(p_1, p_2, p_3;\mu_f) &=& 
2\sum_{k=4} \sum_{\substack{l=4 \\ l\neq k}}\,\, \bigg[ \,
\Big(\mathcal{T}_{12kl1} \,+\,\mathcal{T}_{21kl2}+\mathcal{T}_{kl21k}+\,\mathcal{T}_{kl12k}\Big)
\mathcal{H}_1(\beta_{12kl},0  )
\nonumber  \\[-0.5cm] &&   \hspace{1.8cm}
+\Big(\mathcal{T}_{13kl1} +\,\mathcal{T}_{31kl3}\,+\mathcal{T}_{kl31k}+\,\mathcal{T}_{kl13k}\Big) 
\mathcal{H}_1(\beta_{13kl},0  )
\nonumber  \\[-0.1cm] &&  \hspace{1.8cm}
+\Big(\mathcal{T}_{23kl2}+\,\mathcal{T}_{32kl3}\,+\mathcal{T}_{kl32k}+\,\mathcal{T}_{kl23k}\Big)
\mathcal{H}_1(\beta_{23kl},0  ) \, \bigg]
\nonumber \\[-0.1cm] && \hspace{-3.5cm}
+ 2 \sum_{\substack{4\leq  j<k<l\leq n } }\,\bigg\{ \Big[
\mathcal{T}_{1klj2}+\mathcal{T}_{2klj1}\Big]
\Big(4 \mathcal{H}_2(\beta_{Pklj},-\beta_{12kl},-\beta_{12kl},0,\beta_{12jk}  ) 
-\mathcal{H}_1(\beta_{Pkjl},\beta_{Pljk}  )  \Big)
\nonumber \\[-0.1cm] && \hspace{-1.1cm}
+ \Big[\mathcal{T}_{1jlk2}+ \mathcal{T}_{2jlk1}\Big]
\Big(4\mathcal{H}_2(\beta_{Pjlk},-\beta_{12jl},-\beta_{12jl},0,\beta_{12jk}  )
-\mathcal{H}_1(\beta_{Pjkl},\beta_{Plkj}  )  \Big)
\nonumber \\ && \hspace{-1.1cm}
+\Big[\mathcal{T}_{1jkl2}+\mathcal{T}_{2jkl1}\Big]
\Big(4\mathcal{H}_2(\beta_{Pjkl},-\beta_{12jk},-\beta_{12jk},0,\beta_{12jl}  )
-\mathcal{H}_1(\beta_{Pjlk},\beta_{Pklj}  )   \Big)
\nonumber \\ && \hspace{-1.1cm}
+\Big[\mathcal{T}_{1klj3}+\mathcal{T}_{3klj1}\Big]
\Big(4\mathcal{H}_2(\beta_{Pklj},-\beta_{13kl},-\beta_{13kl},0,\beta_{13jk}  )
-\mathcal{H}_1(\beta_{Pkjl},\beta_{Pljk}  )  \Big)
\nonumber \\ && \hspace{-1.1cm}
+\Big[ \mathcal{T}_{1jlk3}+\mathcal{T}_{3jlk1}\Big]
\Big(4\mathcal{H}_2(\beta_{Pjlk},-\beta_{13jl},-\beta_{13jl},0,\beta_{13jk}  )
-\mathcal{H}_1(\beta_{Pjkl},\beta_{Plkj}  )  \Big)
\nonumber \\ && \hspace{-1.1cm}
+ \Big[\mathcal{T}_{1jkl3}+\mathcal{T}_{3jkl1}\Big]
\Big(4\mathcal{H}_2(\beta_{Pjkl},-\beta_{13jk},-\beta_{13jk},0,\beta_{13jl}  )
-\mathcal{H}_1(\beta_{Pjlk},\beta_{Pklj}  )   \Big)
\nonumber \\ && \hspace{-1.1cm}
+\Big[\mathcal{T}_{2klj3}+\mathcal{T}_{3klj2}\Big]
\Big(4\mathcal{H}_2(\beta_{Pklj},-\beta_{23kl},-\beta_{23kl},0,\beta_{23jk}  )
-\mathcal{H}_1(\beta_{Pkjl},\beta_{Pljk}  )   \Big) 
\nonumber \\ && \hspace{-1.1cm}
+ \Big[\mathcal{T}_{2jlk3}+\mathcal{T}_{3jlk2}\Big]
\Big(4\mathcal{H}_2(\beta_{Pjlk},-\beta_{23jl},-\beta_{23jl},0,\beta_{23jk}  )
-\mathcal{H}_1(\beta_{Pjkl},\beta_{Plkj}  ) \Big)
\nonumber \\[-0.1cm] && \hspace{-1.1cm}
+\Big[ \mathcal{T}_{2jkl3}+ \mathcal{T}_{3jkl2}\Big]
\Big(4\mathcal{H}_2(\beta_{Pjkl},-\beta_{23jk},-\beta_{23jk},0,\beta_{23jl}  )
-\mathcal{H}_1(\beta_{Pjlk},\beta_{Pklj}  )   \Big) \bigg\}
\nonumber \\[-0.2cm] && \hspace{-3.5cm} 
+ 8 \sum_{m=4}^n   \sum_{\substack{4\leq  k<l\leq n\\  k,l\neq m} }\Big[\,
\mathcal{T}_{12lkm}\mathcal{H}_2(\beta_{12kl},\beta_{12lm},0,\beta_{12km},0 )
\nonumber \\[-0.5cm] && \hspace{-0.8cm}
+\mathcal{T}_{12klm}\mathcal{H}_2(\beta_{12kl},\beta_{12km},0,\beta_{12lm},0 )
\nonumber \\[0.2cm] && \hspace{-0.8cm} 
+\mathcal{T}_{13lkm}\mathcal{H}_2(\beta_{13kl},\beta_{13lm},0,\beta_{13km},0  )
\nonumber \\[0.2cm] && \hspace{-0.8cm}
+ \mathcal{T}_{13klm}\mathcal{H}_2(\beta_{13kl},\beta_{13km},0,\beta_{13lm},0  )
\nonumber \\[0.2cm] && \hspace{-0.8cm}
+ \mathcal{T}_{23lkm}\mathcal{H}_2(\beta_{23kl},\beta_{23lm},0,\beta_{23km},0  )
\nonumber \\[0.1cm] && \hspace{-0.8cm}
+\mathcal{T}_{23klm}\mathcal{H}_2(\beta_{23kl},\beta_{23km},0,\beta_{23lm},0  ) \, \Big]
\nonumber  \\[-0.1cm] &&\hspace{-3.5cm}
+ 2 \sum_{l=4}^n  \Big[\,
\mathcal{T}_{13l21}\mathcal{H}_1(\beta_{132l},\beta_{1l23} ) 
+\mathcal{T}_{12l31}\mathcal{H}_1(\beta_{123l},\beta_{1l32} )
+\mathcal{T}_{123l1}\mathcal{H}_1(\beta_{12l3},\beta_{13l2} ) 
\nonumber  \\[-0.1cm] &&   \hspace{-2.2cm}
+\mathcal{T}_{23l12}\mathcal{H}_1(\beta_{231l},\beta_{2l13} ) 
+\mathcal{T}_{21l32}\mathcal{H}_1(\beta_{213l},\beta_{2l31} )
+\mathcal{T}_{213l2}\mathcal{H}_1(\beta_{21l3},\beta_{23l1} ) 
\nonumber  \\[0.3cm] &&   \hspace{-2.2cm}
+\mathcal{T}_{32l13}\mathcal{H}_1(\beta_{321l},\beta_{3l12} )
+\mathcal{T}_{31l23}\mathcal{H}_1(\beta_{312l},\beta_{3l21} )
+\mathcal{T}_{312l3}\mathcal{H}_1(\beta_{31l2},\beta_{32l1} ) 
\nonumber  \\[0.2cm] &&   \hspace{-2.2cm}
+\mathcal{T}_{l231l}\mathcal{H}_1(\beta_{l213},\beta_{l312}  )
+\mathcal{T}_{l132l}\mathcal{H}_1(\beta_{l123},\beta_{l321}  )
+\mathcal{T}_{l123l}\mathcal{H}_1(\beta_{l132},\beta_{l231}  )  \, \Big]
\nonumber \\[-0.1cm] && \hspace{-3.5cm}
+ 8 \hspace{-0.1cm}\sum_{\substack{4\leq   k<l\leq n } }\hspace{-0.1cm}
\Big[\,
\mathcal{T}_{23lk1}\mathcal{H}_2(\beta_{23kl},\beta_{1l23},\beta_{132l},\beta_{1k32},\beta_{123k} )
\nonumber  \\[-0.2cm] &&   \hspace{-1.8cm}
+\mathcal{T}_{23kl1}\mathcal{H}_2(\beta_{23kl},\beta_{1k23},\beta_{132k},\beta_{1l32},\beta_{123l} )
\nonumber \\[0.3cm] && \hspace{-1.8cm} 
+\mathcal{T}_{13lk2}\mathcal{H}_2(\beta_{13kl},\beta_{132l},\beta_{1l23},\beta_{13k2},\beta_{12k3} )
\nonumber  \\[0.3cm] &&   \hspace{-1.8cm}
+\mathcal{T}_{13kl2}\mathcal{H}_2(\beta_{13kl},\beta_{132k},\beta_{1k23},\beta_{13l2},\beta_{12l3} ) 
\nonumber \\[0.3cm] && \hspace{-1.8cm}
+\mathcal{T}_{12lk3}\mathcal{H}_2(\beta_{12kl},\beta_{123l},\beta_{1l32},\beta_{12k3},\beta_{13k2} )
\nonumber  \\[0.2cm] &&   \hspace{-1.8cm}
+\mathcal{T}_{12kl3}\mathcal{H}_2(\beta_{12kl},\beta_{123k},\beta_{1k32},\beta_{12l3},\beta_{13l2} )\, \Big]
\nonumber \\[-0.2cm] && \hspace{-3.5cm}
+8 \sum_{m=4}^n   \sum_{\substack{l =4 \\   l\neq m} }^n\,\,\,\Big[\,
\mathcal{T}_{13l2m}\mathcal{H}_2(\beta_{13l2},\beta_{13lm},0,\beta_{132m},\beta_{1m23}  )
\nonumber \\[-0.4cm] && \hspace{-1.3cm}
+\mathcal{T}_{12l3m}\mathcal{H}_2(\beta_{12l3},\beta_{12lm},0,\beta_{123m},\beta_{1m32}  )
\nonumber \\[0.2cm] && \hspace{-1.3cm}
+\mathcal{T}_{123lm}\mathcal{H}_2(\beta_{123l},\beta_{12m3},\beta_{13m2},\beta_{12lm},0  )\, \Big].
\end{eqnarray}
In the first three sums we recognise the
two-particle collinear-like terms that 
generalise the expression for ${\bf{ \Gamma}}_{\SP, 2}^{5{\rm{T}}}
$ in \Eqn{eq:5.187} to each pair out of the three
particles becoming collinear in the case at hand. Therefore, for these terms
we can follow the manipulations below \Eqn{eq:5.187} to recover the constraints in
equations Eqs.~\eqref{eq:H1constraint} and \eqref{eq:Kminus-constraint} 
that allow to drop these terms from further analysis
as, in the end, they do not  contribute to the splitting amplitude
soft anomalous dimension. We continue with the remaining terms which
are the new type terms that contain 
information on all three particles becoming collinear appearing 
for the first time in the three-particle collinear limit.
We use colour identities, such as 
${\mathcal{T}}_{ijklm} = -{\mathcal{T}}_{jilkm} $
and ${\mathcal{T}}_{ijklm} = - {\mathcal{T}}_{ikjlm} $,
along with kinematic identities, such as
ones in Eqs.~\eqref{eq:H2-sym2}, \eqref{eq:H2-sym3},
and
\begin{eqnarray}\label{eq:H2-sym1}
        \mathcal{H}_2(y_1,y_2,y_3,y_4,y_5)
        &= &
        -\mathcal{H}_2(y_1,y_4,y_5,y_2,y_3)\,,
\end{eqnarray}
to rewrite the expression for ${\bf{ \Gamma}}_{\SP, 3}^{5{\rm{T}}}$ 
as follows 
\begin{eqnarray}\label{eq:5.308-c1} \nn
{\bf{ \Gamma}}_{\SP, 3}^{5{\rm{T}}}(p_1, p_2, p_3) && \\ \nn
&&\hspace{-2.5cm}=\, 2 \sum_{l=4}^n  \Big[\, 
\mathcal{T}_{13l21}\mathcal{H}_1(\beta_{132l},\beta_{1l23} ) 
+\mathcal{T}_{12l31}\mathcal{H}_1(\beta_{123l},\beta_{1l32} )
+\mathcal{T}_{123l1}\mathcal{H}_1(\beta_{12l3},\beta_{13l2} ) \\[-0.1cm] \nn
&& \hspace{-1.1cm}
+\mathcal{T}_{23l12}\mathcal{H}_1(\beta_{231l},\beta_{2l13} ) 
+\mathcal{T}_{21l32}\mathcal{H}_1(\beta_{213l},\beta_{2l31} )
+\mathcal{T}_{213l2}\mathcal{H}_1(\beta_{21l3},\beta_{23l1} ) \\[0.2cm] \nn
&& \hspace{-1.1cm}
+\mathcal{T}_{32l13}\mathcal{H}_1(\beta_{321l},\beta_{3l12} )
+\mathcal{T}_{31l23}\mathcal{H}_1(\beta_{312l},\beta_{3l21} )
+\mathcal{T}_{312l3}\mathcal{H}_1(\beta_{31l2},\beta_{32l1} ) \\[0.1cm] \nn
&& \hspace{-1.1cm}
+\mathcal{T}_{l231l}\mathcal{H}_1(\beta_{l213},\beta_{l312}  )
+\mathcal{T}_{l132l}\mathcal{H}_1(\beta_{l123},\beta_{l321}  )
+\mathcal{T}_{l123l}\mathcal{H}_1(\beta_{l132},\beta_{l231}  ) \, \Big] \\[-0.1cm] \nn
&&\hspace{-2.5cm}
+\, 8 \hspace{-0.1cm}\sum_{\substack{4\leq   k<l\leq n } }\hspace{-0.1cm} \Big[ \, 
\mathcal{T}_{23lk1}
\mathcal{H}_2(\beta_{23kl},\beta_{1l23},\beta_{132l},\beta_{1k32},\beta_{123k}) \\[-0.2cm] \nn
&& \hspace{-0.7cm}
+\mathcal{T}_{23kl1}
\mathcal{H}_2(\beta_{23kl},\beta_{1k23},\beta_{132k},\beta_{1l32},\beta_{123l}) \\[0.1cm] \nn
&& \hspace{-0.7cm}
+\mathcal{T}_{31kl2}
\mathcal{H}_2(\beta_{13kl},\beta_{13k2},\beta_{12k3},\beta_{132l},\beta_{1l23}) \\[0.1cm] \nn
&& \hspace{-0.7cm}
+\mathcal{T}_{31lk2}
\mathcal{H}_2(\beta_{13kl},\beta_{13l2},\beta_{12l3},\beta_{132k},\beta_{1k23}) \\[0.1cm] \nn
&& \hspace{-0.7cm}
+\mathcal{T}_{12lk3}
\mathcal{H}_2(\beta_{12kl},\beta_{123l},\beta_{1l32},\beta_{12k3},\beta_{13k2} ) \\[0.0cm] \nn
&& \hspace{-0.7cm}
+\mathcal{T}_{12kl3}
\mathcal{H}_2(\beta_{12kl},\beta_{123k},\beta_{1k32},\beta_{12l3},\beta_{13l2} ) \, \Big]  \\[-0.3cm] \nn
&& \hspace{-2.5cm}
+\, 8\; \sum_{m=4}^n   \sum_{\substack{l =4 \\   l\neq m} }^n\,\,\, \Big[ \, 
\mathcal{T}_{312lm}
\mathcal{H}_2(\beta_{13l2},\beta_{132m},\beta_{1m23},\beta_{13lm},0) \\[-0.5cm] \nn
&& \hspace{-0.1cm}
+\mathcal{T}_{231lm}
\mathcal{H}_2(\beta_{231l},\beta_{23m1},\beta_{21m3},\beta_{23lm},0) \\[0.1cm] 
&& \hspace{-0.1cm}
+\mathcal{T}_{123lm}
\mathcal{H}_2(\beta_{123l},\beta_{12m3},\beta_{13m2},\beta_{12lm},0) \, \Big].
\end{eqnarray}
Next, we use CICR relations such as $\beta_{l132} = \beta_{1l23}, \beta_{l231} = \beta_{132l}$,
and so on, to collect the colour terms according to the ${\cal{H}}_1$ function they multiply. 
We arrive at 
\begin{eqnarray}\label{eq:5.308-c2} \nn
{\bf{ \Gamma}}_{\SP, 3}^{5{\rm{T}}}(p_1, p_2, p_3) &=&
2 \sum_{l=4}^{n} \bigg[ \Big(  
\mathcal{T}_{13l21} -\mathcal{T}_{23l12} 
+\mathcal{T}_{312l3} \Big) \mathcal{H}_1(\beta_{132l},\beta_{1l23})
- \mathcal{T}_{l123l} \mathcal{H}_1(\beta_{132l},\beta_{1l23}) \bigg] \\[-0.1cm] \nn
&+& 2 \sum_{l=4}^{n} \bigg[ \Big(  
\mathcal{T}_{12l31} - \mathcal{T}_{32l13}
+\mathcal{T}_{213l2} \Big) \mathcal{H}_1(\beta_{123l},\beta_{1l32})
- \mathcal{T}_{l132l} \mathcal{H}_1(\beta_{123l},\beta_{1l32}) \bigg] \\[-0.1cm] \nn
&+& 2 \sum_{l=4}^{n} \bigg[ \Big(  
\mathcal{T}_{21l32} -\mathcal{T}_{31l23}
+\mathcal{T}_{123l1} \Big) \mathcal{H}_1(\beta_{12l3},\beta_{13l2})
- \mathcal{T}_{l231l}\mathcal{H}_1(\beta_{12l3},\beta_{13l2}) \bigg] \\[-0.1cm] \nn
&+& 8 \hspace{-0.1cm}\sum_{\substack{4\leq   k<l\leq n } }\hspace{-0.1cm} \Big[ \,  
\mathcal{T}_{23lk1}
\mathcal{H}_2(\beta_{23kl},\beta_{1l23},\beta_{132l},\beta_{1k32},\beta_{123k} ) \\[-0.1cm] \nn
&&\hspace{1.4cm} +\mathcal{T}_{23kl1}
\mathcal{H}_2(\beta_{23kl},\beta_{1k23},\beta_{132k},\beta_{1l32},\beta_{123l} )\\[0.2cm] \nn
&&\hspace{1.4cm} +\mathcal{T}_{31kl2}
\mathcal{H}_2(\beta_{13kl},\beta_{13k2},\beta_{12k3},\beta_{132l},\beta_{1l23} ) \\[0.2cm] \nn
&&\hspace{1.4cm} +\mathcal{T}_{31lk2}
\mathcal{H}_2(\beta_{13kl},\beta_{13l2},\beta_{12l3},\beta_{132k},\beta_{1k23} ) \\[0.2cm] \nn
&&\hspace{1.4cm} +\mathcal{T}_{12lk3}
\mathcal{H}_2(\beta_{12kl},\beta_{123l},\beta_{1l32},\beta_{12k3},\beta_{13k2} ) \\[0.1cm] \nn
&&\hspace{1.4cm} +\mathcal{T}_{12kl3}
\mathcal{H}_2(\beta_{12kl},\beta_{123k},\beta_{1k32},\beta_{12l3},\beta_{13l2} )\, \Big] \\[-0.1cm] \nn
&+& 8\; \sum_{m=4}^n   \sum_{\substack{l =4 \\   l\neq m} }^n\,\,\,\Big[ \,
\mathcal{T}_{312lm}\mathcal{H}_2(\beta_{13l2},\beta_{132m},\beta_{1m23},\beta_{13lm},0) \\[-0.4cm] \nn
&&\hspace{2.0cm}+\mathcal{T}_{231lm}
\mathcal{H}_2(\beta_{231l},\beta_{23m1},\beta_{21m3},\beta_{23lm},0) \\[0.1cm] 
&&\hspace{2.0cm}+\mathcal{T}_{123lm}
\mathcal{H}_2(\beta_{123l},\beta_{12m3},\beta_{13m2},\beta_{12lm},0 )\, \Big],
\end{eqnarray}
where we have separated the colour terms multiplying the ${\cal{H}}_1$
kinematic function that depend on a repeated rest-of-the-process index
from the ones which only have one. For the case of the latter type of term,
remembering that the CICRs with information on three collinear partons do
not depend on the rest-of-the-process parton and can be moved outside of
the sum, we can directly apply colour conservation to bring the dependence
of these terms to degrees of freedom of particles becoming collinear only. 
These steps are performed explicitly below after we 
deal with the remaining terms with a repeated rest-of-the-process index
that needs to be combined with relevant ${\cal{H}}_2$ terms 
to form stuffle product relations. 
This situation is in direct analogy to the two-particle
collinear limit at three loops, see in particular manipulations
leading to \Eqn{eq:stuffleGen}. We start by using the following
identities on the colour terms in the last three lines of the above equation
\begin{eqnarray}\label{eq:Tijklm-idty}
{\cal{T} }_{ijklm} = {\cal{T} }_{ijmlk} -  {\cal{T} }_{ikmlj},
\qquad \mathcal{T}_{ijklm}=-\mathcal{T}_{jilkm}.
\end{eqnarray}
Then, we can combine the two sums with ${\cal{H}}_2$ into one after relabelling the
summed index $m$ to $k$. We find 
\begin{eqnarray}\label{eq:5.308-c3} \nn
{\bf{ \Gamma}}_{\SP, 3}^{5{\rm{T}}}(p_1, p_2, p_3) &=&
2 \sum_{l=4}^{n} \bigg[ \Big(  
\mathcal{T}_{13l21} -\mathcal{T}_{23l12} 
+\mathcal{T}_{312l3} \Big) \mathcal{H}_1(\beta_{132l},\beta_{1l23})
- \mathcal{T}_{l123l}\mathcal{H}_1(\beta_{132l},\beta_{1l23}) \bigg] \\[-0.1cm] \nn
&+& 2\sum_{l=4}^{n} \bigg[ \Big(  
\mathcal{T}_{12l31} - \mathcal{T}_{32l13}
+\mathcal{T}_{213l2} \Big) \mathcal{H}_1(\beta_{123l},\beta_{1l32})
- \mathcal{T}_{l132l}\mathcal{H}_1(\beta_{123l},\beta_{1l32}) \bigg] \\[-0.1cm] \nn
&+& 2 \sum_{l=4}^{n} \bigg[ \Big(  
\mathcal{T}_{21l32} -\mathcal{T}_{31l23}
+\mathcal{T}_{123l1} \Big)\mathcal{H}_1(\beta_{12l3},\beta_{13l2})
- \mathcal{T}_{l231l}\mathcal{H}_1(\beta_{12l3},\beta_{13l2}) \bigg] \\[-0.1cm] \nn
&+& 8 {\sum_{k=4}^{n}   \sum_{\substack{l =4 \\   l\neq k} }^ {n} }
\,\bigg[ \mathcal{T}_{23kl1} \Big(   
\mathcal{H}_2(\beta_{23kl},\beta_{1k23},\beta_{132k},\beta_{1l32},\beta_{123l}) \\[-0.4cm] \nn 
&& \hspace{2.8cm}
+ \mathcal{H}_2(\beta_{231l},\beta_{23k1},\beta_{21k3},\beta_{23lk},0 ) \\[0.2cm] \nn 
&& \hspace{2.8cm}
+  \mathcal{H}_2(\beta_{13l2},\beta_{132k},\beta_{1k23},\beta_{13lk},0 ) \Big) \\[0.1cm] \nn
&& \hspace{1.5cm}+  
\mathcal{T}_{31kl2} \Big( 
\mathcal{H}_2(\beta_{13kl},\beta_{13k2},\beta_{12k3},\beta_{132l},\beta_{1l23} ) \\[0.2cm] \nn 
&& \hspace{2.8cm}
+\mathcal{H}_2(\beta_{13l2},\beta_{132k},\beta_{1k23},\beta_{13lk},0 ) \\[0.1cm] \nn 
&& \hspace{2.8cm}
+\mathcal{H}_2(\beta_{123l},\beta_{12k3},\beta_{13k2},\beta_{12lk},0 ) \Big) \\[0.1cm] \nn
&& \hspace{1.5cm}+  
\mathcal{T}_{12kl3}\Big(
\mathcal{H}_2(\beta_{12kl},\beta_{123k},\beta_{1k32},\beta_{12l3},\beta_{13l2} ) \\[0.2cm] \nn 
&& \hspace{2.8cm}
+\mathcal{H}_2(\beta_{123l},\beta_{12k3},\beta_{13k2},\beta_{12lk},0 ) \\[0.1cm] 
&& \hspace{2.8cm}
+\mathcal{H}_2(\beta_{231l},\beta_{23k1},\beta_{21k3},\beta_{23lk},0 )\Big)\bigg].
\end{eqnarray}
We now use the identities in \Eqn{eq:Tijklm-idty} to move the
repeated indices appearing in the colour structures in the first three lines
to the third and fourth position, so for example 
\begin{eqnarray}
\mathcal{T}_{l123l} = \mathcal{T}_{l1l32}-  \mathcal{T}_{l2l31}
= - \mathcal{T}_{31ll2} + \mathcal{T}_{32ll1}\,.
\end{eqnarray}
We then have 
\begin{eqnarray}\label{eq:5.308-c4} \nn
{\bf{ \Gamma}}_{\SP, 3}^{5{\rm{T}}}(p_1, p_2, p_3) &=&
2 \sum_{l=4}^{n} \bigg[\,\Big(\mathcal{T}_{13l21} -\mathcal{T}_{23l12} 
+\,\mathcal{T}_{312l3} \Big)\mathcal{H}_1(\beta_{132l},\beta_{1l23}) \\[-0.1cm] \nn
&&\hspace{1.0cm}+ \Big( \mathcal{T}_{12l31} 
-\mathcal{T}_{32l13} +\mathcal{T}_{213l2} \Big)
\mathcal{H}_1(\beta_{123l},\beta_{1l32}) \\[-0.1cm] \nn
&&\hspace{1.0cm}+ \Big(\mathcal{T}_{21l32} 
- \mathcal{T}_{31l23} +\mathcal{T}_{123l1} \Big)
\mathcal{H}_1(\beta_{12l3},\beta_{13l2}) \, \bigg] \\[-0.1cm] \nn
&+&\, 8 {\sum_{k=4}^{n}   \sum_{\substack{l =4 \\   l\neq k} }^ {n} }
\Big[\, \mathcal{H}_2(\beta_{23kl},\beta_{1k23},\beta_{132k},\beta_{1l32},\beta_{123l}) \\[-0.4cm] \nn
&&\hspace{1.5cm}+ \mathcal{H}_2(\beta_{231l},\beta_{23k1},\beta_{21k3},\beta_{23lk},0 ) \\[0.1cm] \nn
&&\hspace{1.5cm}+  \mathcal{H}_2(\beta_{13l2},\beta_{132k},\beta_{1k23},\beta_{13lk},0 ) \, 
\Big] \mathcal{T}_{23kl1} \\[-0.1cm] \nn
&+&\,2 \sum_{l=4}^{n} \Big[\mathcal{H}_1(\beta_{132l},\beta_{1l23}) 
- \mathcal{H}_1(\beta_{123l},\beta_{1l32}) \Big] \mathcal{T}_{23ll1} \\[-0.1cm] \nn
&+&\, 8 {\sum_{k=4}^{n}   \sum_{\substack{l =4 \\   l\neq k} }^ {n} }
\Big[\, \mathcal{H}_2(\beta_{13kl},\beta_{13k2},\beta_{12k3},\beta_{132l},\beta_{1l23} ) \\[-0.2cm] \nn
&&\hspace{1.5cm} +\mathcal{H}_2(\beta_{13l2},\beta_{132k},\beta_{1k23},\beta_{13lk},0 ) \\[0.1cm] \nn
&&\hspace{1.5cm} +\mathcal{H}_2(\beta_{123l},\beta_{12k3},\beta_{13k2},\beta_{12lk},0 ) \,
\Big] \mathcal{T}_{31kl2} \\[-0.1cm] \nn
&+&\, 2 \sum_{l=4}^{n}\Big[\mathcal{H}_1(\beta_{12l3},\beta_{13l2})
+\mathcal{H}_1(\beta_{132l},\beta_{1l23})\Big]\mathcal{T}_{31ll2} \\[-0.1cm] \nn
&+&\,8 {\sum_{k=4}^{n}   \sum_{\substack{l =4 \\   l\neq k} }^ {n} }
\Big[\, \mathcal{H}_2(\beta_{12kl},\beta_{123k},\beta_{1k32},\beta_{12l3},\beta_{13l2} ) \\[-0.2cm] \nn
&&\hspace{1.5cm} +\mathcal{H}_2(\beta_{123l},\beta_{12k3},\beta_{13k2},\beta_{12lk},0 ) \\[0.1cm] \nn
&&\hspace{1.5cm} +\mathcal{H}_2(\beta_{231l},\beta_{23k1},\beta_{21k3},\beta_{23lk},0 ) \,
\Big]\mathcal{T}_{12kl3} \\[-0.1cm] 
&+&\, 2 \sum_{l=4}^{n} \Big[ \mathcal{H}_1(\beta_{12l3},\beta_{13l2})
-\mathcal{H}_1(\beta_{123l},\beta_{1l32}) \Big]\mathcal{T}_{12ll3}\,.
\end{eqnarray}
We now notice that we can form stuffle products
for the colour terms with two rest-of-the-process indices,
just as it was the case for the calculation of the two-particle collinear limit
at three loops in~\Eqn{eq:stuffleGen},
which enables us to apply colour conservation to these terms. For example we have that
\begin{eqnarray}\label{eq:NEWstuffle3}
2{\sum_{k=4}^{n}   \sum_{\substack{l =4 \\   k\neq l} }^ {n} }   {\cal T}_{12kl3}  
+2\sum_{l=4}^{n}{\cal T}_{12ll3} = \sum_{k=4}^{n}\sum_{l=4}^{n} 
\big( {\cal T}_{12kl3} +  {\cal T}_{12lk3} \big)\,.
\end{eqnarray}
For this operation to be possible, the coefficients of the repeated index terms
and the terms with two different rest-of-the-process partons must be identified with
each other in the collinear limit.
Precisely this argument was used to obtain constraints in the two-particle
collinear limit for three loop terms. 
In an analogous way we obtain the following constraint here: 
\begin{eqnarray}\label{eq:E.122}
&&\hspace{0cm} 
4\mathcal{H}_2(\beta_{12kl},\beta_{123k},\beta_{1k32},\beta_{12l3},\beta_{13l2})
+4\mathcal{H}_2(\beta_{123l},\beta_{12k3},\beta_{13k2},\beta_{12lk},0)
\nonumber  \\ &&  \hspace{-0.1cm}    
+\mathcal{H}_2(\beta_{231l},\beta_{23k1},\beta_{21k3},\beta_{23lk},0)
= \mathcal{H}_1(\beta_{12l3},\beta_{13l2})
-\mathcal{H}_1(\beta_{123l},\beta_{1l32})\,,
\end{eqnarray}
which originates from the coefficient of the $\mathcal{T}_{12kl3}$
term. Two more constraints
are obtained from coefficients of the $\mathcal{T}_{31kl2}$
and $\mathcal{T}_{23kl1}$ colour structures, but these are cyclic permutations of 
\Eqn{eq:E.122} and so do not contain additional new information. 
In Section~\ref{sec:three-particle-collinear-fourloop} we argued that
this seemingly new constraint obtained in the three-particle 
collinear limit is in fact the second constraint obtained in the
two-particle collinear limit, namely the one reported in 
\Eqn{eq:Kminus-constraint}. Therefore, no new
constraint appears from the three-particle collinear limit 
even at the four loop order. 

To summarise, we have obtained the result for ${\bf{ \Gamma}}_{\SP, 3}^{5{\rm{T}}}$
starting from generic $n$ amplitudes.
It yields a seemingly new constraint, which we show in the main text to simply be
the one obtained in the two-particle collinear limit \Eqn{eq:Kminus-constraint}.
After the constraint is implemented, colour conservation can be used to obtain 
the final result for ${\bf{ \Gamma}}_{\SP, 3}^{5{\rm{T}}}$, 
which is given in~\Eqn{eq:5.308-c4-main-res}.

We would like to finish with a remark that
if we restricted ourselves at the start to the soft anomalous dimension
with only $n=4$ partons to begin with, we would arrive at the result for
$ {\bf{ \Gamma}}_{\SP, 3}^{5{\rm{T}}}$ 
in \Eqn{eq:5.308-c4} without the double summed terms containing
the dependence on ${\cal{H}}_2$ (since these cannot exist for
$n<5$), and one could directly eliminate $\T_4$ in favour of the sum of
$\T_1,\T_2,$ and $\T_3$, even in the repeated index term. This last 
step is prohibited for amplitudes with $n>4$ for this term alone,
and the purpose of the formation of the stuffle products is to enable the application of 
colour conservation for amplitudes with any $n$. While it is simplest
to obtain $ {\bf{ \Gamma}}_{\SP, 3}^{5{\rm{T}}}$ from
${\bf{ \Gamma}}_{4,5{\rm{T}}-4{\rm{L}}} $, here we show
that the same result is obtained in a calculation with generic
number of external partons $n$. Moreover, it is clear that
the requirement that the splitting amplitude soft anomalous 
dimension is identical regardless of the amplitudes used to 
compute it, is the same as the non-dependence of splitting amplitude
soft anomalous dimension on the rest-of-the-process
degrees of freedom as was already observed below~\Eqn{GammaSpQuad2eHybrid}.

\section{Four-particle collinear limit}
\label{app:4partcol}
In this appendix, we explicitly consider 
the three-loop contribution to the 
four-particle collinear limit 
splitting amplitude soft anomalous dimension. 

We begin with the momentum-dependent 
part of ${\bf{ \Gamma}}_{\SP, 4}^{4{\rm{T}}}$ 
which is obtained by considering
the difference ${\bf{ \Gamma}}_{n,4{\rm{T}}-4{\rm{L}}}(\{\beta\},\alpha_s) - {\bf{ \Gamma}}_{n-3,4{\rm{T}}-4{\rm{L}}}(\{\beta\},\alpha_s)$ as defined in \Eqn{eq:GammaSpQuad4}.
We proceed by
separating the contribution of 
the collinear particles 1, 2, 3,  and 
4 and splitting the sum in 
\Eqn{eq:quardupoleABdef} for the amplitudes with $n$ and $n-3$ external partons accordingly:
\begin{subequations} \label{Aterms-4part}
\bea \label{An-4part}   \nn
{\bf{ \Gamma}}_{n,4{\rm{T}}-4{\rm{L}}}(\{\beta\})
&=& 4 \bigg[
\textcolor{orange}{
\sum_{5\leq i<j<k<l \leq n} \, {\bf a}_{ijkl}^{ }(\{\beta\})}
+\textcolor{red}{\sum_{5\leq j<k<l \leq n} \, {\bf a}_{1jkl}^{ }(\{\beta\}) }
\\ 
&&\hspace{-1.5cm}
+\hspace{-0.1cm}\textcolor{red}{\,\sum_{5\leq j<k<l \leq n} \, {\bf a}_{2jkl}^{ }(\{\beta\})}
+\hspace{-0.1cm}\textcolor{red}{\sum_{5\leq j<k<l \leq n} \, {\bf a}_{3jkl}^{ }(\{\beta\})}
+\hspace{-0.1cm}\textcolor{red}{\sum_{5\leq j<k<l \leq n} \, {\bf a}_{4jkl}^{ }(\{\beta\})}
\nn \\&&\hspace{-1.5cm} 
+\,\sum_{5\leq k<l \leq n} \, {\bf a}_{12kl}^{ }(\{\beta\})
+\sum_{5\leq k<l \leq n} \, {\bf a}_{13kl}^{ }(\{\beta\})
+\sum_{5\leq k<l \leq n} \, {\bf a}_{14kl}^{ }(\{\beta\})
\nn\\ 
&&\hspace{-1.5cm}
+\sum_{5\leq k<l \leq n} \, {\bf a}_{23kl}^{ }(\{\beta\})
+\sum_{5\leq k<l \leq n} \, {\bf a}_{24kl}^{ }(\{\beta\})
+\sum_{5\leq k<l \leq n} \, {\bf a}_{34kl}^{ }(\{\beta\})
\nn\\ 
&&\hspace{-1.5cm}
+ \textcolor{darkgreen}{
\sum_{5\leq  l \leq n} \, {\bf a}_{123l}^{ }(\{\beta\})}
+ \textcolor{darkgreen}{
\sum_{5\leq  l \leq n} \, {\bf a}_{124l}^{ }(\{\beta\})}
+ \textcolor{darkgreen}{
\sum_{5\leq  l \leq n} \, {\bf a}_{134l}^{ }(\{\beta\})}
\nn\\ 
&&\hspace{-1.5cm}
+ \textcolor{darkgreen}{
\sum_{5\leq  l \leq n} \, {\bf a}_{234l}^{ }(\{\beta\})}
+ \textcolor{cPurple}{
 \, {\bf a}_{1234}^{ }(\{\beta\})}
\bigg], \\[0.1cm] \label{Anm3-4part}
{\bf{ \Gamma}}_{n-3,4{\rm{T}}-4{\rm{L}}}(\{\beta\})
&=& 4  \bigg[\textcolor{orange}{
\sum_{5\leq i<j<k<l \leq n} \, {\bf a}_{ijkl}^{ }(\{\beta\})}
+\textcolor{red}{\sum_{5\leq j<k<l \leq n} \, {\bf a}_{Pjkl}^{ }(\{\beta\})} \bigg].
\eea
\end{subequations} 
The terms which do not 
involve any of the collinear or parent partons 
(\textcolor{orange}{orange} terms) cancel each other 
in the difference. 
Using $\T_P=\T_1+\T_2+\T_3+\T_4$
leads to the terms involving one collinear particle in 
${\bf{ \Gamma}}_{n,4{\rm{T}}-4{\rm{L}}}(\{\beta\})$
cancelling with the second term in the ${\bf{ \Gamma}}_{n-3,4{\rm{T}}-4{\rm{L}}}(\{\beta\})$
amplitude (\textcolor{red}{red} terms).  
What remains are the two-particle collinear-like terms
for each pair of the four particles becoming collinear (black terms), 
the \textcolor{darkgreen}{green} terms which
are the three-particle
collinear-like terms out of the four becoming collinear,
and the last term in 
${\bf{ \Gamma}}_{n,4{\rm{T}}-4{\rm{L}}}(\{\beta\})$ 
which captures the full
four-particle collinear limit kinematics. 

Turning our attention to the constant terms, we again use the method of splitting the sums in $ {\bf{ \Gamma}}_{n,4{\rm{T}}-3{\rm{L}}} $ and $ {\bf{ \Gamma}}_{n-3,4{\rm{T}}-3{\rm{L}}} $ in \Eqn{BtermsHybrid}. We have
\begin{subequations}   \label{eq:Bterms-4part}
\bea    \nn
 {\bf{ \Gamma}}_{n,4{\rm{T}}-3{\rm{L}}}(\alpha_s) &=& 2  f(\alpha_s)  \Bigg\{ 
\textcolor{orange}{
\sum_{i = 5}^{n}\sum_{\substack{5\leq j<k \leq n \\ j,k\neq i}}
  \, {\cal T}_{iijk}}
 +\textcolor{blue}{
\sum_{ 5\leq j<k \leq n  } \Big(
{\cal T}_{11jk} +{\cal T}_{22jk} +{\cal T}_{33jk}+{\cal T}_{44jk}  \Big)} \\[-0.1cm] 
&&\hspace{-0.5cm}+\,
\textcolor{red}{\sum_{i = 5}^{n}\sum_{\substack{k = 5, \\ k\neq i}}^{n} 
\Big({\cal T}_{ii1k} +{\cal T}_{ii2k} +{\cal T}_{ii3k}+{\cal T}_{ii4k} \Big)
}+
\sum_{i = 5}^{n} \Big( 
{\cal T}_{112i} + {\cal T}_{221i} + {{\cal T}_{ii12}} + {\cal T}_{113i}
  \nn
 \\[-0.1cm] 
&&\hspace{-0.5cm} + {\cal T}_{331i} + {{\cal T}_{ii13}}  + {\cal T}_{223i} + {\cal T}_{332i} + {{\cal T}_{ii23}}    
+
{\cal T}_{114i} + {\cal T}_{441i} + {{\cal T}_{ii14}}
+
{\cal T}_{224i} + {\cal T}_{442i}
  \nn
 \\[-0.1cm] 
&&\hspace{-0.5cm} 
+ {{\cal T}_{ii24}}
+
{\cal T}_{334i} + {\cal T}_{443i} + {{\cal T}_{ii34}}
\Big) 
+ \textcolor{darkgreen}{ {\cal T}_{1123}+ {\cal T}_{2213} + {\cal T}_{3312} }
+ \textcolor{darkgreen}{ {\cal T}_{1124} }
+ \textcolor{darkgreen}{ {\cal T}_{1134} }
  \nn
 \\[-0.1cm] 
&&\hspace{-0.5cm} 
+ \textcolor{darkgreen}{ {\cal T}_{2214}  }
+ \textcolor{darkgreen}{ {\cal T}_{2234}  }
+ \textcolor{darkgreen}{ {\cal T}_{3314} }
+ \textcolor{darkgreen}{ {\cal T}_{3324} }
+ \textcolor{darkgreen}{ {\cal T}_{4412} }
+ \textcolor{darkgreen}{ {\cal T}_{4413} }
+ \textcolor{darkgreen}{ {\cal T}_{4423} }
 \Bigg\} , \\[0.1cm] 
 {\bf{ \Gamma}}_{n-3,4{\rm{T}}-3{\rm{L}}}(\alpha_s) &=& 2 f(\alpha_s)  \Bigg\{   \textcolor{orange}{\sum_{i = 5}^{n}\sum_{\substack{5\leq j<k \leq n \\ j,k\neq i}} \, {\cal T}_{iijk}}
+\textcolor{blue}{
\sum_{\substack{5\leq j<k \leq n }} {\cal T}_{PPjk}  }
+\textcolor{red}{\sum_{i = 5}^{n} \sum_{\substack{k = 5, \\ k\neq i}}^{n} {\cal T}_{iiPk}}\Bigg\} .
\eea
\end{subequations}
The black terms depend on pairs of particles out of the four  becoming collinear, 
and the \textcolor{darkgreen}{green} terms contain information 
about three out of the four collinear particles. 
When we take the difference, the \textcolor{orange}{orange} and \textcolor{red}{red} contributions exactly cancel each other. The contributions marked in \textcolor{blue}{blue} leave behind some terms. This is because upon 
substituting for the total charge of the parent parton $\T_P=\T_1+\T_2+\T_3+\T_4$
we have 
\begin{eqnarray}\label{eq:teal-fourpartcol}
{\cal T}_{PPjk} &=&   {\cal T}_{11jk} 
+{\cal T}_{22jk}+{\cal T}_{33jk}
+{\cal T}_{44jk}
+{\cal T}_{12jk}+{\cal T}_{12kj}
+{\cal T}_{13jk}+{\cal T}_{13kj}
\nonumber \\ &&
+{\cal T}_{14jk}+{\cal T}_{14kj}
+{\cal T}_{23jk}+{\cal T}_{23kj}
+{\cal T}_{24jk}+{\cal T}_{24kj}
+{\cal T}_{34jk}+{\cal T}_{34kj}\,.
\end{eqnarray}
Combining together the terms from 
Eqs.~\eqref{Aterms-4part} and \eqref{eq:Bterms-4part}
according to definition of the four-parton splitting amplitude soft 
anomalous dimension in \Eqn{eq:GammaSpQuad4} gives the following 
\bea\label{GammaSpQuad4eHybrid} \nn 
{\bf{ \Gamma}}_{\SP, 4}^{4{\rm{T}}}(p_1,p_2, p_3,p_4)
&=&   \sum_{5\leq k<l \leq n} \bigg[
\, 4 {\cal F}^{\rm A}_{12kl}(\{\beta\}) \,{\cal T}_{1kl2} +
\Big( 4 {\cal F}^{\rm S}_{12kl}(\{\beta\}) 
\textcolor{blue}{ -2f(\alpha_s)} \Big) \Big( {\cal T}_{12lk}
+ {\cal T}_{12kl} \Big)  \bigg]
\nn \\ && \hspace{-0.35cm} 
+\sum_{5\leq k<l \leq n} \bigg[
\, 4 {\cal F}^{\rm A}_{13kl}(\{\beta\}) \,{\cal T}_{1kl3} +
\Big( 4 {\cal F}^{\rm S}_{13kl}(\{\beta\}) 
\textcolor{blue}{ -2f(\alpha_s)} \Big) \Big( {\cal T}_{13lk}
+ {\cal T}_{13kl} \Big)  \bigg]
\nn \\ && \hspace{-0.35cm} 
+\sum_{5\leq k<l \leq n} \bigg[
\, 4 {\cal F}^{\rm A}_{14kl}(\{\beta\}) \,{\cal T}_{1kl4} +
\Big( 4 {\cal F}^{\rm S}_{14kl}(\{\beta\}) 
\textcolor{blue}{ -2f(\alpha_s)} \Big) \Big( {\cal T}_{14lk}
+ {\cal T}_{14kl} \Big)  \bigg]
\nn \\ && \hspace{-0.35cm} 
+ \sum_{5\leq k<l \leq n} \bigg[
\, 4 {\cal F}^{\rm A}_{23kl}(\{\beta\}) \,{\cal T}_{2kl3} +
\Big( 4 {\cal F}^{\rm S}_{23kl}(\{\beta\}) 
\textcolor{blue}{ -2f(\alpha_s)} \Big) \Big( {\cal T}_{23lk}
+ {\cal T}_{23kl} \Big)  \bigg]
\nn \\ && \hspace{-0.35cm} 
+ \sum_{5\leq k<l \leq n} \bigg[
\, 4 {\cal F}^{\rm A}_{24kl}(\{\beta\}) \,{\cal T}_{2kl4} +
\Big( 4 {\cal F}^{\rm S}_{24kl}(\{\beta\}) 
\textcolor{blue}{ -2f(\alpha_s)} \Big) \Big( {\cal T}_{24lk}
+ {\cal T}_{24kl} \Big)  \bigg]
\nn \\ && \hspace{-0.35cm} 
+ \sum_{5\leq k<l \leq n} \bigg[
\, 4 {\cal F}^{\rm A}_{34kl}(\{\beta\}) \,{\cal T}_{3kl4} +
\Big( 4 {\cal F}^{\rm S}_{34kl}(\{\beta\}) 
\textcolor{blue}{ -2f(\alpha_s)} \Big) \Big( {\cal T}_{34lk}
+ {\cal T}_{34kl} \Big)  \bigg]
\nn \\  &&\hspace{-2.0cm}
+ 2f(\alpha_s) \bigg[ 
\sum_{i = 5}^{n} \Big( {\cal T}_{112i} + {\cal T}_{221i} 
+ \textcolor{teal}{{\cal T}_{ii12}} 
+ {\cal T}_{113i}+ {\cal T}_{331i} 
+ \textcolor{teal}{{\cal T}_{ii13}}
+ {\cal T}_{114i}+ {\cal T}_{441i} 
+ \textcolor{teal}{{\cal T}_{ii14}}
\nn \\[-0.1cm]  &&\hspace{0.4cm} 
+ {\cal T}_{223i} + {\cal T}_{332i} 
+ \textcolor{teal}{{\cal T}_{ii23}}
+ {\cal T}_{224i} + {\cal T}_{442i} 
+ \textcolor{teal}{{\cal T}_{ii24}}
+ {\cal T}_{334i} + {\cal T}_{443i} 
+ \textcolor{teal}{{\cal T}_{ii34}}
\Big) 
\nn \\[-0.1cm]  &&\hspace{0.4cm} 
+ \textcolor{darkgreen}{ {\cal T}_{1123}+ {\cal T}_{2213} + {\cal T}_{3312} } + \textcolor{darkgreen}{ {\cal T}_{1124} }
+ \textcolor{darkgreen}{ {\cal T}_{1134} }
+ \textcolor{darkgreen}{ {\cal T}_{2214}  }
\nn \\[-0.1cm]  &&\hspace{0.4cm} 
+ \textcolor{darkgreen}{ {\cal T}_{2234}  }
+ \textcolor{darkgreen}{ {\cal T}_{3314} }
+ \textcolor{darkgreen}{ {\cal T}_{3324} }
+ \textcolor{darkgreen}{ {\cal T}_{4412} }
+ \textcolor{darkgreen}{ {\cal T}_{4413} }
+ \textcolor{darkgreen}{ {\cal T}_{4423} } \bigg]
\nn \\[-0.1cm]  &&\hspace{-2.5cm} 
+4 \textcolor{darkgreen}{
\sum_{5\leq  l \leq n} \,\big( {\bf a}_{123l}^{ }(\{\beta\})
+ {\bf a}_{124l}^{ }(\{\beta\})
+ \, {\bf a}_{134l}^{ }(\{\beta\})
+ \, {\bf a}_{234l}^{ }(\{\beta\}) \big)}
+4 \textcolor{cPurple}{
 \, {\bf a}_{1234}^{ }(\{\beta\})}\,.
\eea
We now continue as in the previous cases.
It is immediately possible 
to apply colour conservation to the terms
which contain the rest-of-the-process
dependence only through one index, such as 
${\cal T}_{112i}$.
The resultant expression is similar to~\Eqn{eq:calT-AABi-3pc},
however, since here the sum 
over the rest-of-the-process partons
starts at $i=5$, additional terms can enter 
through colour conservation.  
Using $\sum_{i=1}^{n}\T_i=0$, and applying it to the
rightmost generator we find the following:
\begin{eqnarray}\label{eq:calT-AABi-4pc}
\sum_{i=5}^{n}  \, {\cal T}_{aabi}  &\stackrel{m.c.c}{=}&
- \frac{1}{8}C_A^2 \T_a\cdot \T_b -
\mathcal{T}_{aabb} - \mathcal{T}_{aabc}- \mathcal{T}_{aabd}\,,
\end{eqnarray}
where now the collinear particles 1, 2, 3, and 4 are
generically represented by $a, b, c, d$.   

As in the three-particle collinear limit,
we again see that stuffle products can be formed
when the two-particle collinear limit constraints in 
\Eqn{eq:twocoll-constraint1c-main} are imposed
for each of the two out of the four particles becoming collinear, 
\bea\label{eq:fourcoll-constraint2} 
&& \Big( 4 {\cal F}^{\rm S}_{abkl}(\{\rho\})-2f(\alpha_s) \Big)\bigg|_{p_a|| p_b} \!\!\!\! = 2 f(\alpha_s)\,, 
\qquad (a,b) \in\{ (1,2), \,\,\, (1,3), \,\,(1,4), 
\\[-0.3cm] && \nonumber \hspace{9.1cm}  (2,3), \,\,(2,4), 
\,\,   \,\,(3,4)\}\,.
\eea
The stuffle products are identical to the ones encountered 
previously in \Eqn{eq:stuffle-2.p.c.c} and \Eqn{eq:stuffleGen}, 
aside from the fact that now
the sums start at $5$,
\begin{eqnarray}\label{eq:stuffleGen-4pc}
2\hspace{-0.2cm}\sum_{5\leq k<l \leq n}  \Big( {\cal T}_{ablk} + {\cal T}_{abkl} \Big)
+2\sum_{i=5}^{n}{\cal T}_{abii} = \sum_{k=5}^{n}\sum_{l=5}^{n} 
\big( {\cal T}_{ablk} +  {\cal T}_{abkl} \big).
\end{eqnarray}
This again allows for application of colour conservation, 
leading to appearance of structures encountered in the three-particle 
collinear limit in \Eqn{eq:mcolcons-triplecollinear-2a},
and additional ones with extra particles from the collinear set. 
For example, we have 
\begin{eqnarray}\label{eq:k5l5stuffle-m}
\sum_{k=5}^{n}\sum_{l=5}^{n}  \mathcal{T}_{12kl}&& 
\stackrel{m.c.c}{=}
- {\cal{T}}_{1122} - \frac{1}{8}C_A^2 \T_1\cdot \T_2
- {\cal T}_{1123} + {\cal T}_{1233} - {\cal T}_{2213} 
\nonumber \\[-0.25cm] && \hspace{0.82cm}
- {\cal T}_{1124} + {\cal T}_{1244} - {\cal T}_{2214} 
+ \mathcal{T}_{1234}   + \mathcal{T}_{1243}\,.
\end{eqnarray}
Implementing the two-particle collinear limit type
constraints in \Eqn{eq:fourcoll-constraint2} and
applying colour conservation to resulting stuffle
products as in \Eqn{eq:k5l5stuffle-m}, but not yet
to the terms with a single rest-of-the-process index
in \Eqn{eq:calT-AABi-4pc},
yields the following
\bea\label{GammaSpQuad4eHybrid-d} \nn 
{\bf{ \Gamma}}_{\SP, 4}^{4{\rm{T}}}(p_1,p_2, p_3,p_4)
&=&  
-2f(\alpha_s)
\left(   \frac{1}{8}C_A^2 \T_1\cdot \T_2 +  {\cal{T}}_{1122}
       -\textcolor{red}{ \mathcal{T}_{1234}}   - \textcolor{orange}{ \mathcal{T}_{1243}} \right)
\nn \\ && 
-
2f(\alpha_s)
\left(    \frac{1}{8}C_A^2 \T_1\cdot \T_3+ {\cal{T}}_{1133}
       - \textcolor{red}{ \mathcal{T}_{1324}}   - \textcolor{darkgreen}{\mathcal{T}_{1342}}  \right)
\nn \\ && 
-
2f(\alpha_s)  
\left(   \frac{1}{8}C_A^2 \T_1\cdot \T_4 + {\cal{T}}_{1144}
       - \textcolor{orange}{ \mathcal{T}_{1423}}   -\textcolor{darkgreen}{ \mathcal{T}_{1432}} \right)
\nn \\ && 
-
2f(\alpha_s) 
\left(    \frac{1}{8}C_A^2 \T_2\cdot \T_3+  {\cal{T}}_{2233}
       - \textcolor{purple}{\mathcal{T}_{2314}}   - \textcolor{blue}{ \mathcal{T}_{2341} } \right)
\nn \\ && 
-
2f(\alpha_s)
\left(  \frac{1}{8}C_A^2 \T_2\cdot \T_4 + {\cal{T}}_{2244}
       -\textcolor{cBlue}{  \mathcal{T}_{2413} }   - \textcolor{blue}{ \mathcal{T}_{2431} } \right) 
\nn \\ && 
-2f(\alpha_s) 
\left(   \frac{1}{8}C_A^2 \T_3\cdot \T_4 + {\cal{T}}_{3344}
       -\textcolor{cBlue}{ \mathcal{T}_{3412} }  - \textcolor{purple}{\mathcal{T}_{3421}}  \right)
\nn \\[-0.1cm]  &&
+ 2f(\alpha_s)  
\sum_{i = 5}^{n} \Big( {\cal T}_{112i} + {\cal T}_{221i} 
+ {\cal T}_{113i}+ {\cal T}_{331i} 
+ {\cal T}_{114i}
+ {\cal T}_{441i} 
\nn \\[-0.4cm]  &&\hspace{+2.4cm} 
+ {\cal T}_{223i} + {\cal T}_{332i} 
+ {\cal T}_{224i} + {\cal T}_{442i} 
+ {\cal T}_{334i} + {\cal T}_{443i} 
\Big) 
\nn\\[-0.2cm]  &&
+4 \textcolor{black}{
\sum_{5\leq  l \leq n} \,\Big( {\bf a}_{123l}^{ }(\{\beta\})
+ {\bf a}_{124l}^{ }(\{\beta\})
+ \, {\bf a}_{134l}^{ }(\{\beta\})
+ \, {\bf a}_{234l}^{ }(\{\beta\}) \Big)}
\nn \\[0.01cm]  &&
+4 \textcolor{cPurple}{
 \, {\bf a}_{1234}^{ }(\{\beta\})}\,,
\eea
where the constant colour terms with one repeated collinear
parton index cancelled, i.e., terms of type ${\cal T}_{1134}$
in \textcolor{darkgreen}{green} in \Eqn{GammaSpQuad4eHybrid}, against the same type of term 
from colour conservation applied to terms originating
from stuffle products relations, as
the one in \Eqn{eq:k5l5stuffle-m}.
We have also marked in different colours the pairwise cancellation
between the constant terms with all four different particle
indices. These contributions enter in individual terms
through \Eqn{eq:k5l5stuffle-m} and
if not for their explicit cancellation, these
could have constituted a new type of contribution
appearing for the first time in the four-particle collinear limit
for the splitting amplitude anomalous dimension. 

Now using the results for application of colour conservation 
to terms with a single rest-of-the-process parton index in 
\Eqn{eq:calT-AABi-4pc}, we find 
\bea\label{GammaSpQuad4eHybrid-g} \nn 
{\bf{ \Gamma}}_{\SP, 4}^{4{\rm{T}}}(p_1,p_2, p_3,p_4)
&=&  - \frac{3}{4} f(\alpha_s)
\big( C_A^2 \T_1\cdot \T_2 + 8{\cal{T}}_{1122} \big)
- \frac{3}{4}f(\alpha_s)
\big(   C_A^2 \T_1\cdot \T_3+8{\cal{T}}_{1133} \big)
\nn \\ && \hspace{-0.0cm} 
- \frac{3}{4}f(\alpha_s)  
\big( C_A^2 \T_1\cdot \T_4  - 8{\cal{T}}_{1144}  \big)
 - \frac{3}{4}f(\alpha_s) 
\big(    C_A^2 \T_2\cdot \T_3- 8{\cal{T}}_{2233} \big)
\nn \\ && \hspace{-0.0cm} 
 - \frac{3}{4}f(\alpha_s)
\big(  C_A^2 \T_2\cdot \T_4- 8{\cal{T}}_{2244} \big)
- \frac{3}{4}f(\alpha_s) 
\big(  C_A^2 \T_3\cdot \T_4- 8{\cal{T}}_{3344} \big)
\nn \\  &&\hspace{-0.0cm}
- 4f(\alpha_s) \Big[ 
  \mathcal{T}_{1123}+\mathcal{T}_{1124}
+ \mathcal{T}_{1134}+\mathcal{T}_{2213}
+ \mathcal{T}_{2214}+\mathcal{T}_{2234}
\nn \\  &&\hspace{1.4cm}
+\mathcal{T}_{3312}+\mathcal{T}_{3314}
+\mathcal{T}_{3324}+\mathcal{T}_{4412}
+\mathcal{T}_{4413}+\mathcal{T}_{4423} \Big]
\nn \\[-0.1cm] &&\hspace{-3.9cm} 
+4 \textcolor{darkgreen}{
\sum_{5\leq  l \leq n} \,\big( {\bf a}_{123l}^{ }(\{\beta\})
+ {\bf a}_{124l}^{ }(\{\beta\})
+ \, {\bf a}_{134l}^{ }(\{\beta\})
+ \, {\bf a}_{234l}^{ }(\{\beta\}) \big)}
+4 \textcolor{cPurple}{
 \, {\bf a}_{1234}^{ }(\{\beta\})}\,.
\eea
It remains to apply colour conservation to the 
terms highlighted in \textcolor{darkgreen}{green}, just as in the corresponding
step in the three-particle collinear limit calculation 
that can be found in~\Eqn{eq:m3partcol-pen}, now for all
the combinations of three out of the four particles becoming
collinear. However, new terms connecting more of the
collinear particles arise through colour conservation, see
for example the result in~\Eqn{eq:3partcol-kin-c.c} 
with respect to \Eqn{eq:4partcol-kin-c.c}. The result
of these final manipulations is presented in the main 
text in \Eqn{eq:GammaSpQuad4eHybrid-i}.

\section{Colour identities}\label{colourID}
The first set of identities does not require colour conservation to be proven. 
\be\label{colourId1}
f^{abe}f^{cde}\{\T^a_i,\T^c_i\}\T^d_i\T^b_j = \frac{1}{4}C_A^2\T_i \cdot \T_j\,,
\ee
\be\label{colourId2}
f^{abe}f^{cde}  \T^{a}_{i}\T^{c}_{j}\T^{d}_{i}\T^{b}_{j}= -\frac{1}{8}C_A^2 \T_i \cdot \T_j - \frac{1}{4} f^{abe}f^{cde}\{\T^{a}_{i},\T^{c}_{i}\}\{\T^{d}_{j},\T^{b}_{j}\} \,.
\ee
In addition to these identities one can manipulate terms involving colour generators using colour conservation. The identities in this case depend on the number of colour carrying partons in the process. It is important to stress that the sum of colour generators does not vanish in general, but only when it is acting on a colour singlet object. 
For $n=3$ colour conservation implies $\T_1 + \T_2 + \T_3 = 0$ and one can show the following

\bea \label{eq:colourId3NewInt} \nn
{\cal T}_{1233} = \frac{1}{2} f^{ade}f^{bce}\T^{a}_{1}\T^{b}_{2} \{\T^{c}_{3},\T^{d}_{3}\} 
&=& - \frac{1}{8}C_A^2 \T_1 \cdot \T_2 - \frac{1}{4} f^{ade}f^{bce} 
 \{\T^{a}_{1},\T^{b}_{1}\}  \{\T^{c}_{2},\T^{d}_{2}\}  
 \\ &=& -\frac{1}{8} C_A^2 \, \T_1\cdot \T_2- {\cal T}_{1122}\,,
\eea 
and
\be\label{eq:colourId4}
f^{abe}f^{cde}\{\T^{a}_{1},\T^{c}_{1}\}\T^{b}_{2}\T^{d}_{3} = - \frac{1}{4}C_A^2 \T_1 \cdot \T_2 - \frac{1}{2}f^{abe}f^{cde}\{\T^{a}_{1},\T^{c}_{1}\}\{\T^{b}_{2},\T^{d}_{2}\}\,.
\ee
These relations have been used to obtain \Eqn{eq:GammaSpQuad2d} from \Eqn{eq:GammaSpQuad2c}. However, in the $n=4$ case $\T_1 + \T_2 + \T_3 +\T_4 = 0$. Hence similar expressions become 
\be\label{2.11}
\begin{split}
	f^{abe}f^{cde} \{\T^{a}_{4},\T^{c}_{4}\}\T^{b}_{1}\T^{d}_{2}  = -\frac{1}{4}C_A^2 \T_1 \cdot \T_2 +f^{abe}f^{cde} \bigg\{ \{\T^{a}_{3},\T^{c}_{3}\}\T^{b}_{1}\T^{d}_{2} - 2\T^{c}_{1}\T^{a}_{1}\T^{d}_{2}\T^{b}_{3}\\-2\T^{a}_{2}\T^{c}_{2}\T^{b}_{1}\T^{d}_{3}  - \frac{1}{2} \{\T^{a}_{2},\T^{c}_{2}\}\{\T^{d}_{1},\T^{b}_{1}\}  
	\bigg\}   \,,
\end{split}
\ee
and
\be
\begin{split}
	f^{abe}f^{cde}\{\T^{a}_{1},\T^{c}_{1}\}\T^{b}_{2}\T^{d}_{4}=   - \frac{1}{4}C_A^2\T_1 \cdot \T_2    -\frac{1}{2}f^{abe}f^{cde}\{\T^{a}_{1},\T^{c}_{1}\}\{\T^{b}_{2},\T^{d}_{2}\} \\-f^{abe}f^{cde}\{\T^{a}_{1},\T^{c}_{1}\}\T^{b}_{2}\T^{d}_{3}\,.
\end{split}
\ee
These expressions can be used in the three-particle collinear limit to obtain the constant part of  
${\bf{ \Gamma}}_{\SP, 3}^{4{\rm{T}}}$ in \Eqn{GammaSpQuad3d}.

\subsection{Two-particle collinear limit}
In the two-particle collinear limit, we have the following 
result
\bea \label{eq:D13} 
\sum_{i=3}^{n}  \, {\cal T}_{112i}  &\stackrel{m.c.c}{=}&
- \frac{1}{8}\,C_A^2 \,\T_1\cdot \T_2 -
 \mathcal{T}_{1122}\,,
\eea
which is obtained by 
\bea
 \sum_{i=3}^{n}  \, {\cal T}_{112i}  &=&
 \sum_{i=3}^{n}  f^{ade}f^{bce}(\T_1^{a}\T_1^{b})_+\T_2^{c}\T_i^{d}
 \nn = 
 \frac{1}{2}
   f^{ade}f^{bce}\{\T_1^{a},\T_1^{b}\}\T_2^{c}
 \left(-\T_1^{d} - \T_2^{d}
 \right) 
  \\[0.2cm]  &&
 {=} - \frac{1}{8}C_A^2 \T_1\cdot \T_2 -
 \mathcal{T}_{1122}\,,
\eea
where in the second step we used Eqs.~\eqref{colourId1} and \eqref{colourId2}.
For the terms beginning at four loops we need 
\begin{eqnarray}\label{eq:calD1i22cc}
\sum_{  i = 3}^n       { \bf{\cal{ D}}}^R_{1i22} = 
\sum_{  i = 3}^n       { \bf{\cal{ D}}}^R_{122i} =  
-  { \bf{\cal{ D}}}^R_{1221}     -  { \bf{\cal{ D}}}^R_{1222} =  
-  { \bf{\cal{ D}}}^R_{1122}     -  { \bf{\cal{ D}}}^R_{1222} \,,
\end{eqnarray} 
and
\begin{eqnarray}\label{eq:calD2iikcc}
\sum_{i=3}^n\sum_{\substack{ k=3\\  k \neq i } }^n \, 
{ \bf{\cal{ D}}}^R_{2iik}
= - \sum_{i=3}^{n} { \bf{\cal{ D}}}^R_{12ii} - \sum_{i=3}^{n} 
{ \bf{\cal{ D}}}^R_{22ii}  -\sum_{i=3}^{n} { \bf{\cal{ D}}}^R_{2iii} \,.
\end{eqnarray}

\subsection{Three-particle collinear limit}
For the three-particle collinear limit, \Eqn{eq:D13} has to be
reconsidered since the rest-of-the-process partons sum starts at index $i=4$ with the first three taken up by particles becoming collinear. Therefore, here we need
\bea \label{eq:D13-3pc} 
\sum_{i=4}^{n}  \, {\cal T}_{112i}  &\stackrel{m.c.c}{=}&
- \frac{1}{8}\,C_A^2 \,\T_1\cdot \T_2
- \mathcal{T}_{1122} - \mathcal{T}_{1123}\,,
\eea
which is obtained by 
\bea
 \sum_{i=4}^{n}  \, {\cal T}_{112i}  &&=
 \sum_{i=4}^{n}  f^{ade}f^{bce}(\T_1^{a}\T_1^{b})_+\T_2^{c}\T_i^{d}
 \nn = 
 \frac{1}{2}
   f^{ade}f^{bce}\{\T_1^{a},\T_1^{b}\}\T_2^{c}
 \left(-\T_1^{d} - \T_2^{d} - \T_3^{d}
 \right) 
  \\[0.2cm]  &&
 {=} - \frac{1}{8}C_A^2 \T_1\cdot \T_2 -
 \mathcal{T}_{1122} - \mathcal{T}_{1123}\,,
\eea
where in the second step we used
Eqs.~\eqref{colourId1} and \eqref{colourId2}.
For the calculations of the three-particle collinear limit,
we also need relations such as 
\begin{eqnarray} \label{eq:5.231}\nn
\sum_{4\leq l \leq n} 
\mathcal{T}_{123l}  
&&=  \sum_{4\leq l \leq n} f^{ade}f^{bce}
\T_1^a\T_2^b\T_3^c\T_l^d \stackrel{m.c.c}{=}
-f^{ade}f^{bce}
\T_1^a\T_2^b\T_3^c\big( \T_1^d + \T_2^d+ \T_3^d   \big) 
\\  && =
-f^{ade}f^{bce}
\T_1^a\T_2^b\T_3^c \T_1^d 
-f^{ade}f^{bce}
\T_1^a\T_2^b\T_3^c  \T_2^d 
-f^{ade}f^{bce}
\T_1^a\T_2^b\T_3^c  \T_3^d  \nonumber
\\ \nonumber
&&= - \frac{C_A}{2} \mathcal{T}_{123} + \mathcal{T}_{2213}
- \frac{C_A}{4}\mathcal{T}_{213} - \mathcal{T}_{3312}
+ \frac{C_A}{4} \mathcal{T}_{312}
\\ &&= \mathcal{T}_{2213}  - \mathcal{T}_{3312}  \,.
\end{eqnarray}
For the terms beginning at four loops we need 
\begin{eqnarray}\label{eq:calD1i22-3cc}
\sum_{  i = 4}^n { \bf{\cal{ D}}}^R_{1i22} = 
\sum_{  i = 4}^n { \bf{\cal{ D}}}^R_{122i} =  
-  { \bf{\cal{ D}}}^R_{1221}    
-  { \bf{\cal{ D}}}^R_{1222}
-  { \bf{\cal{ D}}}^R_{1223} =  
-  { \bf{\cal{ D}}}^R_{1122}     
-  { \bf{\cal{ D}}}^R_{1222} 
-  { \bf{\cal{ D}}}^R_{1223} \,,
\end{eqnarray}  
and for terms with a double sum
\begin{eqnarray}\label{eq:calD2iik-3cc}
      \sum_{i=4}^n\sum_{\substack{ k=4\\  k \neq i } }^n \, { \bf{\cal{ D}}}^R_{2iik}
 = - \sum_{i=4}^{n} { \bf{\cal{ D}}}^R_{12ii} - \sum_{i=4}^{n} { \bf{\cal{ D}}}^R_{22ii}- \sum_{i=4}^{n} { \bf{\cal{ D}}}^R_{23ii}  -\sum_{i=4}^{n} { \bf{\cal{ D}}}^R_{2iii}\,, 
 \end{eqnarray}
 and 
 \begin{eqnarray}\label{eq:calD3iik-3cc}
      \sum_{i=4}^n\sum_{\substack{ k=4\\  k \neq i } }^n \, { \bf{\cal{ D}}}^R_{3iik}
 = - \sum_{i=4}^{n} { \bf{\cal{ D}}}^R_{13ii} - \sum_{i=4}^{n} { \bf{\cal{ D}}}^R_{23ii}- \sum_{i=4}^{n} { \bf{\cal{ D}}}^R_{33ii}  -\sum_{i=4}^{n} { \bf{\cal{ D}}}^R_{3iii} \,.
 \end{eqnarray}

\subsection{Four-particle collinear limit}
As in the case of moving from the two-particle to the three-particle collinear limit, now \Eqn{eq:D13-3pc} has to be again
reconsidered. In the four-particle collinear limit, the rest-of-the-process partons sum starts at index $i=5$ with the first four taken up by particles becoming collinear. Therefore, here we need
\bea \label{eq:D13-4pc} 
\sum_{i=4}^{n}  \, {\cal T}_{112i}  &\stackrel{m.c.c}{=}&
- \frac{1}{8}\,C_A^2 \,\T_1\cdot \T_2
- \mathcal{T}_{1122} - \mathcal{T}_{1123}- \mathcal{T}_{1124}
\eea
which is obtained by 
\bea
\sum_{i=5}^{n}\,{\cal T}_{112i}&&=
\sum_{i=5}^{n}  f^{ade}f^{bce}
(\T_1^{a}\T_1^{b})_+\T_2^{c}\T_i^{d}
\nn = 
\frac{1}{2}
f^{ade}f^{bce}\{\T_1^{a},\T_1^{b}\}\T_2^{c}
\left(-\T_1^{d} - \T_2^{d} - \T_3^{d}- \T_4^{d}
\right) 
\\[0.2cm]  &&
{=} - \frac{1}{8}C_A^2 \T_1\cdot \T_2 -
\mathcal{T}_{1122} - \mathcal{T}_{1123}- \mathcal{T}_{1124}\,,
\eea
where in the second step we used Eqs.~\eqref{colourId1} and
\eqref{colourId2}. It can be seen that in the generalisation
to more particles becoming collinear, additional terms with
the last index changed to ones corresponding to more
of the particles becoming collinear appear.

Moreover, the terms appearing first in the three-particle collinear limit of the type below, again receive 
additional contributions through colour conservation. Namely, Eq.~\eqref{eq:5.231} generalises to 
\begin{eqnarray} \label{eq:5.231-four} \nn
  \sum_{5\leq l \leq n} 
\mathcal{T}_{123l}  
&&=  \sum_{5\leq l \leq n} f^{ade}f^{bce}
\T_1^a\T_2^b\T_3^c\T_l^d =
-f^{ade}f^{bce}
\T_1^a\T_2^b\T_3^c\big( \T_1^d + \T_2^d+ \T_3^d+ \T_4^d   \big) 
\\  && \nn
= -f^{ade}f^{bce}
\T_1^a\T_2^b\T_3^c \T_1^d 
-f^{ade}f^{bce}
\T_1^a\T_2^b\T_3^c  \T_2^d 
\\  && \hspace{0.45cm}
-f^{ade}f^{bce}
\T_1^a\T_2^b\T_3^c  \T_3^d  \nonumber
-f^{ade}f^{bce}
\T_1^a\T_2^b\T_3^c  \T_4^d   
\\ \nn && 
= -T_{1231} - T_{1232} - T_{1233}- T_{1234}
\\ \nn && 
= - \frac{C_A}{2} \mathcal{T}_{123} 
+ \mathcal{T}_{2213} - \frac{C_A}{4}\mathcal{T}_{213} 
- \mathcal{T}_{3312} + \frac{C_A}{4} \mathcal{T}_{312}
- \mathcal{T}_{1234} 
\\  &&\stackrel{ }{=}
\mathcal{T}_{2213}  - \mathcal{T}_{3312}
- \mathcal{T}_{1234} \,. 
\end{eqnarray}
It is straight forward to generalise this type of terms
for the case of more particles becoming collinear, 
we can see that simply additional terms where the 
index of particle 4 will be replaced by 5, 6, and so on, will appear in addition to the ones present above.

\section{Three-particle collinear limit constraints on \texorpdfstring{$F_{\rm{h}3}$}{}} 
\label{sec:Fh3Constraints}

In Section~\ref{sec:three-particle-collinear-massive}, we deduced a relation valid in the triple collinear limit between the function~$F_{\rm{h}3}$ of \Eqn{eq:Massive-aijkI} describing the singularities associated with interactions between a single massive particle and three massless ones, and ${\cal F}$ of \Eqn{eq:GammaF}, corresponding to the massless case. The purpose of this appendix is to show in detail that these two anomalous dimensions are directly related in the triple collinear limit. 

We start by using the Jacobi identity 
\begin{equation}
\label{JacobiI123_}
\mathcal{T}_{I132}=
\mathcal{T}_{I231}
+\mathcal{T}_{I123}\,,
\end{equation}
which we may reduce the set of independent colour structures in \Eqn{eq:GammaF} to just two. 
These can be expressed~\cite{Almelid:2017qju} as antisymmetric and symmetric component under a given swap of the Wilson-line velocities, as quoted in Eqs.~(\ref{eq:quadrupole-ab-defshu}) and (\ref{quadrupole-AS-defsh}). Given the identical colour structure, the same rearrangement of the soft anomalous dimension can be obtained for one-mass case of Eq.~(\ref{eq:adm-massive}).

We will show in what follows that the antisymmetric and symmetric parts of $F_{\rm{h}3}$ are related in the triple-collinear limit to the respective combinations of ${\cal F}$.
To this end we first use the Jacobi identity in~\Eqn{JacobiI123_} to express \Eqn{TripleCollinearMassiveConstraints} as 
\begin{eqnarray}
\label{TripleCollinearMassiveConstraintsAgain}
0\,\,&=&\,\,-\Bigg\{2{\mathcal{T}}_{I231}
\Big(
F_{{\rm{h}}3}\left(r_{12I},r_{13I},r_{23I}  \right)
-4{\cal F} (\beta_{12l3},\beta_{13l2}) \Big)
 \\ \nonumber && \hspace{.7cm}
+ \left[{\mathcal{T}}_{I132}
+{\mathcal{T}}_{I123}
+{\mathcal{T}}_{I231}\right]
\Big( 
-F_{{\rm{h}}3}\left(r_{23I},r_{21I},r_{13I}  \right)
+4{\cal F}  (\beta_{1l32},\beta_{123l}) \Big)
 \\ \nonumber &&
 \hspace{.7cm}
 \left.
+ \left[
{\mathcal{T}}_{I132}
+{\mathcal{T}}_{I123}
-{\mathcal{T}}_{I231}
\right] \Big(
-F_{{\rm{h}}3}\left(r_{32I},r_{31I},r_{21I}  \right)
 + 4  \,{\cal F}  (\beta_{1l23},\beta_{132l})  \Big)\Bigg\}\right\vert_{p_1||p_2||p_3}\,,
\end{eqnarray}
and then collect the coefficients of ${\mathcal{T}}_{123I}$, which is antisymmetric under swapping $2\leftrightarrow  3$, and ${\mathcal{T}}_{213I}
+{\mathcal{T}}_{312I}$, which is symmetric under this swap.
In the second and third lines of Eq.~(\ref{TripleCollinearMassiveConstraintsAgain}) we have also used Eqs.~(\ref{calFAntisymm}) and (\ref{Fh3Antisymm}) to swap the (first two) arguments of both~${\cal F}$ and~$F_{\rm{h}3}$ to obtain an overall minus sign.
Collecting the colour structures which are symmetric and antisymmetric with respect to swapping the partons $2$ and $3$ we then obtain:
\begin{align}
\label{TripleCollinearMassiveConstraintsIntermediateFormCollectColour}
\begin{split}
0\,\,&=\,\,{\mathcal{T}}_{I321}\Bigg\{
\Big(
2F_{{\rm{h}}3}\left(r_{12I},r_{13I},r_{23I}  \right)
-8{\cal F} (\beta_{12l3},\beta_{13l2}) \Big)
\\[-0.2cm]&\hspace*{80pt}
+\Big( 
-F_{{\rm{h}}3}\left(r_{23I},r_{21I},r_{13I}  \right)
+4{\cal F}  (\beta_{1l32},\beta_{123l}) \Big)
\\[-0.2cm]&\hspace*{80pt}
+\Big(
+F_{{\rm{h}}3}\left(r_{32I},r_{31I},r_{21I}  \right)
 - 4  \,{\cal F}  (\beta_{1l23},\beta_{132l})  \Big)
 \Bigg\}
 \\[-0.2cm] & \hspace*{0pt}
+ \left[{\mathcal{T}}_{I132}+{\mathcal{T}}_{I123}\right]
\Bigg\{
\Big( 
F_{{\rm{h}}3}\left(r_{23I},r_{21I},r_{13I}  \right)
-4{\cal F}  (\beta_{1l32},\beta_{123l}) \Big)
 \\[-0.3cm] &\hspace*{90pt}
+ \Big(
F_{{\rm{h}}3}\left(r_{32I},r_{31I},r_{21I}  \right)
 - 4  \,{\cal F}  (\beta_{1l23},\beta_{132l})  \Big)\Bigg\}\,,
 \end{split}
\end{align}
where in the last two lines we absorbed the minus sign inside the brackets to change the overall sign of the kinematic functions.
Here we can readily identify the antisymmetric and symmetric kinematically-dependent functions, separately for ${\cal F}$ and for $F_{\rm{h}3}$. These read:
\begin{subequations}
\label{quadrupole-AS-defsh_} 
    \begin{align}
{\cal F}_{1l23}^{\rm A}(\{\beta\}) &=
{\cal F}(\beta_{1l23},\beta_{132l})
-{\cal F}(\beta_{1l32},\beta_{123l})
+ 2{\cal F}(\beta_{12l3},\beta_{13l2})\,, \\[0.1cm]
{\cal F}_{1l23}^{\rm S}(\{\beta\}) &= 
 {\cal F}(\beta_{1l23},\beta_{132l})
+{\cal F}(\beta_{1l32},\beta_{123l})
 \,.
\end{align}
\end{subequations}
Therefore, the contribution due to the massless part of the soft anomalous dimension corresponds to 
  \begin{align}
{\bf a}_{1l23}(\{\beta\})={\mathcal{T}}_{l321}{\cal F}_{1l23}^{\rm A}(\{\beta\}) 
+\left[{\mathcal{T}}_{l132}+{\mathcal{T}}_{l123}\right]
{\cal F}_{1l23}^{\rm S}(\{\beta\})\, ,
\end{align}
which can be directly identified as ${\bf a}_{ijkl}(\{\beta\})$ in~\Eqn{eq:quadrupole-ab-defshu} along with the functions in~\Eqn{quadrupole-AS-defsh} with the identification of $(i,j,k,l)\to (1,l,2,3)$.
Similarly, on the one-mass side we have
\begin{subequations}
\label{Fh3quadrupole-AS-defsh_} 
\begin{align}
F^{\rm A}_{{\rm{h}}3, {1I23}}
(\{r\})
&\equiv
F_{{\rm{h}}3}\left(r_{32I},r_{31I},r_{21I} \right)  
-F_{{\rm{h}}3}\left(r_{23I},r_{21I},r_{13I}\right)
+
2F_{{\rm{h}}3}\left(r_{12I},r_{13I},r_{23I}  \right)
\\
F^{\rm S}_{{\rm{h}}3, {1I23}}
(\{r\})&\equiv 
F_{{\rm{h}}3}\left(r_{32I},r_{31I},r_{21I}\right)
+F_{{\rm{h}}3}\left(r_{23I},r_{21I},r_{13I}
  \right)\,,
\end{align}
\end{subequations}
with which Eq.~(\ref{eq:Massive-aijkI}) may be expressed as
\begin{align}\label{eq:ah1l23}
{\bf a}^h_{1I23}(\{r\})=
{\mathcal{T}}_{I321}
F^{\rm A}_{{\rm{h}}3, 1I23}\left(\{r\} \right)
+ \left[{\mathcal{T}}_{I132}+{\mathcal{T}}_{I123}\right]
F^{\rm S}_{{\rm{h}}3, 1I23}\left(\{r\} \right) \,.
\end{align}
in direct analogy with \Eqn{eq:quadrupole-ab-defshu}.
With these identifications in place, the expression in~\Eqn{TripleCollinearMassiveConstraintsIntermediateFormCollectColour}, can be compactly written as
\begin{align}
\label{TripleCollinearMassiveConstraintsFinalForm}
\begin{split}
0\,\,&=\,\,
\Bigg\{
{\mathcal{T}}_{I321}
\Big( 
F^{\rm A}_{{\rm{h}}3, 1I23}\left(\{r\} \right)
-4{\cal F}^{\rm A}_{1I23}  (\{\beta\}) \Big)
 \\ & 
 \left.
 \hspace{.8cm}
+ \left[{\mathcal{T}}_{I132}+{\mathcal{T}}_{I123}\right]
\Big(F^{\rm S}_{{\rm{h}}3, 1I23}\left(\{r\} \right)
-4{\cal F}^{\rm S}_{1I23}  (\{\beta\}) \Big)\Bigg\}\right\vert_{p_1||p_2||p_3}.
 \end{split}
\end{align}
Since now we have a basis of colour structures, it can be therefore concluded that the three-particle collinear limit constraint on the function $F_{\rm{h}3}$ takes the form:
\begin{subequations}
\label{TripleCollinearMassiveConstraintsFinalFormLim}
    \begin{align}
\lim_{p_1||p_2||p_3}F^{\rm A}_{{\rm{h}}3, {1I23}}\left(\{r\} \right)
&=\left.
4{\cal F}^{\rm A}_{1l23}  (\{\beta\}) \right\vert_{p_1||p_2||p_3}\,,
\\
\lim_{p_1||p_2||p_3}F^{\rm S}_{{\rm{h}}3, {1I23}}\left(\{r\} \right)
&=\left.4{\cal F}^{\rm S}_{1l23}  (\{\beta\})\right\vert_{p_1||p_2||p_3}\,,
\end{align}
\end{subequations}
thus directly relating $F_{\rm{h}3}^{\rm A/S}$ to its massless counterpart ${\cal F}^{\rm A/S}$. 
Comparing Eqs.~(\ref{Fh3quadrupole-AS-defsh_}) and (\ref{quadrupole-AS-defsh_}) we note the correspondence of the arguments between these two functions. Considering for example 
$F_{{\rm{h}}3}\left(r_{31I},r_{32I},r_{21I}  \right)$ we see that the first and second arguments, normalised by the third, corresponds directly to the two arguments of 
${\cal F}(\beta_{132l},\beta_{1l23})$:
\begin{align}
\label{eq:r_CICRs}
\begin{split}
\frac{r_{31I}}{r_{21I}}&= 
\frac{p_3\cdot p_1 \,   }{p_3\cdot p_I \,  }
\frac{  \, p_2\cdot p_I}{p_2\cdot p_1 \,   }
= \frac{p_2\cdot p_l  \,   }{p_1\cdot  p_2 \,  }
\frac{p_1 \cdot  p_3 \, }{ p_l\cdot p_3  \,   }
= \rho_{132l}
\,, \\
\frac{r_{32I}}{r_{21I}} &= 
\frac{p_3\cdot p_2 \, }{p_3\cdot p_I \, }
\frac{p_1\cdot p_I \, }{p_2\cdot p_1 \,  }
= \frac{p_2\cdot p_3 \, }{p_1\cdot  p_2\, }
\frac{p_1\cdot p_l \, }{p_3\cdot p_l \,  }
\,= \rho_{1l23}
\,.
\end{split}
\end{align}
where $r_{abI}$ are defined in \Eqn{eq:r-kin} while the massless cross ratios are defined in Eqs.~(\ref{eq:CICR}) and (\ref{eq:lnCICR}). 

\bibliography{IR-Collinear-v15}

\end{document}